%% file: ANA-HIGG-2019-13-PAPER.tex
\documentclass[cernpreprint, atlasdraft=false, texlive=2020, UKenglish, texmf, texmf]{atlasdoc}
 
\usepackage{atlaspackage}
\usepackage{atlasbiblatex}
\usepackage{multirow}
\usepackage[snippets,process]{atlasphysics}

\graphicspath{{logos/}{figures/}}
 
\usepackage{ANA-HIGG-2019-13-PAPER-defs}

\AtlasTitle{Measurements of the Higgs boson inclusive and differential fiducial cross-sections in the diphoton decay channel with \(pp\) collisions at \(\sqrt{s} = 13\,\text{TeV}\) with the ATLAS detector}
 
\AtlasVersion{2.9}
 
\AtlasAbstract{A measurement of inclusive and differential fiducial cross-sections for the production of the Higgs boson decaying into two photons is performed using \(139\,\text{fb}^{-1}\) of proton--proton collision data recorded at \(\sqrt{s} = 13\,\text{TeV}\) by the ATLAS experiment at the Large Hadron Collider.
The inclusive cross-section times branching ratio, in a fiducial region closely matching the experimental selection, is measured to be \(67\pm 6\)\,fb, which is in agreement with the state-of-the-art Standard Model prediction of \(64\pm 4\)\,fb.
Extrapolating this result to the full phase space and correcting for the branching ratio, the total cross-section for Higgs boson production is estimated to be \(58\pm 6\)\,pb.
In addition, the cross-sections in four fiducial regions sensitive to various Higgs boson production modes and differential cross-sections as a function of either one or two of several observables are measured.
All the measurements are found to be in agreement with the Standard Model predictions.
The measured transverse momentum distribution of the Higgs boson is used as an indirect probe of the Yukawa coupling of the Higgs boson to the bottom and charm quarks.
In addition, five differential cross-section measurements are used to constrain anomalous Higgs boson couplings to vector bosons in the Standard Model effective field theory framework.
}

\AtlasRefCode{HIGG-2019-13}
 
\PreprintIdNumber{CERN-EP-2021-227}

\arXivId{2202.00487}
 
\HepDataRecord{111349}
 
\AtlasJournal{\JHEP}
\AtlasJournalRef{\JHEP 08 (2022) 027}
\AtlasDOI{10.1007/JHEP08(2022)027}

\hypersetup{pdftitle={ATLAS document},pdfauthor={The ATLAS Collaboration}}
 
\begin{document}
 
\maketitle
 
\tableofcontents

\clearpage
\section{Introduction}
 
Following the discovery of the Higgs boson (\(H\)) in July 2012 by the ATLAS~\cite{HIGG-2012-27} and CMS~\cite{CMS-HIG-12-028} collaborations, extensive studies were carried out to precisely measure the properties of the new particle.
These studies yielded measurements of the different Higgs boson production mechanisms in several decay modes~\cite{HIGG-2014-06, CMS-HIG-14-009, HIGG-2015-07, HIGG-2016-25, HIGG-2016-22, HIGG-2016-29,HIGG-2017-02,HIGG-2014-14,HIGG-2016-21,CMS-HIG-16-041,CMS-HIG-16-041, CMS-HIG-17-011,CMS-HIG-17-028,HIGG-2017-14,STDM-2017-09,HIGG-2017-07,HIGG-2016-07,HIGG-2018-04,HIGG-2017-06,HIGG-2018-13,HIGG-2016-33,HIGG-2017-11, CMS-HIG-19-015, HIGG-2018-29}, constraints on the spin and parity of the Higgs boson~\cite{HIGG-2013-17,  CMS-HIG-14-018}, and measurements of the Higgs boson mass, which average to \mbox{\(\mH = \SI[separate-uncertainty,multi-part-units=single]{125.09(24)}{\GeV}\)}~\cite{CMS-HIG-14-042}.
 
The results of these measurements were found to be consistent with the Standard Model (SM) predictions. These measurements were carried out using data from proton--proton (\pp) collisions at the Large Hadron Collider (LHC) using the full \RunOne\ data sets at centre-of-mass energies of \(\sqrt{s}=\SI{7}{\TeV}\) and \SI{8}{\TeV}, as well as partial and full \RunTwo\ data sets at \(\sqrt{s}=\SI{13}{\TeV}\).
 
Among the possible studies of the properties of the Higgs boson, a particular interest is covered by the measurement of its production cross-section in fiducial regions for which the final-state particles are limited to a specific volume of the phase space defined by the detector acceptance, thus minimising the physics assumptions that would be needed for the extrapolation to the full phase space. In addition, the measured cross-sections are inclusive instead of being split by production process, thus further minimising SM assumptions.
The measurements are corrected for detector response effects (unfolding) and yield results that can be compared directly with any current or future theoretical predictions.
 
This paper presents measurements of fiducial inclusive and differential cross-sections in the \Hgg\ decay channel.
The signature of the Higgs boson in the diphoton final state is a narrow resonance rising above a smooth background in the diphoton invariant mass (\mgg) distribution with a width consistent
with the detector resolution.
Despite the small branching ratio for Higgs boson decay into two photons, \((2.27\pm0.07)\times 10^{-3}\) for \(\mH=\SI{125.09}{\GeV}\)~\cite{deFlorian:2016spz}, the Higgs boson signal can be extracted thanks to the excellent photon reconstruction and identification efficiency of the ATLAS detector.
All the measurements are performed assuming a Higgs boson mass of \SI{125.09}{\GeV}, and are compared with SM predictions.
Unless explicitly stated otherwise, all results include the Higgs boson branching ratio to two photons.
 
The measurements presented in this paper follow previous fiducial cross-section measurements using the full \RunOne\ data set~\cite{HIGG-2013-10} and a partial \RunTwo\ data set~\cite{HIGG-2016-21}. Those measurements showed agreement with SM predictions, and the measured differential cross-sections were used to constrain anomalous coupling of the Higgs boson to other SM particles in an effective field theory (EFT) framework~\cite{Contino:2013kra}.
The study described in this paper relies on the full \RunTwo\ \pp\ collision data set collected at \(\sqrt{s}=\SI{13}{\TeV}\) at the LHC by the ATLAS detector between 2015 and 2018, corresponding to an integrated luminosity of \SI{139.0}{\femto\barn^{-1}}.
This data set is approximately four times larger than the one used in the previous ATLAS publication, and hence it significantly reduces the statistical uncertainty of the measurements. In addition, the latest ATLAS developments in the reconstruction and identification of the various physics objects used in the analysis, namely updates to photon reconstruction and identification and jet reconstruction, are employed. Furthermore, the paper presents measurements of new kinematical observables and new fiducial regions sensitive to the various Higgs boson production modes, including regions that are sensitive to potential beyond-the-SM (BSM) effects.
 
The paper also reports a measurement of the total Higgs boson production cross-section in the full phase space, as well as two interpretations of the measured fiducial differential cross-sections: constraints on Higgs boson Yukawa couplings to charm and bottom quarks from the differential cross-section as a function of the diphoton transverse momentum, \ptgg, and constraints on anomalous couplings of the Higgs boson to other SM particles from a combined fit including several kinematic observables in an EFT approach.
 
The paper is organised as follows. Section~\ref{sec:detector} provides a brief description of the ATLAS detector. The data and Monte Carlo samples used are described in Section~\ref{sec:samples}. An overview of the event reconstruction and selection is given in Section~\ref{sec:reco}. Section~\ref{sec:intro_xs} summarises the definition of the fiducial regions and the differential cross-sections measured by the analysis. The details of the signal and background modelling and of how these models are fitted to the data are provided in Section~\ref{sec:SandB}. A summary of the different theoretical and experimental systematic uncertainties is given in Section~\ref{sec:sysunc}. The measured cross-sections are presented in Section~\ref{sec:xsec_results}. The interpretations of the measured cross-sections are presented in Section~\ref{sec:interpretations}. The summary and conclusion are given in Section~\ref{sec:conclusion}.
 
\section{ATLAS detector}
\label{sec:detector}
The ATLAS detector~\cite{PERF-2007-01} is a multipurpose particle detector
with a forward--backward symmetric cylindrical geometry\footnote{
ATLAS uses a right-handed coordinate system with its origin at the nominal interaction point (IP)
in the centre of the detector and the \(z\)-axis along the beam pipe.
The \(x\)-axis points from the IP to the centre of the LHC ring,
and the \(y\)-axis points upwards.
Cylindrical coordinates \((r,\phi)\) are used in the transverse plane, \(\phi\) being the azimuthal angle around the \(z\)-axis.
The pseudorapidity is defined in terms of the polar angle \(\theta\) as \(\eta = -\ln \tan(\theta/2)\).
When dealing with massive particles, the rapidity \(y=(1/2)\ln[(E+p_z)/(E-p_z)]\) is used, where \(E\) is the energy and \(p_z\) is the \(z\)-component of the momentum.
Angular distance is measured in units of \(\Delta R = \sqrt{(\Delta\eta)^{2} + (\Delta\phi)^{2}}\).}
and a near \(4\pi\) coverage in solid angle. It consists of an inner tracking detector (ID) surrounded by a thin superconducting solenoid, which provides a \SI{2}{\tesla} axial magnetic field, electromagnetic (EM) and hadronic calorimeters, and a muon spectrometer (MS).
 
The inner tracking detector covers the pseudorapidity range \(|\eta| < 2.5\).
It consists of a silicon pixel detector, including the insertable B-layer~\cite{ATLAS-TDR-2010-19, PIX-2018-001} installed before Run 2, a silicon microstrip detector, and a straw-tube tracking detector featuring transition radiation to aid in the identification of electrons.
 
Lead/liquid-argon (LAr) sampling calorimeters provide electromagnetic energy measurements
with high granularity. In the region up to \(|\eta|=2.5\) they are segmented longitudinally into three layers and they are complemented with an additional thin LAr presampler covering \(|\eta| < 1.8\) to correct for energy loss in material upstream of the calorimeters.
A steel/scintillator-tile hadron calorimeter covers the central pseudorapidity range (\(|\eta| < 1.7\)).
The endcap and forward regions are instrumented up to \(|\eta| = 4.9\) with LAr calorimeters for both the EM and hadronic energy measurements.
 
The calorimeters are surrounded by the muon spectrometer, which has
three large air-core toroidal superconducting magnets with eight coils each.
The field integral of the toroid magnets ranges between \num{2.0} and \SI{6.0}{\tesla\metre} across most of the detector.
The muon spectrometer includes a system of precision tracking chambers and fast detectors for triggering with a coverage of \(|\eta| < 2.7\).
 
Events are selected using a first-level trigger implemented in custom electronics, which reduces the event
rate to a maximum of \SI{100}{\kilo\hertz} using a subset of detector information. Software algorithms with access
to the full detector information are then used in the high-level trigger to  yield a recorded event rate of
about \SI{1}{\kilo\hertz}~\cite{TRIG-2016-01}.
 
An extensive software suite~\cite{ATL-SOFT-PUB-2021-001} is used in the reconstruction and analysis of real and simulated data, in detector operations, and in the trigger and data acquisition systems of the experiment.
\section{Data and simulation samples}
\label{sec:samples}
 
The results presented in this paper are based on the full \RunTwo\ proton--proton collision data at \(\sqrt{s}=\SI{13}{\TeV}\) recorded by the ATLAS detector between 2015 and 2018. Only data collected while all the detector components were operational are used~\cite{DAPR-2018-01}. The integrated luminosity is \SI{139.0}{\femto\barn^{-1}} and the average number of interactions per bunch-crossing is
\(\left<\mu\right> =34\), varying from 24 in 2015--2016 to 38 in 2017 and 36 in 2018.
 
Events were selected with a trigger requiring at least two photon candidates with energies greater than 35 and \SI{25}{\GeV}, respectively~\cite{TRIG-2018-05}. In addition, loose photon identification requirements were applied by the trigger in 2015--2016 and were tightened in 2017 and 2018 to cope with the higher instantaneous luminosity. Once the full diphoton event selection described in Section~\ref{sec:reco} is applied, the average trigger efficiency for \Hgg\ events is found to be greater than 99\% for the 2015--2016 data-taking period, and greater than 98\% for the 2017--2018 data-taking period.
 
Monte Carlo (MC) event generators were used to generate signal samples for the main Higgs boson production modes: gluon--gluon fusion (ggF), vector-boson fusion (VBF), Higgs boson production in association with a vector boson \(V=W,Z\) (\(VH\)) or with a top-quark pair (\ttH), a bottom-quark pair (\bbH) or a single top quark (\(tH\), with either an additional \(W\) boson, \(tWH\), or an additional \(b\)-quark and light quark, \(tHqb\)).
About 40~million events were produced with the nominal set-up and almost twice as many with alternative set-ups used to estimate modelling uncertainties, as described later.
The samples are normalised to the latest available calculations of the corresponding SM production cross-sections.
The normalisation of all Higgs boson samples also accounts for the \Hgg\ branching ratio of 0.227\%, calculated with HDECAY~\cite{Djouadi:1997yw,Spira:1997dg,Djouadi:2006bz} and \PROPHECY~\cite{Bredenstein:2006ha,Bredenstein:2006rh,Bredenstein:2006nk}.
 
The events were generated using \POWHEGBOX[v2]~\cite{Alioli:2010xd,Nason:2004rx,Frixione:2007vw}
and the \PDFforLHC[15] PDF set~\cite{Butterworth:2015oua},
except for \(tH\) production events, which were generated with \MGNLO[2.6]~\cite{Alwall:2014hca} and the \NNPDF[3.0nlo] PDF set.
The mass of the Higgs boson was set to \(\mH=\SI{125}{\GeV}\), while the width was set to \(\Gamma_H=\SI{4.07}{\MeV}\).
For all signal samples, the parton-level events from the generator were interfaced to \PYTHIA[8]~\cite{Sjostrand:2014zea} for the parton shower and the modelling of the underlying event, with parameter values set according to the \AZNLO tune~\cite{STDM-2012-23} for ggF, VBF and \(VH\) production, and to the A14 tune~\cite{ATL-PHYS-PUB-2014-021} for the others.
 
Higgs boson production via ggF was generated at next-to-next-to-leading-order (NNLO) accuracy in
QCD using \POWHEGBOX[v2]~\cite{Hamilton:2013fea,Hamilton:2015nsa} and the NNLO family of \PDFforLHC[15] PDFs.
The simulation achieves NNLO accuracy for arbitrary inclusive \(gg\to H\) observables by reweighting the Higgs boson
rapidity spectrum in \textsc{Hj}-\MINLO~\cite{Hamilton:2012np,Campbell:2012am,Hamilton:2012rf} to that of HNNLO~\cite{Catani:2007vq},
and the total cross-section is normalised to a prediction calculated at next-to-next-to-next-to-leading-order (N\({}^3\)LO) accuracy in QCD and
has next-to-leading-order (NLO) electroweak (EW) corrections applied~\cite{deFlorian:2016spz,Anastasiou:2016cez,Anastasiou:2015ema,Dulat:2018rbf,Harlander:2009mq,Harlander:2009bw,Harlander:2009my,Pak:2009dg,Actis:2008ug,Actis:2008ts,Bonetti:2018ukf,Bonetti:2018ukf}.
 
Higgs boson production via VBF was simulated with \POWHEGBOX[v2]~\cite{Nason:2009ai} using the \PDFforLHC[15nlo] PDF set.
The generation is accurate to NLO in QCD and the total cross-section is normalised to a calculation with full NLO QCD and EW corrections and approximate-NNLO QCD
ones~\cite{Ciccolini:2007jr,Ciccolini:2007ec,Bolzoni:2010xr}.
 
Higgs boson production via \(VH\) was simulated using
\POWHEGBOX[v2] and the \PDFforLHC[15nlo] PDF set.
The generation has NLO accuracy in QCD for \(q\bar{q}/qg\to VH\) events with up to one extra jet in the event,
while the loop-induced \(gg\to ZH\) process was generated separately at leading order in QCD.
The two samples are normalised to cross-sections calculated at NNLO in QCD with NLO electroweak corrections for \(q\bar{q}/qg \to VH\) and at NLO
and next-to-leading-logarithm (NLL) accuracy in QCD for \(gg \to ZH\)~\cite{Ciccolini:2003jy,Brein:2003wg,Brein:2011vx,Altenkamp:2012sx,Denner:2014cla,Brein:2012ne,Harlander:2014wda}.
 
The production of \ttH events was modelled using
\POWHEGBOX[v2]~\cite{Frixione:2007nw,Hartanto:2015uka}, while \(tHqb\) and \(tWH\) events were generated using \MGNLO[2.6.0] and 2.6.2~\cite{Alwall:2014hca}, respectively.
In these samples, the decays of bottom and charm hadrons were performed by \EVTGEN[1.6.0]~\cite{Lange:2001uf}.
Events in the \(tH\) samples originating from \ttH production were removed using the diagram removal scheme~\cite{Frixione:2008yi,Demartin:2016axk}.
The cross-section used to normalise the \ttH sample is calculated at NLO in QCD and electroweak couplings~\cite{deFlorian:2016spz,Beenakker:2002nc,Dawson:2003zu,Zhang:2014gcy,Frixione:2014qaa}, while those used to normalise the \(tH\) samples are calculated at NLO in QCD~\cite{Demartin:2015uha,Demartin:2016axk}.
 
Events from \bbH production were generated with \POWHEGBOX at NLO in QCD with the \NNPDF[3.0] PDF set~\cite{Ball:2014uwa}.
The \bbH sample contains additional NLO electroweak corrections, accounting for the treatment of the quark mass effects.
The sample is normalised with the cross-section calculation obtained by matching the five-flavour scheme cross-section accurate to NNLO in QCD with the four-flavour scheme cross-section accurate to NLO in QCD~\cite{Dawson:2003kb,Dittmaier:2003ej,Harlander:2011aa}, using the Santander scheme~\cite{Harlander:2011aa}.
 
Additionally, alternative signal samples were generated in order to estimate uncertainties related to the modelling of the parton shower or of the matrix element and, in particular, of extra jet radiation in ggF.
For the estimation of the uncertainties related to the modelling of the parton shower, for the ggF, VBF, \(VH\), \(tH\) (\ttH) samples, the same events from the matrix element generator of the nominal signal samples were showered with \HERWIG[7.1.3] (\HERWIG[7.0.4])~\cite{Bahr:2008pv,Bellm:2015jjp} instead of \PYTHIA, using the H7UE set of tuned parameters~\cite{Bellm:2015jjp}.
For the estimation of the uncertainties related to the matrix element calculation, alternative ggF, \(VH\) and \ttH (VBF) samples were produced with \MGNLO interfaced to \PYTHIA[8] (\HERWIG[7.1.3]). The alternative ggF sample is at NLO QCD accuracy for zero, one and two additional partons merged using the FxFx merging scheme~\cite{Frederix:2012ps}.
 
The generated Higgs boson events were passed through a \GEANT~\cite{Agostinelli:2002hh} simulation of the ATLAS detector~\cite{SOFT-2010-01} and reconstructed with the same software used for the data~\cite{ATL-SOFT-PUB-2021-001}.
 
The main background originates from continuum diphoton production. Prompt diphoton MC events were generated with \SHERPA[2.2]~\cite{Bothmann:2019yzt}, using the \NNPDF[3.0nnlo] PDF set and the dedicated set of tuned parton-shower parameters developed by the \SHERPA authors.
In this set-up, NLO-accurate matrix elements for up to one parton, and LO-accurate matrix elements for up to three partons were calculated with the Comix~\cite{Gleisberg:2008fv} and \OPENLOOPS~\cite{Buccioni:2019sur,Cascioli:2011va,Denner:2016kdg} libraries.
They were matched with the \SHERPA parton shower~\cite{Schumann:2007mg} using the \MEPSatNLO prescription~\cite{Hoeche:2011fd,Hoeche:2012yf,Catani:2001cc,Hoeche:2009rj} with a dynamic merging cut~\cite{Siegert:2016bre} of \SI{10}{\GeV}.
Photons were required to be isolated according to a smooth-cone isolation criterion~\cite{Frixione:1998jh}.
Due to the large size of the non-resonant diphoton sample (around one~billion generated events), needed for an accurate modelling of the background shape, these events were passed through a fast parametric simulation of the ATLAS detector response~\cite{SOFT-2010-01}.
Smaller backgrounds from non-prompt photons are studied using control regions in data as described in the following.

\PileupSnippet\
Events in the simulation are weighted in order to reproduce the distribution of the number of interactions per bunch crossing observed in real collisions. In addition, simulated events are corrected to reflect the momentum scales and resolutions as well as the trigger, reconstruction, identification and isolation efficiencies measured in data for all the objects used in the analysis.
\section{Event reconstruction and selection}
\label{sec:reco}
 
This section describes how the events and the objects used in the analysis are reconstructed and selected. In addition to photons, these objects include jets, \bjets, leptons and missing transverse momentum since several fiducial regions and differential cross-sections are defined using these additional objects.
 
\subsection{Photon reconstruction and identification}
\label{subsec:photon_reco}
Photon candidates are reconstructed~\cite{EGAM-2018-01} from dynamic, variable-size topological clusters of cells with significant energy in the EM calorimeter~\cite{ATL-PHYS-PUB-2017-022} and from potentially matching tracks reconstructed in the inner detector.
The photon candidates are classified as converted if two tracks forming a conversion vertex, or one track with the signature of an electron track but without hits in the innermost pixel layer, are matched to the cluster; otherwise they are labelled as unconverted.
Photon candidates are required to have pseudorapidity \(|\eta| < 2.37\), excluding the transition region between the barrel and endcap calorimeter, \(1.37 < |\eta| < 1.52\). In this acceptance region, the high granularity of the first sampling layer of the EM calorimeter allows efficient discrimination between isolated photons and closely spaced photon pairs from meson decays.
The photon candidate energy is calibrated using the procedure described in Ref.~\cite{PERF-2017-03}.
 
Photon candidates are selected by an identification algorithm based on one-dimensional selection criteria on multiple shower-shape variables related to the energy deposition by the candidate in the calorimeters.
Two working points of the identification algorithm are defined in order to reduce the contamination from the background, primarily associated with diphoton decays of neutral hadrons in jets~\cite{PERF-2017-02}. The \textit{loose} working point, with a nominal efficiency above 98\%, uses the lateral and longitudinal shape of the electromagnetic shower in the second layer of the calorimeter, together with the fraction of the shower's energy deposited in the hadronic calorimeter. The \textit{tight} selection adds information from the finely segmented first sampling layer of the EM calorimeter, and imposes requirements tighter than those of the \textit{loose} working point on shower shapes in other layers. The criteria are tuned separately for unconverted and converted photons in several pseudorapidity regions and as a function of the photon transverse energy \et~\cite{EGAM-2018-01}. The efficiency of the tight selection for prompt photons increases with $\pt$ from about 85\% for $\pt=25$~GeV to about 92\% (98\%) for unconverted (converted) photon candidates with $\pt$ above a few hundred GeV.
 
To further reject the hadronic jet background, photon candidates are required to be isolated from any significant activity in the calorimeter and tracking detector. A calorimeter-based isolation variable is defined as the sum of the transverse energy of positive-energy topological clusters contained within a cone of \(\Delta R = 0.2\) around the photon candidate, after removing the transverse energy of the photon candidate. The pile-up and underlying-event contributions are removed by using an ambient energy correction computed from low-\pt\ jets in the events~\cite{Cacciari:2008gn,Cacciari:2009dp,PERF-2013-04,STDM-2010-08,EGAM-2018-01}. A track-based isolation observable is computed as the scalar sum of the transverse momenta of tracks within a \(\Delta R=0.2\) cone around the photon candidate. Tracks are required to have \(\pT > \SI{1}{\GeV}\) and to originate from the selected diphoton vertex, defined in Section~\ref{subsec:vertex}. For converted photon candidates, the tracks associated with the conversion are not considered.
Isolation requirements that scale with the transverse energy \et\ of the candidate are applied to the selected photons. Photons are considered to be isolated if the calorimeter-based isolation is less than 6.5\% of the photon \et\ and if the track-based isolation is less than 5\% of the photon \et.
 
\subsection{Event selection and identification of the diphoton primary vertex}
\label{subsec:vertex}
An initial preselection retains events with at least two photon candidates with \(\et>\SI{25}{\GeV}\) satisfying the \textit{loose} identification criteria.
Among all photon candidates passing this requirement, the two with the highest \et\ values are retained for further analysis.
 
The primary vertex of the event is then selected from among all the reconstructed vertices, using a neural-network algorithm based on track and primary vertex information, as well as the directions of the two selected photons measured in the calorimeter and inner detector~\cite{HIGG-2013-08,}. The algorithm was optimised to distinguish between hard vertices from gluon--gluon fusion signal events and ones from pile-up interactions. The direction of each photon candidate is recomputed with respect to the selected primary vertex. In MC simulations, this leads to an improvement of 8\% in the inclusive case, compared to the default primary vertex algorithm of ATLAS which retains the vertex candidate with the largest sum of squared transverse momenta of the associated tracks. Agreement between data and simulation was checked with \Zee\ events, using only the electrons' calorimeter energy clusters, and not their tracks, as input.
 
The event is finally selected if the leading- and subleading-\et\ photon candidates have \(\et / \mgg >0.35\) and 0.25, respectively, if they fulfil the \textit{tight} identification criteria and the calorimetric and track-based isolation requirements, and
if their invariant mass is in the range \SIrange[range-phrase = --]{105}{160}{\GeV}.
 
The number of selected events in the full \RunTwo\ data set is \num{1178855}. The reconstruction efficiency estimated from simulated \Hgg\ events with respect to the full phase space is 36\%.
A shift in the simulated Higgs boson mass corresponding to the precision of the Higgs boson measurement
has a negligible impact on the signal acceptance.
 
\subsection{Reconstruction and selection of hadronic jets, leptons and missing transverse momentum}
Jets, electrons and muons are also considered in events passing the diphoton selection described above. Hadronic \(\tau\)-lepton decays are also reconstructed as jets.
 
Jet clustering uses the anti-\(k_t\) algorithm~\cite{Cacciari:2008gp,Fastjet} with a radius parameter \(R = 0.4\). Differently from the previous analysis, the inputs come from a particle-flow algorithm which combines information from the tracker and the calorimeters~\cite{PERF-2015-09}. The resulting jets exhibit improved energy and angular resolution, reconstruction efficiency, and pile-up stability compared to jets reconstructed using only the calorimeter information. Jets must satisfy \(|y|<4.4\) and \(\pT>\SI{30}{\GeV}\). In order to suppress jets coming from pile-up interactions, a jet-vertex tagger (JVT) multivariate discriminant~\cite{PERF-2014-03} is applied for jets within the tracking acceptance (\(|\eta|<2.5\)) and \(\pT < \SI{60}{\GeV}\).
 
Jets with \(|\eta| < 2.5\) containing \(b\)-hadrons (\bjets) are identified using the DL1r \btag\ algorithm with the 70\% efficiency working point~\cite{FTAG-2018-01}.
 
Electron candidates are reconstructed by matching tracks in the inner detector with variable-size topological clusters of cells with significant energy in the EM calorimeter formed with the same algorithm as in the photon reconstruction. Tracks are required to be consistent with the diphoton vertex using their
longitudinal (\(z_0\)) and transverse (\(d_0\)) impact parameters. In particular, tracks must satisfy \(|z_0\sin\theta| < \SI{0.5}{\milli\meter}\) and a transverse impact parameter significance \(|d_0/\sigma(d_0)| < 5\), where \(\theta\) is the track's angle with respect to the beam axis and \(\sigma(d_0)\) is the uncertainty of \(d_0\). In addition, electron candidates are selected using a likelihood-based identification method (\textit{medium} working point) combining both track and calorimeter information and are required to satisfy isolation criteria based on the calorimeter and track information (\textit{Fixed-cut loose} working point), detailed in Ref.~\cite{EGAM-2018-01}. Electron candidates are preselected by requiring \(\pT > \SI{10}{\GeV}\) and \(|\eta|<2.47\), excluding the transition region between the barrel and endcap sections of the EM calorimeter \((1.37 < |\eta|< 1.52)\). In the measurement, only electrons with \(\pt > \SI{15}{\GeV}\) are considered.
 
Muon candidates are reconstructed by matching tracks from the MS and ID subsystems~\cite{MUON-2018-03}. Muon candidates without an ID track but whose MS track is compatible with the interaction point, in the pseudorapidity range of \(2.5 < |\eta| < 2.7\), are also considered. Muon tracks must satisfy \(|z_0 \sin\theta| < \SI{0.5}{\milli\meter}\) and \(|d_0/\sigma(d_0)| < 3\). Muon candidates are required to have \(\pT > \SI{15}{\GeV}\) and must satisfy \textit{medium} identification requirements. Muons are also required to satisfy isolation criteria based on calorimeter and track information (\textit{PflowLoose} working point), detailed in Ref.~\cite{MUON-2018-03}.
 
To avoid double-counting of reconstructed objects, an overlap removal procedure based on the angular distance \(\Delta R_y=\sqrt{(\Delta y)^2 + (\Delta\phi)^2}\) is applied. First, electrons overlapping with the selected photons (\(\Delta R_y< 0.4\)) are removed. Next, jets overlapping with the selected photons (\(\Delta R_y < 0.4\)) and preselected electrons (\(\Delta R_y < 0.2\)) are removed. After that, electrons overlapping with the remaining jets (\(\Delta R_y < 0.4\)) are removed. Finally, muons overlapping with the selected photons or jets (\(\Delta R_y < 0.4\)) are removed.
 
The missing transverse momentum is computed as the negative vector sum of the transverse momenta of the
selected photon, electron, muon and jet candidates, as well as the transverse momenta of remaining low-\pt\ particles, estimated using tracks associated with the diphoton primary vertex but not with any of the selected objects~\cite{PERF-2016-07}.
\section{Fiducial phase space and differential observables}
\label{sec:intro_xs}
 
In this paper, inclusive and differential cross-sections are measured in various fiducial regions.
As described in detail in Section~\ref{sec:SandB}, the \Hgg\ signal is extracted in each fiducial region or bin of a differential distribution using a signal-plus-background fit to the corresponding diphoton invariant mass spectrum.
The cross-sections are then computed from the signal yields, by correcting the signal yields for the effects of detector inefficiency and resolution, and accounting for the integrated luminosity of the data set.
 
Most of the measurements are performed in a baseline `diphoton' fiducial region which closely matches the selection requirements of the reconstructed photons, described in Section~\ref{subsec:photon_reco}, in order to minimise model-dependent acceptance extrapolations.
The diphoton fiducial region is defined by the presence of two isolated photons in the final state with transverse momenta greater than 35\% and 25\% of the diphoton invariant mass, respectively. Each photon is required to have an absolute pseudorapidity \(|\eta|<1.37\) or \(1.52<|\eta|<2.37\).
The photons are required to be isolated in order to reduce hadronic activity, and the isolation energy must be less than 5\% of the photon's transverse momentum.
The isolation energy is defined as the scalar sum of the transverse momenta of stable charged particles (with a mean lifetime \(c\tau > \SI{10}{\milli\meter}\)) with \(\pT>\SI{1}{\GeV}\) within a \(\Delta R=0.2\) cone around the photon direction. This isolation criterion is chosen so that it matches the detector isolation requirement in order to reduce model dependence that is introduced when assuming SM composition for response matrices built from the different Higgs boson production modes.
The acceptance of the diphoton fiducial region with respect to the full phase space is 50\% and the efficiency of the selection criteria described in Section~\ref{sec:reco} for signal events in the fiducial volume is close to 70\%.
 
In addition, subsets of the diphoton baseline fiducial region are defined, providing a variety of phase-space regions sensitive to particular Higgs boson production modes.
The definitions of these subregions, as well as of some of the observables for the differential measurements, are based on the following particle-level selections:
 
\begin{itemize}
\item \textbf{Leptons} are defined from all electrons and muons that are not produced during hadronisation. The prompt leptons are dressed by adding the four-momenta of stable photons within \(\Delta R <0.1\). Selected electrons (muons) are required to pass the kinematic selection of \(\pT > \SI{15}{\GeV}\) and \(|\eta|<2.47\) (\(|\eta|<2.7\) for muons). Electrons are rejected if they pass through the barrel--endcap transition region \(1.37<|\eta|<1.52\), or if their distance from a selected photon is \(\Delta R<0.4\). No isolation requirement is applied.
 
\item \textbf{Jets} are defined by clustering all stable particles using the \antikt\ algorithm with a radius parameter \(R=0.4\). The clustering algorithm excludes prompt leptons and Higgs boson decay products. Selected jets are required to have transverse momentum \(\pt>\SI{30}{\GeV}\) and rapidity \(|y|<4.4\). Selected jets are required to be well separated from photons with \(\pT > \SI{15}{\GeV}\) (\(\Delta R>0.4\)) and electrons (\(\Delta R>0.2\)). Leading and sub-leading jets are defined as the ones with the largest and second-largest transverse momenta.
 
\item \textbf{\bjets} are defined from selected central jets (\(|\eta|<2.5\)) if there is a \(b\)-hadron with \(\pT > \SI{5}{\GeV}\) within \(\Delta R=0.4\) of the jet axis.
 
\item \textbf{Missing transverse momentum (\met)} is defined as the vector sum of the transverse momenta of all neutrinos that do not originate from the decay of a hadron.
\end{itemize}
 
\begin{figure}[htbp]
\centering
\includegraphics[width=0.8\textwidth]{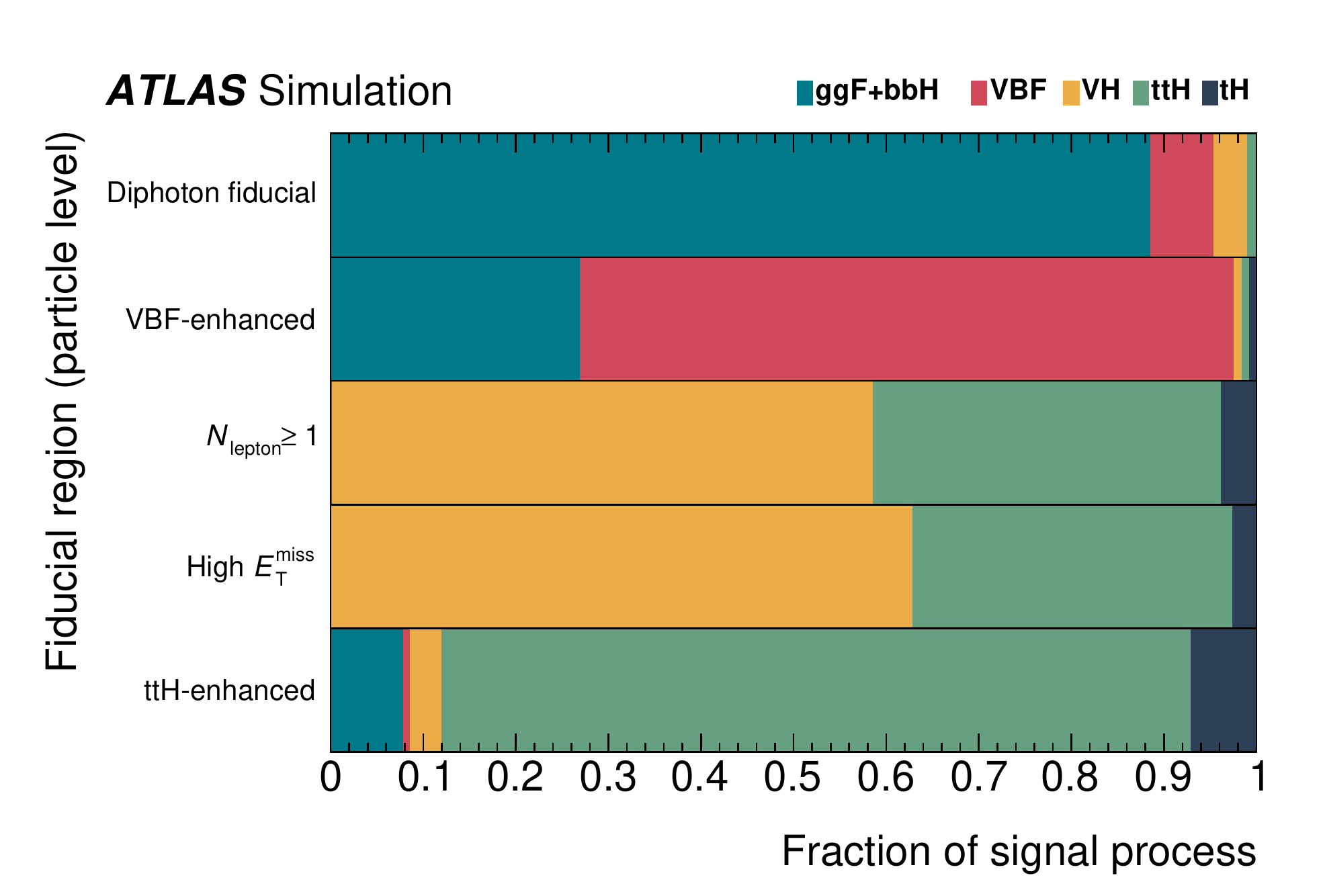}
\caption{The expected production mode composition of Higgs boson events in each fiducial region at particle level, as estimated from simulated SM Higgs boson events with \(\mH=\SI{125.09}{\GeV}\).}
\label{fig:purity_fiducial}
\end{figure}

The following fiducial subregions of the diphoton baseline region, with larger sensitivity to specific Higgs production modes, are defined:
\begin{itemize}
\item\textbf{VBF-enhanced}: a region composed of events with at least two jets, where the two leading jets have a large invariant mass, \(\mjj \geq \SI{600}{\GeV}\), and large rapidity separation, \(\dyjj \geq 3.5\).
\item\textbf{\(N_\text{lepton} \geq 1\)}: a region composed of events containing at least one additional charged lepton (electron or muon) with transverse momentum \(\pt^{\ell} > \SI{15}{\GeV}\). This region is sensitive to the \(VH\), \ttH and \(tH\) production modes.
\item\textbf{High \MET}: a region composed of events with large missing transverse momentum (\(\met > \SI{80}{\GeV}\)) and a diphoton transverse momentum (\(\ptgg>\SI{80}{\GeV}\)). This region is sensitive to the \(VH\) and \ttH\ production mechanisms and to BSM effects such as weakly interacting dark matter particles.
\item\textbf{\ttH-enhanced}: a region composed of events with at least one \bjet\ and either no leptons and at least four jets, or at least one lepton and at least three jets. This region is mostly sensitive to \ttH, but also to \(tH\) production.
\end{itemize}
The acceptance of these fiducial regions with respect to the full phase space of Higgs bosons decaying to $\gamma\gamma$ varies between 0.2\% for the High \met\ region and 1.2\% for the VBF-enhanced region.
Figure~\ref{fig:purity_fiducial} shows the expected contributions of signal events from the different SM Higgs boson production modes to the various fiducial regions. The fiducial regions are not orthogonal and non-negligible overlap exists between the \(N_\text{lepton}\geq 1\), High \met and \ttH-enhanced regions. The largest overlap is between the first two, where 42\% of the events in the High \met fiducial region are also in the \(N_\text{lepton}\geq 1\) fiducial region.

Differential cross-sections are measured as a function of one or two of several observables in the diphoton fiducial region. In addition, a limited set of observables is studied in the VBF-enhanced fiducial region.
The measured observables probe the Higgs boson kinematics and decay information, and also associated jet activity.
The list of observables and a brief motivation for them is outlined below.
 
\begin{itemize}
\item Diphoton kinematic variables:
\begin{itemize}
\item The transverse momentum \ptgg\ and the rapidity \ygg\ of the diphoton system describe the fundamental kinematics of the Higgs boson.
The low-\pt\ region of the Higgs boson is sensitive to the bottom- and charm-quark Yukawa couplings~\cite{Bishara:2016jga}, and also to QCD resummation effects. The high-\pt\ region, on the other hand, is sensitive to the Higgs boson's coupling to the top quark, and to BSM scenarios where heavy resonances in the loops can further boost the Higgs boson~\cite{Grojean:2013nya, Biekoetter:2014jwa,Mimasu:2015nqa,Grazzini:2016paz,Banfi:2018pki}. The Higgs boson rapidity, \ygg, is sensitive to light-quark Yukawa couplings~\cite{Soreq:2016rae} in addition to the parton distribution functions (PDFs) of the colliding protons.
\item The relative transverse momenta of the leading and sub-leading photons, \(\ptgOne/\mgg\) and \(\ptgTwo/\mgg\), probe the kinematics of the Higgs boson decay.
\end{itemize}
 
\item Jet multiplicities:
\begin{itemize}
\item The cross-sections as a function of the inclusive and exclusive jet multiplicities, \Njets\ (for jets with \(\pt^j > \SI{30}{\GeV}\)) are sensitive to the different Higgs boson production mechanisms and to QCD modelling of the gluon--gluon fusion production mode.
 
\item The cross-section as a function of the number of \bjets, \Nbjets, is sensitive to Higgs boson production in association with heavy-flavour particles. The measurement is carried out in the diphoton fiducial region requiring at least one central jet\footnote{Central jets are defined as having \(|\eta| < 2.5\), matching the acceptance of the inner detector.} with \(\pt^j > \SI{30}{\GeV}\). In addition, in order to suppress the \ttH\ contribution, a veto on electrons and muons is employed.
 
\end{itemize}
 
\item \(\geq 1\)-jet observables:
\begin{itemize}
\item The transverse momentum of the leading jet, \ptj[1], and of the scalar sum of the transverse momenta of all jets in the events, \HT, probe the perturbative QCD modelling and are sensitive to the relative contributions of the different Higgs production mechanisms.
\item The transverse momentum \(p_{\text{T}}^{\gamma\gamma j}\) and the invariant mass \(m_{\gamma\gamma j}\) of the system made by the two leading photons and the leading jet are sensitive to resummation effects.
\item Beam-thrust-like variables \(\tau_{C,j1}\) and \(\Sigma\tau_{C,j}\), with \(\tau_{C,j}\) for a given jet \(j\) being defined as:
\begin{equation}\nonumber
\tau_{C,j} = \frac{m_\mathrm{T}}{2\cosh(y_j-y_{\gamma\gamma})},~~~~~m_\mathrm{T}=\sqrt{\pt^2+m^2},
\end{equation}
where \(m\) is the jet mass and \(y_j\) is the jet rapidity. The \(\tau_{C,j1}\) observable is defined as the highest value of \(\tau_{C,j}\) among all jets in the event,  whereas \(\Sigma\tau_{C,j}\) is the scalar sum of \(\tau\) for all jets with \(\tau > \SI{5}{\GeV}\).
For large jet rapidities, \(\tau\) corresponds to the small light-cone component of the jet, \(p_j^{+}=E_j - |p_{z,j}|\),
whereas the sum corresponds to the beam-thrust global event-shape variable measured in the diphoton rest frame~\cite{Stewart_2010}.
\end{itemize}
 
\item The transverse momentum distribution of the diphoton system in events with a veto on the transverse momentum of the hardest accompanying jet, \(\pt^{\gamma\gamma\text{, jet veto}}\), provides insights into jet-veto resummation~\cite{Monni:2019yyr}.
This cross-section is measured using different jet vetoes: \(\pt^j > 30,~40,~50,~\SI{60}{\GeV}\).
 
\item \(\geq 2\)-jet observables:
\begin{itemize}
\item The dijet invariant mass, \mjj, and the signed dijet azimuthal angle separation, \dphijjsig, are sensitive to the VBF production mechanism.
The sign of \dphijjsig\ is determined by ordering the jets in decreasing rapidity,\footnote{This definition of \dphijjsig\ is invariant under a redefinition of the ordering by choosing the opposite beam axis as detailed in Ref.~\cite{Klamke:2007cu}} making this observable sensitive to the CP properties of the Higgs boson's couplings to gluons and weak vector bosons~\cite{Klamke:2007cu,Plehn:2001nj}.
The azimuthal angle between the dijet and diphoton systems, \dphiggjj, is sensitive to the VBF production mechanism and can be used to distinguish it from ggF events with at least two jets. Only the leading and sub-leading jets are considered.
\item The transverse momentum of the system made by the two leading photons and the two leading jets, \ptggjj, is sensitive to additional jet activity in the event.
\end{itemize}
 
\item VBF-enriched phase space observables:
 
Several fiducial differential cross-sections are measured in the VBF-enriched fiducial region as a function of observables sensitive to the kinematic features of the VBF production mode. These observables include: (i) the transverse momentum of the leading jet, \ptj[1]; (ii) the signed dijet azimuthal angle separation, \dphijjsig, which helps disentangle CP effects originating from gluon--gluon fusion and VBF; (iii) the pseudorapidity of the diphoton system relative to the average rapidity of the two leading jets, \(|\eta^{*}|\), as its shape differs between ggF and VBF and hence can be used to disentangle the two production modes; (iv) the transverse momentum of the diphoton-plus-dijet system~\ptggjj, which also helps to disentangle ggF and VBF, in addition to being sensitive to additional jet activity.
 
\end{itemize}
 
For each observable, the binning was designed to have an expected signal significance of close to or greater than \(2\sigma\) and to minimise migrations of signal events between bins.
The definition of the binning is summarised in Table~\ref{tab:binning_1d}.
 
\begin{table}[htbp]
\centering
\caption{Bin ranges for the differential cross-section measurements. Transverse momenta, invariant masses and the \(\HT\) and \(\tau\) variables (defined in the text) are given in \GeV. For the jet multiplicities, the values reported in the table are the definition of the bins.}
\resizebox{\textwidth}{!}{
\begin{tabular}{lcc}
\toprule
Variable & Bin Edges & \(N_{\text{bins}}\) \\
\midrule
\ptgg & 0, 5, 10, 15, 20, 25, 30, 35, 45, 60, 80, 100, 120, 140, 170, 200, 250, 300, 450, 650, 13000 & 20~~ \\
\ygg & 0, 0.15, 0.3, 0.45, 0.6, 0.75, 0.9, 1.2, 1.6, 2.0, 2.5 & 10~~ \\
\(\ptgOne/\mgg\) & 0.35, 0.45, 0.5, 0.55, 0.6, 0.65, 0.75, 0.85, 0.95, 10 & 9 \\
\(\ptgTwo/\mgg\) & 0.25, 0.35, 0.4, 0.45, 0.5, 0.55, 0.65, 0.75, 0.85, 10 & 9 \\
\midrule
\Njets & 0, 1, 2, \(\geq\)3 & 4 \\
\Nbjets & \(\Njets^\text{central} = 0\) or \(N_\text{lep}>0\), \(\Nbjets = 0\), \(\geq 1\) & 3 \\
\midrule
\ptj[1] & 30, 60, 90, 120, 350, 13000 & 5 \\
\HT & 30, 60, 140, 200, 500, 13000 & 5 \\
\ptggj  & 0, 30, 60, 120, 13000 & 4 \\
\mggj & 120, 220, 300, 400, 600, 900, 13000 & 6 \\
\maxtau & 0, 5, 15, 25, 40, 13000  & 5 \\
\sumtau  & 5, 15, 25, 40, 80, 13000  & 5 \\
\midrule
\(\pT^{\gamma\gamma\text{, jet veto 30 GeV}}\) & 0, 5, 10, 15, 20, 30, 40, 50, 100, 13000 & 9\\
\(\pT^{\gamma\gamma\text{, jet veto 40 GeV}}\) & 0, 5, 10, 15, 20, 30, 40, 50, 60, 100, 13000 & 10~~ \\
\(\pT^{\gamma\gamma\text{, jet veto 50 GeV}}\) & 0, 5, 10, 15, 20, 30, 40, 50, 60, 70, 100, 13000 & 11~~ \\
\(\pT^{\gamma\gamma\text{, jet veto 60 GeV}}\) & 0, 5, 10, 15, 20, 30, 40, 50, 60, 70, 80, 100, 13000 & 12~~ \\
\midrule
\mjj & 0, 120, 450, 3000, 13000 & 4 \\
\dphijj & \(-\pi\), \(-\frac{\pi}{2}\), 0, \(\frac{\pi}{2}\), \(\pi\) & 4 \\
\(\pi - \absdphiggjj\) & 0, 0.15, 0.65, \(\pi\) & 3 \\
\ptggjj & 0, 30, 60, 120, 13000 & 4 \\
\midrule
VBF-enhanced: \ptj[1] & 30, 120, 13000 & 2 \\
VBF-enhanced: \dphijjsig & \(-\pi\), \(-\frac{\pi}{2}\), 0, \(\frac{\pi}{2}\), \(\pi\) & 4 \\
VBF-enhanced: \(|\eta^{*}|\) & 0, 1, 2, 10 & 3 \\
VBF-enhanced: \ptggjj & 0, 30, 13000 & 2 \\
\bottomrule
\end{tabular}
}
\label{tab:binning_1d}
\end{table}
 
Fiducial differential cross-sections are also measured in two-dimensional combinations of some of these observables, providing deeper insight into event kinematics and correlations across observables.
The full list and the binning are summarised in Table~\ref{tab:binning_2D}.
 
\begin{table}[htbp]
\caption{Binning for the double-differential cross-section measurements. Transverse momenta and \(\tau\) variables (defined in the text) are given in GeV.}
\centering
\resizebox{\textwidth}{!}{
\begin{tabular}{lccc}
\toprule
Variable & \multicolumn{2}{c}{Bin Edges} & \(N_{\text{bins}}\)     \\
\midrule
\ptgg\ vs \ygg & 0.0 < \ygg < 0.5 &  \ptgg: 0, 45, 120, 350 & \multirow{4}{*}{12}~~\\
& 0.5 < \ygg < 1.0 &  \ptgg: 0, 45, 120, 350 & \\
& 1.0 < \ygg < 1.5 &  \ptgg: 0, 45, 120, 350 & \\
& 1.5 < \ygg < 2.5 &  \ptgg: 0, 45, 120, 350 & \\
\midrule
\((\ptgOne+\ptgTwo)/\mgg\) vs \((\ptgOne-\ptgTwo)/\mgg\) & \(0.6 < (\ptgOne+\ptgTwo)/\mgg \leq 0.8\) & \((\ptgOne-\ptgTwo)/\mgg\): 0, 0.3 & \multirow{3}{*}{8} \\
& \(0.8 < (\ptgOne+\ptgTwo)/\mgg \leq 1.1\) & \((\ptgOne-\ptgTwo)/\mgg\): 0, 0.05, 0.1, 0.2, 0.8 & \\
& \(1.1 < (\ptgOne+\ptgTwo)/\mgg \leq 4\)~~~ & \((\ptgOne-\ptgTwo)/\mgg\): 0, 0.3, 0.6, 4 & \\
\midrule
\ptgg\ vs \ptggj& \(\Njets = 0\) &  \ptgg: 0, 350 & \multirow{4}{*}{9} \\
& \(0 < \ptggj \leq 30\) & \ptgg: 0, 100, 350 & \\
& \(30 < \ptggj \leq 60\)~~  & \ptgg: 0, 45, 120, 350 &  \\
& \(60 < \ptggj \leq 350 \) & \ptgg: 0, 80, 250, 450 & \\
\midrule
\ptgg\ vs \maxtau&  \(\Njets = 0\) & \ptgg: 0, 350 & \multirow{5}{*}{9} \\
& \(0 < \maxtau \leq 15\) & \ptgg: 0, 100, 350 & \\
& \(15 < \maxtau \leq 25\)~~ & \ptgg: 0, 120, 350 & \\
& \(25 < \maxtau \leq 40\)~~ & \ptgg: 0, 200, 350 & \\
& \(40 < \maxtau \leq 400\) & \ptgg: 0, 250, 650 & \\
 
\midrule
VBF-enhanced:  \ptj[1] vs \dphijjsig & \(-\pi < \dphijjsig < 0\) & \ptj[1]: 30, 120, 500 & \multirow{2}{*}{4} \\
& \,~~\(0 < \dphijjsig < \pi\) & \ptj[1]: 30, 120, 500 & \\
\bottomrule
\end{tabular}
}
\label{tab:binning_2D}
\end{table}
\section{Signal and background modelling of the diphoton mass spectrum}
\label{sec:SandB}
The Higgs boson signal yield is measured using an unbinned maximum-likelihood fit to the diphoton invariant mass spectrum in the fiducial regions and in each bin of the differential distributions. The fit is performed in the invariant mass range \(\SI{105}{\GeV}<\mgg<\SI{160}{\GeV}\). This range is chosen to be wide enough to allow a reliable determination of the background shape from the data, while being narrow enough to limit the uncertainties from the choice of background parameterisation.
The fit model (detailed in Section~\ref{sec:measurement}) is the sum of the two analytic functions describing signal and background components.
The signal and background shapes are modelled as described below.
 
\subsection{Signal model}
 
The Higgs boson signal manifests itself as a narrow peak in the \mgg\ spectrum.
The signal distribution is empirically modelled as a double-sided Crystal Ball function consisting of a Gaussian
central part and a power-law tail on each side.
The Gaussian core of the Crystal Ball function is parameterised by the peak position \((\mH + \Delta \mu_\mathrm{CB})\) and width \((\sigma_\mathrm{CB})\).
The non-Gaussian contributions to the mass resolution arise mostly from photons converted to electrons with at least one electron losing a significant fraction of its energy through bremsstrahlung in the inner-detector material.
 
The parametric form of the double-sided Crystal Ball function can be found in Ref.~\cite{HIGG-2016-21}.
The parameters of the model except its normalisation are determined through fits to the
simulated signal samples, taking into account all production modes according to their expected contributions.
To take into account the different values of the Higgs boson mass assumed in the analysis (\(\mH= \SI{125.09}{\GeV}\)) and in the MC event samples (\(\mH = \SI{125}{\GeV}\)), a shift of \SI{90}{\MeV} is applied to the position of the signal peak.
 
\begin{figure}[htbp]
\centering
\includegraphics[width=0.6\textwidth]{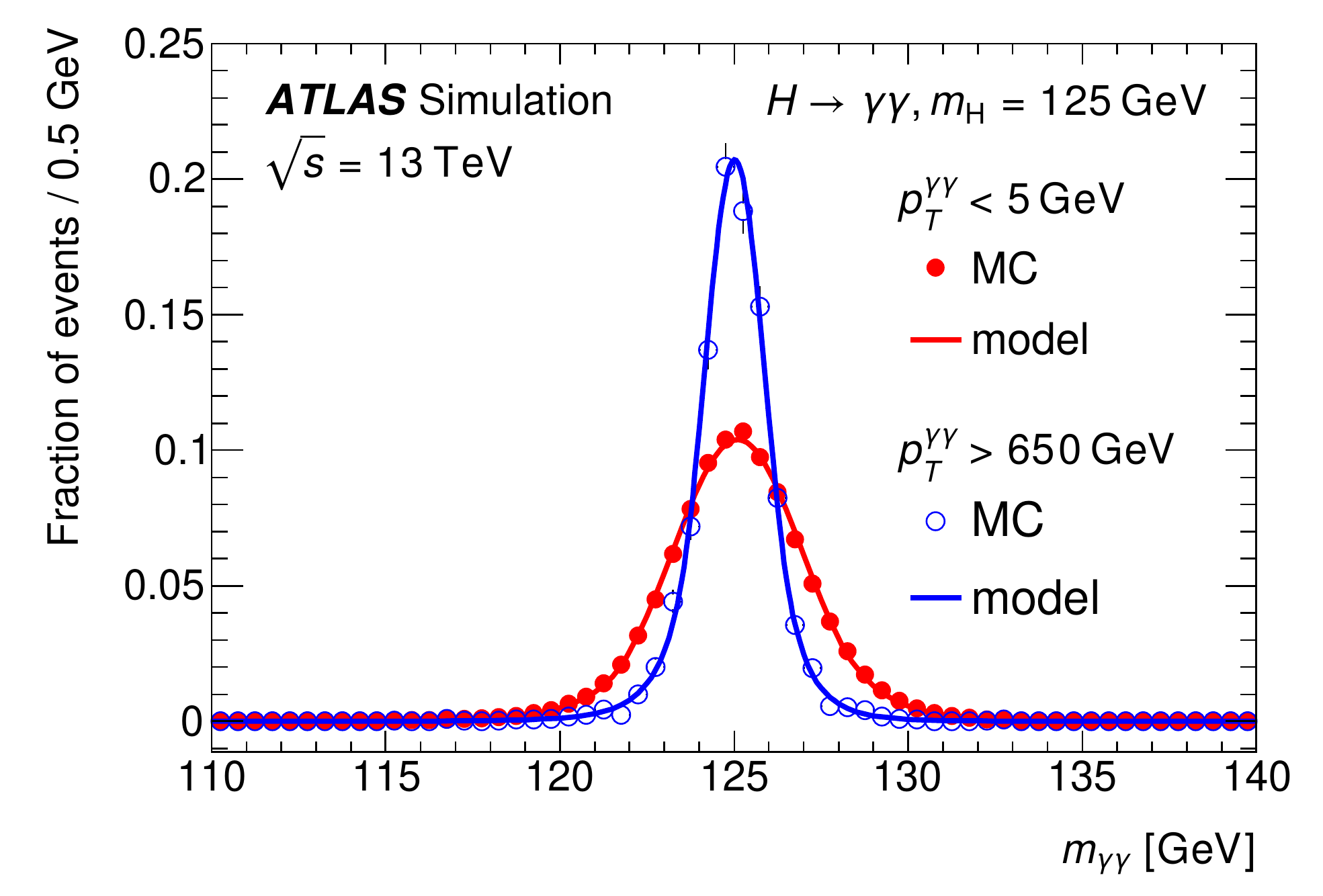}
\caption{
Signal \mgg\ model in the lowest and highest \ptgg\ bins considered. The two fitted models (solid curves) are compared with the \mgg\ distributions of the signal MC events in the lowest (filled markers) and highest (open markers) \ptgg\ bins. The resolution, evaluated as half the width of the narrowest interval containing 68.3\% of the simulated events, varies between \SI{1.0}{\GeV} and \SI{1.9}{\GeV}.
}\label{fig:dscb}
\end{figure}

The parameterisation is derived separately for each bin considered in the cross-section measurement.
As an example of the signal model, Figure~\ref{fig:dscb} shows the parameterisations in the lowest and the highest \ptgg\ bin considered in the measurement.
The \mgg\ resolution for the signal is evaluated as half the width of the narrowest interval containing 68.3\% of the distribution. In the inclusive case, it corresponds to \SI{1.9}{\GeV}. The resolution in the bins of the single-differential cross-sections ranges from a minimum of \SI{1.0}{\GeV} in the highest \ptgg\ bin, to \SI{2.2}{\GeV} in the bins corresponding to the diphoton rapidity region \(1.2<\ygg<2.0\).

\subsection{Background model}
\label{sec:background}
The background in the Higgs boson signal extraction fit is modelled analytically. The functional form chosen for the background model is based on background templates built using simulations and data control-regions, and these are also used in the estimation of related systematic uncertainties. This section details the choice of background model.
 
The main sources of background are the non-resonant production of prompt and isolated diphotons (\(\gamma\gamma\)) and the \(\gamma j\) and \(jj\) processes where one or two hadronic jets are misidentified as photons. For each bin of the analysis, the \mgg\ distribution of the background falls smoothly and is described by an empirically chosen function. The parameters of these functions are fitted to data, but the functional forms are chosen from dedicated studies.
The fraction contributed by each background component is measured in data using a two-dimensional double-sideband method~\cite{STDM-2011-05} for each bin of a given observable. In this method, the identification and isolation requirements for the photons are loosened. The events are then separated into 16 orthogonal regions, depending on whether one or both photons satisfy or fail the nominal identification and/or isolation requirements as described in Section~\ref{subsec:photon_reco}. The \(\gamma\gamma\), \(\gamma j\) and \(jj\) yields after the nominal selection are obtained by solving a system of equations using the observed yields in the 16 regions and the photon identification and isolation efficiencies estimated from MC simulation as inputs. The systematic uncertainties in the measured background fractions are due
to the definition of the control regions arising from the different inverted photon identification and isolation criteria, resulting in a systematic uncertainty that dominates the total uncertainty.
The fractions contributed by the \(\gamma\gamma\), \(\gamma j\) and \(jj\) background sources after the inclusive diphoton selection are \SI{75\pm 4}{\percent}, \SI{22\pm 3}{\percent} and \SI{3\pm 1}{\percent}, respectively. The \(\gamma\gamma\) fraction changes smoothly across the bins of the differential measurements and ranges from 66\% to 92\%.
 
To study the modelling of the background, a background template of the diphoton invariant mass is built as the sum of the \(\gamma\gamma\) and the \(\gamma j\) components. Adding the \(jj\) component or the contribution due to cases where the two photons originate from two different pile-up interactions was found to have a negligible impact on the results.
For each bin, the background template is built by summing the two considered components according to the relative fractions measured with data. The \(\gamma\gamma\) template is a histogram determined from the events of the simulation described in Section~\ref{sec:samples} that pass the full event selection. The template of the reducible \(\gamma j\) component is obtained from data in control regions formed by inverting the tight photon identification requirement, while still imposing the loose identification requirement, on any of the two photons in the final state. The total background template is then normalised to match the data entries in the mass sidebands (i.e.\ excluding the range \(\SI{123}{\GeV}< \mgg < \SI{127}{\GeV}\)).
 
Several functional forms were considered for the modelling of the background template distributions, including an exponential function of a polynomial of first to fourth order, a power law of first or second order, and a Bernstein polynomial of third, fourth or fifth order. The choice of functional form for the background modelling is based on the estimated potential bias in the fitted signal yield (spurious signal) and the goodness of the fit. Both criteria are evaluated using the background template. The spurious signal is estimated as the maximum of the absolute value of the fitted signal yield in successive fits to background templates, using a signal model with resonant mass scanning the range between \SI{123}{\GeV} and \SI{127}{\GeV}. The spurious signal must be less than 20\% of the background uncertainty or less than 10\% of the expected signal yield.
In addition, the goodness of the fit of the functional form to the background template is evaluated with a \(\chi^2\) test and the relative \(p\)-value is required to be larger than 1\%.
If more than one function fulfils the criteria, the one with fewer degrees of freedom is chosen.
The value of the spurious signal is considered as a systematic uncertainty of the signal yield due to the background modelling.
 
Due to the finite size of the simulation samples used to build the background templates, large statistical fluctuations are often observed. These fluctuations can adversely affect the estimation of the spurious signal, particularly when they occur in the Higgs signal window between \SI{123}{\GeV} and \SI{127}{\GeV}. These fluctuations would nominally be interpreted as a spurious signal, hence resulting in an overestimation of the background model's systematic uncertainty.
Given the computational limitations in generating larger data sets, an alternative approach is employed. The background templates are smoothed using Gaussian process regression (GPR)~\cite{Frate:2017mai}, using the Gibbs Kernel. The GPR approach suppresses statistical fluctuations in the background templates, without biasing the shape of wider features in the template.
The GPR approach was validated using pseudo-experiments in which sets of pseudo-data, with different statistics, were generated from known functions, using the functional forms considered to model the background distribution, and the bias from the GPR smoothing was tested.
The pseudo-experiments show that the smoothing procedure maintains the underlying shape of the \mgg\ distribution, hence introducing no significant bias in the spurious signal. Overall, use of the GPR approach resulted in an average reduction of the spurious signal by 30\%, with the largest improvements seen in the low-yield bins, and little to no change seen in high-yield bins. This is expected given the statistical fluctuations in the low-yield templates, and hence GPR provides an accurate estimate of the spurious signal that is due to real shape mis-modelling. This is in contrast to the high-yield templates, where the spurious-signal estimates from the GPR-smoothed template and the original template are compatible.
 
The selected background functions are further validated in the \mgg\ data sideband regions by means of likelihood-ratio tests that check the hypothesis that a function with an additional degree of freedom is not needed to better describe the distribution. The following test statistic is computed:
\begin{equation}\nonumber
\lambda_{1,2}= -2 \log (L_1 / L_2),
\end{equation}
where \(L_1\) and \(L_2\) are the likelihood with the nominal background model and an alternative one with an additional degree of freedom, respectively. The function with the extra degree of freedom is chosen if the probability of obtaining a \(\lambda_{1,2}\) value higher than the observed one computed under the nominal-background hypothesis, \(P(\lambda_{1,2} \geq \lambda_{1,2}^\text{obs})\), is less or equal to \(0.05\). The procedure is then repeated with higher degrees of freedom until \(P(\lambda_{1,2} \geq \lambda_{1,2}^\text{obs})\) exceeds 0.05.
Overall, approximately 6\% of the total number
of different differential observable bins and fiducial regions considered for the measurement required an increase of the number of degrees of freedom of the background model by one unit.

\subsection{Measurement procedure}
\label{sec:measurement}
 
Reconstructed events passing the event selection are assigned to one of several different bins, each corresponding to a fiducial bin to be measured, called reco-bins. The classification is performed using reconstructed quantities following the same definition as at particle level. When measuring the cross-section for the fiducial regions which are subsets of the diphoton region, an additional reco-bin is filled with events passing the diphoton selection but failing the full selection corresponding to that region in order to help the fit constrain systematic uncertainties. For example, when measuring the cross-section in the VBF-enhanced fiducial region, an additional reco-bin named `anti-VBF', filled with events passing the analysis selection but not passing the definition of the VBF region at detector level, is added.
Similarly, for the differential cross-sections involving the presence of at least one or at least two jets, one or two additional reco-bins are filled with events passing the selection with zero jets or exactly one jet.
In addition, when applicable, underflow and overflow reco-bins are considered.
 
\begin{figure}[htbp]
\centering
\subfloat[]{\includegraphics[width=0.45\textwidth]{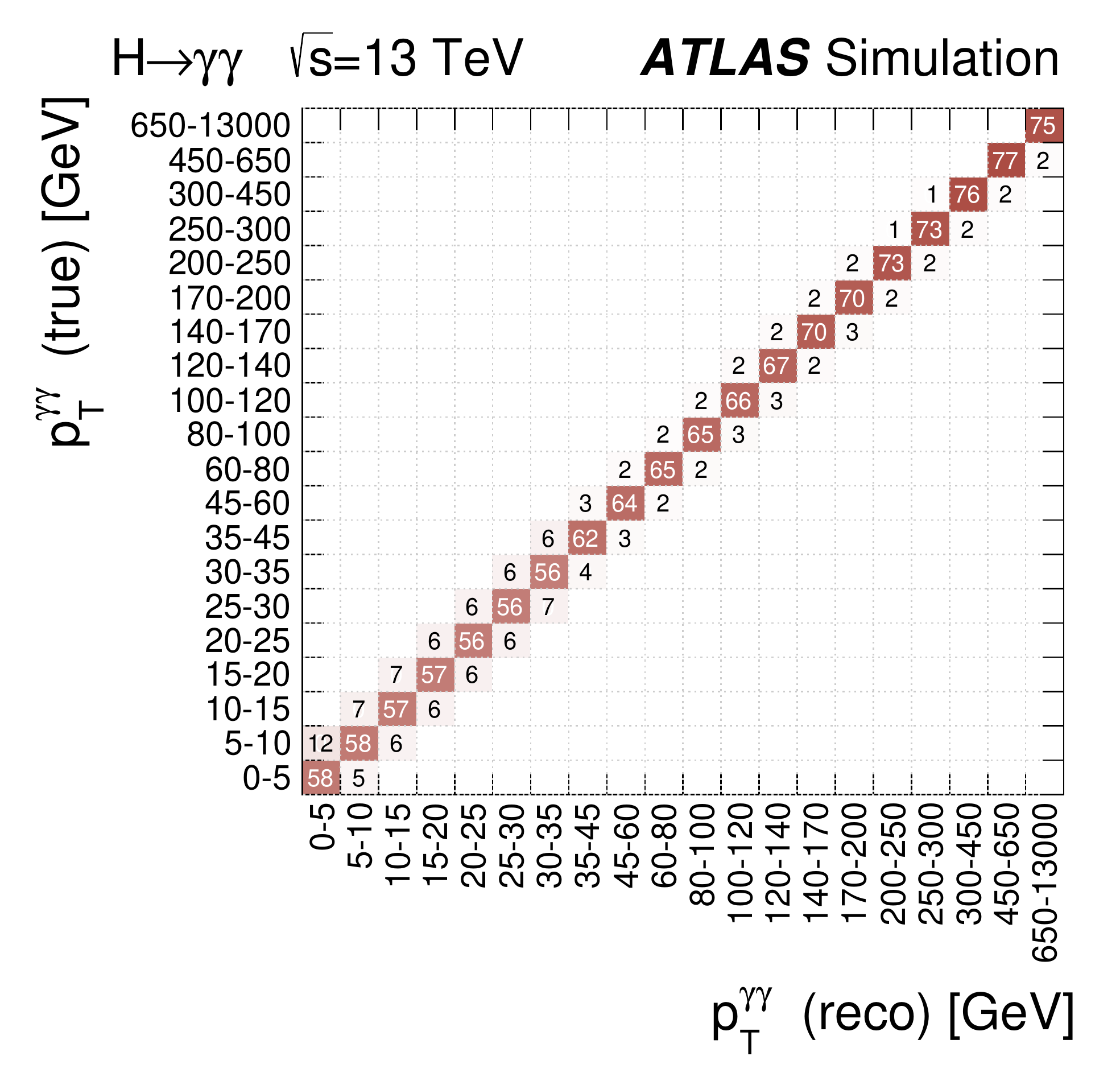}\label{fig:response_example_a}}
\subfloat[]{\includegraphics[width=0.45\textwidth]{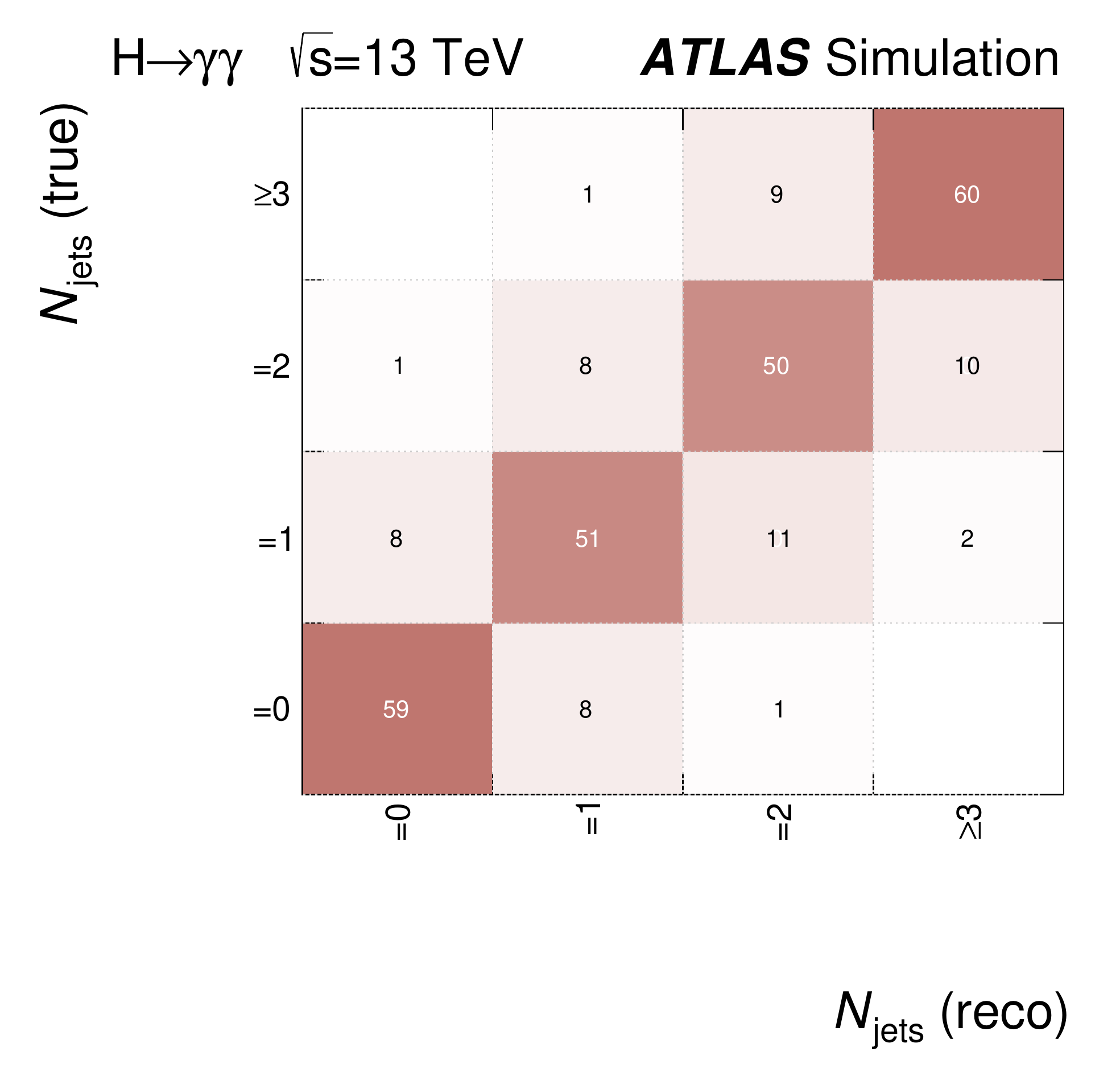}\label{fig:response_example_b}}
\caption{Response matrices (in \%) evaluated from MC simulations of Higgs bosons decaying into two photons for \protect\subref{fig:response_example_a} the transverse momentum of the diphoton system \ptgg\ and \protect\subref{fig:response_example_b} the number of jets \Njets. Each matrix element is the probability for a signal event generated in a fiducial truth-bin to be selected in a reconstructed reco-bin. Values smaller than 1\% (after rounding) are not shown.}
\label{fig:response_example}
\end{figure}

Differently from the previous analysis~\cite{HIGG-2016-21}, the measurement of the fiducial cross-section as a function of a certain particle-level observable is performed in a single step through a simultaneous fit of the \mgg\ distribution in all the reco-bins for the corresponding detector-level observable. For each reco-bin \(r\), the number of selected events originating from Higgs boson diphoton decays \(N_r^{(H)}\) is parameterised as a function of the cross-sections under study through the response matrix \(R\):
 
\begin{equation}
N_r^{(H)} = \frac{1}{C_r^\text{fid}}\left[\sum_t L\times(\sigma_{t} \times B_{\gamma\gamma}) \times R_{t,r} \right],
\label{eq:unfold1}
\end{equation}
where the sum inside the square brackets represents the number of fiducial signal events reconstructed in the reco-bin \(r\). In this formula, \(L\) is the integrated luminosity, and \((\sigma_t \times B_{\gamma\gamma})\) is the fiducial Higgs boson production cross-section in the `truth'-bin \(t\) times the Higgs boson diphoton branching ratio, where `truth' refers to information from the MC generator's event record. The matrix element \(R_{t,r}\) of the response matrix describes the probability to reconstruct a signal event originating from a particle-level truth-bin \(t\) in a reconstructed detector-level reco-bin \(r\). As a consequence of carefully choosing the binning of the observables under study to minimise migrations between bins, the response matrices for all the variables considered are well-conditioned, with a condition number\footnote{The condition number of a matrix is defined as the ratio of the maximum and minimum singular values of the matrix itself. In the case of a response matrix, the condition number is related to how much an unfolding procedure relying on such a matrix is sensitive to statistical fluctuations in the input.} that ranges from 1.1 for quantities with very small migrations, such as \(|y_{\gamma\gamma}|\), to 2.1 for quantities involving jets for which the migrations are larger.
As an example, Figure~\ref{fig:response_example} shows the response matrices for \ptgg\ and \Njets. The slightly decreasing values of the diagonal elements of the \ptgg\ response matrix up to $\ptgg=45~\GeV$ reflects the dependence of the photon identification efficiency on the single-photon \ET~\cite{PERF-2017-02}.
The factor \(C_r^\text{fid}\) is equal to the fraction of selected events in the signal simulation that originate from events within the fiducial volume. It is in general close to one, and corrects for events that pass the selection but are outside of the fiducial region. It also removes a small fraction (around 0.4\%) of reconstructed \(H\to f\bar{f}\gamma\) Dalitz decays, where \(f\) is any fermion except a top quark, that are present in the MC samples showered with \PYTHIA[8]. For example, the value for the diphoton inclusive fiducial cross-section \(C_r^\text{fid}\) is 98\%. This correction factor and the response matrix are estimated from the SM Higgs boson Monte Carlo simulations taking into account all the production modes and their SM cross-sections.
 
For each reco-bin, the \mgg\ distribution is modelled as a mixture of the signal model and the background model taking into account their yields. The yield of the background, as well as its shape parameters, are free in the fit and uncorrelated among the reco-bins. The measurement of \((\sigma_t \times B_{\gamma\gamma})\) is performed with a simultaneous maximum-likelihood fit of all the reco-bins of a given observable.
 
Systematic uncertainties, described in Section~\ref{sec:sysunc}, are incorporated into the likelihood as nuisance parameters, each one constrained by a probability density function. Gaussian constraints are used for the peak position of the signal model and for the spurious-signal uncertainty. Log-normal constraints are used for the uncertainties related to the yield and for the signal mass resolution.
\section{Systematic uncertainties}
\label{sec:sysunc}
The measurements presented in the paper are affected by several systematic uncertainties. Most of the uncertainties fall in one of the following categories and are described in the sections below: (i) uncertainties affecting the modelling of the signal and background shapes, and (ii) experimental and theoretical uncertainties affecting the response matrices. Two additional uncertainties affecting the measured cross-sections are those in the luminosity \(L\) of the analysed data set (which enters Eq.~(\ref{eq:unfold1}) directly) and the branching ratio of the Higgs boson Dalitz decays (which enters Eq.~(\ref{eq:unfold1}) through the correction factor \(C_r^\text{fid}\)).
 
The uncertainty in the combined 2015--2018 integrated luminosity is 1.7\%. It is derived from the calibration of the luminosity scale using \(x\)--\(y\) beam-separation scans, following a methodology similar to that detailed in Ref.~\cite{DAPR-2013-01}, and using the LUCID-2 detector for the baseline luminosity measurements~\cite{LUCID2}.
 
Some of the Higgs boson Dalitz decays pass the analysis selection and the yields have to be corrected to subtract this contribution as explained in the previous section.
Since the branching ratio of Higgs boson Dalitz decay is poorly known~\cite{HIGG-2018-43}, a 100\% uncertainty is assigned to the Dalitz contribution; this results in a small uncertainty corresponding to about 0.4\% of the measured cross-section in the fiducial inclusive region.
 
\subsection{Systematic uncertainties in the signal and background \mgg\ models}
\label{subsec:sysunc_shape}
As described in Section~\ref{sec:SandB}, the Higgs boson cross-sections are estimated from a fit to the diphoton invariant mass spectrum. Therefore, the cross-sections are affected by uncertainties in the signal \mgg model introduced by photon energy scale and resolution uncertainties, as well as by uncertainties in the background model related to choosing a particular analytic function to represent it.
 
The photon energy scale uncertainties shift the peak position, whereas the photon energy resolution uncertainties affect the signal width, broadening or narrowing it.
These systematic uncertainties are estimated from \(Z\to ee\) decays as detailed in Ref.~\cite{EGAM-2018-01}. Photon energy resolution uncertainties dominate the Higgs boson signal shape uncertainties, and typically change the width of the signal distribution by between 5\% and 27\%. These uncertainties increase with \pt\ of the individual photons and the impact on the final measurement reaches 16\% at \(\ptgg > \SI{650}{\GeV}\). The photon energy scale uncertainties, on the other hand, have a smaller impact, varying the peak position by between 0.2\% and 0.5\%.
An additional systematic uncertainty in the peak position is included to account for the uncertainty (\SI{0.24}{\GeV}) in the Higgs boson mass~\cite{HIGG-2014-14}.
 
As detailed in Section~\ref{sec:background}, choosing an analytic function to model the background diphoton invariant mass distribution in the fit can cause deviations of the true background distribution from the nominal model to be identified as a spurious signal. The spurious signal estimated from fits to the background \mgg\ templates is assigned as the systematic uncertainty in the signal yield originating from the choice of background model. In the likelihood it is implemented as an additional signal component in each reco-bin. The uncertainty is considered to be uncorrelated between different reco-bins.
 
\subsection{Experimental and theoretical uncertainties affecting the response matrices}
\paragraph{Experimental uncertainties} The response matrices \(R\) are affected by different experimental uncertainties in the calibration and identification of photons, jets, and leptons. They can change the total signal yield, migrations across reco-bins, and migrations into and out of the fiducial acceptance:
\begin{itemize}
 
\item \textbf{Diphoton trigger efficiency} The efficiency of the diphoton trigger is estimated using the bootstrap method~\cite{TRIG-2016-01} in data and simulation, with an uncertainty close to 1.0\%.
 
\item \textbf{Vertex selection efficiency} This uncertainty reflects the difference in the selection of the primary vertex using the neural network algorithm between data and the simulation. This uncertainty is estimated using \(Z\to e^+ e^-\) events after removing the electron tracks. The resulting uncertainty is found to be \({<}0.3\%\).
 
\item \textbf{Photon identification efficiency} This uncertainty is evaluated by varying the efficiency scale factors between data and the simulation, measured using three data-driven techniques as detailed in Ref.~\cite{EGAM-2018-01}, within their uncertainties. This uncertainty is estimated to be 1.8\% for the inclusive fiducial region, and decreases to 1\% with increasing \ptgg as the uncertainties of the single-photon scale factors decrease.
 
\item  \textbf{Photon isolation efficiency} Similarly to the photon identification efficiency, this uncertainty is evaluated by varying the track and calorimeter isolation scale factors within their uncertainties as detailed in Ref.~\cite{EGAM-2018-01}. This amounts to an uncertainty of 1.6\% for the inclusive fiducial region, and varies as function of \ptgg, increasing to 2\% for the very high \ptgg region.
 
\item \textbf{Photon energy scale and resolution} In addition to affecting the signal invariant mass distribution, as detailed in Section~\ref{subsec:sysunc_shape}, the photon energy scale and resolution uncertainties also have an effect on the response matrices which results from migrations across bin boundaries and across the boundaries of the fiducial region. The magnitude of this uncertainty is at the per-mille level, reaching 0.2\% for the highest \ptgg bins.
 
\item \textbf{Modelling of pile-up in the simulation} This uncertainty is derived by varying the average number of interactions per bunch crossing, \avgmu, in the simulation by an amount consistent with data. This yields an uncertainty of 1.5\% for the inclusive fiducial region, and increases with \pt and jet activity to 6\% for the highest jet multiplicity.
 
\item \textbf{Jet energy calibration and jet selection} These uncertainties affect only jet-related measurements. Jet energy calibration uncertainties reflect the remaining differences in the jet energy scale and resolution between data and the simulation as estimated through the \pt-balance technique in \(Z\)+jets, \(\gamma\)+jet, and dijet events, as detailed in Ref.~\cite{JETM-2018-05}. Jet selection uncertainties are related to the efficiency of both the jet-vertex tagger and the forward jet-vertex tagger used at higher \(|\eta|\). The typical size of the jet-related uncertainties ranges from 5\% for topologies with low jet multiplicities to 24\% for the highest jet multiplicities.
 
\item \textbf{Lepton uncertainties}
Lepton uncertainties account for the uncertainties in the reconstruction, identification and isolation efficiencies of electrons~\cite{EGAM-2018-01} and muons~\cite{MUON-2018-03}.
They are obtained from dilepton decays of \(Z\) bosons and \(J/\psi\) mesons collected in \RunTwo, using a tag-and-probe technique. The typical size of these uncertainties is
about 0.6\% for electrons and about 0.5\% for muons, and their effect is significant only for the cross-section measurement in the lepton-enhanced fiducial region.
 
\item \textbf{\MET uncertainties}
Uncertainties related to the reconstruction and calibration of missing transverse momentum are estimated by propagating uncertainties associated with the energy scales and resolutions of photons, jets and leptons in addition to uncertainties from unassociated charged-particle tracks. This results in a cross-section measurement uncertainty of 13\% in the High \MET fiducial region.
 
\item \textbf{\btag uncertainties} These uncertainties are associated with the efficiency of the \btag algorithm and affect only measurements where \btag is required. They correspond to a maximum uncertainty of 4\% for the highest \Nbjets\ bin. They are determined for jets containing the decay of a \bquark, using \(t\bar{t}\) events in \SI{13}{\TeV} data and the method outlined in Ref.~\cite{PERF-2016-05}.
\end{itemize}
 
\paragraph{Theoretical uncertainties} In addition to the previous experimental uncertainties, the following theoretical uncertainties affect the response matrices:
\begin{itemize}
\item \textbf{Signal composition uncertainty} The response matrices used in the fit to account for detector effects in each distribution are built considering all the production modes of the Higgs boson. The simulated samples are then combined assuming SM cross-sections, and hence model dependence can be introduced if the matrices vary significantly between production modes. Therefore, a modelling uncertainty is estimated by varying the cross-section of each production mode within its measured uncertainty~\cite{HIGG-2018-57}. The resulting uncertainties are quite small, reaching at most 1\% for the highest jet multiplicities.
 
\item \textbf{Modelling of the matrix element generator} This uncertainty results from the bias estimated by using an alternative matrix element generator (\MGNLO) and using it to unfold the predictions from the nominal matrix element generator (\POWHEGBOX). Both matrix element generators are interfaced with the \PYTHIA[8] parton shower. The response matrices are built using all production modes. The relative difference between the two response matrices is then included in the fit likelihood as a nuisance parameter, resulting in a small uncertainty (1\%--2\%) in the cross-sections for the high-\pt regions and the highest jet multiplicities. In addition, a robustness check was performed by reweighting the default simulation using the EFT model that gives the largest variation (see Section~\ref{sec:eft} for more details), and using the nominal response matrix. This check resulted in negligible non-closure at the level of at most $0.5\%$.
 
\item\textbf{Modelling of the parton shower, underlying event, and hadronisation} This uncertainty arises from the relative change in the response matrices when switching the parton showering algorithm from \PYTHIA[8] to \HERWIG[7]. These uncertainties are typically small, reaching 1.6\% for the highest jet multiplicities.
 
\end{itemize}
 
Uncertainties in the response matrix due to the scale or PDF variations are negligible and not considered further.
For the total cross-section in the full phase space, an additional uncertainty of 2.9\% in the \Hgg\ branching ratio is included in the measurement.
\section{Cross-section results and comparison with theoretical predictions}
\label{sec:xsec_results}
This section presents the measured fiducial inclusive cross-sections and a subset of the differential cross-sections described in Section~\ref{sec:intro_xs}, following the fit procedure detailed in Section~\ref{sec:measurement}. The measurements are compared with one or more theoretical predictions, which are described in Section~\ref{subsec:theoryunc}.
 
\subsection{Theoretical predictions}
\label{subsec:theoryunc}
The measured cross-sections are compared with several theoretical predictions described below, the nominal one being that from the fully simulated MC samples scaled to their latest cross-section calculations, called `default simulation' in the following.
The difference between the nominal prediction and other predictions, except the \textsc{proVBF} predictions used for comparison with the VBF observables, is only in the calculation of the ggF component.
The alternative predictions were either provided by their corresponding authors or calculated by using their tools and the set-up recommended by the authors themselves. For the inclusive parton-level predictions, acceptance corrections derived from the default simulation are applied.
 
\paragraph{Default simulation}
 
The default simulated signal samples described in Section~\ref{sec:samples} are used for the different processes.
The uncertainties in the predictions are computed as follows:
 
\begin{itemize}
\item Uncertainties from the choice of the PDF set and \alphas\ are evaluated using the \PDFforLHC[15] error PDF set, which takes into account 30 variations of NNLO (ggF) or NLO (other modes) PDFs and two variations of \alphas, following the \PDFforLHC\ recommendations~\cite{Butterworth:2015oua}. The PDF uncertainties are treated as fully correlated across production modes, given that the eigen-variations are completely independent of the physics process.
 
\item Perturbative uncertainties for ggF, VBF, \(VH\), \ttH are estimated bin-by-bin using the simplified template cross-sections (STXS) stage 1.2 uncertainty scheme ~\cite{deFlorian:2016spz}. The scheme includes various normalisation and migration uncertainties in fine binning corresponding to the STXS 1.2 granularity. For ggF, the scheme defines 18 sources of uncertainty, which are added in quadrature: two accounting for yield uncertainties related to the total cross-section, two for migration uncertainties related to splitting the phase space by jet multiplicity, one accounting for the treatment of \(m_t\), and the remaining sources account for migrations across the different STXS bin boundaries defined as a function of various observables including the Higgs boson transverse momentum, the Higgs-plus-jet transverse momentum and the dijet invariant mass. The same scheme is also defined for $gg\to ZH$. For the other production modes, a similar set of nuisance parameters accounting for migrations across the different STXS 1.2 bins is used. For \bbH\ and \(tH\) production modes, the perturbative uncertainties are estimated as an envelope of the scale variations available in \POWHEGBOX.
 
\item The Higgs to diphoton branching ratio uncertainty from Ref.~\cite{deFlorian:2016spz} is also included.
\end{itemize}
 
In addition to the default simulation, several theory predictions for the various measured cross-sections are compared with data. A summary of the uncertainties in the new predictions is reported in Appendix~\ref{sec:aux theory preds unc}. An overview of the different predictions is given below:
 
\paragraph{MATRIX+RadISH} The \textsc{MATRIX+RadISH} interface~\cite{Kallweit:2020gva} combines fully differential cross-sections at NNLO accuracy in QCD through \textsc{MATRIX}~\cite{Grazzini:2017mhc,Grazzini:2015wpa} with all-order resummation through \textsc{RadISH}~\cite{Monni:2016ktx,Bizon:2017rah} for various \(2\to 1\) and \(2\to 2\) colour-singlet production processes. \textsc{MATRIX+RadISH} is used to perform the double-differential resummation of the transverse momentum of the colour singlet and of the leading jet at next-to-next-to-leading-logarithm (NNLL) accuracy~\cite{Monni:2019yyr}.

\paragraph{RadISH+NNLOjet} A prediction for the transverse momentum distribution of the Higgs boson decay products with fiducial cuts has been calculated within the \textsc{RadISH} framework~\cite{Monni:2016ktx,Bizon:2017rah}, matched to the Higgs~\(+\geq 1\)-jet NNLO-accurate QCD calculation at large Higgs boson \pt from \textsc{NNLOjet}~\cite{Chen:2016zka,Chen:2014gva}. The N\(^3\)LL\(^\prime\) calculation includes a resummation correction of linear fiducial power terms at the same accuracy with respect to N\(^3\)LL accuracy~\cite{Bizon:2018foh,Re:2021con}.
 
\paragraph{\(\SHERPA+\)MCFM+\OPENLOOPS}
\SHERPA~\cite{Bothmann:2019yzt} predictions were produced using version 2.2.11. The predictions are accurate to NLO in QCD for Higgs~\(+ \ge 0, \ge 1, \ge 2, \ge3\)~jets, with the fourth jet being accurate to leading order (LO) in QCD and subsequent jets produced by the parton shower (with leading-logarithmic accuracy). The Higgs~\(+\ge2\)-jets matrix elements were produced using MCFM~\cite{Campbell:2010ff}, and Higgs~\(+\ge3\)-jets matrix elements were calculated using the \OPENLOOPS~\cite{Buccioni:2019sur,Cascioli:2011va,Denner:2016kdg} libraries. They are matched with the \SHERPA\ parton shower~\cite{Schumann:2007mg} using the \MEPSatNLO\ prescription~\cite{Hoeche:2011fd,Hoeche:2012yf,Catani:2001cc,Hoeche:2009rj}. \SHERPA[2.2.11] uses an improved parton clustering procedure resulting in a reduction in the predicted cross-sections and a reduction in the uncertainties in phase-space regions with multiple jets.

\paragraph{\textsc{ResBos2}}
\textsc{ResBos2}~\cite{Balazs:1997xd,Wang:2012xs} provides predictions for inclusive Higgs production via gluon--gluon fusion calculated at N\(^3\)LL+NNLO accuracy~\cite{Wang:2012xs,Glosser:2002gm}.
For inclusive Higgs-plus-jet production, the \textsc{ResBos2} program uses the transverse-momentum-dependent (TMD) resummation formalism as proposed in the Collins 2011 scheme~\cite{Collins:2011zzd}. The prediction is made to NLL+NLO accuracy and matched to the NLO calculation~\cite{Sun:2016kkh}.
The jet-veto results for the \textsc{ResBos2} code are taken as the difference between the inclusive Higgs and Higgs-plus-jet results.
 
\paragraph{\scetlib::qT}
Predictions were obtained using the \scetlib::qT module~\cite{scetlib, Billis:2021ecs, Ebert:2020dfc, Billis:2019vxg, Ebert:2017uel} including the resummation of logarithms at small \(\pT < \mH\) for the usual leading-power in \(\pT/\mH\) contributions as well as all fiducial power corrections (induced by the fiducial cuts)~\cite{Ebert:2020dfc}.
\scetlib resummation achieves N\(^3\)LL\('\) accuracy, including the complete three-loop corrections in the small-\(\pT\) limit~\cite{Li:2016ctv, Billis:2019vxg, Ebert:2020yqt}.
For other photonic observables, the required corrections for matching to N\(^3\)LO are not available, so they are computed at N\(^3\)LL\(+\)NNLO accuracy in QCD.
Only the dominant top-quark loop contributions are included in the predictions for the differential fiducial cross-sections.
 
\paragraph{\scetlib::pTj1}
The predictions for \ptj[1] are obtained using the \scetlib::pTjet module~\cite{scetlib, Stewart:2013faa, Michel:2018hui}.
The predictions are for \(gg\to H\) in the narrow-width limit with
\(\mH = \SI{125.09}{GeV}\). The \ptj[1] spectrum is computed to NNLL\('+\)NNLO accuracy for the
dominant top-quark Yukawa coupling \(y_t^2\) contribution in the rescaled EFT limit~\cite{Stewart:2013faa}.
 
\paragraph{SCET+MG5 (NNLL\('\)+NNLO)}
The predictions are obtained for the 0-jet ggF cross-section using the rapidity-dependent jet veto observable \maxtau\ at NNLL\('\)+NNLO accuracy~\cite{tauC}. The predictions are obtained with a resummation of \(\ptj/\mH\) with NNLL\('\) accuracy.
 
\paragraph{STWZ, BLPTW}
The perturbative STWZ, BLPTW predictions~\cite{Stewart:2013faa, Boughezal:2013oha} include NNLL\('\)+NNLO resummation in QCD for the $\pT$ of the leading jet , combined with a NLL\('\)+NLO resummation in QCD for the subleading jet. The numerical predictions for $\sqrt{s}=13$~TeV are taken from Ref.~\cite{deFlorian:2016spz}.
This prediction is shown for the inclusive zero-, one- and two-jet cross-sections as well as for the exclusive zero- and one-jet cross-sections.
 
\paragraph{proVBF NNLO}
The \textsc{proVBF}~\cite{proVBFNNLO} program calculates the fully differential NNLO corrections to vector-boson fusion Higgs boson production. This is achieved with a `projection-to-Born' method that combines an inclusive NNLO calculation in the structure-function approach with a suitably factorised NLO VBF Higgs plus 3-jet calculation.
 
\paragraph{GoSam}
\textsc{GoSam}~\cite{Cullen:2011ac,Cullen_2014} provides the fixed-order loop contributions accurate at NLO in QCD in the inclusive $\text{Higgs} + 0, 1, 2, 3$ jets regions. The real-emission contributions at fixed order in QCD are provided by \SHERPA~\cite{Bothmann:2019yzt}.

\subsection{Inclusive fiducial cross-section measurements}
\begin{figure}[htbp]
\centering
\includegraphics[width=0.6\textwidth]{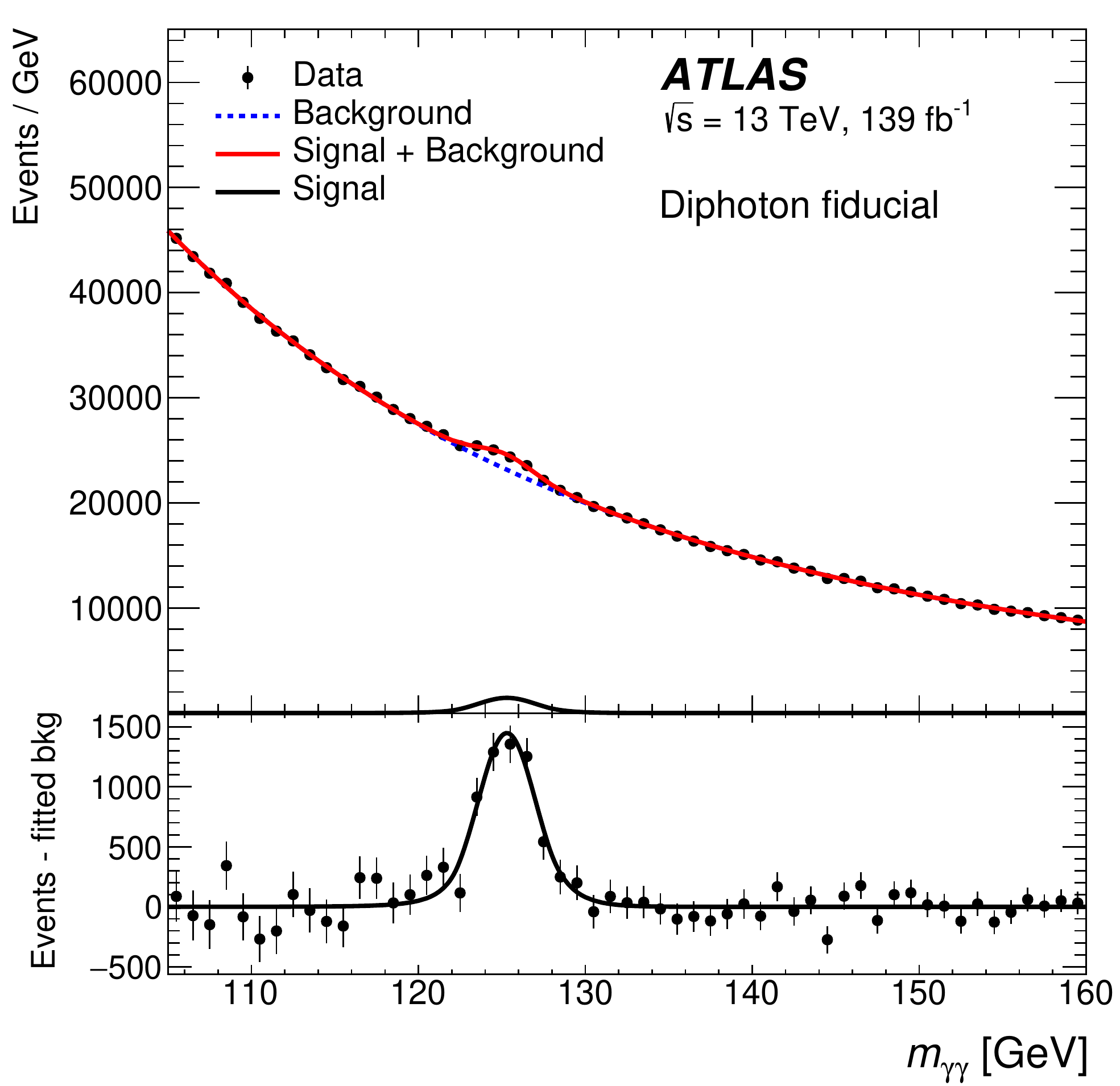}
\caption{The diphoton invariant mass spectrum of events passing the selection. The solid red curve shows the fitted signal-plus-background model with the Higgs boson mass constrained to \SI[multi-part-units=single]{125.09 \pm 0.24}{\GeV}.
The bottom pad shows the residuals between the data and the background component of the fitted model.}
\label{fig:myy_incl}
\end{figure}

\begin{figure}[htbp]
\centering
\subfloat[]{\includegraphics[width=0.5\textwidth, page=2]{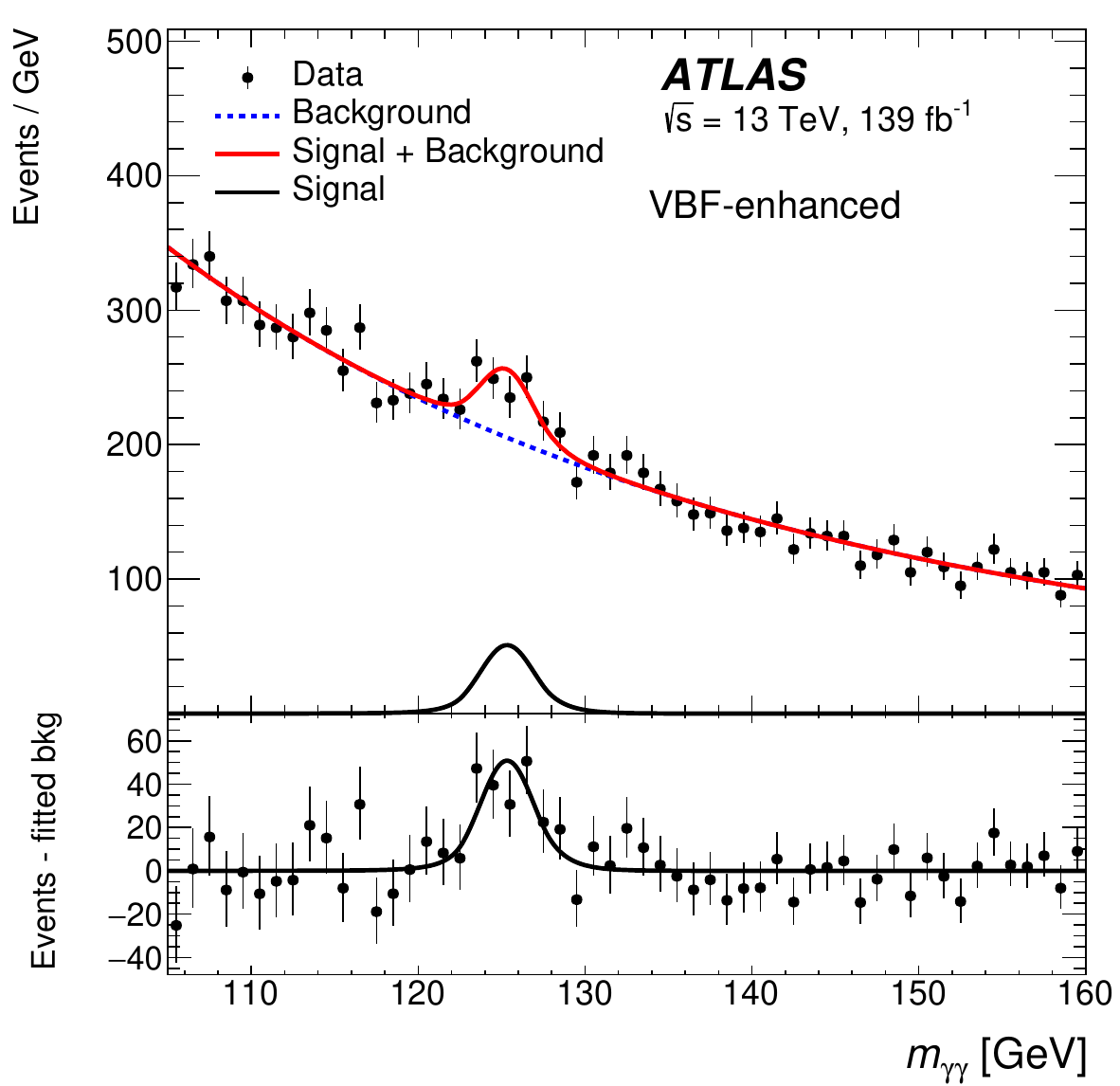}\label{fig:myy_fid_a}}
\subfloat[]{\includegraphics[width=0.5\textwidth, page=2]{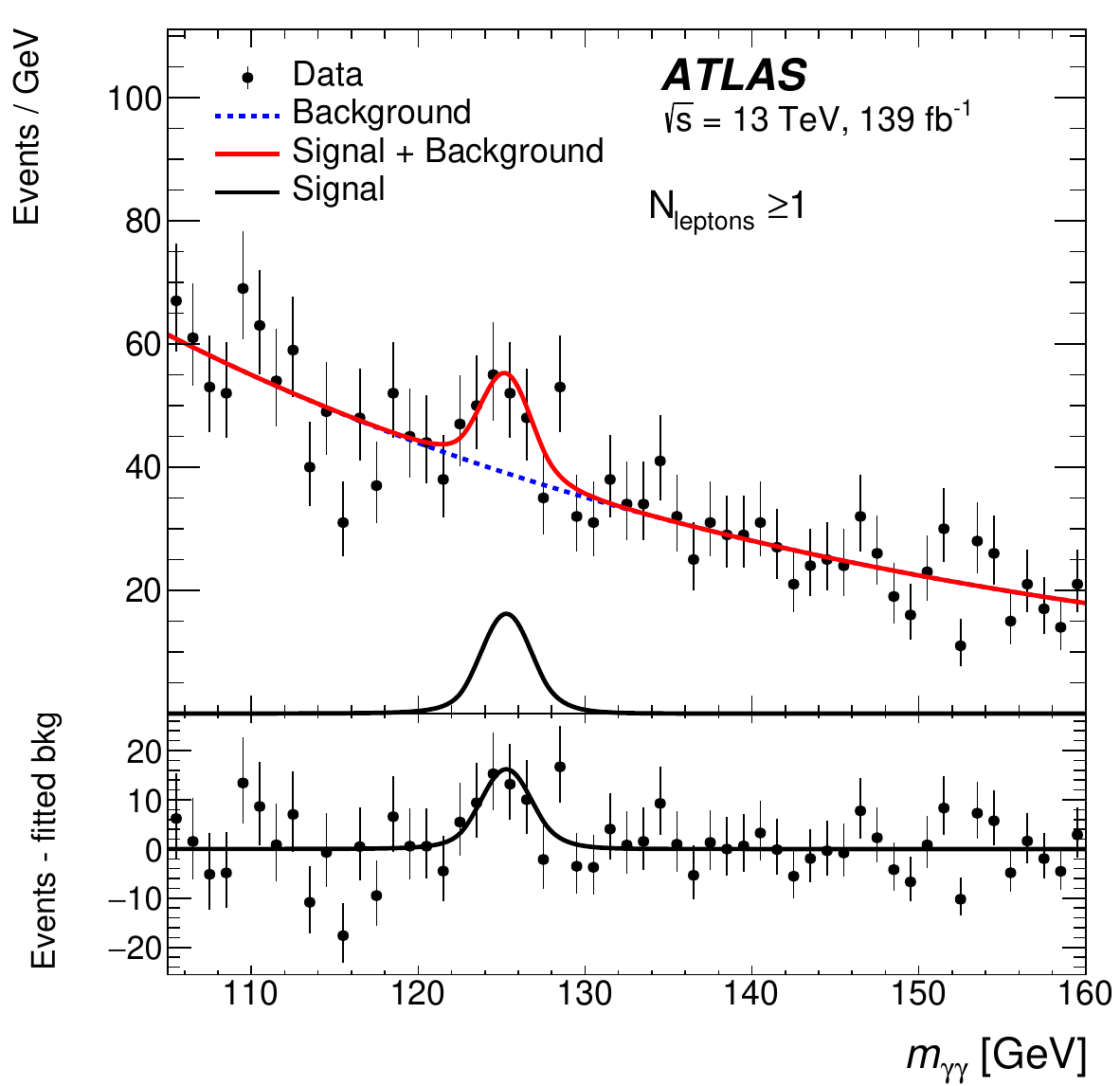}\label{fig:myy_fid_b}} \\
\subfloat[]{\includegraphics[width=0.5\textwidth, page=2]{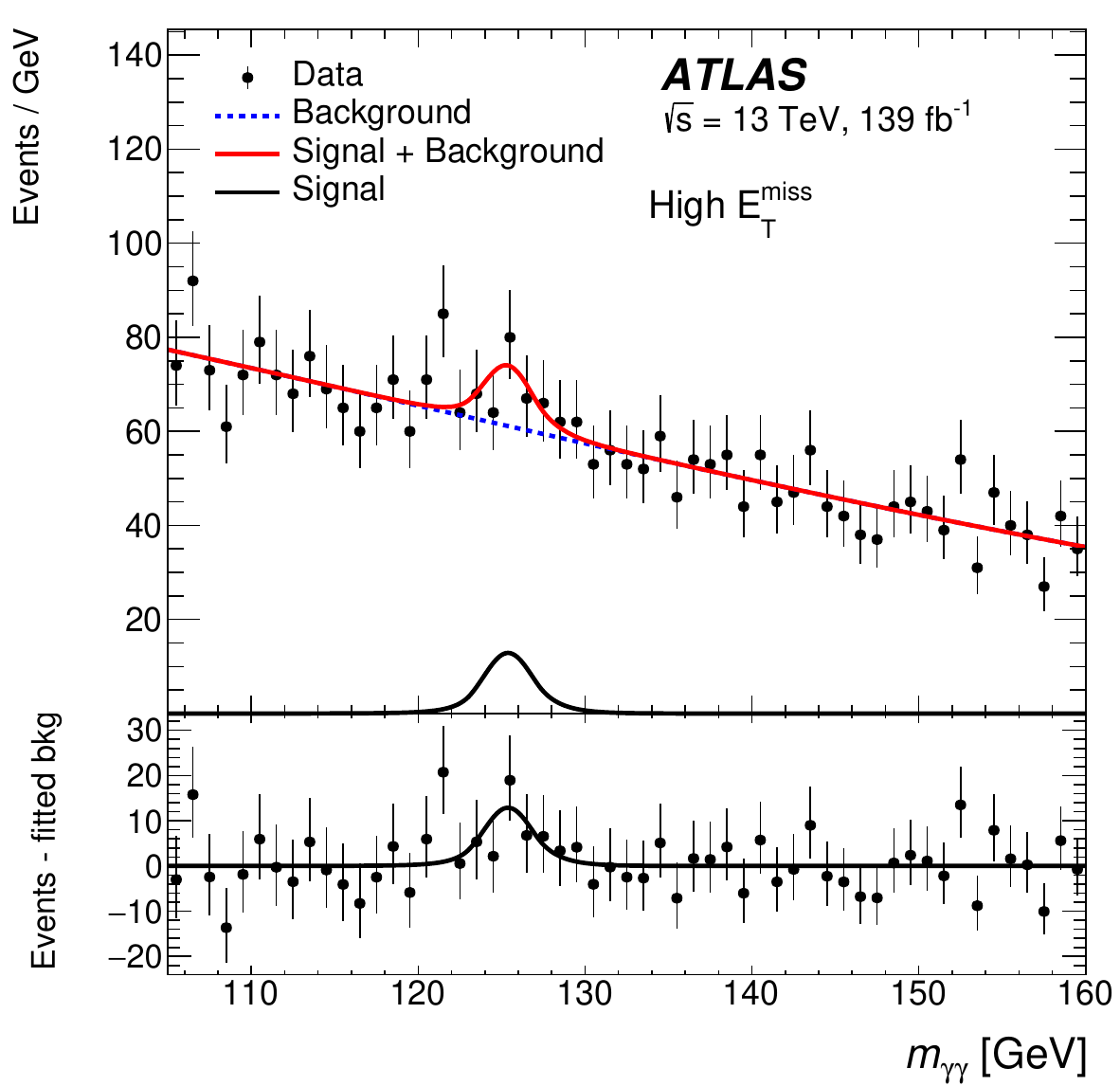}\label{fig:myy_fid_c}}
\subfloat[]{\includegraphics[width=0.5\textwidth, page=2]{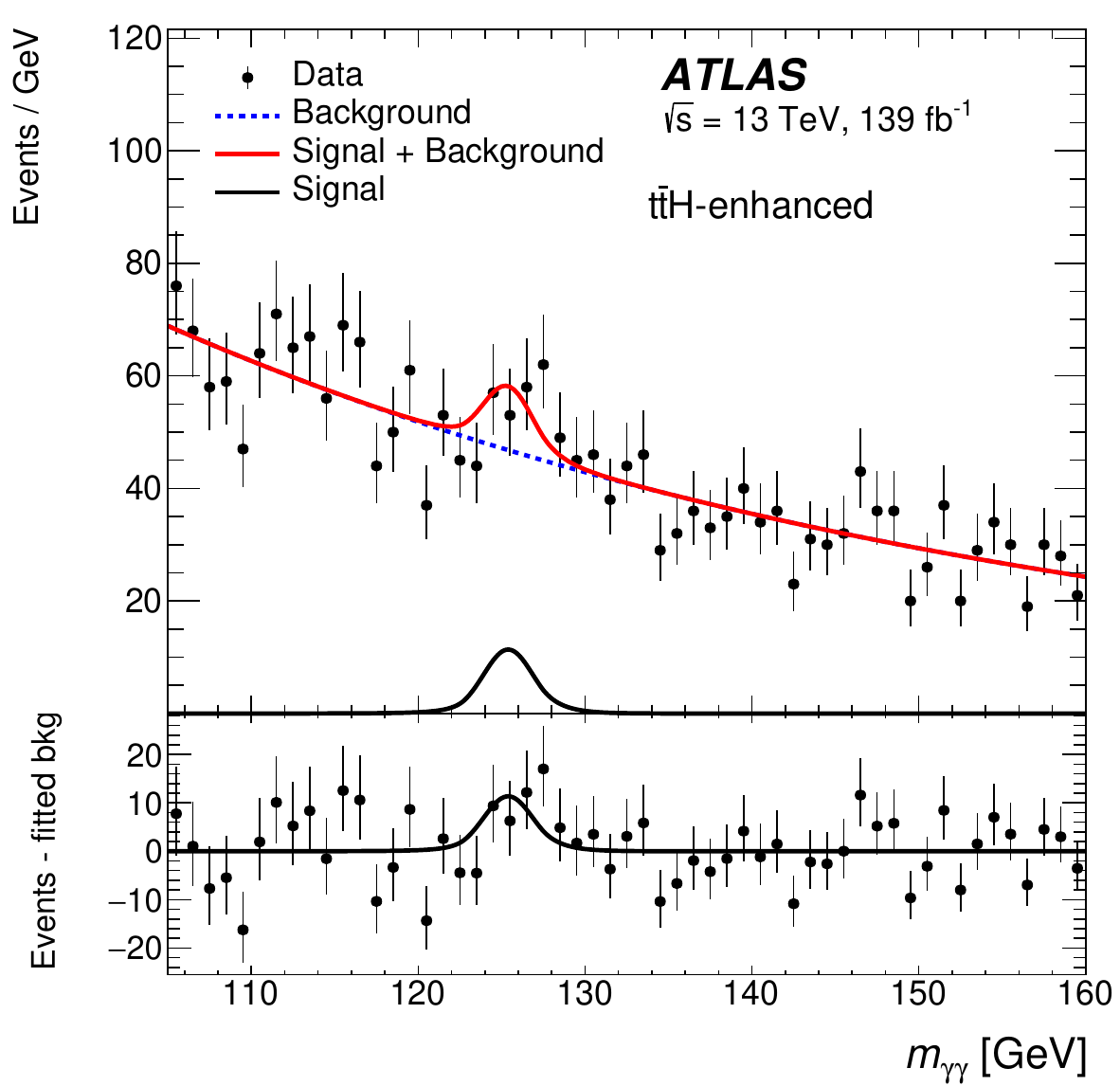}\label{fig:myy_fid_d}}
\caption{The diphoton invariant mass spectrum of events passing the selection for the region \protect\subref{fig:myy_fid_a} VBF-enhanced, \protect\subref{fig:myy_fid_b} \(N_\text{lepton}\geq 1\), \protect\subref{fig:myy_fid_c} High \MET and \protect\subref{fig:myy_fid_d} \ttH-enhanced. The solid red curve shows the fitted signal-plus-background model with the Higgs boson mass constrained to \SI[multi-part-units=single]{125.09 \pm 0.24}{\GeV}. The bottom pad shows the residuals between the data and the background component of the fitted model.}
\label{fig:myy_fid}
\end{figure}

The observed \mgg distribution in data for the photons passing the selection is shown in Figure~\ref{fig:myy_incl}. The figure also shows the signal-plus-background (S+B) fit of the data with the model described in Section~\ref{sec:SandB} for the measurement of the cross-section in the diphoton fiducial region.
Similarly, Figure~\ref{fig:myy_fid} shows the \mgg distribution and the corresponding S+B fit for the events passing the requirements corresponding to the other fiducial regions.
 
\begin{figure}[htbp]
\centering
\includegraphics[width=0.8\textwidth]{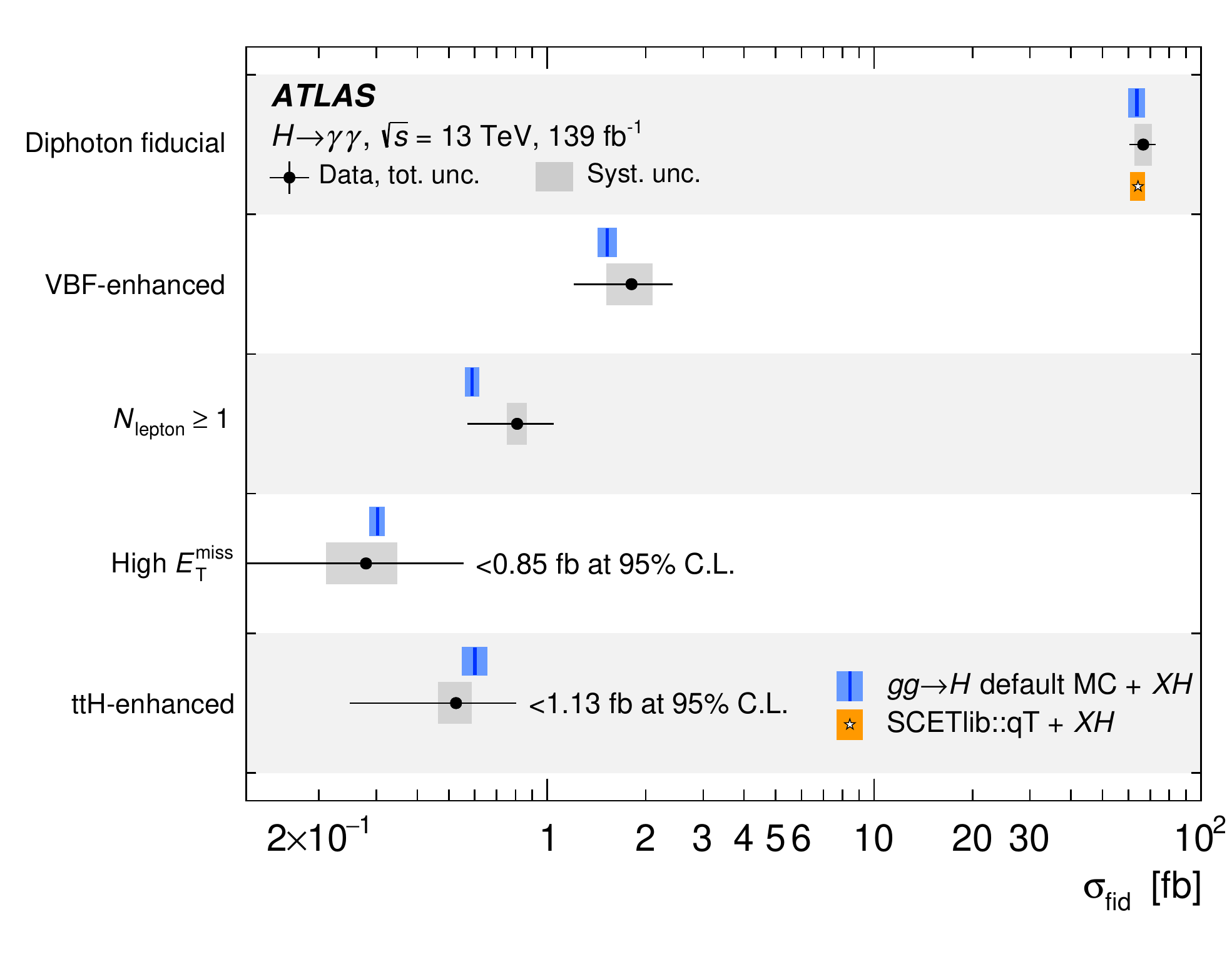}
\caption{Particle-level cross-sections times branching ratio in the five fiducial regions. The data are shown as filled (black) circles. The error bar on each measured cross-section represents the total uncertainty in the measurement, with the systematic uncertainty shown as a dark grey rectangle. The default prediction with its uncertainty is superimposed. \(XH\) indicates all the Higgs production modes except for ggF. Upper limits at 95\% CL are shown for fiducial regions with observed significance below $3\sigma$.}
\label{fig:xsec_fid}
\end{figure}

The measured particle-level cross-sections of the \(\pp\to H\to \gamma\gamma\) process in the fiducial regions defined in Section~\ref{sec:intro_xs} are summarised in Table~\ref{tab:xsec_fid} and in Figure~\ref{fig:xsec_fid}. Upper limits at 95\% confidence level (CL) were set on fiducial cross-sections with observed significance below $3\sigma$, using the CL\(_{\text{s}}\) procedure~\cite{Read:2002hq}.
The cross-section in the inclusive fiducial region is also compared with \scetlib::qT predictions, where the efficiency of the particle-level photon isolation requirement (98\%) is corrected for using the default simulation. \scetlib::qT directly accounts for the photon kinematic acceptance and relevant resummation corrections, resulting in the most accurate prediction of the cross-section of \SI[multi-part-units=single]{64.2 \pm 3.4}{\femto\barn}~\cite{Billis:2021ecs}, compared to a measured fiducial cross-section of \(\sigma_\textrm{fid} = 67\pm5~\text{(stat.)}\pm4~\text{(sys.)}~\si{\femto\barn}=\SI[multi-part-units=single]{67 \pm 6}{\femto\barn}\).
 
The total cross-section times branching ratio in the full phase space is also reported. It was computed in the same way as the fiducial cross-sections, from the fit to the inclusive sample of all the selected candidates. In the fit, the full acceptances from the full phase space to the selected one are used to replace those from the fiducial ones. Taking into account the Higgs diphoton branching ratio, the measured total Higgs boson production cross-section is \(\sigma_\text{tot}=58 \pm 4~\text{(stat.)} \pm 4~\text{(sys.)}~\si{\pico\barn} = \SI[multi-part-units=single]{58\pm 6}{\pico\barn}\) compared to the SM prediction of \SI[multi-part-units=single]{55.6 \pm 2.7}{\pico\barn}.
 
The uncertainty in each measured cross-section is dominated by the statistical component. Usually the spurious signal is the largest source of systematic uncertainty, typically followed by the photon energy resolution uncertainty. For fiducial regions with jet requirements, such as the VBF-enhanced one, jet energy scale and resolution uncertainties are the dominant source of systematic uncertainty, accounting for approximately 40\% of the total relative uncertainty, whereas the uncertainty in the \met\ reconstruction is important in the measurement of the cross-section of the High \met fiducial region, accounting for approximately 13\% of the total relative uncertainty.
Table~\ref{tab:syst_summary_diff} summarises the systematic uncertainties and shows their relative impact on the inclusive fiducial cross-section.
 
\begin{table}[htbp]
\centering
\caption{Particle-level cross-sections times branching ratio in the five fiducial regions, together with the total cross-section times branching ratio. The measured values with their statistical and systematic uncertainties are compared with the expected uncertainties and the default SM predictions. Upper-limits at 95\% CL are shown for fiducial regions with observed significance below $3\sigma$.
The last column shows the probabilities from a \(\chi^2\) compatibility test between the fitted cross-sections and the SM prediction. The \(\chi^2\) is computed using the full set of uncertainties in the data and in the prediction.}
\resizebox{\textwidth}{!}{
 
\begin{tabular}{lS@{\(\;\pm\;\)}S@{\(\;\pm\;\)}SS@{\(\;\pm\;\)}SSr}
\toprule
Fiducial region           & \multicolumn{3}{c}{Measured [\si{\femto\barn}]} & \multicolumn{2}{c}{SM prediction [\si{\femto\barn}]} & {95\% $\text{CL}_\text{s}$ upper limit [\si{\femto\barn}]} & \(p\)-value                                                    \\
\cmidrule(lr){2-4}
&                                                 & stat                                                 & sys     & \multicolumn{2}{c}{} &    &                       \\
\midrule
Diphoton                  & 67                                              & 5                                                    & 4       & 64                   & 4     &{-}& \SI{69}{\percent} \\
VBF-enhanced              & 1.8                                             & 0.5                                                  & 0.3     & 1.53                 & 0.10  &{-}& \SI{64}{\percent} \\
\(N_\text{lepton}\geq 1\) & 0.81                                            & 0.23                                                 & 0.06    & 0.59                 & 0.03  &{-}& \SI{36}{\percent} \\
High \met                 & 0.28                                            & 0.27                                                 & 0.07    & 0.302                & 0.017 & 0.85  & \SI{93}{\percent} \\
\ttH-enhanced             & 0.53                                            & 0.27                                                 & 0.06    & 0.60                 & 0.05  &1.13 & \SI{79}{\percent} \\
\midrule
Total                     & 132                                             & 10                                                   & 8       & 126                  & 7     &{-}& \SI{69}{\percent} \\
\bottomrule
\end{tabular}
}
\label{tab:xsec_fid}
\end{table}
 
For each measured cross-section, a \(\chi^2\) test is used to evaluate the \(p\)-value for compatibility between the measurement and the default SM prediction. The \(\chi^2\) is computed using the full set of uncertainties in the fitted cross-section and the theory uncertainties in the default SM prediction. The \(p\)-values computed in this way are listed in Table~\ref{tab:xsec_fid}, showing good agreement between the measurements and predictions.
 
\begin{table}[t]
\centering
\caption{Breakdown of the uncertainties in the inclusive diphoton fiducial cross-section measurement.}
\begin{tabular}{lc}
\toprule
Source & {Uncertainty [\%]} \\
\midrule
Statistical uncertainty                   & 7.2  \\
Systematic uncertainties                  & 6.0\\
~~~Background modelling (spurious signal) & 3.6 \\
~~~Photon energy scale \& resolution      & 3.3 \\ 
~~~Photon selection efficiency            & 2.5 \\
~~~Luminosity                             & 1.7 \\
~~~Pile-up modelling                      & 1.3 \\
~~~Trigger efficiency                     & 1.0 \\
~~~Theoretical modelling                  & 0.4 \\ 
\midrule
Total                                     & 9.4 \\
\bottomrule
\end{tabular}
\label{tab:syst_summary_diff}
\end{table}
 
\FloatBarrier
\subsection{Differential fiducial cross-section measurements}
\begin{figure}[htb!]
\begin{center}
\subfloat[]{\includegraphics[width=0.5\textwidth]{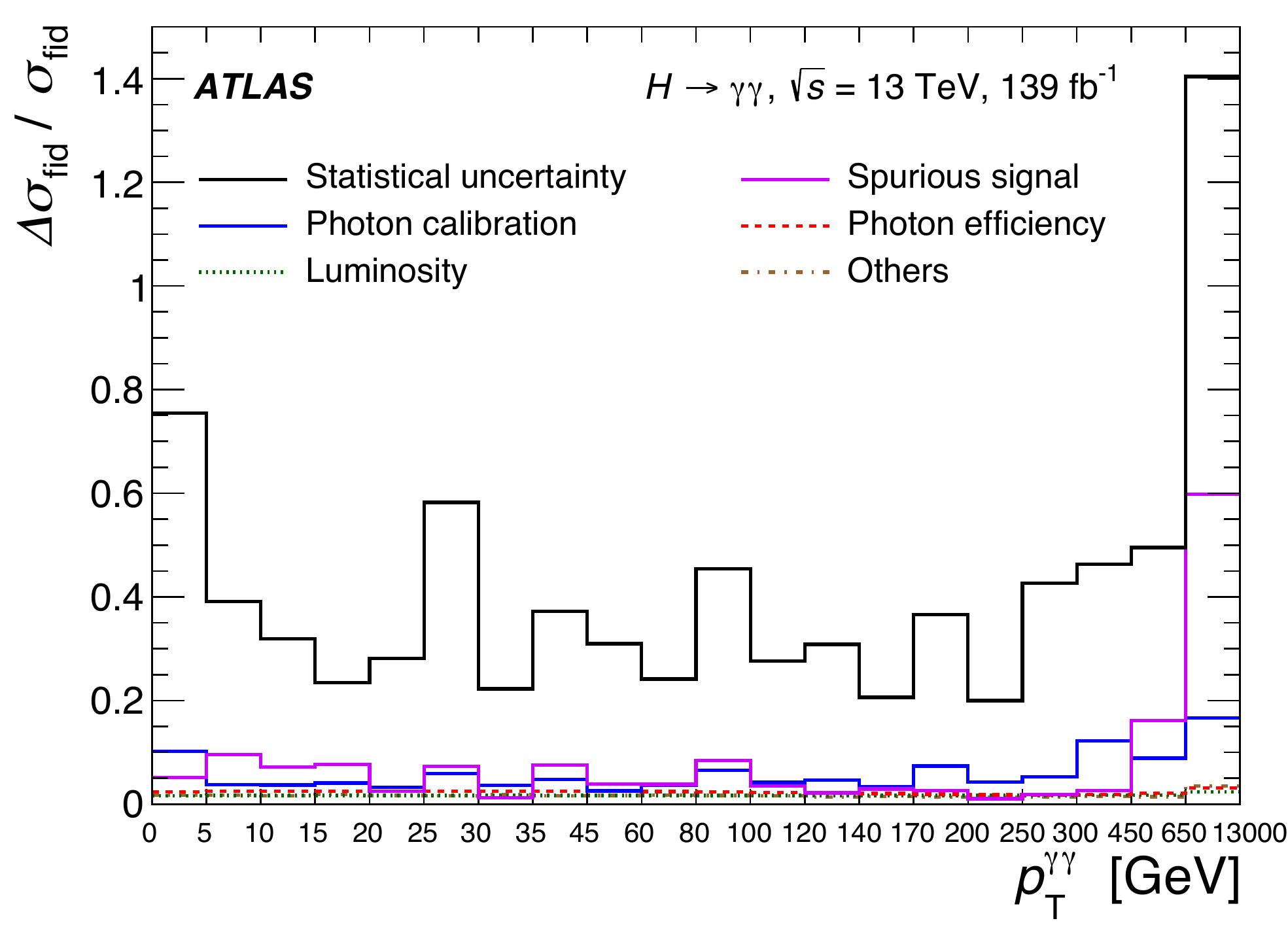}\label{fig:unc_breakdown_1}}
\subfloat[]{\includegraphics[width=0.5\textwidth]{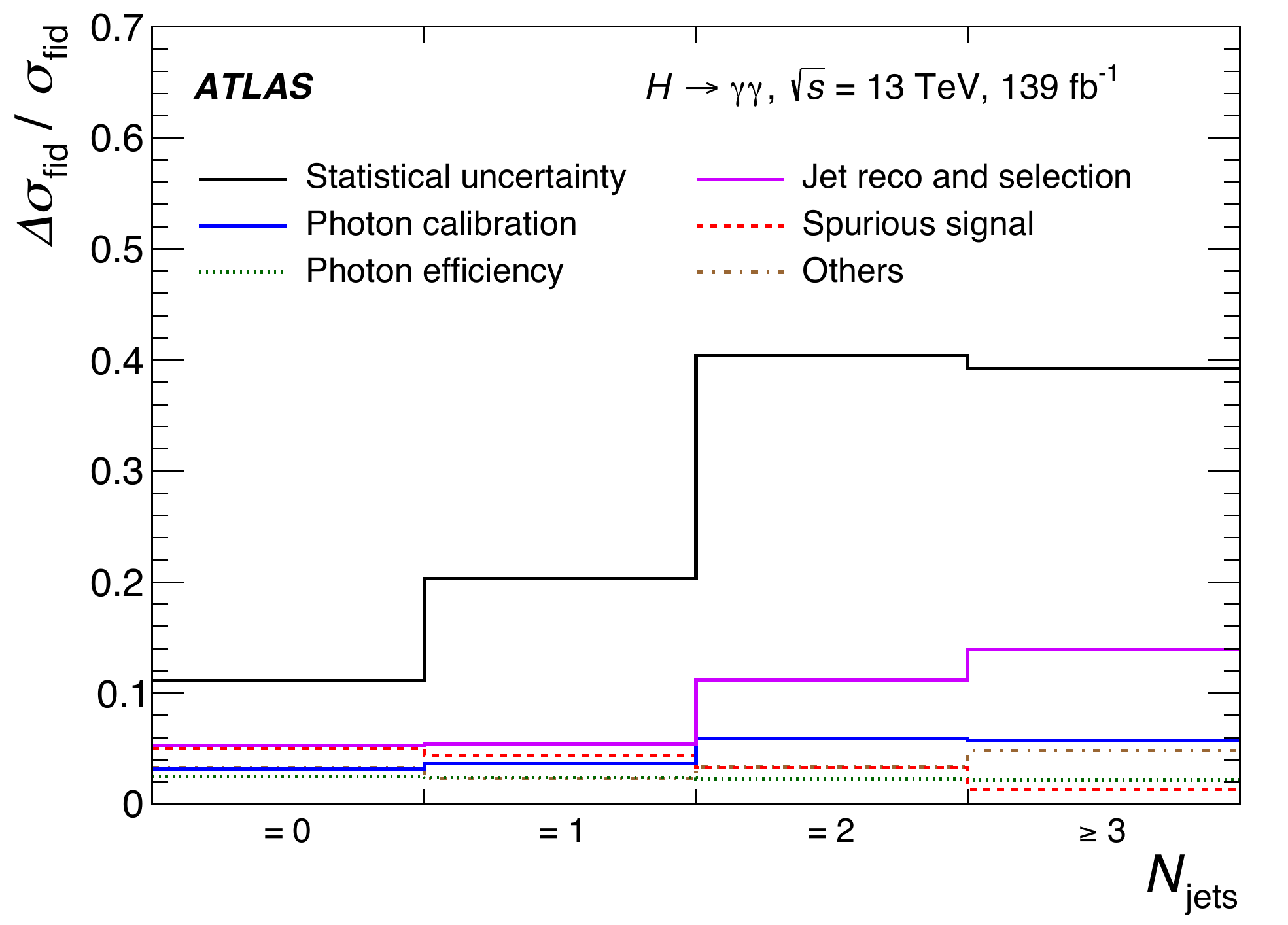}\label{fig:unc_breakdown_2}}\\
\end{center}
\caption{Summary of the uncertainties in the differential cross-section measurement for \protect\subref{fig:unc_breakdown_1} \ptgg, \protect\subref{fig:unc_breakdown_2} \Njet. The five leading uncertainties are shown separately, while all other uncertainties (labelled as `Others' in the figures) are summed in quadrature and shown as a single contribution.}
\label{fig:unc_breakdown}
\end{figure}

A subset of the measured differential fiducial cross-sections for the observables under study are reported. The corresponding correlation matrices for these observables are reported in Appendix~\ref{sec:aux diff xs corr}. Additional differential cross-section measurements are presented in Appendix~\ref{sec:aux diff xs}. All the results are compared with the default SM prediction.
All differential measurements are limited by the statistical uncertainties. As an example, Figure~\ref{fig:unc_breakdown} shows the breakdown of the uncertainties for \ptgg\ and \Njet. The five leading sources of systematic uncertainty are shown in each plot along with the statistical uncertainty. The plots show that the leading systematic uncertainties are from the spurious signal and photon calibration for photon observables, while they are from jet energy calibration and selection uncertainties for jet observables. Important systematic uncertainties also originate from the photon selection efficiency uncertainties.
 
\paragraph{Diphoton kinematics differential cross-sections}
 
\begin{figure}[htb]
\centering
\subfloat[]{ \includegraphics[width=0.5\textwidth]{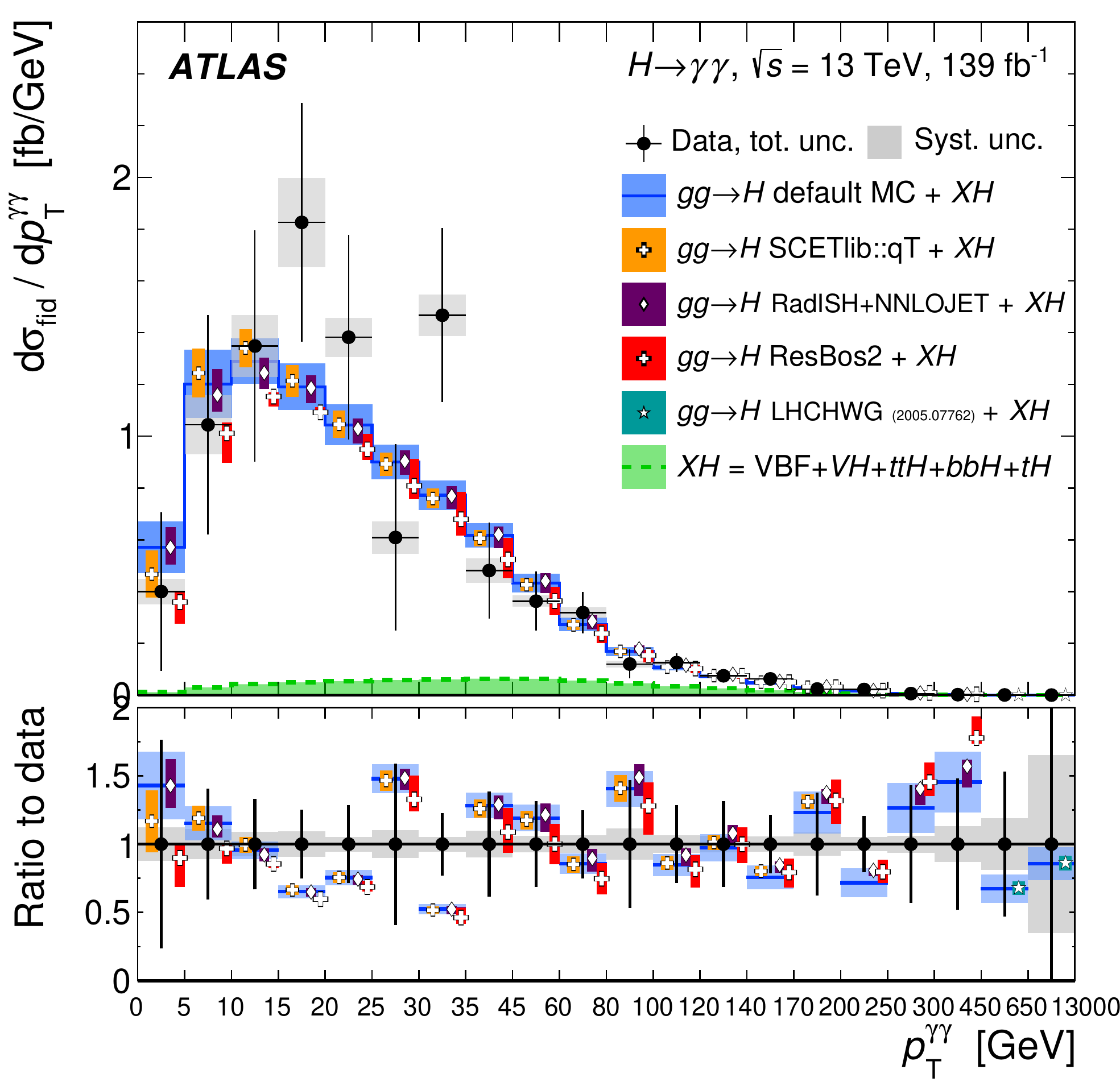}\label{fig:data_unfolded_xsections_matrixinversion_1D_photon_0_a}}
\subfloat[]{\includegraphics[width=0.5\textwidth]{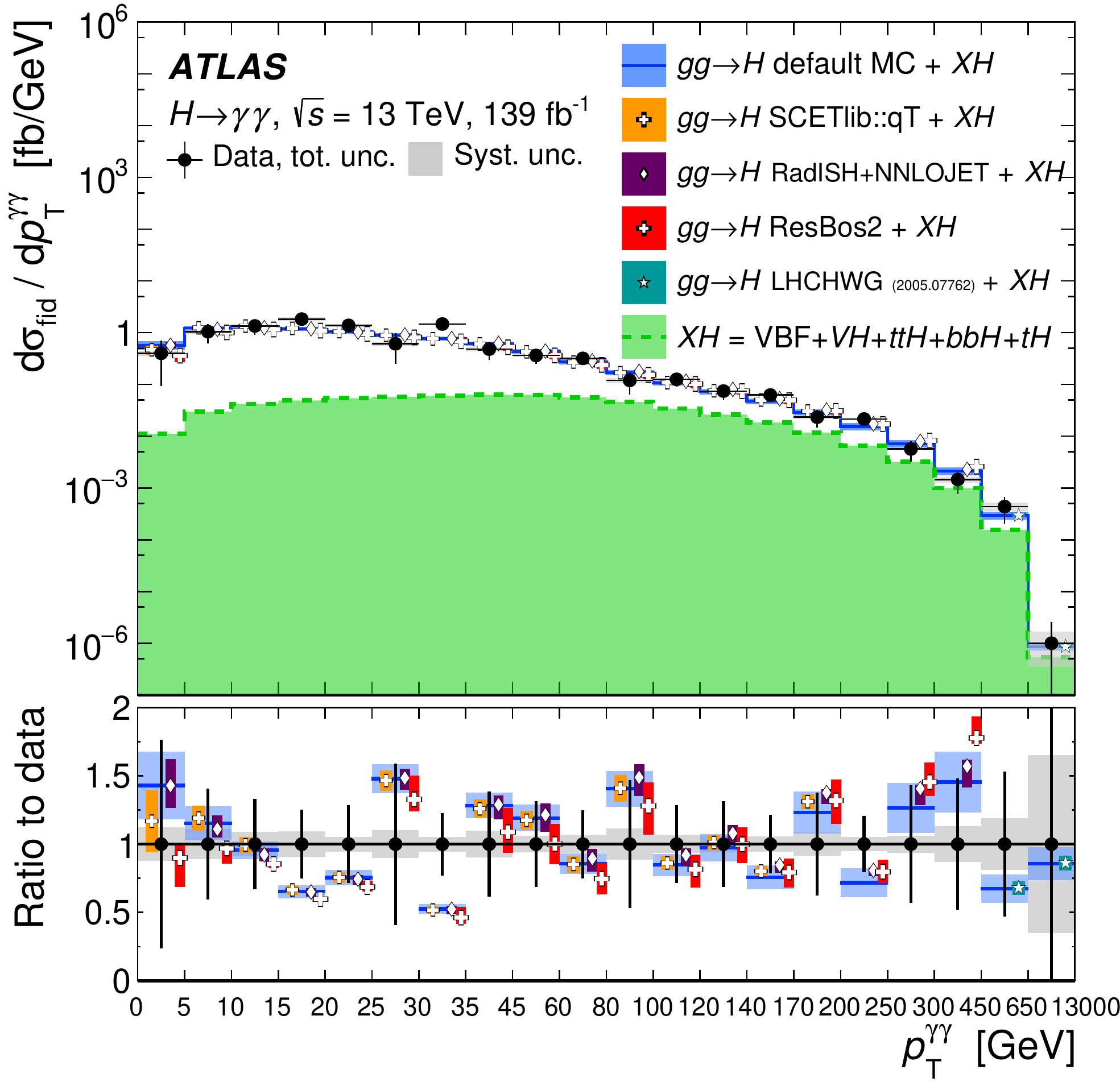}\label{fig:data_unfolded_xsections_matrixinversion_1D_photon_0_b}}
\caption{Particle-level fiducial differential cross-sections times branching ratio for the diphoton variable \ptgg\ in \protect\subref{fig:data_unfolded_xsections_matrixinversion_1D_photon_0_a} linear and \protect\subref{fig:data_unfolded_xsections_matrixinversion_1D_photon_0_b} logarithmic scale. The measured cross-sections are compared with several predictions changing the ggF components as described in the text: the default simulation, \scetlib::qT (up to \SI{200}{\GeV}), \textsc{RadISH+NNLOjet} (up to \SI{450}{\GeV}), \textsc{ResBos2} (up to \SI{450}{\GeV}) and LHCHWG (for the two highest \pt\ bins). Total uncertainties are indicated by the error bars on the data points, while the systematic uncertainties are indicated by the boxes. The uncertainties in the predictions are indicated with shaded bands. The bottom panel shows the predicted values from the top panel divided by data.}
\label{fig:data_unfolded_xsections_matrixinversion_1D_photon_0}
\end{figure}

Figure~\ref{fig:data_unfolded_xsections_matrixinversion_1D_photon_0} shows the measured differential cross-sections probing \ptgg.
The measured differential cross-section is in good agreement with the benchmark default simulation, and is statistically limited. The measured differential cross-sections in bins of \ptgg include a new measurement in the boosted region \(\ptgg>\SI{350}{\GeV}\) using kinematic ranges similar to those used in a search for highly boosted \Hbb decays by CMS~\cite{CMS-HIG-19-003}. This region is of interest given its sensitivity to BSM effects.
Our measurement is in agreement with the SM predictions, albeit the large uncertainties for \(\ptgg>\SI{650}{\GeV}\). For \(\ptgg>\SI{450}{\GeV}\), the measured cross-section is compared to the state-of-the-art predictions from the LHC Higgs Working Group (LHCHWG)~\cite{becker2021precise}, which match the predictions from the default simulation but provide an improved estimation of the uncertainties.
 
Using the CL\(_{\text{s}}\) procedure~\cite{Read:2002hq}, 95\% CL upper limits of 3.1 and 5.8 were set on the ratio \(\sigma^{\text{observed}}/\sigma^{\text{SM}}\) for \(450 < \ptgg < \SI{650}{\GeV}\) and \(\ptgg>\SI{650}{\GeV}\), respectively. These correspond to upper limits on the cross-section times branching ratio of
0.18~\si{\femto\barn} (0.06~\si{\femto\barn}) for \(450 < \ptgg < \SI{650}{\GeV}\) (\(\ptgg>\SI{650}{\GeV}\)).
These limits are a significant improvement on the upper limits from measurements in the \Hbb\ channel~\cite{ATLAS:2021tbi}, with the caveat that for \(\ptgg>\SI{650}{\GeV}\) the photon isolation criteria in the fiducial selection reject events with \(\ptgg>\SI{1.25}{\TeV}\).
 
For the lower \pt\ range, the measured \ptgg\ distribution is compared with \textsc{RadISH+NNLOjet}, \scetlib and \textsc{ResBos2} theoretical predictions. The first two are accurate to N\(^3\)LL\('\) in resummation accuracy, whereas \textsc{ResBos2} is accurate to N\(^3\)LL, but all are in good agreement with the data within the statistical uncertainty.
 
\begin{figure}[htb]
\centering
\subfloat[]{\includegraphics[width=0.5\textwidth]{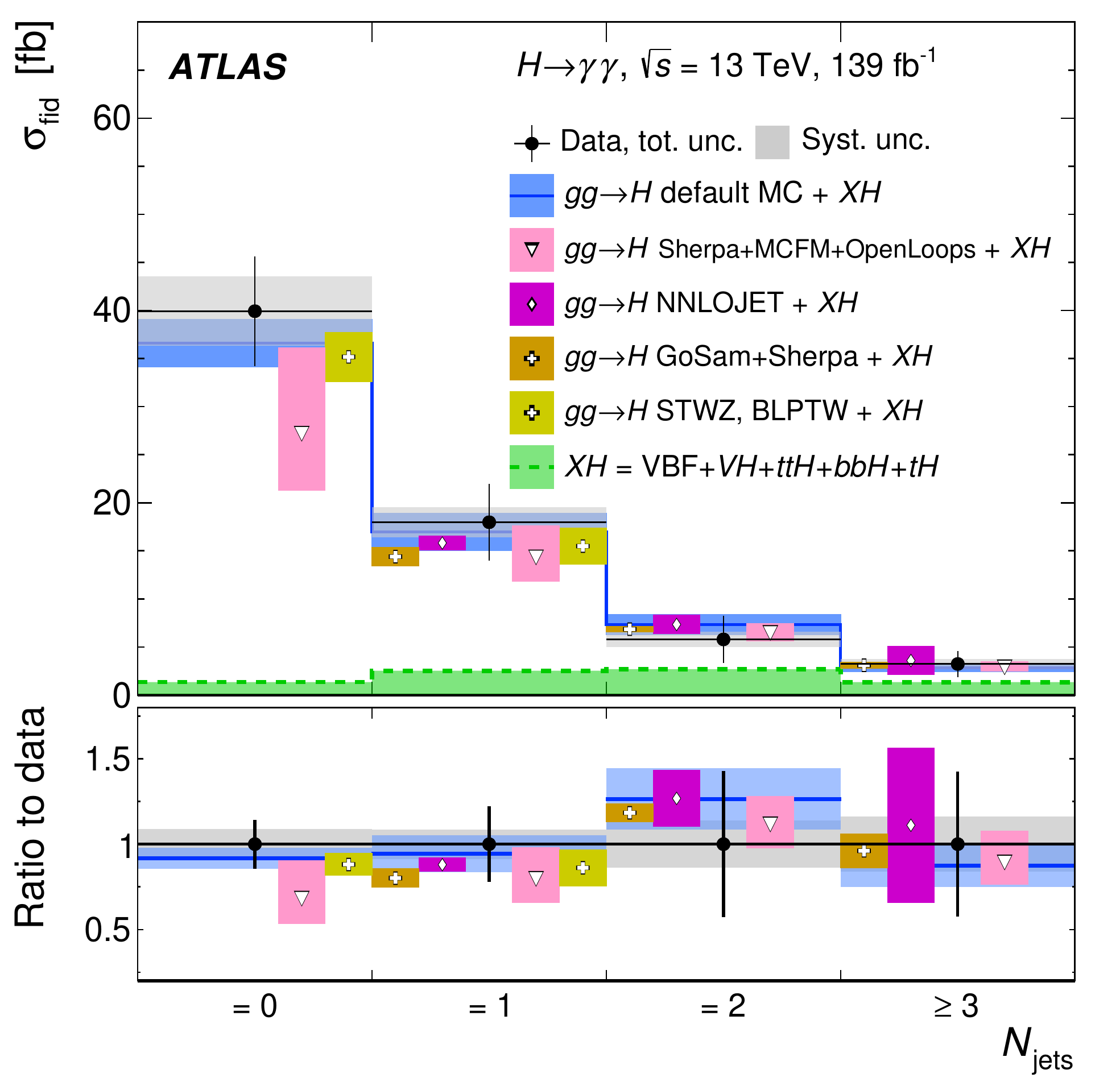}\label{fig:data_unfolded_xsections_matrixinversion_1D_njet_1}}
\subfloat[]{\includegraphics[width=0.5\textwidth]{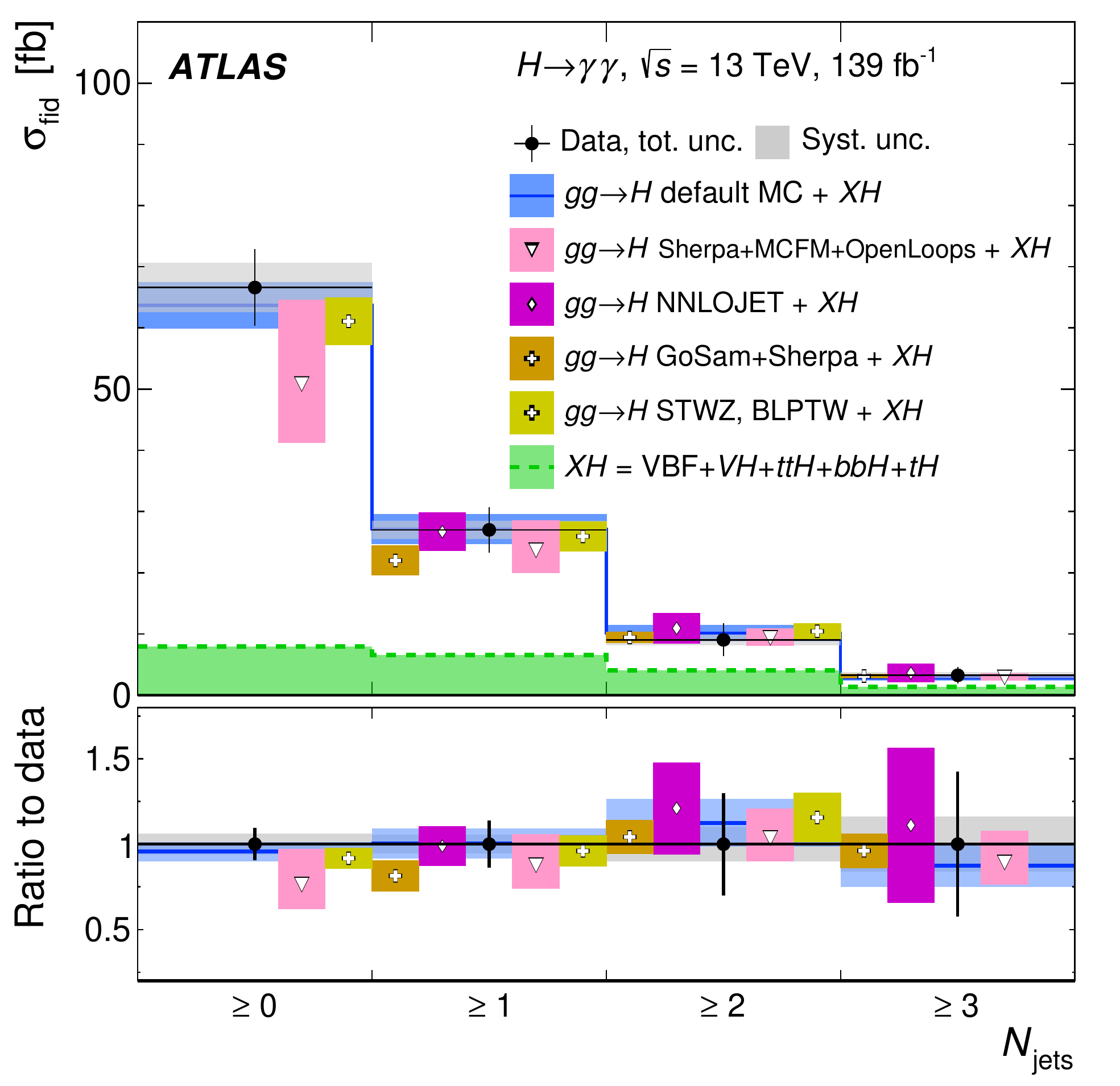}\label{fig:data_unfolded_xsections_matrixinversion_1D_njet_3}}\\
\caption{Particle-level fiducial differential cross-sections times branching ratio for \protect\subref{fig:data_unfolded_xsections_matrixinversion_1D_njet_1} the exclusive jet multiplicity \Njets\ and \protect\subref{fig:data_unfolded_xsections_matrixinversion_1D_njet_3} the inclusive jet multiplicity. The \textsc{NNLOJET} and \textsc{GoSam+Sherpa} predictions are available only for the \(\geq 1\) jet phase space. The STWZ, BLPTW predictions are available only for the exclusive 0, 1-jet bins and the inclusive $\geq 0$, $\geq 1$, $\geq 2$-jet bins.}
\label{fig:data_unfolded_xsections_matrixinversion_1D_njet}
\end{figure}

\begin{figure}[htb]
\centering
\includegraphics[width=0.5\textwidth]{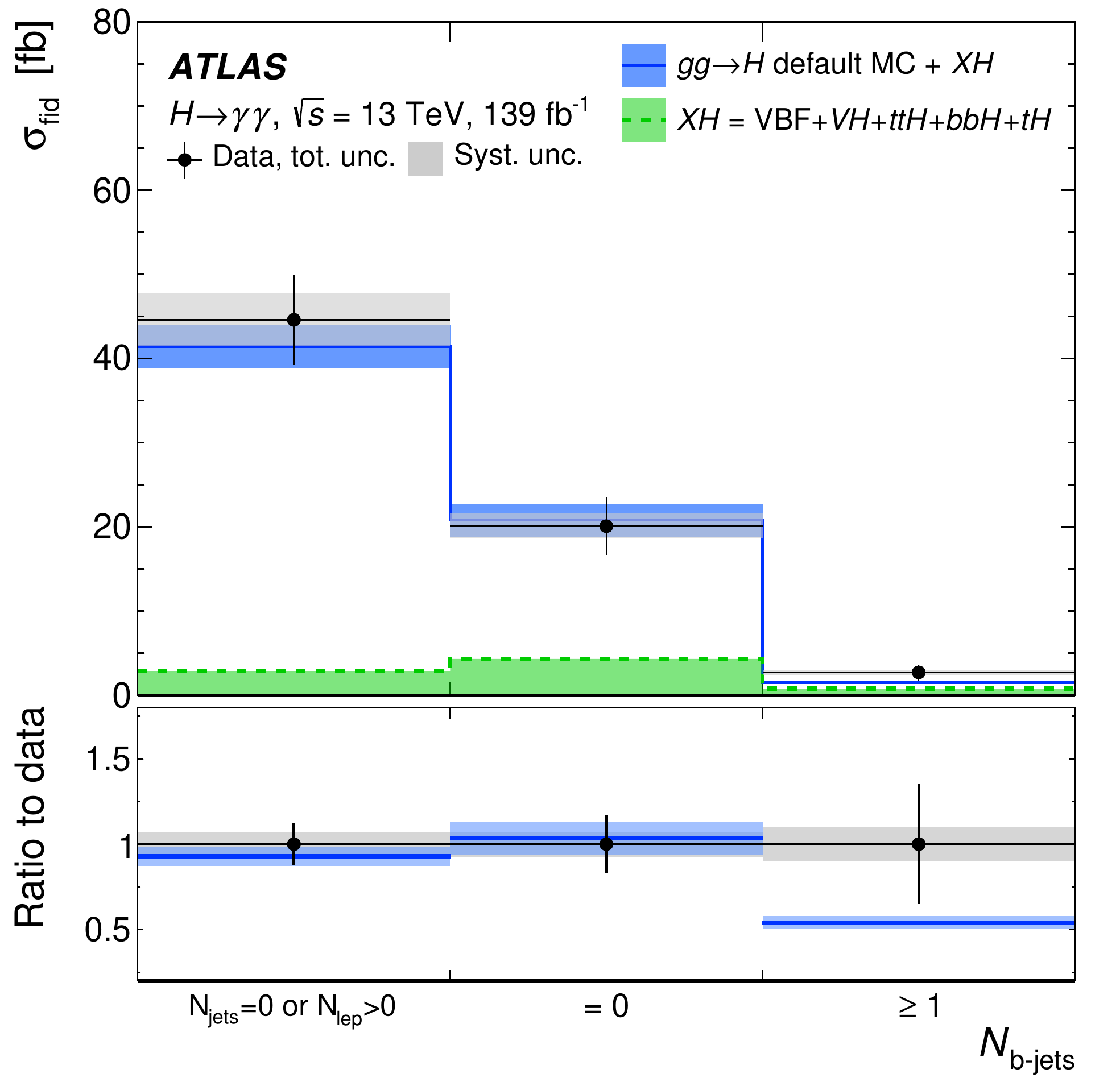}
\caption{Particle-level fiducial differential cross-sections times branching ratio for the \bjet\ multiplicities variable \Nbjets.
The first bin includes events with no central jets or at least one lepton,
while the two other bins contain events with zero or at least one $b$-jet
in the remaining part of the diphoton fiducial phase space.}
\label{fig:data_unfolded_xsections_matrixinversion_1D_nbjet}
\end{figure}

\paragraph{Jet multiplicities}
Measured cross-sections with respect to exclusive and inclusive jet multiplicity are shown in Figure~\ref{fig:data_unfolded_xsections_matrixinversion_1D_njet}, while the \bjets multiplicity dependence is shown in Figure~\ref{fig:data_unfolded_xsections_matrixinversion_1D_nbjet}.
The measured cross-sections are compared with various predictions at different orders in QCD accuracy. Good agreement is observed between the measured \Njets\ and \Nbjets\ distributions and the corresponding predictions. For \Njets, the predictions vary significantly in their uncertainties among the different bins since they vary in their order of QCD accuracy. This is most evident for \textsc{NNLOjet} predictions~\cite{Chen:2016zka,Chen:2014gva} which is an \NNLO\ prediction for \(H +\geq 1\) jet, and hence a leading-order prediction for the \(\geq3\)-jet bin, yielding a larger uncertainty. The \textsc{Sherpa+MCFM+OpenLoops} and \textsc{GoSam} predictions are at \NLO\ for the different bins with \(\geq 1\) jet, and hence has a smaller uncertainty for the highest jet multiplicity. The \(\geq 3\)-jet bin from the default simulation is produced solely by the parton shower and thus the uncertainty estimate is unreliable. The uncertainties in the different predictions for the differential cross-sections in bins of exclusive \Njets\ are underestimated as the exclusive-jet requirement results in a severe restriction of the phase space that is not taken into account in the formalism of these predictions.
 
\begin{figure}[htb]
\centering
\includegraphics[width=0.5\textwidth]{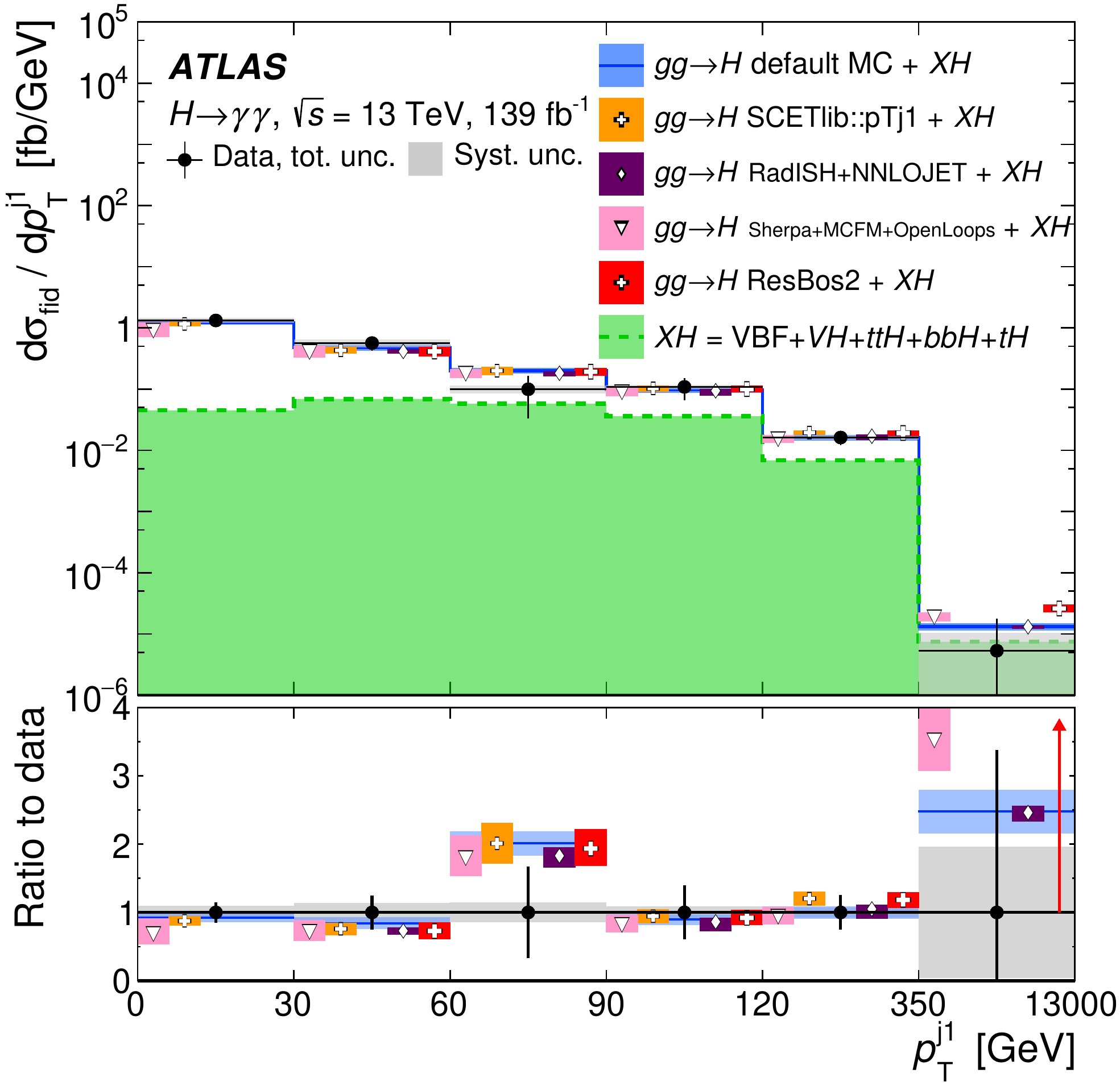}
\caption{Particle-level fiducial differential cross-sections times branching ratio for \ptj[1]. The \textsc{ResBos2} and \textsc{RadISH+NNLOJET} predictions for \ptj[1] are available only for $\ptj[1]>\SI{30}{\GeV}$, whereas \scetlib is available up to \SI{350}{\GeV}.}
\label{fig:data_unfolded_xsections_matrixinversion_1D_1jet_1}
\end{figure}

\begin{figure}[htb]
\centering
\includegraphics[width=0.5\textwidth]{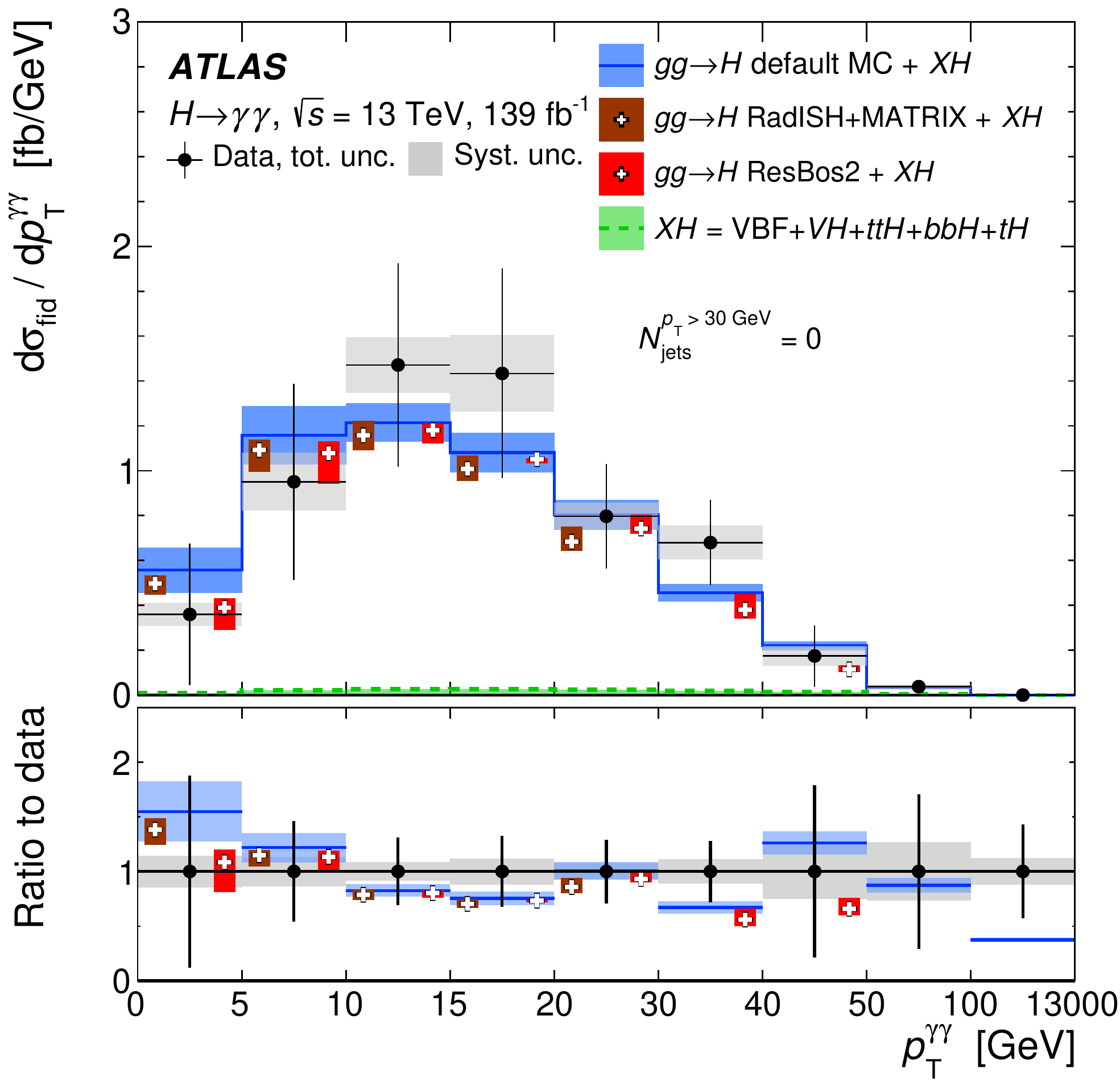}
\caption{Particle-level fiducial differential cross-sections times branching ratio for \ptgg\ with a $p_\text{T}^{j}>\SI{30}{\GeV}$ jet veto. The \textsc{ResBos2} predictions are available up to \SI{50}{\GeV}. The \textsc{RadISH+Matrix} predictions are available up to \SI{30}{\GeV}.}
\label{fig:data_unfolded_xsections_matrixinversion_1D_JV_1}
\end{figure}

\paragraph{\(\geq\) 1-jet differential cross-sections}
Figure~\ref{fig:data_unfolded_xsections_matrixinversion_1D_1jet_1} shows the measured differential cross-section for \ptj[1].
The \ptj\ distribution covers the same kinematic range as the Higgs boson \ptgg\ measurement, but coarser bins were chosen at low \pt, with the \scetlib and \textsc{RadISH+NNLOjet} predictions providing the greatest accuracy (NNLO) among the different predictions.
Figure~\ref{fig:data_unfolded_xsections_matrixinversion_1D_JV_1} shows \ptgg\ with a jet veto for $p_\text{T}^{j}>\SI{30}{\GeV}$.
The measured cross-sections are compared with the default Monte Carlo predictions and with the resummed predictions from \textsc{RadISH+MATRIX} and \textsc{ResBos2}, which carry out the jet-veto resummation at NNLL accuracy. The predictions are considered accurate up to \SI{10}{\GeV} above the jet-veto threshold. The current data uncertainty does not allow detailed conclusions to be drawn for various predictions, but future comparisons with improved precision will allow refinements in similar resummation calculations.
 
\begin{figure}[htb]
\centering
\subfloat[]{\includegraphics[width=0.5\textwidth]{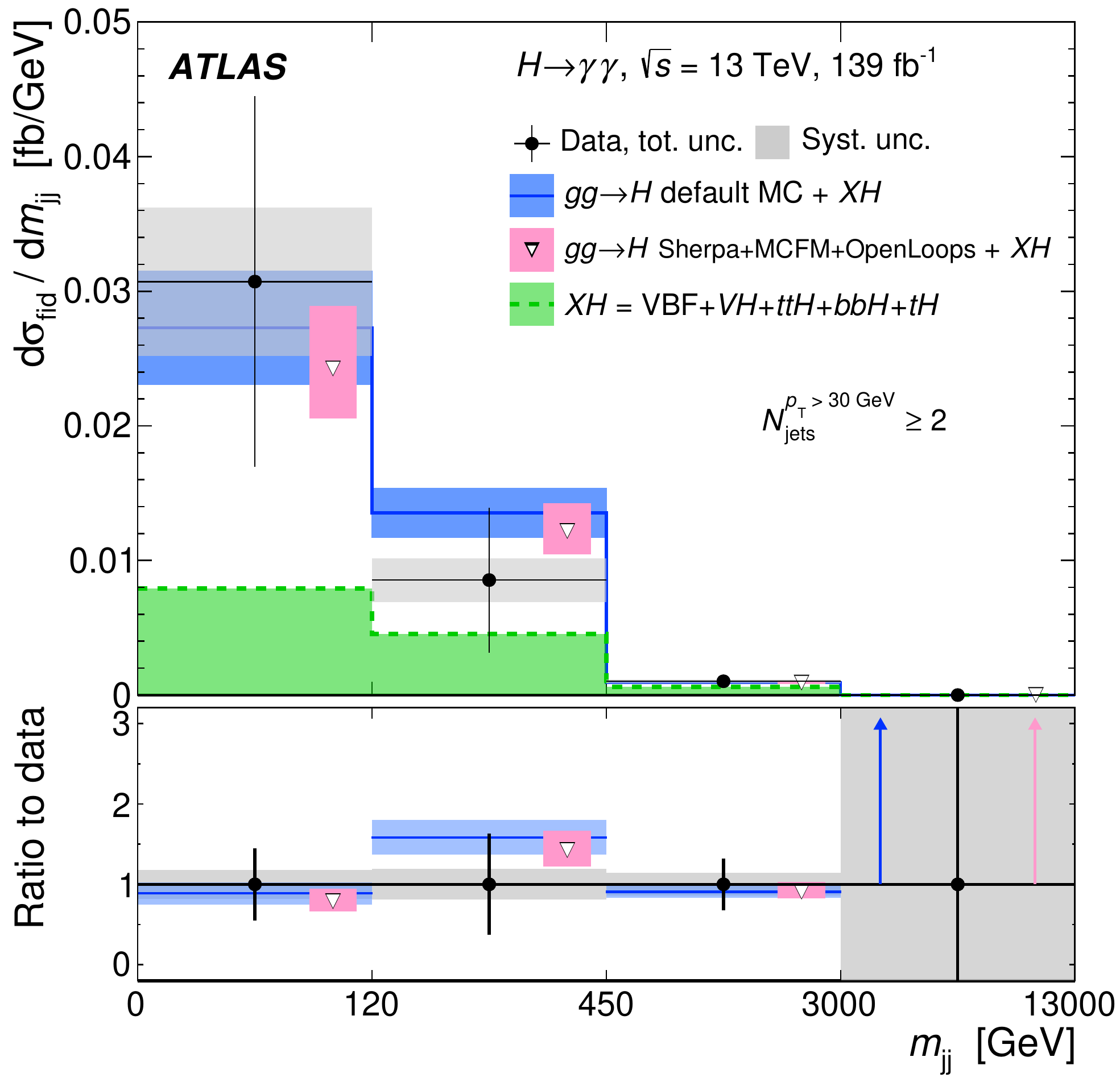}\label{fig:data_unfolded_xsections_matrixinversion_1D_2jet_a}}
\subfloat[]{\includegraphics[width=0.5\textwidth]{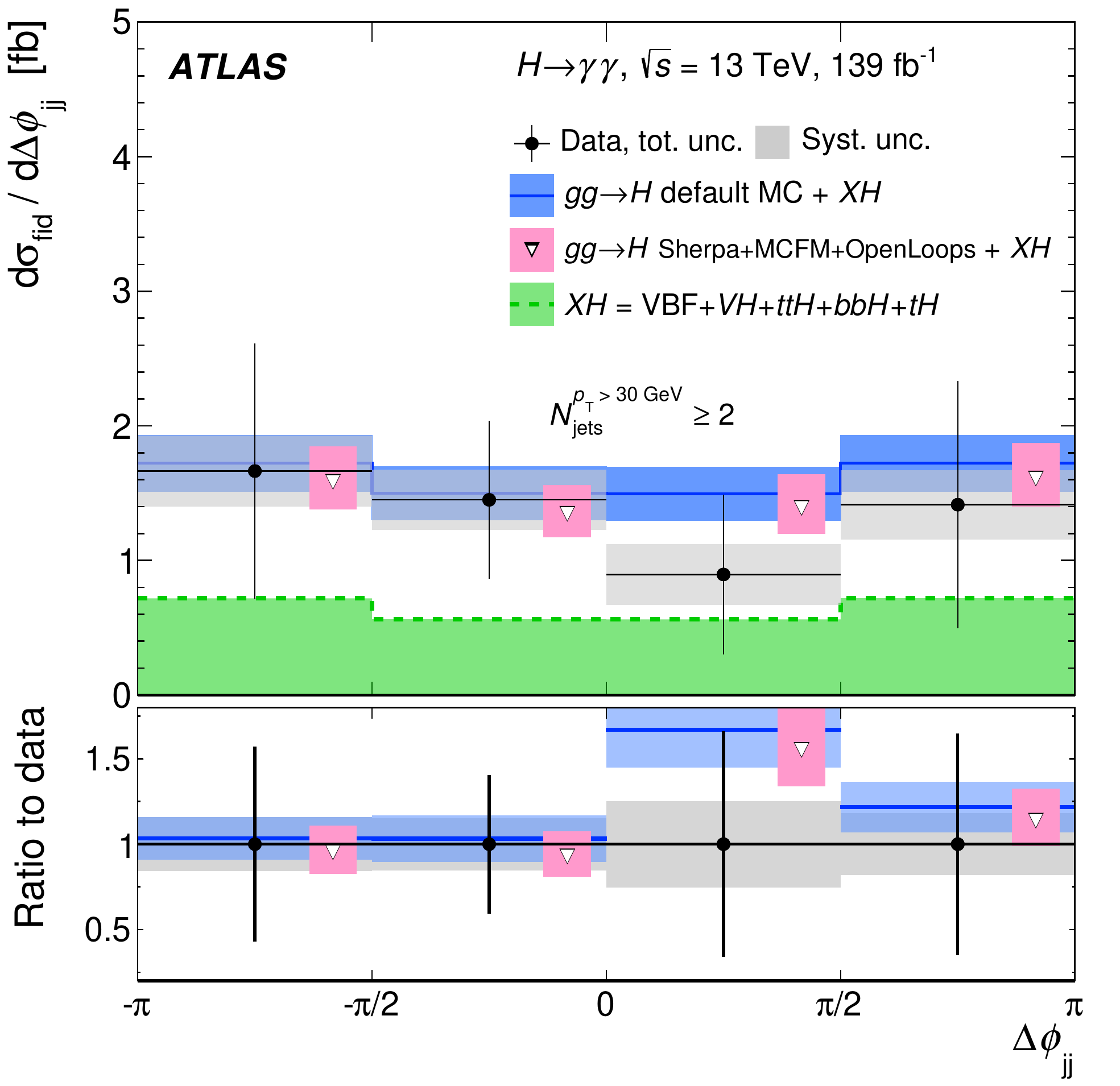}\label{fig:data_unfolded_xsections_matrixinversion_1D_2jet_c}}\\
 
\caption{Particle-level fiducial differential cross-sections times branching ratio for the variables \protect\subref{fig:data_unfolded_xsections_matrixinversion_1D_2jet_a} \mjj\ and \protect\subref{fig:data_unfolded_xsections_matrixinversion_1D_2jet_c} \dphijj\ in the diphoton baseline fiducial region.}
\label{fig:data_unfolded_xsections_matrixinversion_1D_2jet_1}
\end{figure}

\paragraph{\(\geq\) 2-jet differential cross-sections}
Figure~\ref{fig:data_unfolded_xsections_matrixinversion_1D_2jet_1} shows the differential cross-sections for the variables \mjj\ and  \dphijj.
The \mjj\ and \dphijj\ distributions are compared with \SHERPA\ predictions that are of NLO accuracy for this jet multiplicity, whereas the default simulation is accurate only to leading order.
Good agreement is observed between data and the predictions, including the default simulation.
In the highest \mjj\ bin, which is more sensitive to VBF production, the data are in agreement with the predictions within the uncertainty of the measurement.
The \dphijj\ distribution, which has sensitivity to the CP properties of the Higgs boson, is in good agreement with the expected shape in the SM.
 
\begin{figure}[htb]
\centering
\includegraphics[width=0.7\textwidth]{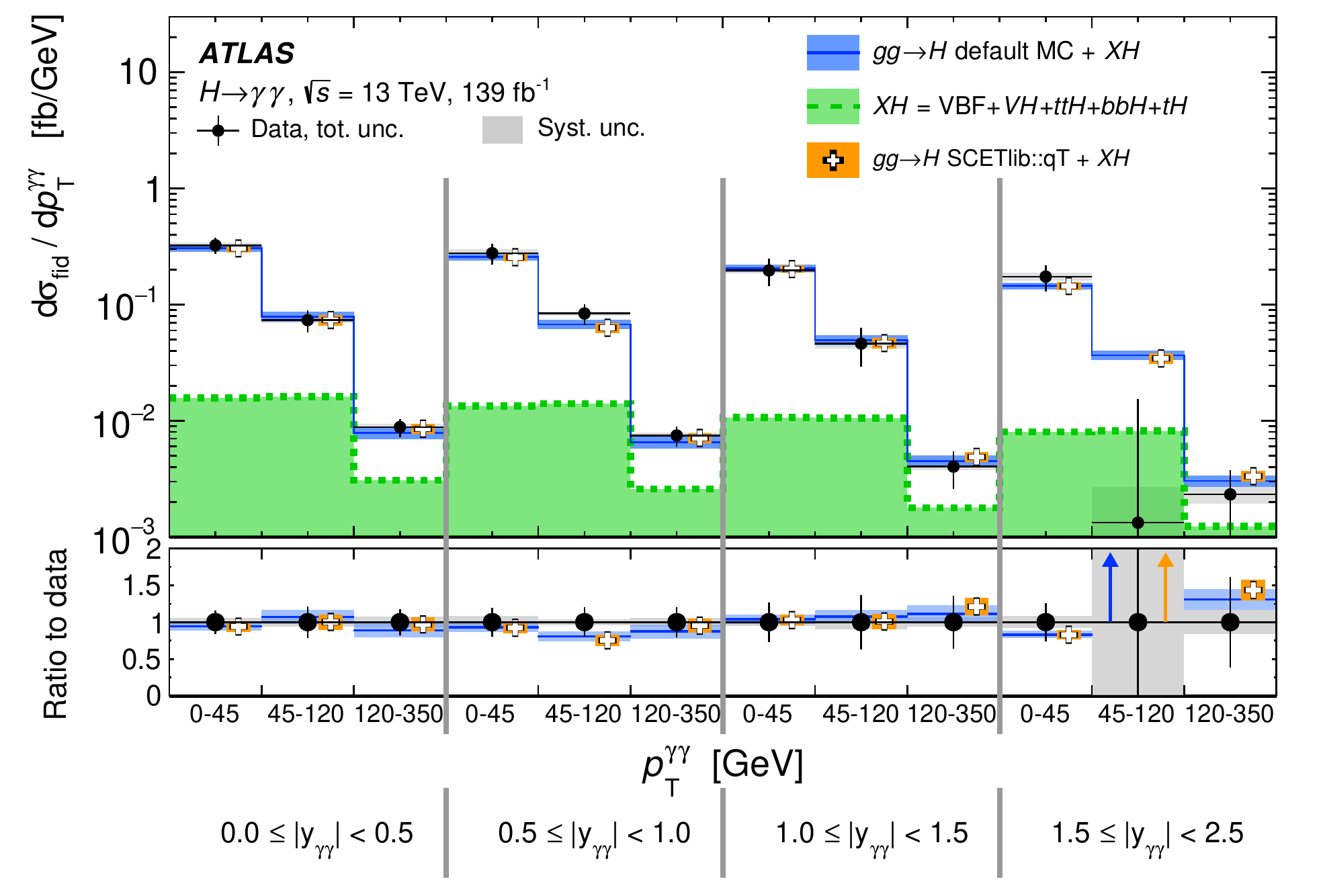}
\caption{Double-differential particle-level fiducial cross-sections times branching ratio of \ptgg\ in bins of \ygg.}
\label{fig:data_unfolded_xsections_matrixinversion_2D_1}
\end{figure}

\paragraph{Double-differential cross-sections}
Figure~\ref{fig:data_unfolded_xsections_matrixinversion_2D_1} shows the double-differential cross-section for  \ptgg\ vs \ygg. Overall, good agreement is observed between data and predictions, with \scetlib providing a more accurate description than the default simulation.
 
\begin{figure}[htbp]
\centering
\includegraphics[width=0.5\textwidth]{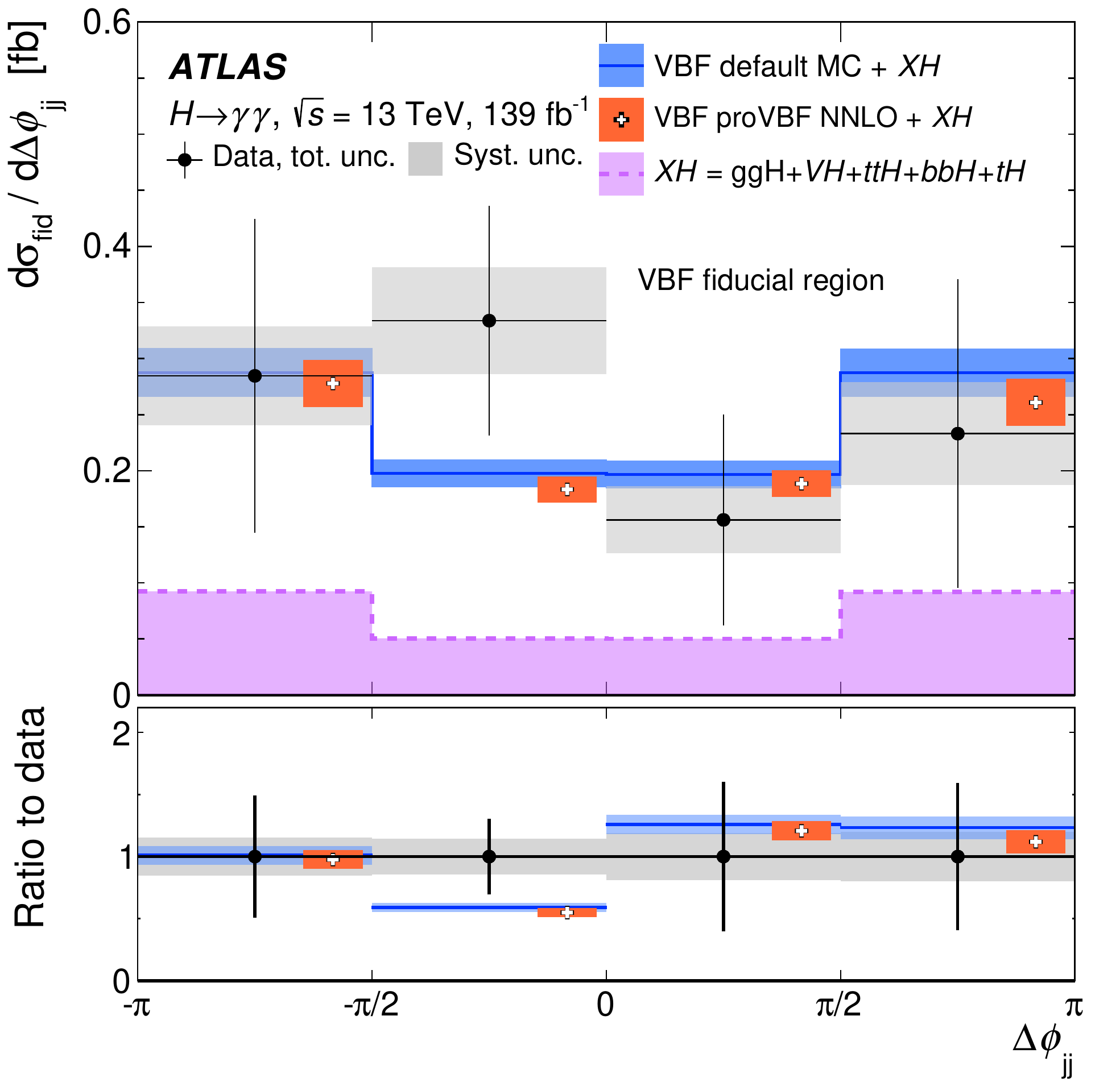}
 
\caption{Particle-level fiducial differential cross-sections times branching ratio for \dphijj\ in the VBF-enhanced fiducial region.}
\label{fig:data_unfolded_xsections_matrixinversion_VBF_1}
\end{figure}

\paragraph{Cross-sections in the VBF-enhanced phase space}
Figure~\ref{fig:data_unfolded_xsections_matrixinversion_VBF_1} shows the differential cross-section in the VBF-enhanced phase space for \dphijj. Overall, good agreement is observed between the data and the default simulation prediction and the \textsc{proVBF} prediction, which is at higher-order accuracy in QCD.

\FloatBarrier
 
Similarly to the inclusive cross-sections, a \(\chi^2\) test was used to evaluate \(p\)-values for the compatibility of the measured differential cross-sections and the predictions. For the uncertainty in the theoretical predictions, the correlation is ignored since it is not available for most of the predictions.
Using the default simulation prediction, it was estimated that neglecting this correlation can change the \(p\)-value by few percent.
As indicated in Table~\ref{tab:chi2Data}, the measurements are compatible with the SM for all the predictions.
In addition, it was checked that when fitting different differential cross-sections relative to the same fiducial region the integrals of the signal yields in the bins are compatible.
 
\begin{table}[htbp]
\setlength{\tabcolsep}{2pt}
\centering
\caption{The \(p\)-values obtained with a \(\chi^2\) compatibility test between the fitted cross-sections and the SM predictions for each differential distribution. The \(\chi^2\) is computed using the full set of uncertainties for the data, including their correlation, and for the SM predictions. The correlation of the SM predictions are neglected as most of the predictions are provided without this information.
}
\label{tab:chi2Data}
\clearpage
\resizebox{\textwidth}{!}{
\begin{tabular}{lrrrrrrrrrrr}
\toprule
\multirow{2}{*}{Variable}   & \multicolumn{11}{c}{\(p\)-value}                                                                                                                                          \\
\cmidrule(lr){2-12}
& default                          & \texttt{RadISH}  & NNLOJet & STWZ  & MATRIX & \SHERPA & \GOSAM & \scetlib & \texttt{TAUC} & \texttt{ResBos2} & \textsc{proVBF} \\
&                                  & \texttt{NNLOJET} &         & BLPTW &        &         &        &          &               &                  &                 \\
\midrule
\relptgOne                  & 56\%                             & --                               & --      & --      & --     & --      & --     & 58\%     & --            & 32\%             & --              \\
\relptgTwo                  & 93\%                             & --                               & --      & --      & --     & --      & --     & 49\%     & --            & 2\%              & --              \\
\ptgg                       & 86\%                             & 68\%                             & --      & --      & --     & --      & --     & 78\%     & --            & 54\%             & --              \\
\ygg                        & 76\%                             & --                               & --      & --      & --     & --      & --     & 78\%     & --            & 66\%             & --              \\
\ptj[1]                     & 78\%                             & 77\%                             & --      & --      & --     & 47\%    & --     & 48\%     & --            & 38\%             & --              \\
\Njets                      & 95\%                             & --                               & 90\%    & 56\%    & --     & 59\%    & 84\%   & --       & --            & --               & --              \\
\Nbjets                     & 60\%                             & --                               & --      & --      & --     & --      & --     & --       & --            & --               & --              \\
\ptggj                      & 81\%                             & --                               & --      & --      & --     & 68\%    & --     & --       & --            & 78\%             & --              \\
\mggj                       & 95\%                             & --                               & --      & --      & --     & 95\%    & --     & --       & --            & --               & --              \\
\maxtau                     & 27\%                             & --                               & --      & --      & --     & --      & --     & --       & 11\%          & 13\%             & --              \\
\sumtau                     & 39\%                             & --                               & --      & --      & --     & --      & --     & --       & --            & --               & --              \\
\HT                         & 46\%                             & --                               & --      & --      & --     & 51\%    & --     & --       & --            & --               & --              \\
\mjj                        & 79\%                             & --                               & --      & --      & --     & 81\%    & --     & --       & --            & --               & --              \\
\dphijj                     & 91\%                             & --                               & --      & --      & --     & 95\%    & --     & --       & --            & --               & --              \\
\absdphiggjj                & 83\%                             & --                               & --      & --      & --     & 88\%    & --     & --       & --            & --               & --              \\
\ptggjj                     & 99\%                             & --                               & --      & --      & --     & 100\%   & --     & --       & --            & --               & --              \\
\ptggjv[30]                 & 84\%                             & --                               & --      & --      & 83\%   & --      & --     & --       & --            & 83\%             & --              \\
\ptggjv[40]                 & 95\%                             & --                               & --      & --      & 45\%   & --      & --     & --       & --            & 83\%             & --              \\
\ptggjv[50]                 & 88\%                             & --                               & --      & --      & 35\%   & --      & --     & --       & --            & 30\%             & --              \\
\ptggjv[60]                 & 67\%                             & --                               & --      & --      & 52\%   & --      & --     & --       & --            & 42\%             & --              \\
\ptgg vs \ygg               & 75\%                             & --                               & --      & --      & --     & --      & --     & 78\%     & --            & --               & --              \\
\ptgg vs \maxtau            & 39\%                             & --                               & --      & --      & --     & --      & --     & --       & --            & --               & --              \\
\ptgg vs \ptggj             & 96\%                             & --                               & --      & --      & --     & --      & --     & --       & --            & --               & --              \\
\reldiffptgg vs    & 81\% & \multirow{2}{*}{--} & \multirow{2}{*}{--}      & \multirow{2}{*}{--}      & \multirow{2}{*}{--}     & \multirow{2}{*}{--}      & \multirow{2}{*}{--}     & \multirow{2}{*}{77\%}     & \multirow{2}{*}{--}            & \multirow{2}{*}{--}               & \multirow{2}{*}{--}              \\
\relsumptgg &                             &                                &       &       &      &       &      &        &             &                &               \\
VBF $|\eta^*|$              & 94\%                             & --                               & --      & --      & --     & --      & --     & --       & --            & --               & 70\%            \\
VBF \dphijj                 & 68\%                             & --                               & --      & --      & --     & --      & --     & --       & --            & --               & 65\%            \\
VBF \ptj[1]                 & 77\%                             & --                               & --      & --      & --     & --      & --     & --       & --            & --               & 70\%            \\
VBF \ptggjj                 & 89\%                             & --                               & --      & --      & --     & --      & --     & --       & --            & --               & 74\%            \\
VBF \ptj[1] vs \dphijj      & 76\%                             & --                               & --      & --      & --     & --      & --     & --       & --            & --               & 74\%            \\
 
\bottomrule
\end{tabular}
}
\clearpage
\end{table}
\section{Interpretations of the measured differential cross-sections}
\label{sec:interpretations}
The fiducial cross-section measurements, shown in Section~\ref{sec:xsec_results}, are largely model independent. This allows direct comparisons with various theory predictions as well as interpretations in alternative theoretical frameworks. In this section, the measured cross-sections are used to constrain \(b\)- and \cquark Yukawa coupling modifiers relative to the SM, \kappab and \kappac, detailed in Section~\ref{subsec:YukawaInterpretation} and to probe physics beyond the SM via the effective field theory approach, detailed in Section~\ref{sec:eft}.

\subsection{Limits on the \(b\)- and \(c\)-quark Yukawa couplings using the Higgs boson \pt spectrum}
\label{subsec:YukawaInterpretation}
 
The Higgs boson \pt spectrum is sensitive to the Yukawa couplings of the Higgs boson to the \(b\)- and \cquarks. This sensitivity is driven by quark-initiated (\(q\bar{q}\) and \(qg\)) production of the Higgs boson and the contributions of \(b\)- and \cquarks to the loop-induced ggF production.
Direct observations of the Higgs boson coupling to \bquarks~\cite{HIGG-2018-04,CMS-HIG-18-016} provided stringent constraints on its possible modification with respect to the SM, whereas current searches for Higgs boson decays to charm final states~\cite{HIGG-2021-12,HIGG-2016-23,CMS:2022psv} still allow for a relatively large modification of the \cquark coupling. This paper presents an indirect method~\cite{HIGG-2018-29} to probe the \(b\)- and \(c\)-coupling modifiers, \kappab and \kappac, through the measured \ptgg spectrum, which has the advantage of not being limited by the tagging efficiency for jets originating from \(b\)- and \(c\)-quarks. The current uncertainties from direct searches are approximately 20\% for \kappab~\cite{HIGG-2018-57,CMS-HIG-17-031}, whereas for \kappac the most stringent limit is $1.1<|\kappa_c|<5.5$~\cite{CMS:2022psv}.
 
Modifications of the coupling strength to \(b\)- and \cquarks would impact the ggF and quark-initiated production modes, thus resulting in changes in both the normalisation and the shape of the \ptgg\ spectrum. In addition, the branching ratio for the \Hgg\ decay would be affected by changes in the \Hgg decay width and in the total Higgs boson decay width, induced by anomalous values of \kappab\ or \kappac\ leading to deviations in the $H\to b\bar{b}$ or $H\to c\bar{c}$ partial widths. Two different fitting strategies are presented to provide limits on \kappab and \kappac with an increasing level of model dependency.
In the first case, only the shape of the measured \ptgg spectrum is considered, whereas the second case also considers normalisation changes due to the cross-section variations in addition to the variations of the \Hgg\ partial decay width and the total Higgs boson width. All the other Higgs boson production modes remain unchanged with \kappab and \kappac variations, and their contributions are taken from the default simulation.
 
The predictions for \kappab and \kappac modifications of ggF production are computed with \scetlib~\cite{scetlib, Billis:2021ecs}, detailed in Sections~\ref{subsec:theoryunc} and Appendix \ref{sec:aux theory preds unc}, including also the bottom- and charm-quark loop contributions.
For the SM, in all the \ptgg bins, the dominant contribution to the ggF cross-section is given by the top-quark loop. The interference between the top-quark gluon-fusion loop and the  \(b\)- and  \(c\)-quark gluon-fusion loops is comparatively small, but not negligible, and negative for \(\ptgg<\SI{100}{\GeV}\). The contributions of \(b\)- and \(c\)-quark gluon-fusion processes and the interference between them are found to be very small. For values of $|\kappab|$ or $|\kappac|$ significantly different from one, the impact of the interference terms  of the \(b\)- or \(c\)-quark gluon-fusion loops can be much larger.
 
Predictions for quark-initiated \(b\bar{b}\to H\) and \(c\bar{c}\to H\) production modes are computed with \MGNLO[2.7.3], including the higher-order contributions \(bg\to H b\) and \(cg\to H c\), using a dedicated PDF set from Ref.~\cite{Bonvini:2016fgf}. \PYTHIA[8] with the A14 tune~\cite{ATL-PHYS-PUB-2014-021} is used for the simulation of the parton shower, hadronisation and underlying event, as well as the Higgs boson decay.
The inclusive \(b\bar{b}\to H\) and \(c\bar{c}\to H\) cross-sections are then normalised to the state-of-the-art \NNLO\ computations available in Refs.~\cite{Bonvini:2016fgf,Harlander_2016}.
The uncertainties due to missing higher-order QCD terms are estimated from simultaneous variations of the renormalisation and factorisation scales around their central values by factors of \(1/2\) and \(2\). The uncertainty from the choice of the \FXFX\ merging scale is estimated by varying the nominal scale (\SI{40}{\GeV}) by factors of \(1/2\) and \(2\). Among these variations, only the downward variation has considerable impact on the \ptgg\ spectrum.
The PDF uncertainty for \(c\bar{c}\rightarrow H\) production is based on the standard deviation computed using the 100 eigen-variations included in the PDF set, in addition to \(\alphas\) uncertainties.
For  \(b\)-quark-initiated production, the PDF-induced uncertainty in the \(\bbbar\to H\) predictions is obtained from variations of the \(b\)-quark pole mass and the threshold above which the \bquark PDF is non-zero~\cite{Bonvini:2016fgf}.
 
Variations in the \Hgg\ branching ratio from modifications of \kappab and \kappac are estimated using \mbox{HDECAY}~\cite{hdecay,hdecay2}. This includes variations in the partial \Hgg\ decay width, in addition to changes in the total Higgs boson width dominated by decay width modifications from \Hbb and \(H\to c\bar{c}\).
 
\begin{figure}[htbp]
\centering
\subfloat[]{\includegraphics[width=0.45\textwidth]{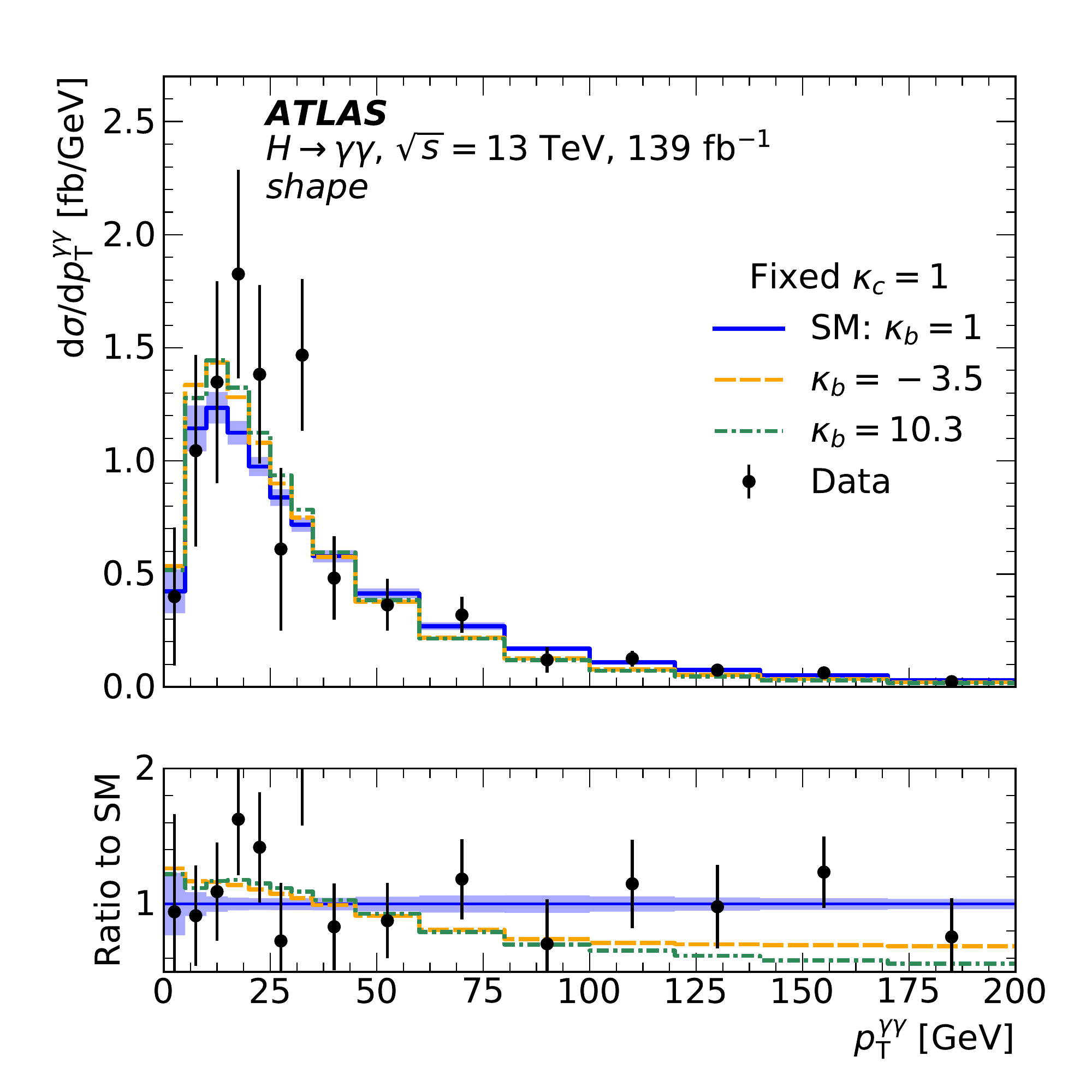}\label{fig:yukawa_data_comp_shape_a}}
\subfloat[]{\includegraphics[width=0.45\textwidth]{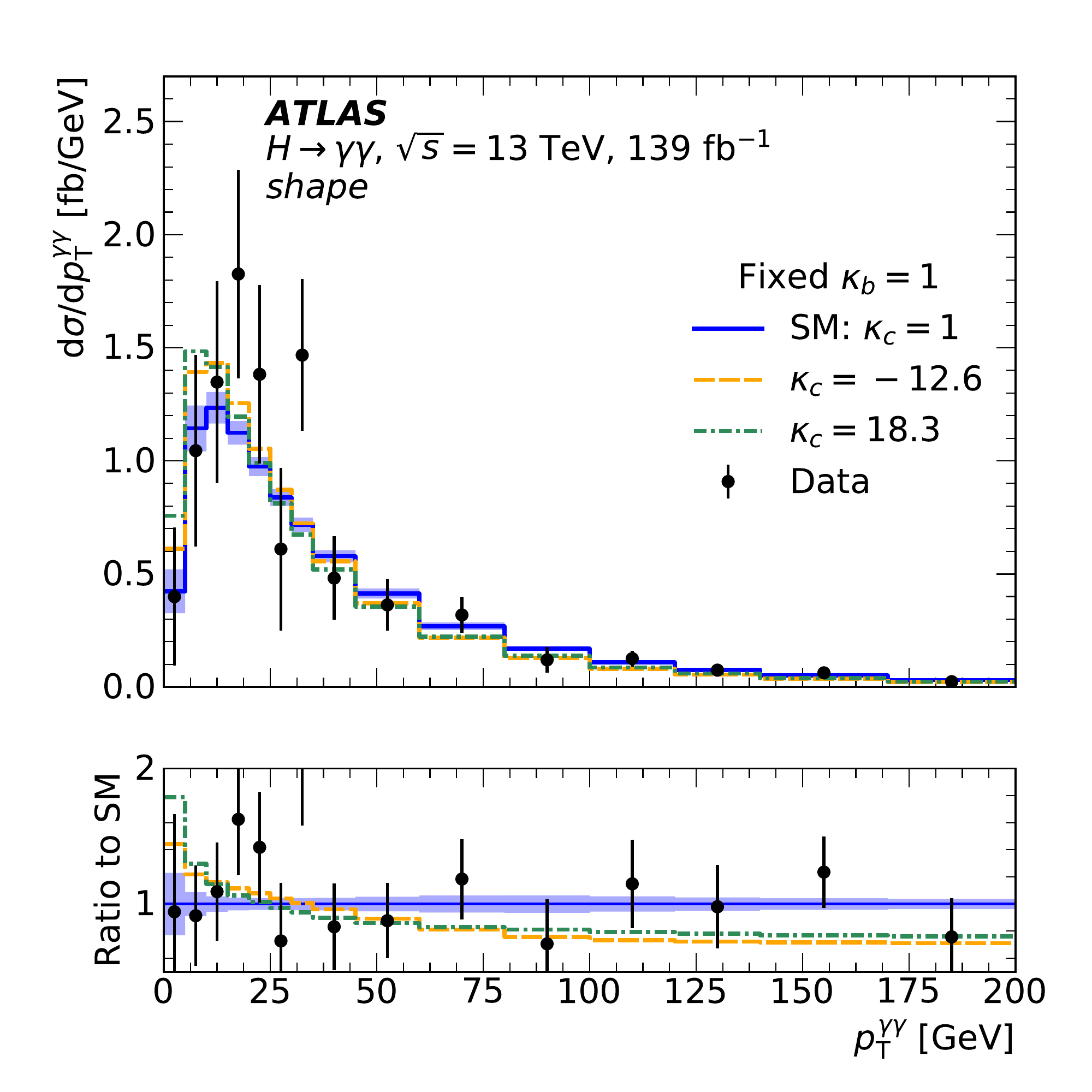}\label{fig:yukawa_data_comp_shape_b}} \\
 
\caption{The observed fiducial differential cross-section times branching ratio for \ptgg compared with the predictions for different values of \protect\subref{fig:yukawa_data_comp_shape_a} \kappab and \protect\subref{fig:yukawa_data_comp_shape_b} \kappac corresponding to the upper (in green) and lower (in orange) limits at 95\% CL for the shape-only fitting strategy. The SM prediction is shown as a blue line with the theoretical uncertainties of the SM prediction as a filled area. The bottom panels show the ratios of the data and the different predictions to the SM prediction.}
\label{fig:yukawa_data_comp_shape}
\end{figure}

\begin{figure}[htbp]
\centering
\subfloat[]{\includegraphics[width=0.45\textwidth]{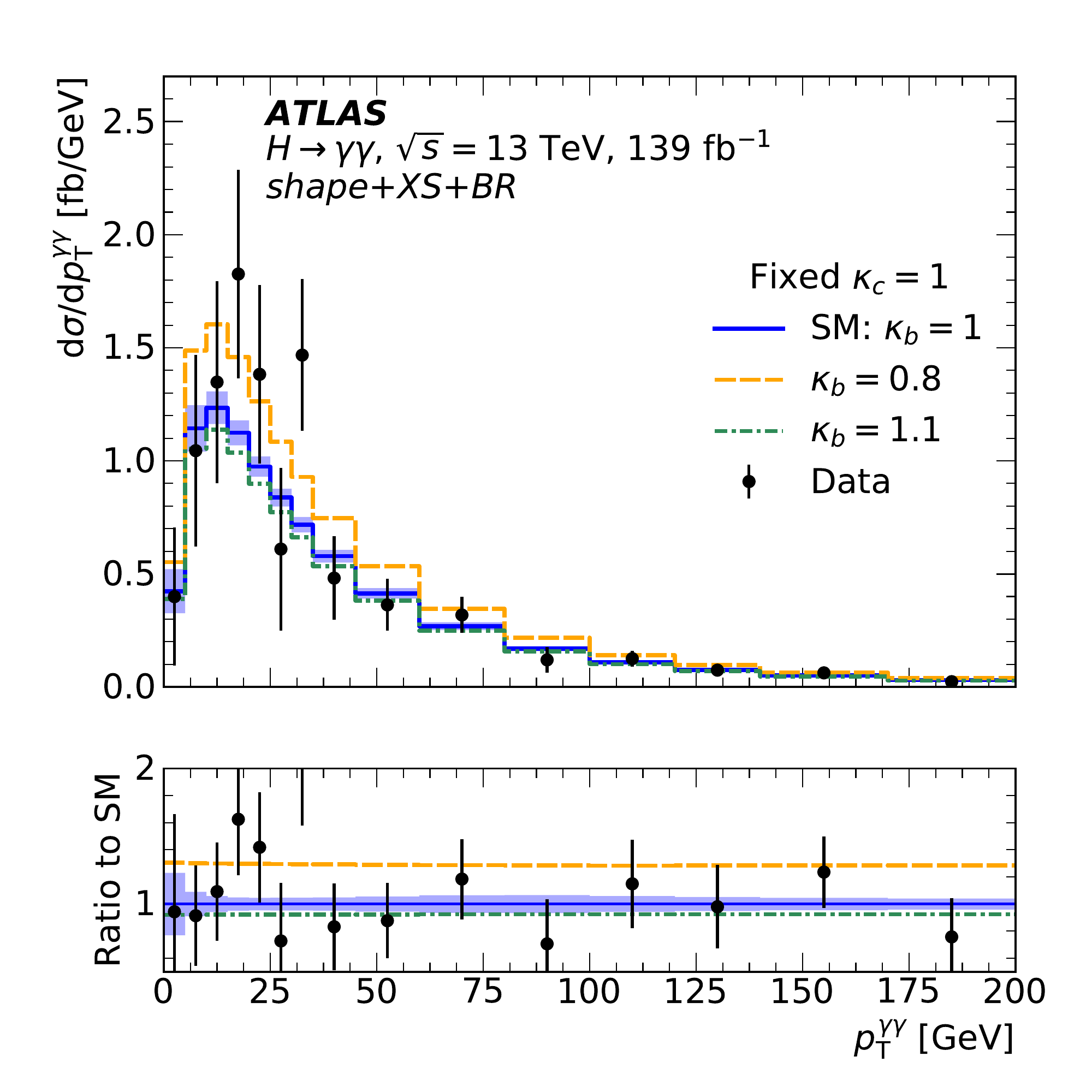}\label{fig:yukawa_data_comp_shape_xs_BR_a}}
\subfloat[]{\includegraphics[width=0.45\textwidth]{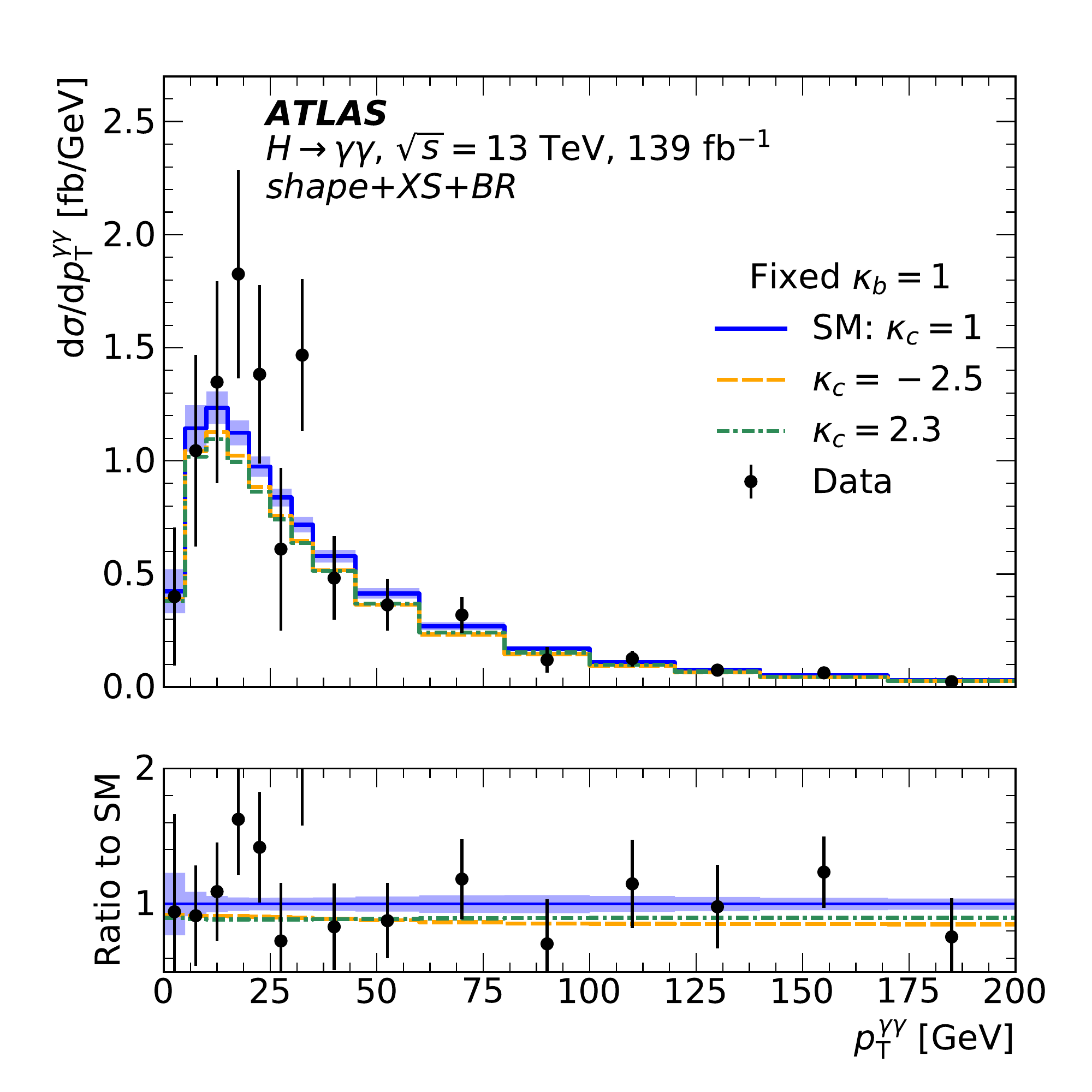}\label{fig:yukawa_data_comp_shape_xs_BR_b}} \\
 
\caption{The observed fiducial differential cross-section times branching ratio for \ptgg compared with the predictions for different values of \protect\subref{fig:yukawa_data_comp_shape_xs_BR_a} \kappab and \protect\subref{fig:yukawa_data_comp_shape_xs_BR_b} \kappac corresponding to the upper (in green) and lower (in orange) limits at 95\% CL for the shape and normalisation fitting strategy (with `XS+BR' denoting `cross-section and branching ratio'). The SM prediction is shown as a blue line with the theoretical uncertainties of the SM prediction as a filled area. The bottom panels show the ratios of the data and the different predictions to the SM prediction.}
\label{fig:yukawa_data_comp_shape_xs_BR}
\end{figure}

\begin{figure}[htbp]
\centering
\subfloat[]{\includegraphics[width=0.45\textwidth]{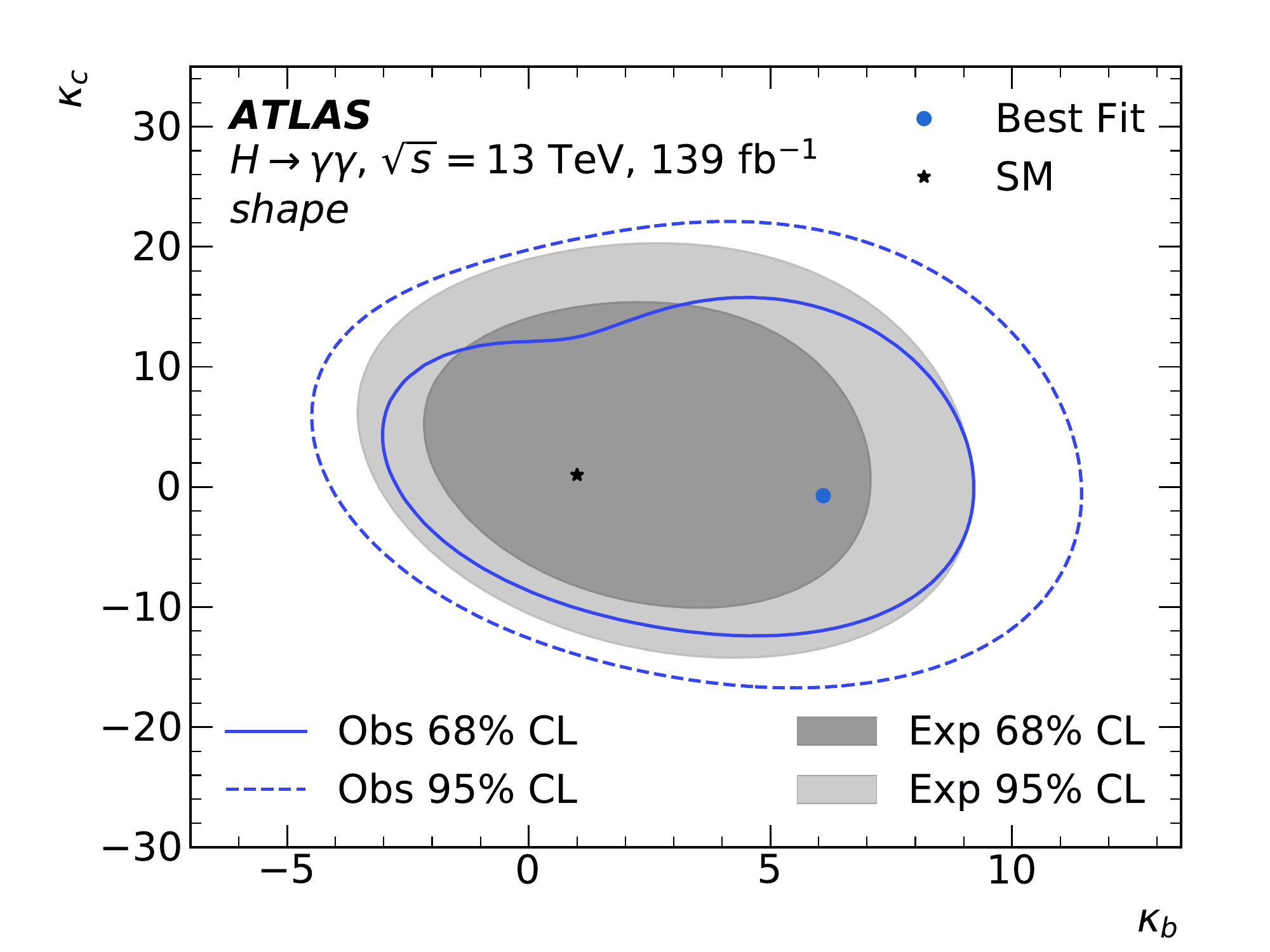}\label{fig:yukawa_2Dlimits_obs_a}}
\subfloat[]{\includegraphics[width=0.45\textwidth]{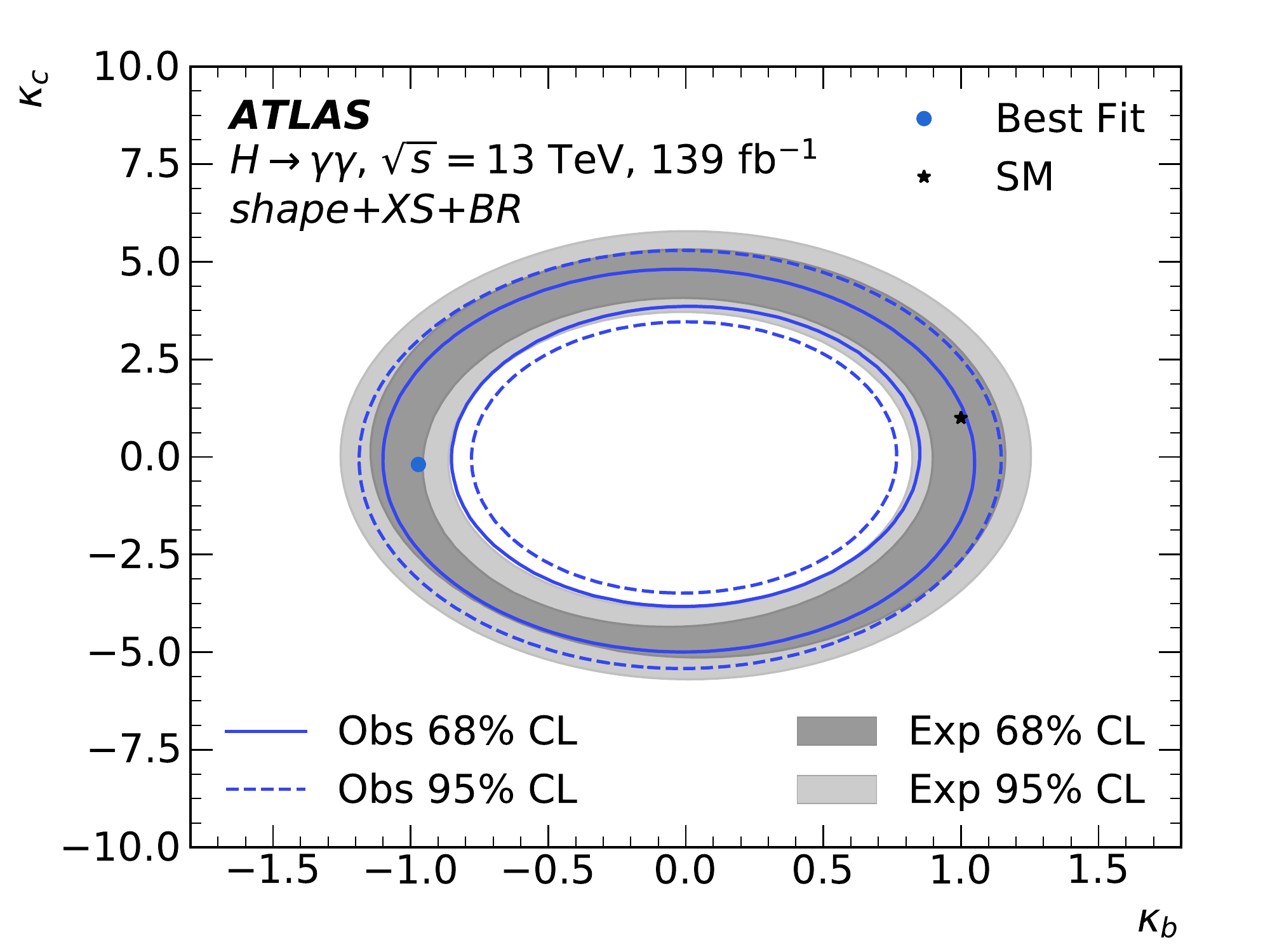}\label{fig:yukawa_2Dlimits_obs_b}}
\caption{Observed and expected 2D limits on \kappab and \kappac when considering modifications of \protect\subref{fig:yukawa_2Dlimits_obs_a} the shape and \protect\subref{fig:yukawa_2Dlimits_obs_b} the shape and normalisation (with `XS+BR' denoting `cross-section and branching ratio') at 68\% and 95\% CL.}
\label{fig:yukawa_2Dlimits_obs}
\end{figure}

The statistical interpretation of the \ptgg\ distribution to set limits on the values of \kappab and \kappac is performed with the profile likelihood method.
The full measurement likelihood for the \ptgg distribution is compared with the predictions parameterised as a function of \kappab and \kappac.
The 95\% CL limits are computed using a test statistic based on a ratio of profiled likelihoods~\cite{Cowan:2010js}.
The measured differential cross-section is used in the range of \ptgg\ up to \SI{200}{\GeV}, which is the region most sensitive to variations of \kappab and \kappac.
The fits were performed with the Higgs boson coupling to the top quark fixed to the SM value (\(\kappa_{t}=1\)). All other Higgs boson couplings are assumed to have SM values ($\kappa=1$).
The observed and expected 95\% confidence intervals are shown in Table~\ref{tab:YbYc_limits} for the different fitting strategies.
The table shows that stricter limits can be computed by including both shape and normalisation variations.
The limits on a given \(\kappa\) parameter are determined while fixing the other one to its SM value \(\kappa=1\). The shape-only limits on \kappab are more stringent than those on \kappac, due to the larger contribution from \(b\bbar\)-initiated process and the \kappab term in the ggF loop. The limits on \kappac are mostly driven by \(c\bar{c}\)-initiated processes. Figures~\labelcref{fig:yukawa_data_comp_shape,fig:yukawa_data_comp_shape_xs_BR} show data compared with predictions for two values of \kappab and \kappac corresponding to the upper and lower limits at 95\% CL from the different fitting strategies.
In addition, two-dimensional limits are derived by simultaneously varying \kappab and \kappac, and the contours for these limits are shown in Figure~\ref{fig:yukawa_2Dlimits_obs}.
In this case the goodness of the fit was computed with a \(\chi^2\) test, resulting in a \(p\)-value of 0.45 when using only the shape and 0.40 when also using the normalisation.
 
The observed \kappab\ and \kappac\ limits are comparable with the limits reported in Ref.~\cite{CMS-HIG-17-028}, which follows a similar approach when interpreting the Higgs boson differential cross-sections but uses multiple decay channels, and about a quarter of the integrated luminosity compared to this paper.
When also using the normalisation information, the observed \kappab\ limits are comparable with those from the direct searches, while the allowed interval for \kappac\ is about two times smaller than that from the direct searches.
In addition to the limits reported below, an additional check was performed by allowing $\kappa_t$ to float in the fit within the limits of the latest $H\to\gamma\gamma$ couplings measurement~\cite{HIGG-2014-06}. This had a negligible effect on the shape-only limits on \kappab and \kappac, whereas the absolute values of the shape-plus-normalisation limits on \kappab and \kappac increased considerably (up to 40\%).

\begin{table}[htbp]
\centering
\caption{Observed and expected allowed ranges at 95\% CL of modifications of the \(b\)- and \(c\)-quark Yukawa couplings to the Higgs boson, \kappab\ and \kappac. The limits on a given \(\kappa\) parameter are computed while fixing the other one to its SM value (\(\kappa=1\)). The table shows the confidence intervals for \kappab\ and \kappac\ using shape-only and using shape and normalisation variations of the SM expectation.}
\begin{tabular}{p{12em}cll}
\toprule
Fit set-up                                                                 & \(\kappa\) & Observed 95\% CL                 & Expected 95\% CL                 \\
\midrule
\multirow{2}{*}{Shape-only}                                               & \kappac    & \([-12.6,~18.3]\)                     & \([-10.1,~17.3]\)                    \\
& \kappab    & \([-3.5,~10.3]\)                      & \([-2.5,~8.1]\)                      \\
\midrule
 
\multirow{2}{12em}{Shape+normalisation (with branching ratio variations)} & \kappac    & \([-2.5,~2.3]\)                      & \([-3.0,~3.1]\)                      \\
& \kappab    & \([-1.1, -0.8]~\cup~[0.8,~1.1]\) & \([-1.2,-0.9]~\cup~[0.8,~1.2]\) \\
\bottomrule
\end{tabular}
\label{tab:YbYc_limits}
\end{table}

\subsection{Limits on anomalous Higgs boson interactions using the Effective Field Theory approach}
\label{sec:eft}
 
The strength and tensor structure of the Higgs boson interactions are investigated following an effective field theory approach. In this approach, an effective Lagrangian~\cite{Contino:2013kra} is defined by the SM Lagrangian, \({\cal L}_\textrm{SM}\) supplemented by additional dimension-6 operators, \({\cal O}^{(6)}_i\) specified by
\begin{equation}\nonumber{\cal L}_\textrm{EFT} = {\cal L}_\textrm{SM}  + \sum_i \frac{c_i}{\Lambda^{2}}{\cal O}^{(6)}_i,
\end{equation}
where the \(c_i\) specify the strengths of the new interactions and are known as the \emph{Wilson coefficients}, and \(\Lambda\) is the scale of new physics.
Contributions from new physics to the differential cross-sections are then probed as non-zero values of the Wilson coefficients of the dimension-6 operators.
Non-zero values of these Wilson coefficients can modify the event rates and the kinematic properties of the Higgs boson, and associated jet spectra, from those predicted by the SM.
 
Contributions from dimension-5 and dimension-7 operators are excluded by assuming lepton and baryon number conservation.
Operators with dimension-8 or higher are neglected as their effects are suppressed by at least \(1/\Lambda^2\) relative to dimension-6 operators.
From the available bases for parameterising the dimension-6 operators, the Warsaw basis of the Standard Model EFT (SMEFT) Lagrangian~\cite{Grzadkowski:2010es,Brivio:2017vri} is used to probe Higgs boson interactions with gauge bosons.
 
Limits on the Wilson coefficients are obtained using a simultaneous fit to five measured fiducial differential cross-sections, including their correlations, in the following variables: \ptgg, \Njets, \mjj, \dphijj\ and \ptj.
 
In the SMEFT formulation, the following operators are considered:
\begin{eqnarray*}{\cal L}_\textrm{eff}^\textrm{SMEFT} \supset  &  \CHG{\cal O}'_{g} + \CHW{\cal O}'_{HW} + \CHB{\cal O}'_{HB} + \CHWB{\cal O}'_{HWB}\\
& \quad + \CHGt \widetilde{\cal O}'_{g} + \CHWt \widetilde{\cal O}'_{HW} + \CHBt \widetilde{\cal O}'_{HB} + \CHWBt \widetilde{\cal O}'_{HWB}\,,
\end{eqnarray*}
The coefficient \CHG\ and its CP-odd counterpart \CHGt\ determine the strength of operators that affect ggF production, while  \CHW, \CHB, \CHWB\ and their corresponding CP-odd counterparts,
\CHWt, \CHBt, \CHWBt, correspond to operators that impact VBF and \(VH\) production and the Higgs boson decay to photons.
 
The effective Lagrangian is implemented in \FEYNRULES~\cite{Alloul:2013naa} within the \textsc{SMEFTsim} package~\cite{SMEFTsim}. The implementation uses \(U(3)^5\) flavour symmetry with non-SM CP-violating phases and the \(\alpha\) scheme that uses \(\alpha_{_\textrm{EW}}\), \mZ, and \(G_\textrm{F}\) as the input parameters for the electroweak sector.
Event generation was performed using \MADGRAPH[2.7.3]~\cite{Alwall:2014hca} for ggF, VBF and \(VH\) production modes with leading-order matrix elements. The event generation was performed with the BSM scale set to \(\Lambda=\SI{1}{\TeV}\).
The ggF Higgs boson events were generated with up to two additional partons in the final state and were merged using the MLM matching scheme to create the full final state~\cite{Mangano:2006rw}.
The remaining Higgs boson production modes, i.e.\ \ttH\ and \bbH, are fixed to their SM expectation.
 
For each production mode, events were generated using the \NNPDF[2.3lo] PDF set~\cite{Ball:2014uwa} and the A14 tune~\cite{ATL-PHYS-PUB-2014-021}.
The parton-level events were then passed to \PYTHIA[8] 
for parton showering, hadronisation and underlying-event simulation.
The analysis \textsc{Rivet}~\cite{Buckley:2010ar} routine is used to apply the fiducial selections and calculate the observables to obtain the differential cross-section predictions.
 
It is assumed that higher-order QCD and electroweak corrections are the same for leading-order SM predictions and leading-order predictions that contain contributions from new physics.
Therefore, to obtain predictions at a given value of the Wilson coefficient, \(c_i\), the following formula is used:
\begin{equation*}
\left(\frac{\textrm{d} \sigma}{\textrm{d}X}\right)_{c_i} = \sum_j \left( \frac{\textrm{d}\sigma_j}{\textrm{d}X} \right)^\textrm{SM} \times \left.\left(
\frac{\textrm{d}\sigma_j}{\textrm{d}X} \right)_{c_i}^\textrm{MG5} \middle/ \left( \frac{\textrm{d}\sigma_j}{\textrm{d}X} \right)_{c_i=0}^\textrm{MG5}\right.,
\end{equation*}
where the summation \(j\) is over the different Higgs boson production mechanisms, `MG5' labels the \MADGRAPH\ predictions and `SM' labels the default SM predictions.
The `MG5' cross-sections at any given value of the Wilson coefficients are obtained using a multidimensional interpolation with the \textsc{Professor} method~\cite{Buckley:2009bj}. The interpolation relies on BSM predictions generated at benchmark non-zero Wilson coefficients. The `MG5' cross-section at a given value of a Wilson coefficient can be separated into three components:
\begin{equation}\nonumber
\sigma \propto |{\cal M}_{\text{EFT}}|^2 = |{\cal M}_{\text{SM}}|^2 + 2Re({\cal M}_\mathrm{SM}^{*}{\cal M}_{\text{d6}}) + |{\cal M}_{\text{d6}}|^2 ~~,
\end{equation}
where the first term is the dimension-4 squared matrix element for the SM, independent of the Wilson coefficients ${c_{i}}$, the second term represents the interference between the SM operators and the dimension-6 operators, which is of order \(c_{i}/\Lambda^{2}\) and therefore linear in ${c_{i}}$, and the last term is the squared matrix element for the dimension-6 operators, which is of order  \(c_{i}^2/\Lambda^{4}\) and thus quadratic in ${c_{i}}$.
For small values of the Wilson coefficients, the interference term, \({\sim}c_i/\Lambda^2\), is the dominant BSM contribution to the cross-section.
For CP-odd operators, the separation of the SM--BSM interference components from the pure BSM components allows purely CP-violating effects to be probed.
Samples of approximately \num{400000} events were generated for each production mode for different values of each Wilson coefficient and used to derive the parameterisation of the observables as a function of $c_i$.
 
Non-zero values of the Wilson coefficients cause changes in the \Hgg\ partial width and the Higgs boson total decay width, and therefore in the \Hgg\ branching ratio. The modifications of the \(H\to\gamma\gamma\) partial width and the Higgs boson total decay width, including both the interference-only and the interference-plus-quadratic terms, were computed using \MADGRAPH.
It was verified that when neglecting quadratic terms the results agree with those obtained using the linear coefficients of the interference term provided in Ref.~\cite{Brivio:2019myy}.
 
The combined effect of the modifications induced by the Wilson coefficients $c_i$ in both the cross-section and the branching ratio ($B$) leads to the following expansion, as a function of $c_i/\Lambda^2$, of the product $\sigma\times B$:
\begin{equation}\nonumber
\sigma \times B = \sigma_\mathrm{SM}B_\mathrm{SM} + \left(\sigma _\mathrm{SM} B_\mathrm{INT} + \sigma_\mathrm{INT} B_\mathrm{SM}\right) + \left( \sigma _\mathrm{SM} B_\mathrm{BSM} + \sigma_\mathrm{BSM} B_\mathrm{SM}\right) + \mathcal{O}(c_i^3/\Lambda^6),
\end{equation}
where the product $\sigma_\mathrm{SM}B_\mathrm{SM}$ of the SM cross-sections and branching ratio, independent of $c_i$, is modified by the first term in parentheses, linear in $c_i$, which results from the SM--BSM interference effects on either the cross-section ($\sigma_\mathrm{INT}$) or the branching ratio ($B_\mathrm{INT}$), and by the second term in parentheses, quadratic in $c_i$, arising from the product of the SM cross-section and the BSM term of the branching ratio and vice versa.
Following the recommendations from the authors of Ref.~\cite{Brivio:2019myy}, quadratic terms of the form $\sigma _\mathrm{INT} B_\mathrm{INT}$ are excluded, since otherwise contributions from dimension-8 operators would need to be included as well in order to guarantee to the renormalisability of the processes under study.
 
\begin{figure}[htbp!]
\centering
\subfloat[]{\includegraphics[width=0.5\textwidth]{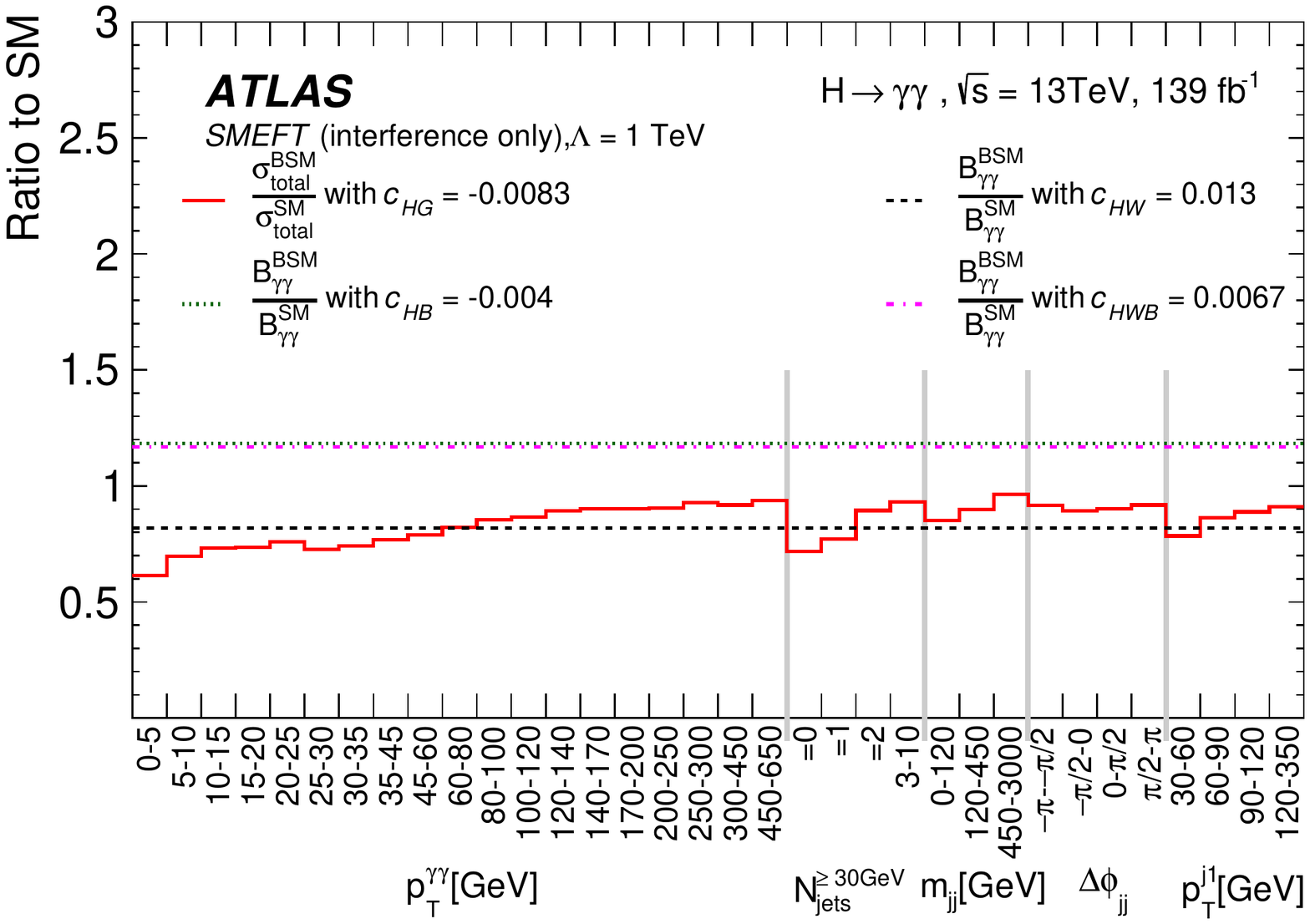}\label{fig:eftmodificationsSMEFT_a}}
\subfloat[]{\includegraphics[width=0.5\textwidth]{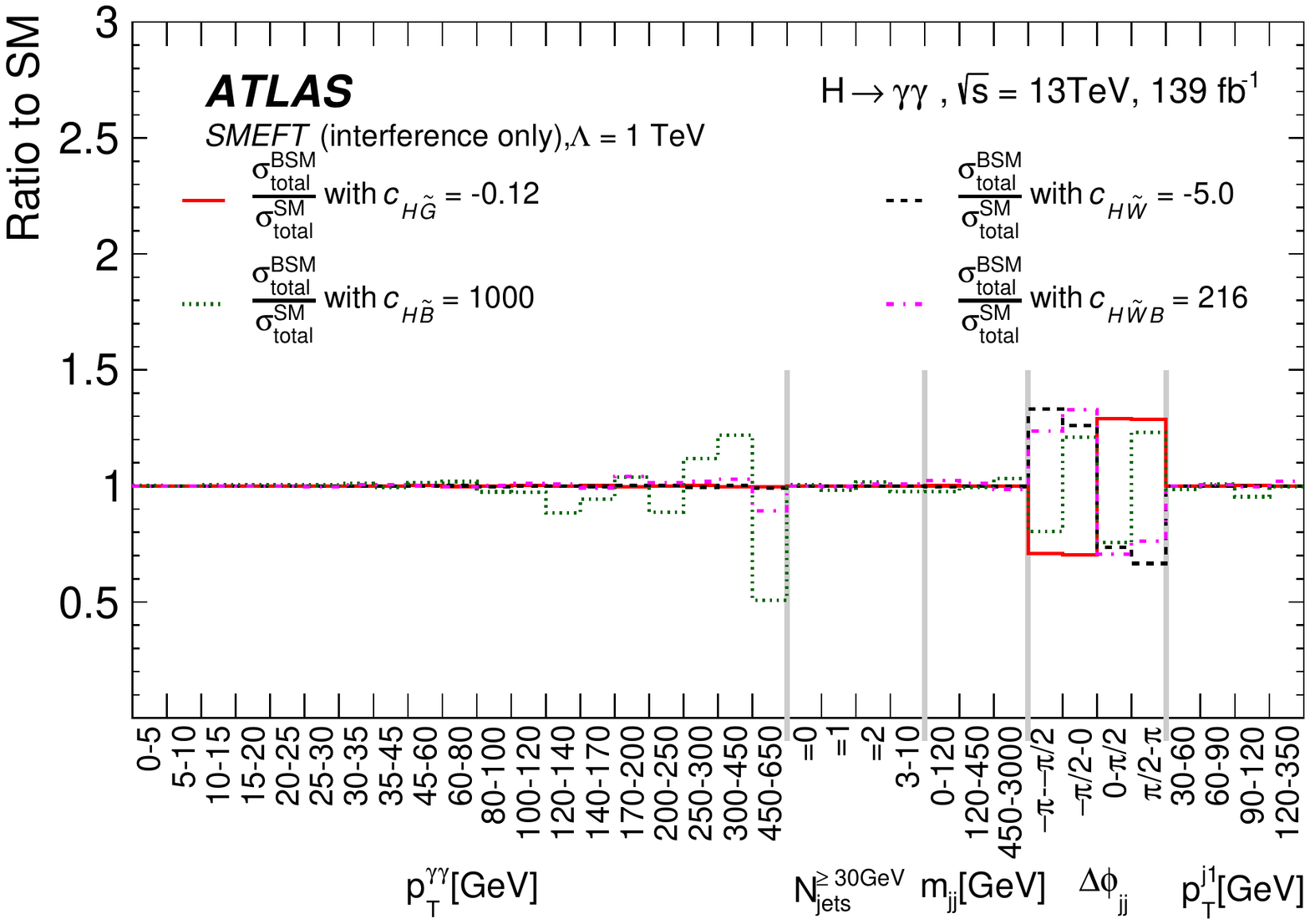}\label{fig:eftmodificationsSMEFT_b}}
\caption{The effect on the five differential distributions used in the analysis
of \protect\subref{fig:eftmodificationsSMEFT_a} the CP-even coefficients \CHG, \CHB, \CHW, \CHWB\ and \protect\subref{fig:eftmodificationsSMEFT_b} the CP-odd coefficients
\CHGt, \CHBt, \CHWt, \CHWBt\ of the SMEFT effective Lagrangian for values of the coefficients close to the expected limits. The \CHB, \CHW, \CHWB\ variations at the expected limits affect mainly the $H\to\gamma\gamma$ branching ratio with negligible effects on the cross-section. The effect is shown at a new-physics scale \(\Lambda=\SI{1}{\TeV}\).	\label{fig:eftmodificationsSMEFT}}
\end{figure}

Figure~\ref{fig:eftmodificationsSMEFT} shows modifications of the differential cross-sections used for benchmark non-zero values of SMEFT Wilson coefficients.
The coefficient \CHG\ and its CP-odd counterpart \CHGt\ affect ggF production while \CHB, \CHW, \CHWB\ and their CP-odd counterparts affect VBF+\(VH\) production.
The main effect of \CHB, \CHW\ and also \CHWB, however, is on the \Hgg\ decay rate, impacting the overall normalisation.
The CP-odd coefficients, as seen in Figure~\ref{fig:eftmodificationsSMEFT}, exhibit sensitivity only to the \dphijj\ observable when only the interference term is considered~\cite{Azatov:2016sqh}.
 
\paragraph{Statistical interpretation}
Limits on Wilson coefficients are set by constructing a likelihood function which is defined, up to a constant normalisation factor, as
 
\begin{equation}\nonumber
L =  \exp \left[-\frac{1}{2} \left( \sigma_\text{obs} -  \sigma_\text{pred} \right)^\text{T} C^{-1} \left( \sigma_\text{obs} - \sigma_\text{pred}  \right) \right] \, ,
\end{equation}
where \(\sigma_\text{obs}\) and \(\sigma_\text{pred}\) are \(k\)-dimensional vectors from the measured and predicted differential cross-sections of the five analysed observables, with \(k=34\) equal to the total number of bins of the five distributions used in the fit, \(C = C_\text{stat} + C_\text{syst} + C_\text{theo}\) is the \(k\times k\) total covariance matrix defined as the sum of the statistical, systematic and theoretical covariances. The overflow bins for \ptgg, \mjj\ and \ptj[1] are not used in the limit-setting fit as they extend beyond the assumed new-physics scale \(\Lambda=\SI{1}{\TeV}\).
 
\begin{figure}[!htbp]
\centering
\includegraphics[width=1.0\textwidth,page=1]{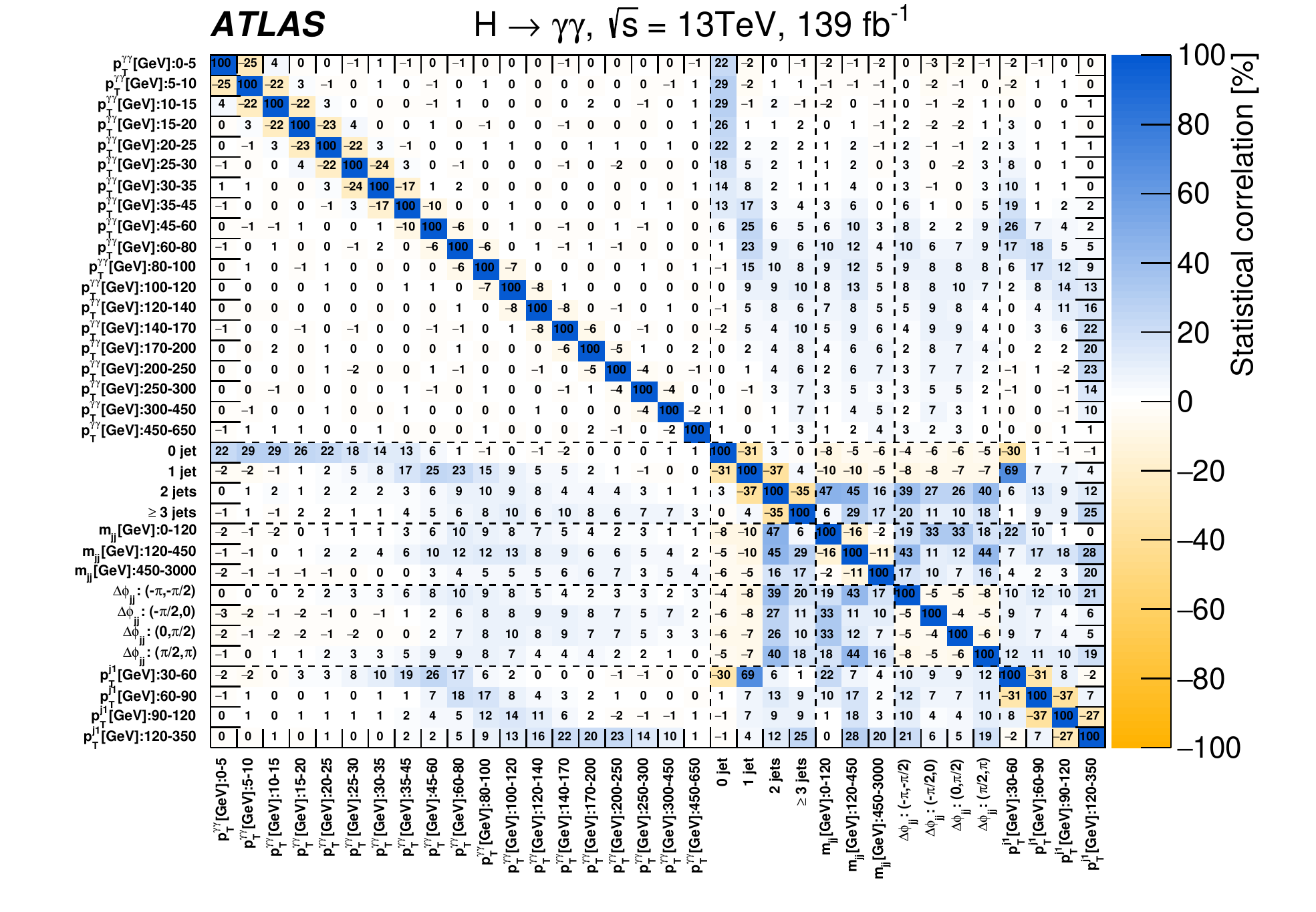}
\caption{The observed statistical correlations, evaluated with a bootstrapping technique, between \ptgg, \Njets, \mjj, \dphijj\ and \ptj\ are shown}
\label{fig:eftCorrelations}
\end{figure}

The statistical covariance matrix is obtained with a bootstrapping technique and the resulting correlation matrix shown in Figure~\ref{fig:eftCorrelations}.
The matrix provides a measure of the statistical correlations between cross-section bins because the same events in data will populate the different observables used in the fit.
 
The covariance matrices for systematic and theoretical uncertainties are constructed from the uncertainties listed in Section~\ref{sec:sysunc}.
Theoretical uncertainties are considered for the different production modes using the default SM MC simulation to estimate the effect of QCD scale and PDF variations, detailed in Section~\ref{subsec:theoryunc}, and are considered to be independent of new physics.
Identical sources are assumed to be fully correlated across bins and variables. In addition, nuisance parameters are included in the fit to account for limited MC sample size, typically affecting the highest \ptgg\ and \mjj\ bins.
In what follows, the likelihood function is numerically maximised to determine \(L_\textrm{max}\) and
confidence intervals for one or several Wilson coefficients are determined via
\begin{align*}
1 - \text{CL} = \int_{x}^{\infty} \, \text{d} x' \, f(x') \, ,
\end{align*}
where $f(x')$ denotes the distribution of the test statistic $x'= -2\ln ( L(c_{i}') / L_\textrm{max})$,
$L(c_i')$ denotes the likelihood value evaluated for a given Wilson coefficient value $c_i'$, and
$x= -2 \ln (L(c_i)/L_\textrm{max})$ determines the value $c_i$ that bounds the confidence interval
at the desired CL.
 
The coverage of 68\% and 95\% CL limits using the likelihood ratio scan was validated using pseudo-experiments.
 
\begin{figure}[!htbp]
\centering
\subfloat[]{\includegraphics[width=0.5\textwidth,page=1]{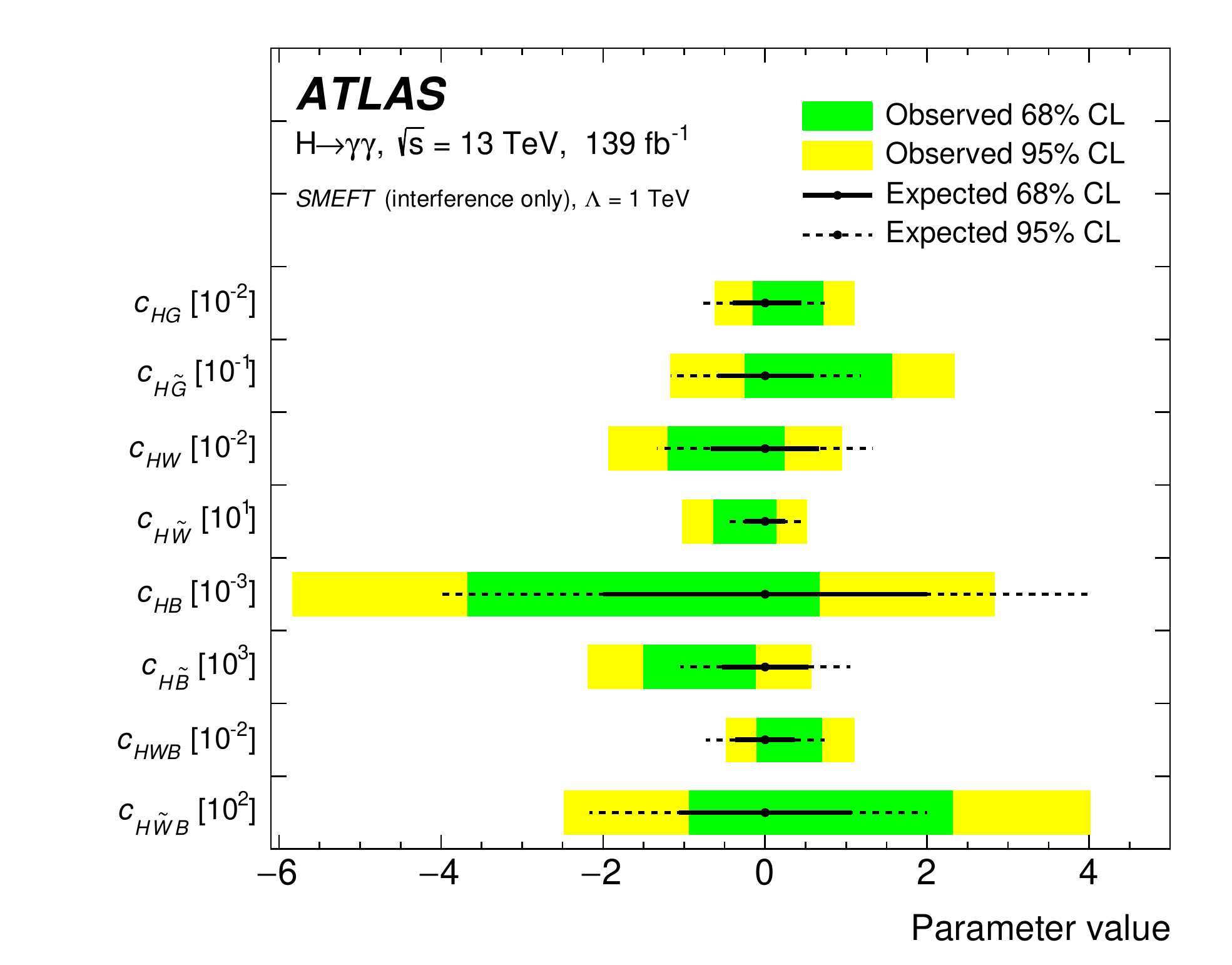}\label{fig:SMEFT1dIndividual_a}}
\subfloat[]{\includegraphics[width=0.5\textwidth,page=1]{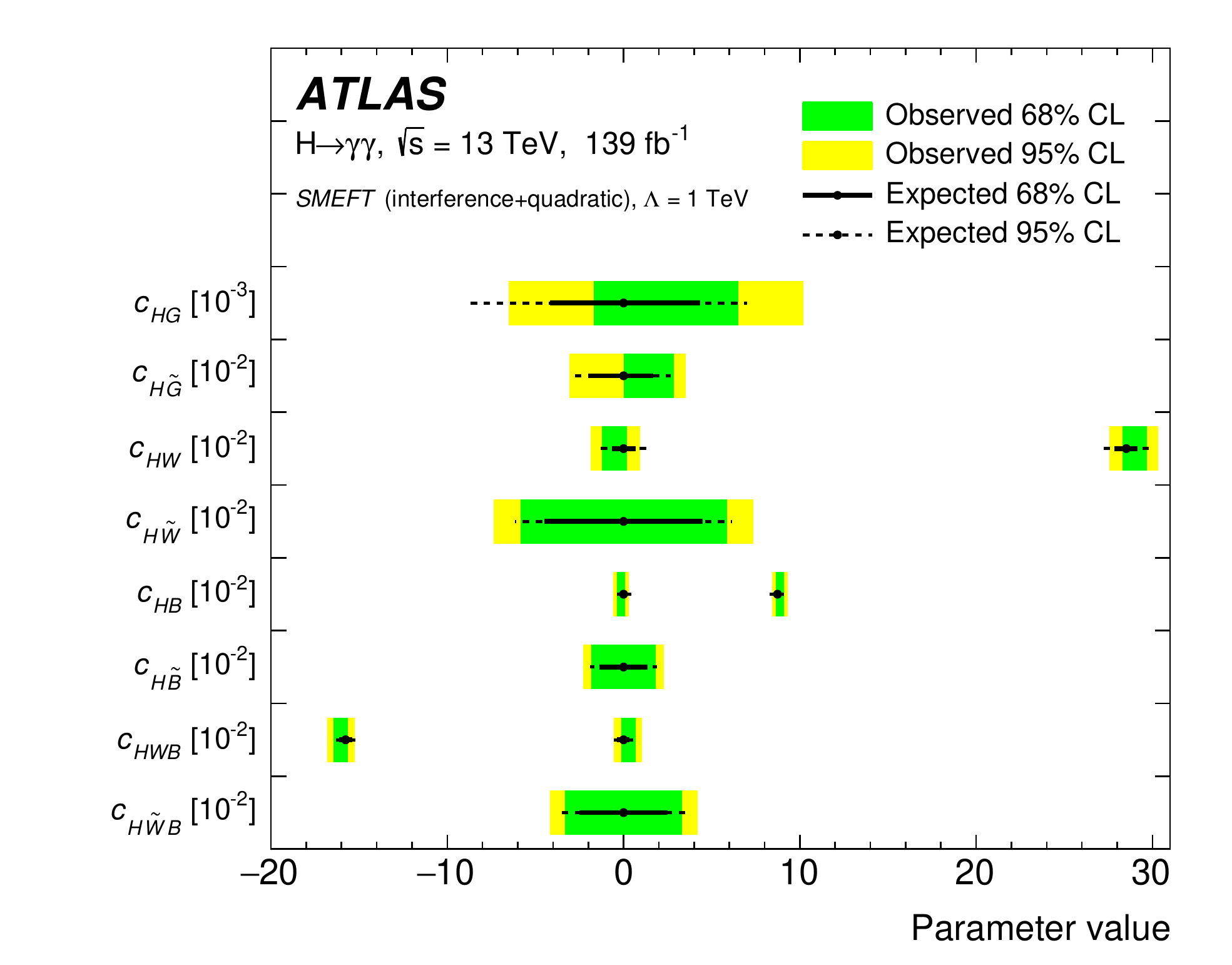}\label{fig:SMEFT1dIndividual_b}}
\caption{Observed and expected 68\% and 95\% CL limits on SMEFT Wilson coefficients using \protect\subref{fig:SMEFT1dIndividual_a} SM and dimension-6 operators interference-only terms and \protect\subref{fig:SMEFT1dIndividual_b} including quadratic dimension-6 terms. Limits are derived by fitting one Wilson coefficient at a time while setting the other coefficients to zero. The limits are computed at a new-physics scale \(\Lambda=\SI{1}{\TeV}\). \label{fig:SMEFT1dIndividual}}
\end{figure}

In Table~\ref{tab:1dSMEFT}, the observed 95\% CL limits are shown for the considered Wilson coefficients. The limits are also summarised in Figure~\ref{fig:SMEFT1dIndividual}.
The limits are obtained assuming that all Wilson coefficients other than the one quoted are zero. The limits are provided for two scenarios, one where predictions are obtained using only the interference term, and the other where both the interference term and the quadratic term are included.
Since the interference terms dominate the predicted cross-sections, the limits in the two approaches are very similar for coefficients of CP-even operators.
Significant differences emerge for the CP-odd ones, for which the interference term cross-section vanishes for CP-even observables, and the sensitivity to pure CP-violating effects is obtained through the \dphijj\ observable.
The results place stringent limits on all CP-even operators, as they affect mainly the normalisation of the five distributions through either the production cross-section, as for \CHG, or the \Hgg\ branching ratio, as for \CHW, \CHB\ and \CHWB.
The results show that the current \dphijj\ measurement can only constrain the \CHGt\ and \CHWt\ coefficients, as the cross-section is dominated by ggF and VBF (which is dominated by \(WW\) fusion). In contrast, the very loose limits on \CHBt\ and \CHWBt\ indicate a breakdown of the EFT regime and unitarity constraints, and the lack of sensitivity to these coefficients at the current measurement accuracy. In addition, two-dimensional limits are derived, allowing two Wilson coefficients (a CP-even coefficient and its CP-odd counterpart) to vary simultaneously, using the interference-only cross-section and including the quadratic dimension-6 cross-section, and these are shown in Appendix~\ref{sec:aux eft 2d}.
 
\begin{table}[htbp!]
\centering
\caption{The 95\% CL observed limits on the \CHG, \CHW, \CHB, \CHWB\ Wilson coefficients of the SMEFT basis and their CP-odd counterparts using interference-only terms and using both the interference and quadratic terms.
Limits are derived by fitting one Wilson coefficient at a time while setting the other coefficients to zero. The limits are computed at a new-physics scale \(\Lambda=\SI{1}{\TeV}\).
\label{tab:1dSMEFT} }{\begin{tabular}{lcc}
\toprule
Coefficient & 95\% CL, interference-only terms & 95\% CL, interference and quadratic terms                  \\
\midrule
\CHG        & \([-6.1,11.0]\times 10^{-3}\)    & \([-6.5,10.2]\times 10^{-3}\)                               \\
\CHGt       & \([-0.12, 0.23]\)                & \([-3.1,3.5]\times 10^{-2} \)                              \\
 
\CHW        & \([-1.9,0.9]\times 10^{-2} \)    & \([-1.8,1.0]\times 10^{-2} \cup [0.28, 0.30]\)             \\
\CHWt       & \([-10.2,5.2]\)                  & \([-7.3,7.3]\times 10^{-2}\)                               \\
 
\CHB        & \([-5.8,2.8]\times 10^{-3}\)     & \([-5.5,3.0]\times 10^{-3} \cup [8.4, 9.3]\times 10^{-2}\) \\
\CHBt       & \([-21.8,5.7]\times 10^{2}\)     & \([-2.3,2.3]\times 10^{-2}\)                               \\
 
\CHWB       & \([-5.2,10.7]\times 10^{-3}\)    & \([-0.17,-0.15] \cup [-5.5, 9.8]\times 10^{-3}\)          \\
\CHWBt      & \([-2.5,4.0]\times 10^{2}\)      & \([-4.0,4.0]\times 10^{-2}\)                               \\
\bottomrule
\end{tabular}}
\end{table}
 
\FloatBarrier
\section{Summary and conclusions}
\label{sec:conclusion}
Measurements of Higgs boson fiducial cross-sections in the diphoton decay channel are performed using \pp\ collision data recorded by the ATLAS experiment at the LHC, assuming the Higgs boson mass to be \SI{125.09}{\GeV}.
The data were taken at a centre-of-mass energy of \(\sqrt{s}=\SI{13}{\TeV}\) and correspond to the full \RunTwo\ data set with an integrated luminosity of \SI{139}{\femto\barn^{-1}}.
 
The measurements are performed in a diphoton fiducial region requiring two isolated photons with transverse momentum greater than 35\% and 25\% of the diphoton invariant mass,
and with \(|\eta|<2.37\), excluding the region of \(1.37 < |\eta|<1.52\).
The inclusive fiducial cross-section times branching ratio is measured to be
\begin{align*}
\sigma_\textrm{fid} = 67\pm5~\text{(stat.)}\pm4~\text{(sys.)}~~\si{\femto\barn},
\end{align*}
which is in agreement with the Standard Model prediction of \SI[multi-part-units=single]{64\pm 4}{\femto\barn}. The measurement has a total relative uncertainty of 10\% with nearly equal contributions from the statistical and the systematic uncertainties.
The inclusive fiducial cross-section is also extrapolated to the full phase space, leading to a total Higgs production cross-section of \(58 \pm 4~\text{(stat.)} \pm 4~\text{(sys.)}~\si{\pico\barn}\), in agreement with the SM prediction of \SI[multi-part-units=single]{55.6 \pm 2.7}{\pico\barn}.
 
In addition, cross-section measurements are reported in various fiducial regions probing Higgs boson production from vector-boson fusion or associated with large missing transverse momentum, leptons or top quarks. The measured cross-sections times branching ratio for the these fiducial regions are:
\begin{align*}
\sigma_\text{VBF-enhanced}       & = 1.8~~\pm0.5~~~\text{(stat.)} \pm0.3~~~\text{(sys.)}~~\si{\femto\barn}, \\
\sigma_{N_{\text{lepton} \geq 1}} & = 0.81 \pm0.23~\text{(stat.)} \pm0.06~\text{(sys.)}~~\si{\femto\barn}, \\
\sigma_{\text{High \met}}             & = 0.28  \pm0.27~\text{(stat.)} \pm0.07~\text{(sys.)}~~\si{\femto\barn}, \\
\sigma_\text{\ttH-enhanced}        &= 0.53 \pm0.27~\text{(stat.)} \pm0.06~\text{(sys.)}~~\si{\femto\barn}, \\
\end{align*}
which show no significant deviation from the Standard Model predictions. The fiducial cross-sections for different inclusive and exclusive jet multiplicities are also measured and compared with different state-of-the-art Standard Model predictions.
 
Twenty differential cross-sections and four double-differential cross-sections are reported for events belonging to the inclusive diphoton fiducial region, as a function of kinematic variables of the diphoton system or of jets produced in association with the Higgs boson. These cross-sections are sensitive to the different Higgs boson production kinematics, jet kinematics, spin, and CP quantum numbers of the Higgs boson. Among the measured cross-sections is a new measurement of the cross-section in the high Higgs boson transverse momentum region, providing the strongest limits to date for the Higgs boson production cross-section above \SI{450}{\GeV}. The reported cross-sections include new measurements in regions of the phase space probing jet-veto resummation effects. In addition, four differential cross-sections and one double-differential cross-section were measured for events belonging to the VBF-enhanced region, probing VBF kinematics and CP properties. All the measured differential cross-sections are compared with various Standard Model predictions, and do not exhibit significant deviations from them.
 
The measured differential cross-sections as a function of \ptgg\ were used to derive limits on the bottom- and charm-quark Yukawa coupling modifiers, \kappab\ and \kappac. These limits were derived using \ptgg\ distribution shape and normalisation variations.
This analysis sets a 95\% CL allowed range \([-3.5,~10.3]\) for \kappab, and \([-12.6,~18.3]\) for \kappac, using only the shape of the \ptgg\ distribution. More stringent constraints were derived using shape-and-normalisation information.
 
The strength and tensor structure of the Higgs boson interactions was investigated using five measured differential cross-sections as functions of \ptgg, \Njets, \mjj, \dphijj\ and \ptj\ in the effective field theory framework. In this framework, the SM Lagrangian is complemented with additional CP-even and CP-odd dimension-6 operators in the SMEFT Warsaw basis. Given the level of agreement between the measured cross-sections and the SM predictions, stringent limits were placed on the CP-even Wilson coefficients. Looser limits were placed on the CP-odd Wilson coefficients that only cause shape modifications of the CP-sensitive \dphijj\ distribution.


\section*{Acknowledgements}

We thank CERN for the very successful operation of the LHC, as well as the
support staff from our institutions without whom ATLAS could not be
operated efficiently.
 
We acknowledge the support of
ANPCyT, Argentina;
YerPhI, Armenia;
ARC, Australia;
BMWFW and FWF, Austria;
ANAS, Azerbaijan;
SSTC, Belarus;
CNPq and FAPESP, Brazil;
NSERC, NRC and CFI, Canada;
CERN;
ANID, Chile;
CAS, MOST and NSFC, China;
Minciencias, Colombia;
MEYS CR, Czech Republic;
DNRF and DNSRC, Denmark;
IN2P3-CNRS and CEA-DRF/IRFU, France;
SRNSFG, Georgia;
BMBF, HGF and MPG, Germany;
GSRI, Greece;
RGC and Hong Kong SAR, China;
ISF and Benoziyo Center, Israel;
INFN, Italy;
MEXT and JSPS, Japan;
CNRST, Morocco;
NWO, Netherlands;
RCN, Norway;
MEiN, Poland;
FCT, Portugal;
MNE/IFA, Romania;
JINR;
MES of Russia and NRC KI, Russian Federation;
MESTD, Serbia;
MSSR, Slovakia;
ARRS and MIZ\v{S}, Slovenia;
DSI/NRF, South Africa;
MICINN, Spain;
SRC and Wallenberg Foundation, Sweden;
SERI, SNSF and Cantons of Bern and Geneva, Switzerland;
MOST, Taiwan;
TAEK, Turkey;
STFC, United Kingdom;
DOE and NSF, United States of America.
In addition, individual groups and members have received support from
BCKDF, CANARIE, Compute Canada and CRC, Canada;
COST, ERC, ERDF, Horizon 2020 and Marie Sk{\l}odowska-Curie Actions, European Union;
Investissements d'Avenir Labex, Investissements d'Avenir Idex and ANR, France;
DFG and AvH Foundation, Germany;
Herakleitos, Thales and Aristeia programmes co-financed by EU-ESF and the Greek NSRF, Greece;
BSF-NSF and GIF, Israel;
Norwegian Financial Mechanism 2014-2021, Norway;
NCN and NAWA, Poland;
La Caixa Banking Foundation, CERCA Programme Generalitat de Catalunya and PROMETEO and GenT Programmes Generalitat Valenciana, Spain;
G\"{o}ran Gustafssons Stiftelse, Sweden;
The Royal Society and Leverhulme Trust, United Kingdom.
 
The crucial computing support from all WLCG partners is acknowledged gratefully, in particular from CERN, the ATLAS Tier-1 facilities at TRIUMF (Canada), NDGF (Denmark, Norway, Sweden), CC-IN2P3 (France), KIT/GridKA (Germany), INFN-CNAF (Italy), NL-T1 (Netherlands), PIC (Spain), ASGC (Taiwan), RAL (UK) and BNL (USA), the Tier-2 facilities worldwide and large non-WLCG resource providers. Major contributors of computing resources are listed in Ref.~\cite{ATL-SOFT-PUB-2021-003}.
 
\clearpage
\appendix
\part*{Appendix}
\addcontentsline{toc}{part}{Appendix}
\section{Correlation matrices for differential cross-section measurements}
\label{sec:aux diff xs corr}
In this section, the correlation matrices for the differential cross-section measurements presented in Section~\ref{sec:xsec_results} are shown.
 
\begin{figure}[htbp]
\centering
\subfloat[]{\includegraphics[width=0.48\textwidth]{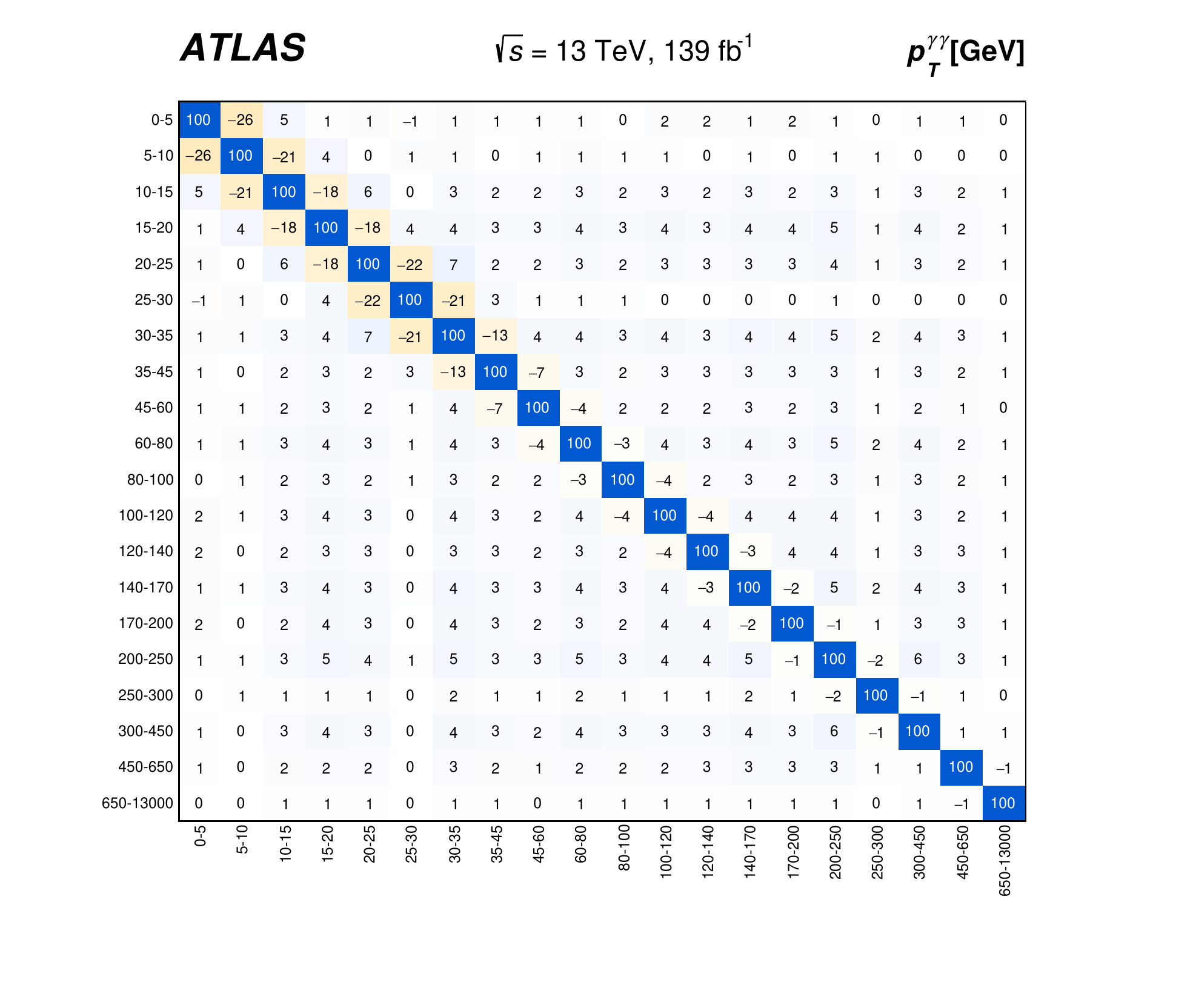}\label{fig:data_unfolded_xsections_matrixinversion_1D_photon_0_c}}
\subfloat[]{\includegraphics[width=0.48\textwidth]{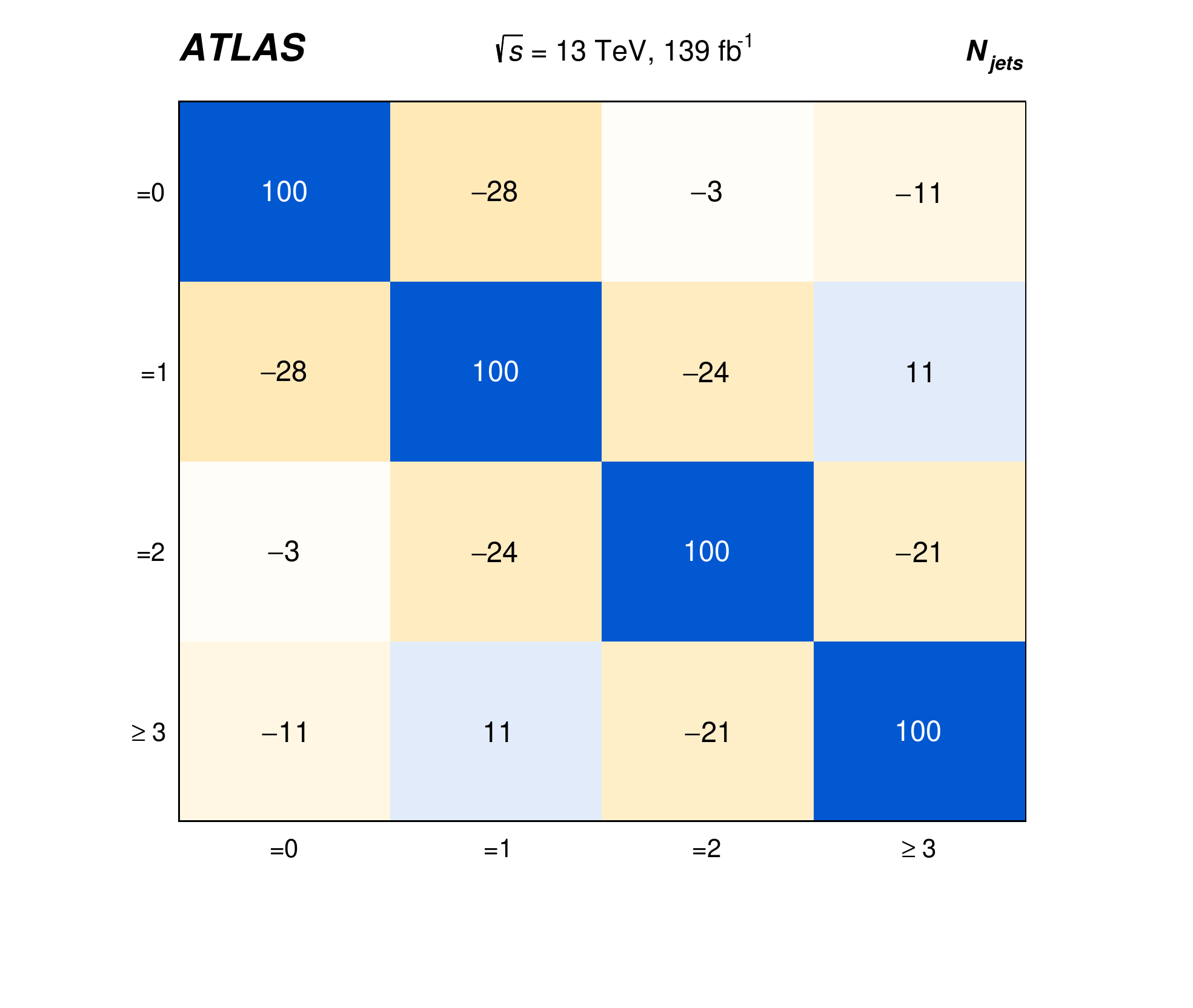}\label{fig:data_unfolded_xsections_matrixinversion_1D_njet_2}} \\
\subfloat[]{\includegraphics[width=0.48\textwidth]{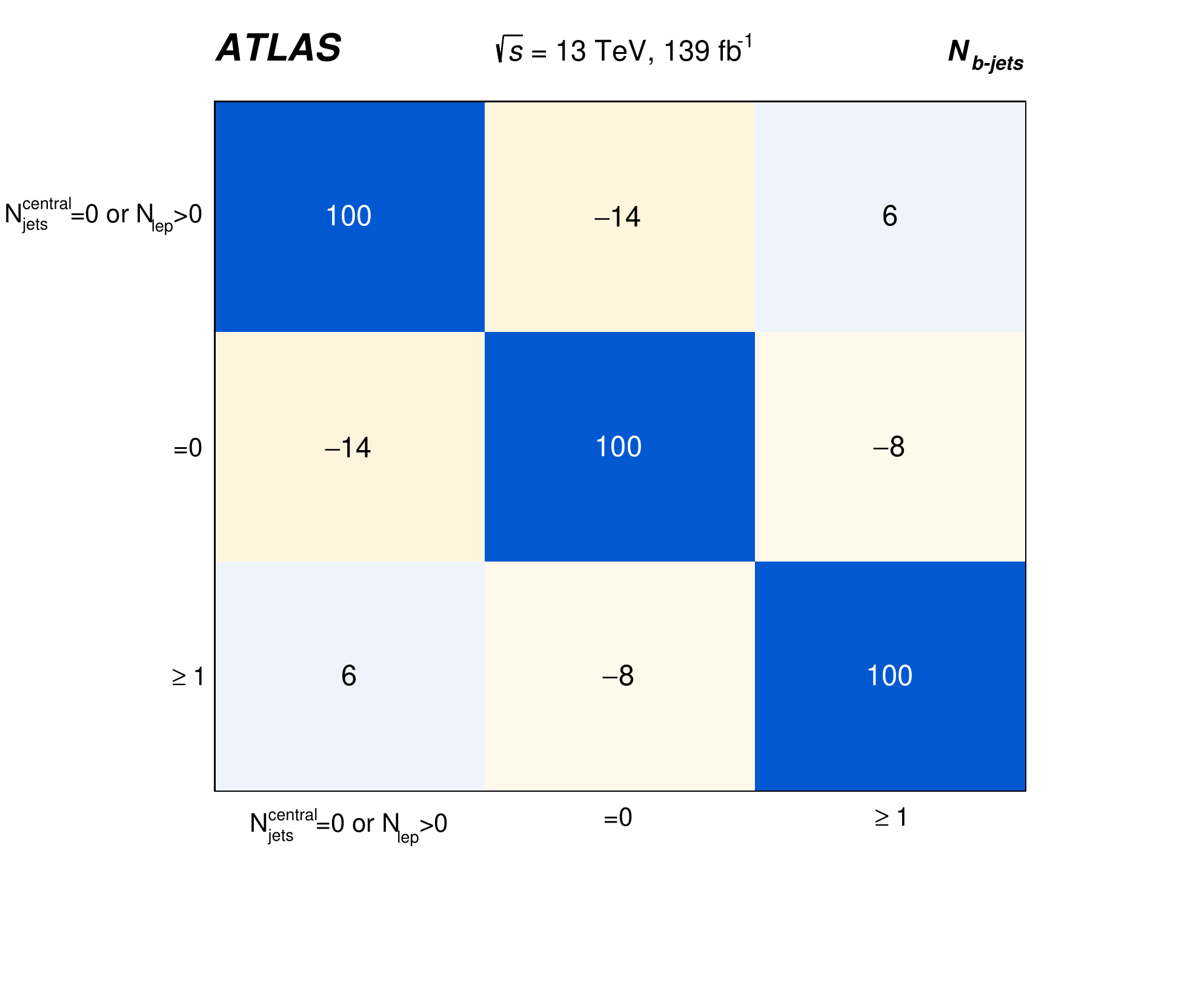}\label{fig:data_unfolded_xsections_matrixinversion_1D_nbjet_b}}
\subfloat[]{\includegraphics[width=0.48\textwidth]{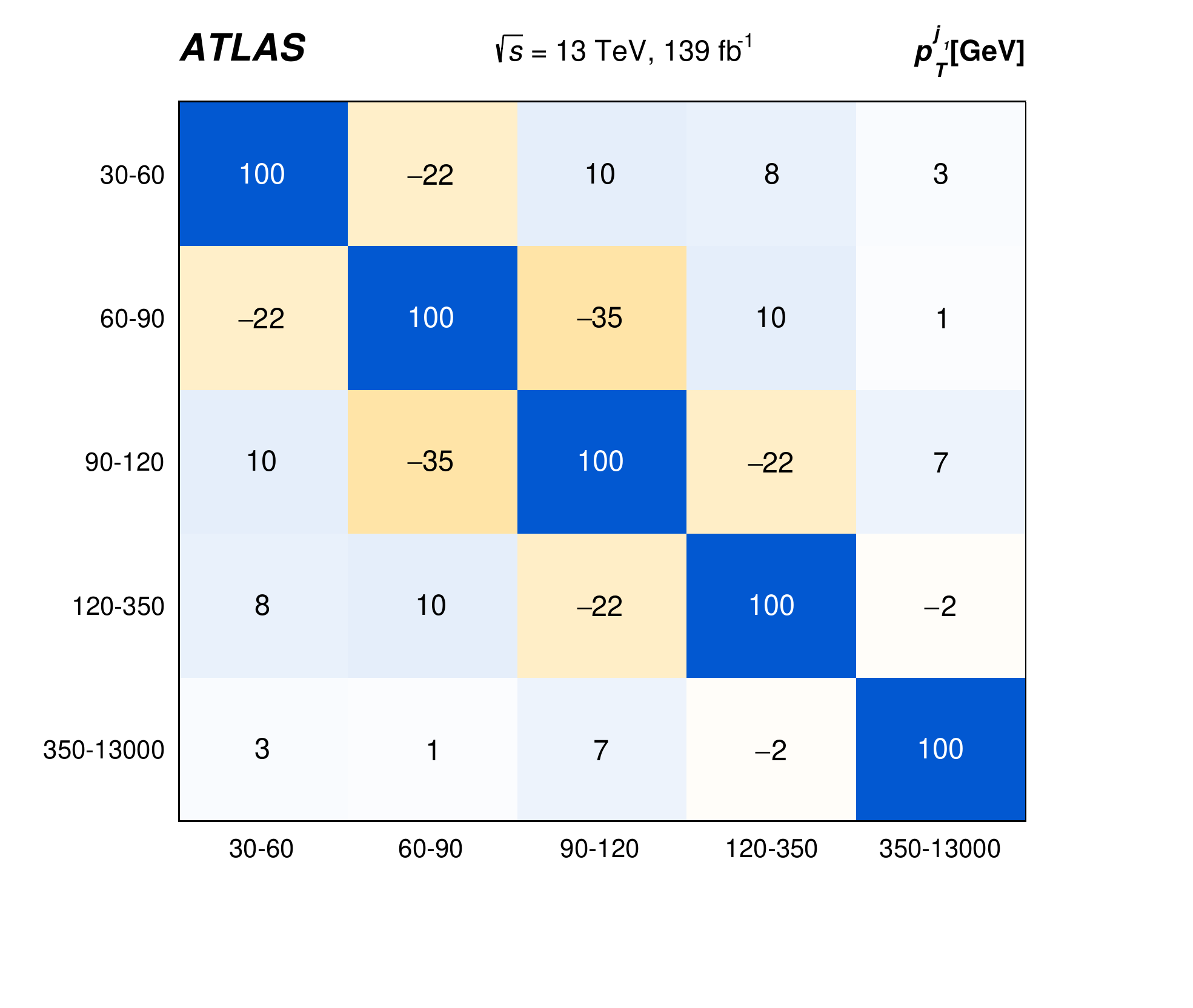}\label{fig:data_unfolded_xsections_matrixinversion_1D_1jet_1_b}}
\caption{Correlation matrices for the differential cross-section variables: \protect\subref{fig:data_unfolded_xsections_matrixinversion_1D_photon_0_c} \ptgg, \protect\subref{fig:data_unfolded_xsections_matrixinversion_1D_njet_2} exclusive \Njets\ distributions for all jets with $p_T^{j}>30~\GeV$, \protect\subref{fig:data_unfolded_xsections_matrixinversion_1D_nbjet_b} $b$-jets multiplicity variable \Nbjets (see caption of Figure~\ref{fig:data_unfolded_xsections_matrixinversion_1D_nbjet} for more details) and \protect\subref{fig:data_unfolded_xsections_matrixinversion_1D_1jet_1_b} \ptj[1].
}
\label{fig:data_unfolded_xsections_matrices_1}
\end{figure}

\begin{figure}[htbp]
\centering
\subfloat[]{\includegraphics[width=0.48\textwidth]{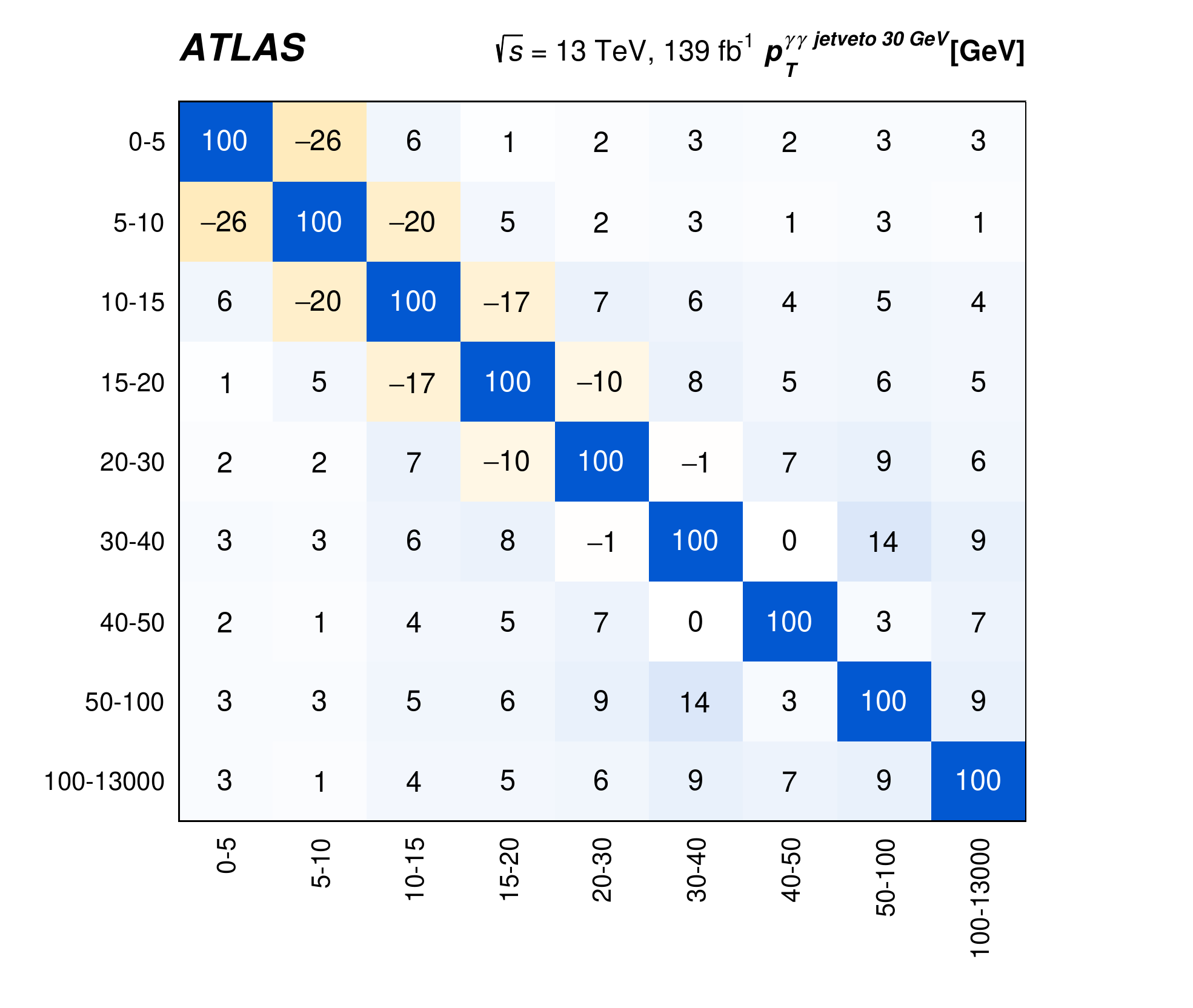}\label{fig:data_unfolded_xsections_matrixinversion_1D_JV_1_b}}
\subfloat[]{\includegraphics[width=0.48\textwidth]{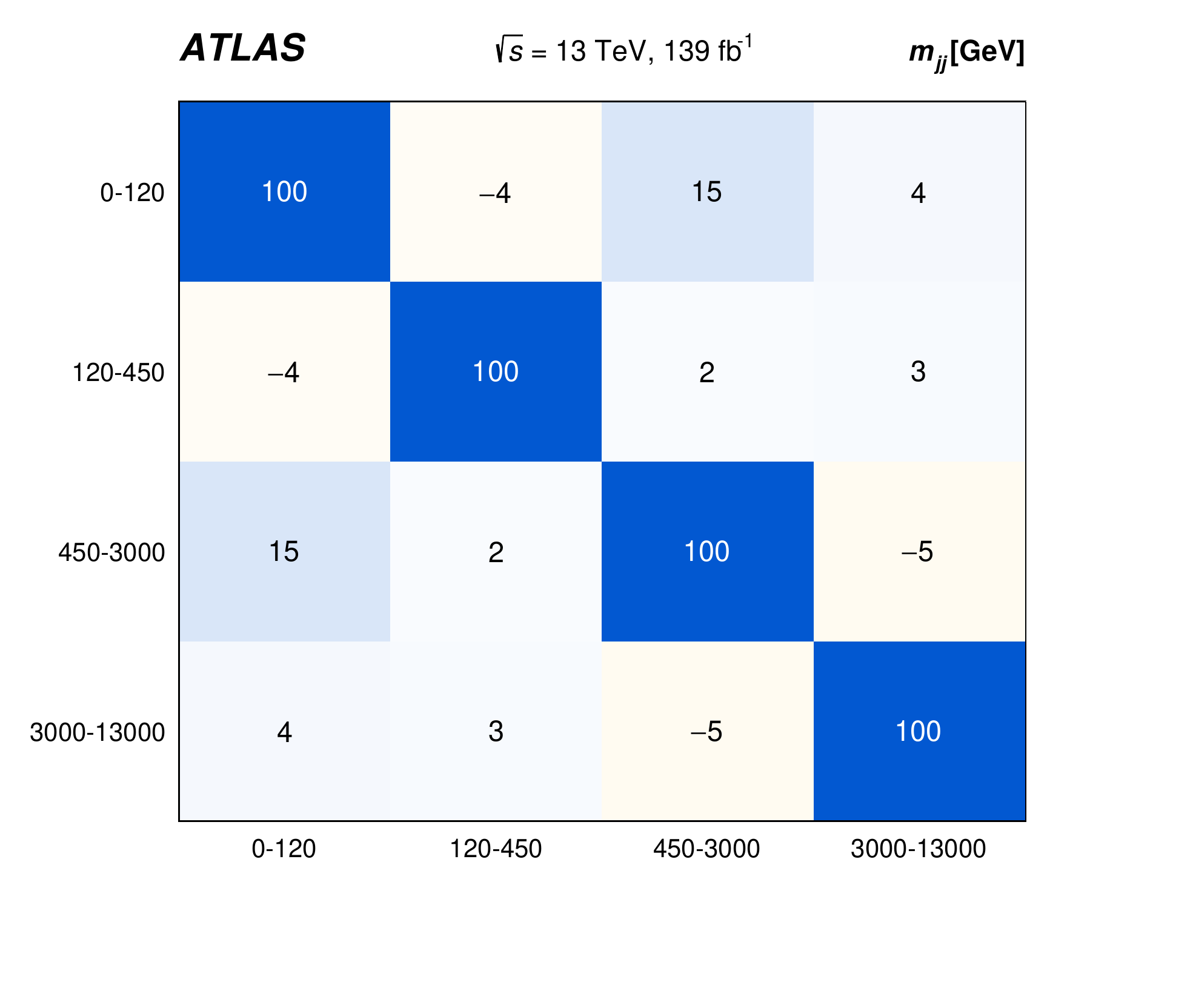}\label{fig:data_unfolded_xsections_matrixinversion_1D_2jet_b}} \\
\subfloat[]{\includegraphics[width=0.48\textwidth]{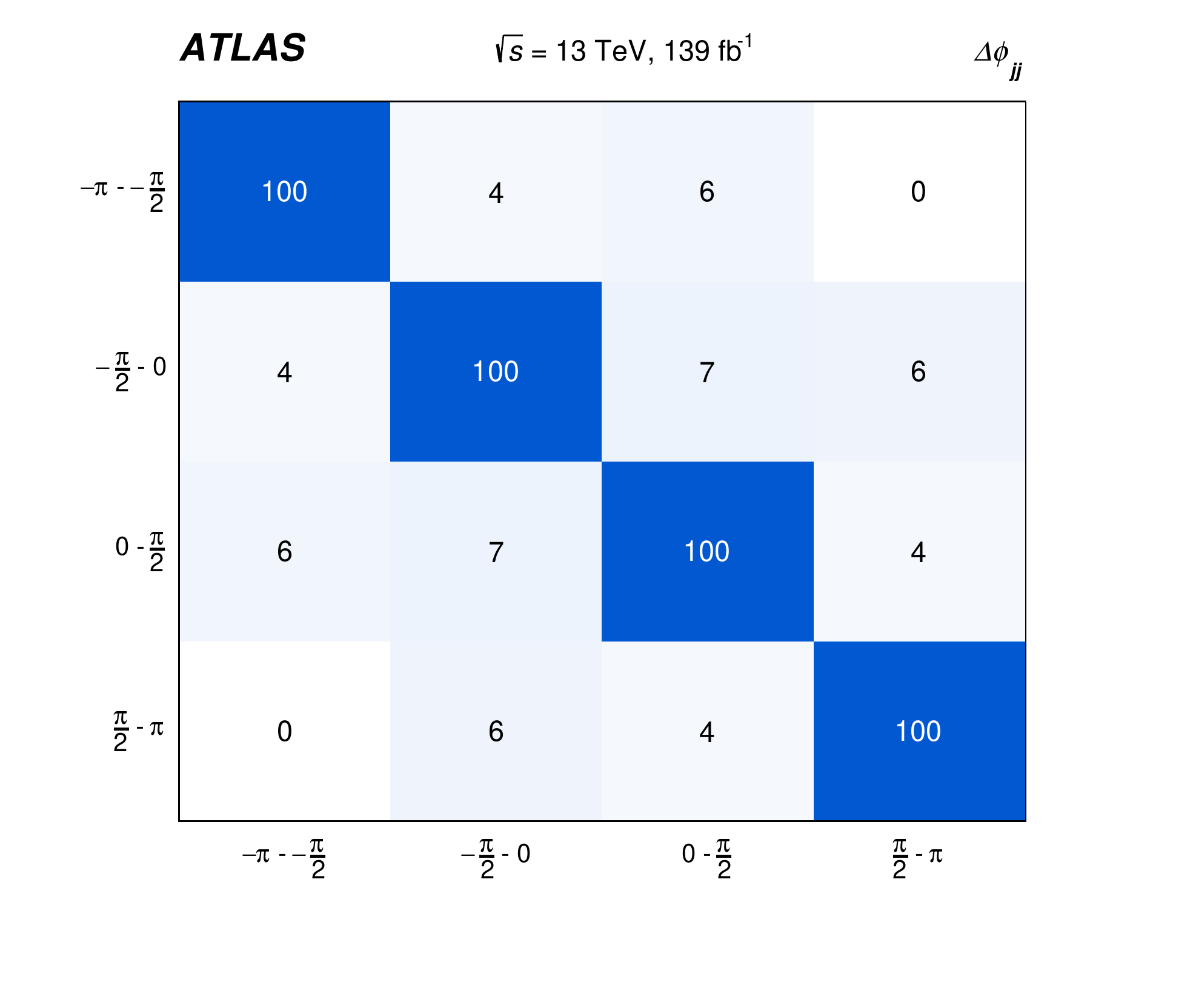}\label{fig:data_unfolded_xsections_matrixinversion_1D_2jet_d}}
\subfloat[]{\includegraphics[width=0.48\textwidth]{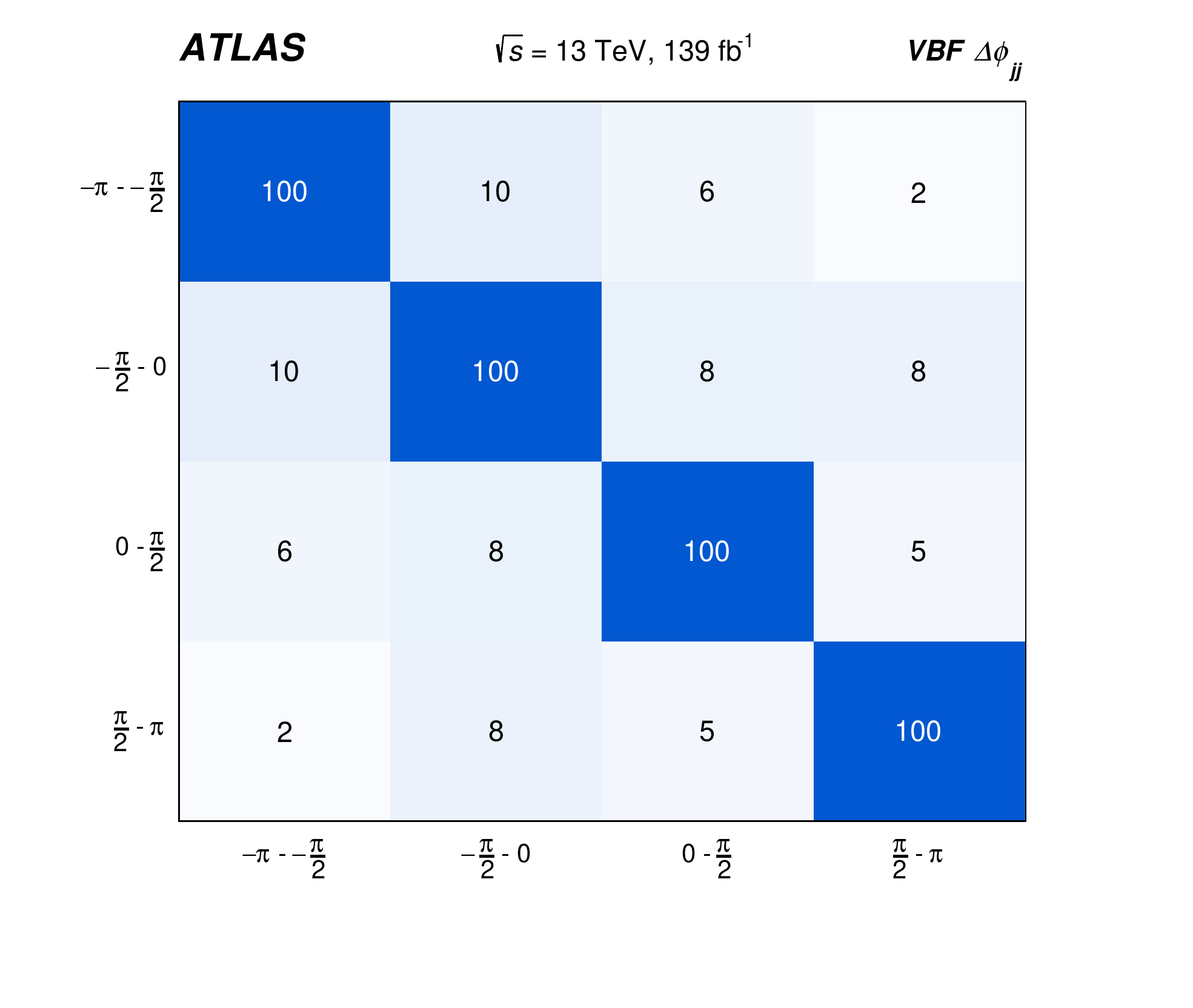}\label{fig:data_unfolded_xsections_matrixinversion_VBF_1_d}}\\
\caption{Correlation matrices for the differential cross-sections variables:  \protect\subref{fig:data_unfolded_xsections_matrixinversion_1D_JV_1_b} \ptgg\ with a \SI{30}{\GeV} jet veto,
\protect\subref{fig:data_unfolded_xsections_matrixinversion_1D_2jet_b} \mjj, \protect\subref{fig:data_unfolded_xsections_matrixinversion_1D_2jet_d} \dphijj and \protect\subref{fig:data_unfolded_xsections_matrixinversion_VBF_1_d} VBF-enhanced \dphijj.
}
\label{fig:data_unfolded_xsections_matrices_2}
\end{figure}
 
\begin{figure}[htb]
\centering
\includegraphics[width=0.48\textwidth]{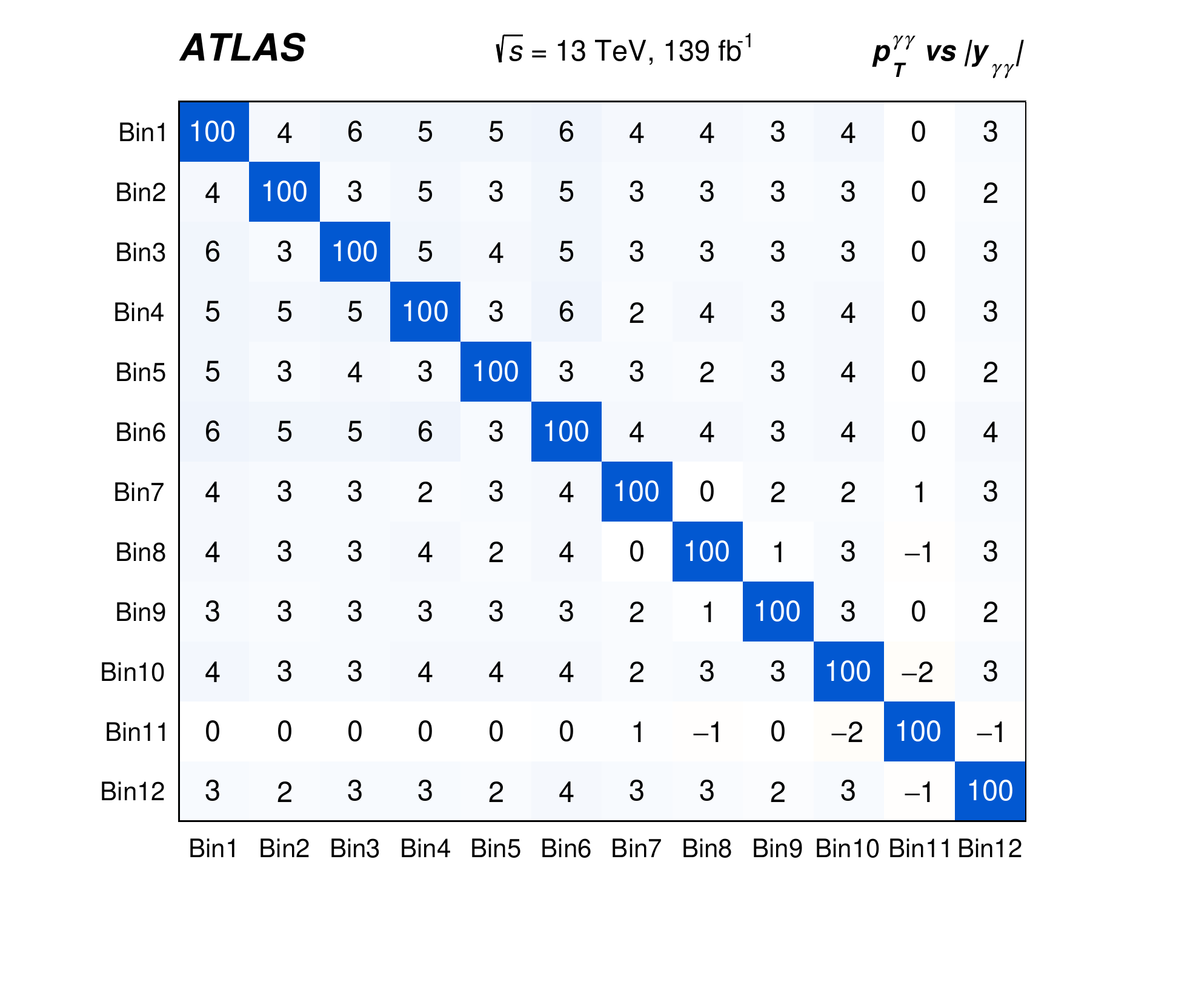}
\caption{Correlation matrix for the double-differential distribution \ptgg\ vs \ygg. The order of the bins in the correlation plots is the same as in the plots with the values of the cross-sections.}
\label{fig:data_unfolded_xsections_matrices_3}
\end{figure}
 
\FloatBarrier
\section{Additional differential cross-section measurements}
\label{sec:aux diff xs}
\paragraph{Diphoton kinematics differential cross-sections}
 
\begin{figure}[htbp]
\centering
\subfloat[]{\includegraphics[width=0.5\textwidth]{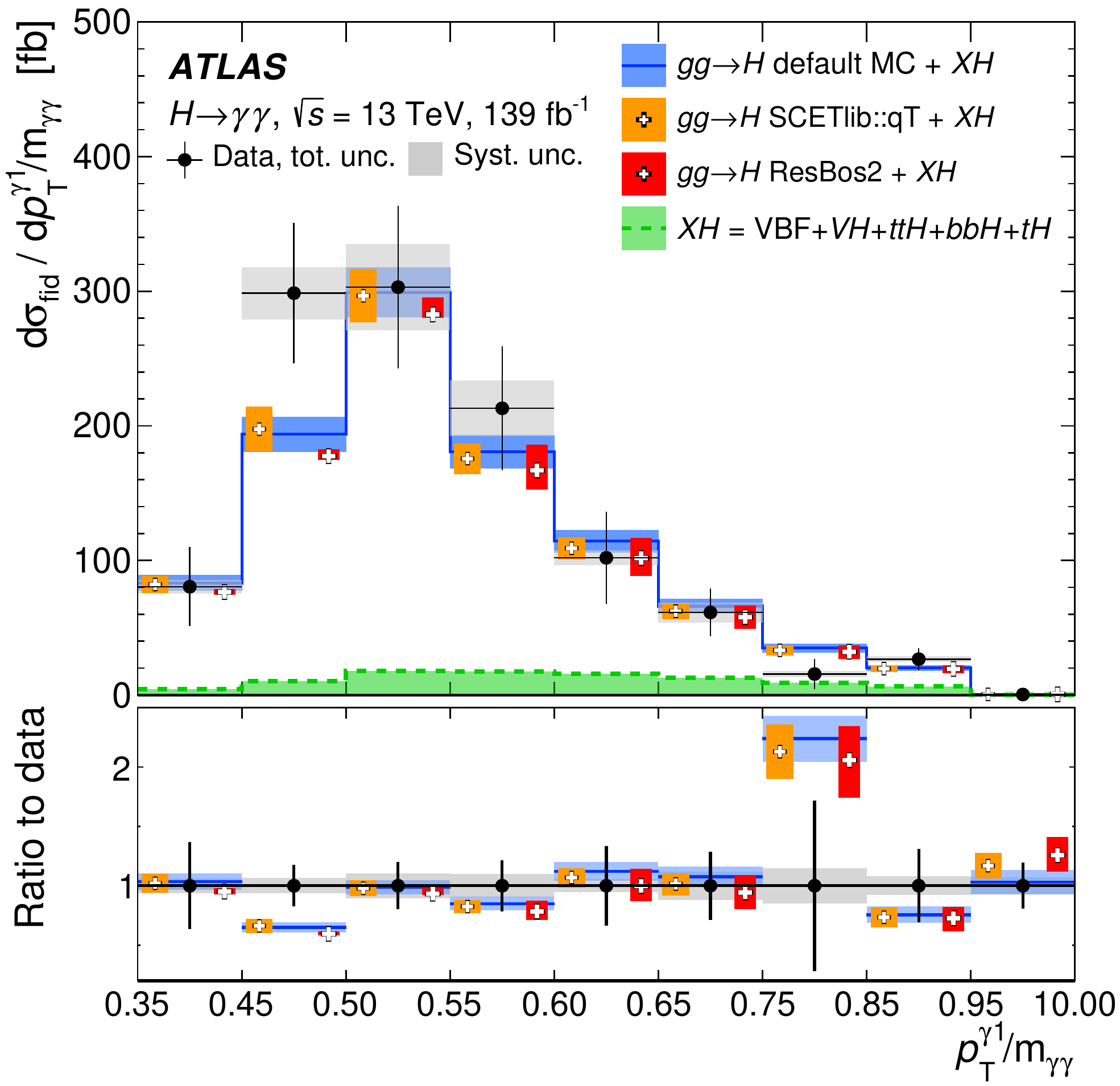}\label{appfig:data_unfolded_xsections_matrixinversion_1D_photon_2_a}}
\subfloat[]{\includegraphics[width=0.5\textwidth]{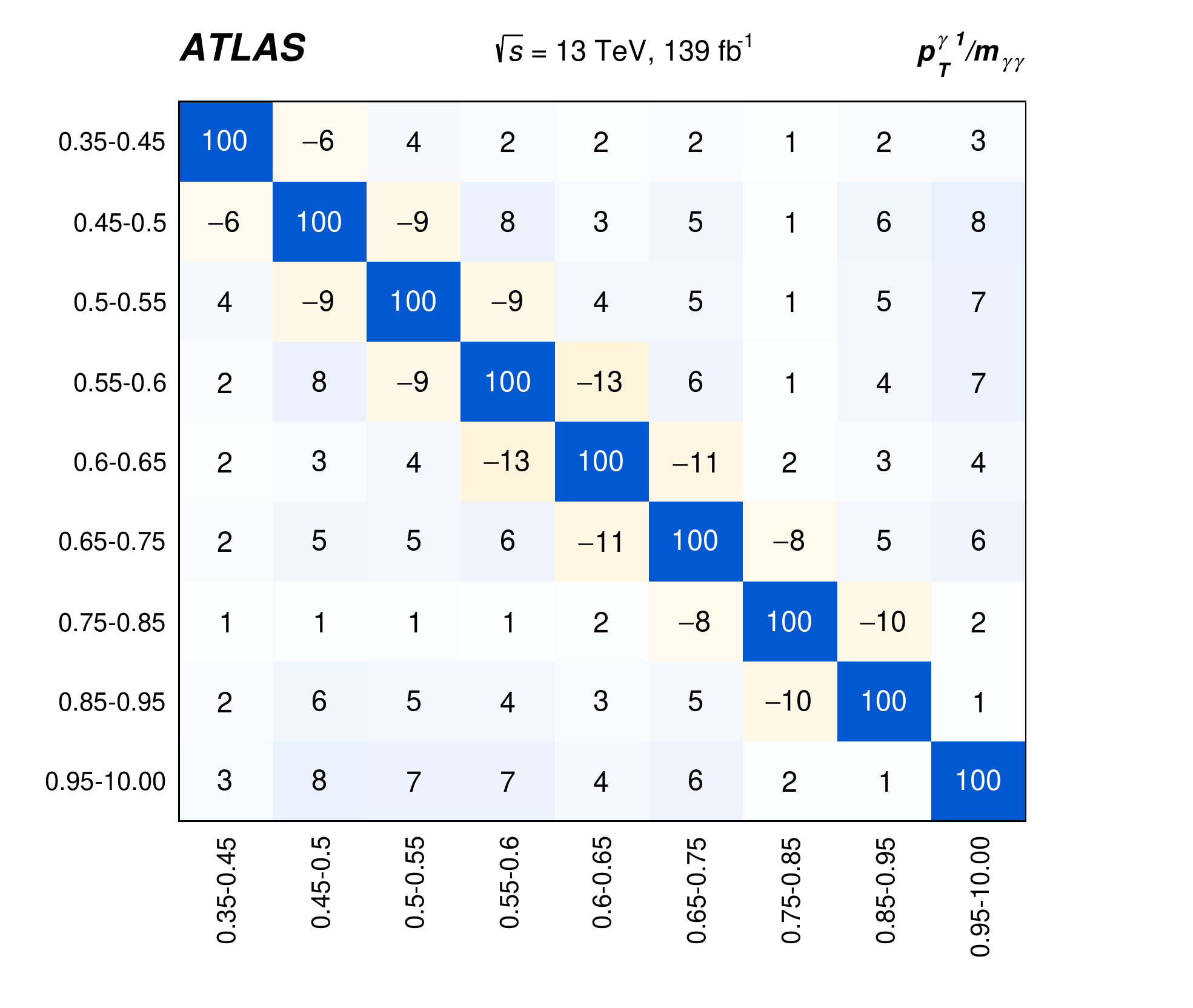}\label{appfig:data_unfolded_xsections_matrixinversion_1D_photon_2_b}} \\
\subfloat[]{\includegraphics[width=0.5\textwidth]{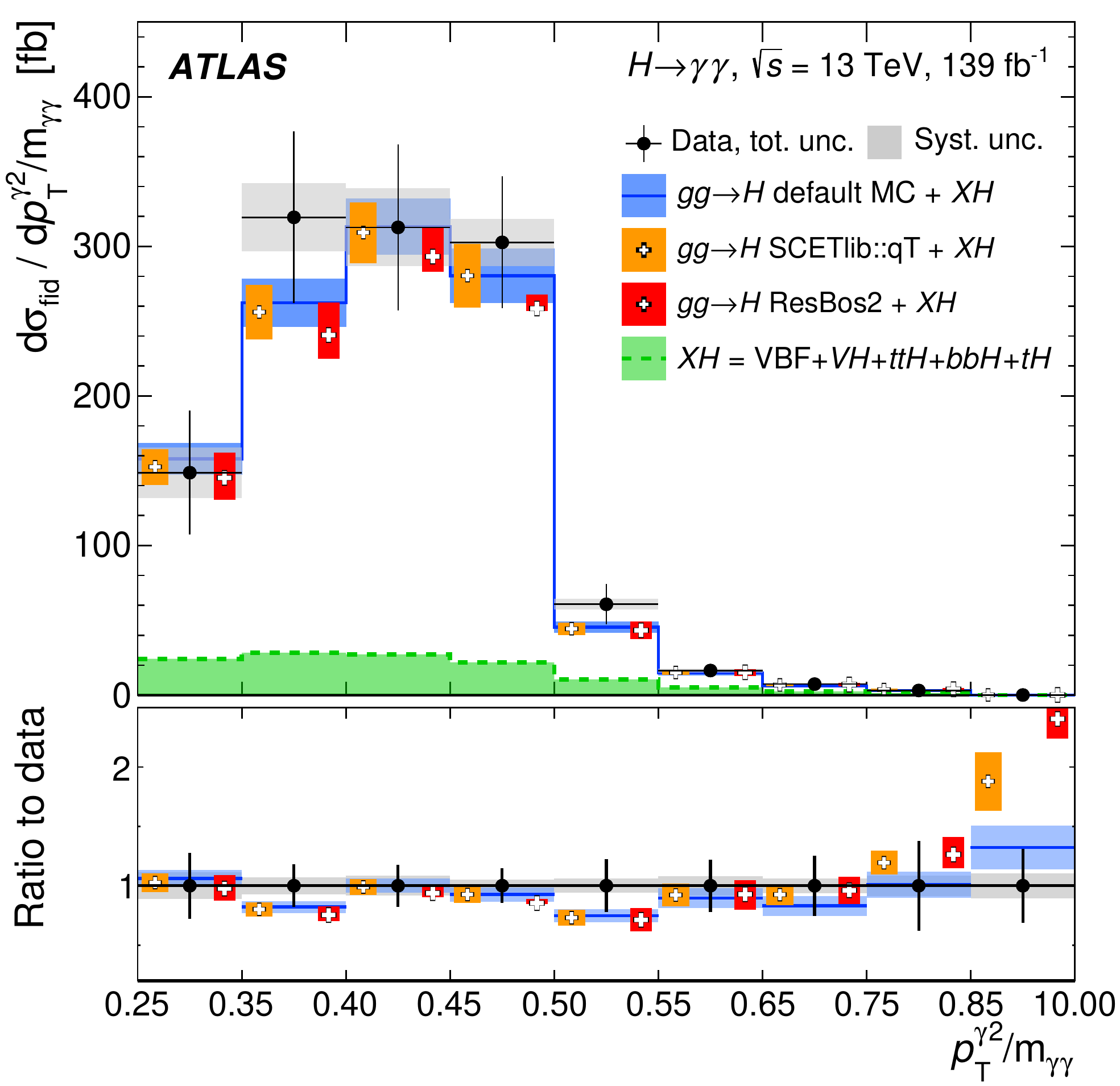}\label{appfig:data_unfolded_xsections_matrixinversion_1D_photon_2_c}}
\subfloat[]{\includegraphics[width=0.5\textwidth]{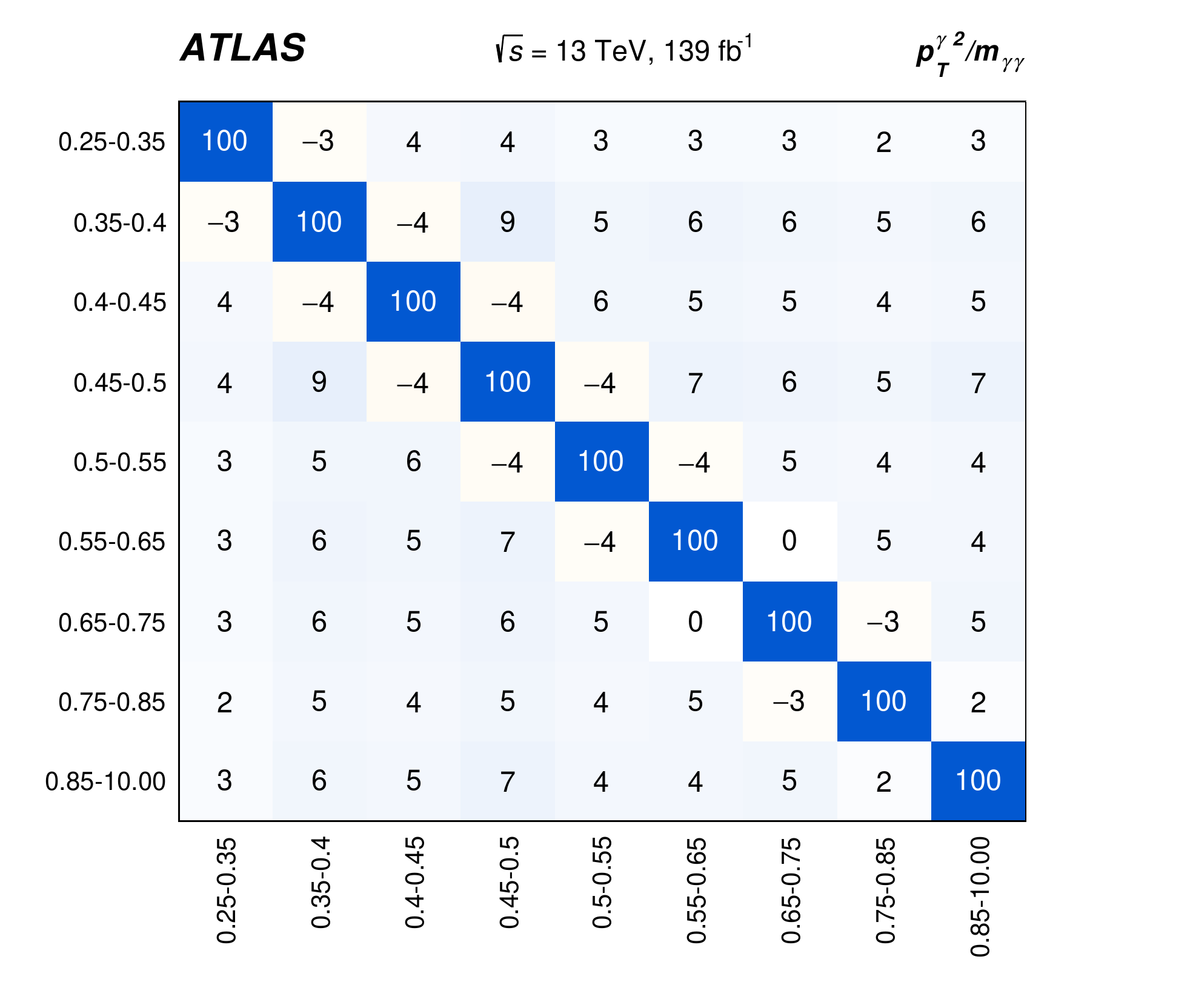}\label{appfig:data_unfolded_xsections_matrixinversion_1D_photon_2_d}}
\caption{Particle-level fiducial differential cross-sections times branching ratio for the photon variables \protect\subref{appfig:data_unfolded_xsections_matrixinversion_1D_photon_2_a} \relptgOne\ and \protect\subref{appfig:data_unfolded_xsections_matrixinversion_1D_photon_2_c} \relptgTwo\ together with the corresponding correlation matrices (\protect\subref{appfig:data_unfolded_xsections_matrixinversion_1D_photon_2_b} and \protect\subref{appfig:data_unfolded_xsections_matrixinversion_1D_photon_2_d}).}
\label{appfig:data_unfolded_xsections_matrixinversion_1D_photon_2}
\end{figure}

\begin{figure}[htb]
\centering
\subfloat[]{\includegraphics[width=0.5\textwidth]{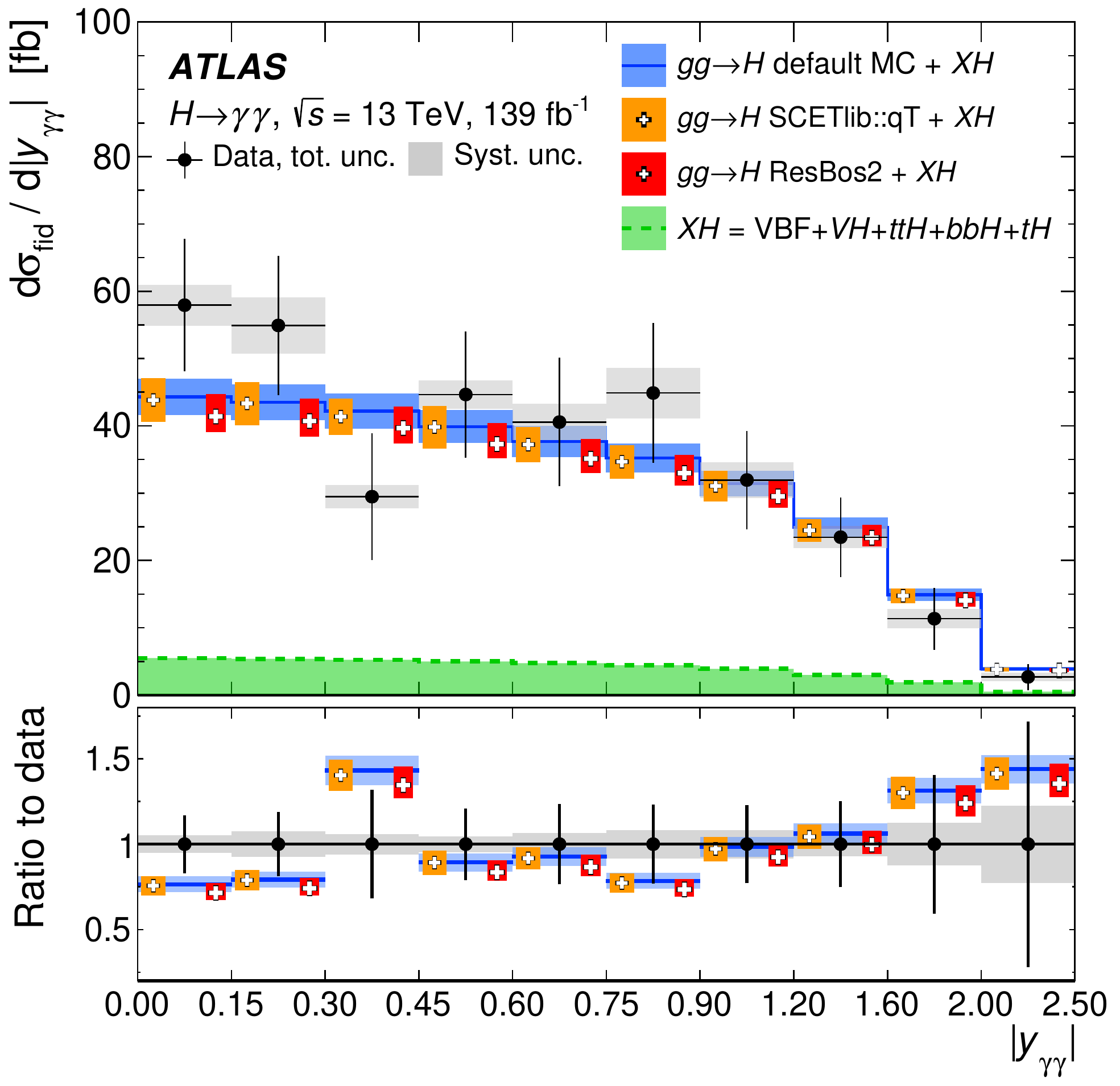}\label{appfig:data_unfolded_xsections_matrixinversion_1D_photon_1_a}}
\subfloat[]{\includegraphics[width=0.5\textwidth]{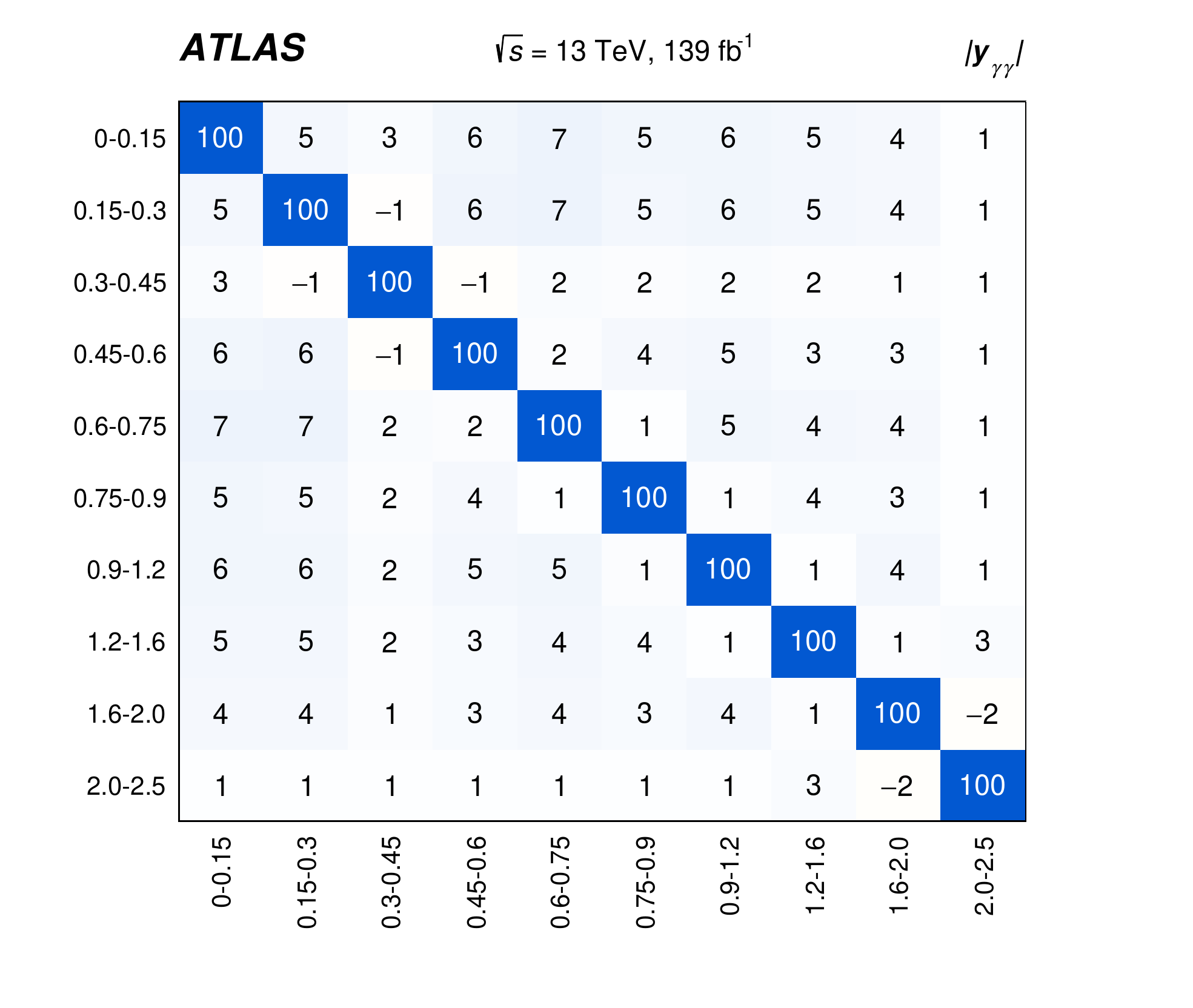}\label{appfig:data_unfolded_xsections_matrixinversion_1D_photon_1_b}} \\
\caption{\protect\subref{appfig:data_unfolded_xsections_matrixinversion_1D_photon_1_a} Particle-level fiducial differential cross-sections times branching ratio for the diphoton rapidity \ygg\ together with \protect\subref{appfig:data_unfolded_xsections_matrixinversion_1D_photon_1_b} the corresponding correlation matrix.}
\label{appfig:data_unfolded_xsections_matrixinversion_1D_photon_1}
\end{figure}

Figures~\labelcref{appfig:data_unfolded_xsections_matrixinversion_1D_photon_2,appfig:data_unfolded_xsections_matrixinversion_1D_photon_1} show the measured differential cross-sections probing the Higgs kinematic variables: \relptgOne, \relptgTwo and \ygg. The measurements are statistically limited, and show good agreement between data and the default simulation, with shape differences at the higher ends of \relptgOne and \relptgTwo distributions between the additional theoretical predictions and the default simulation. For the \ygg\ distribution, similarly good agreement is observed in the full range.
 
\paragraph{\(\geq\) 1-jet differential cross-sections}
 
\begin{figure}[htb!]
\centering
\subfloat[]{\includegraphics[width=0.5\textwidth]{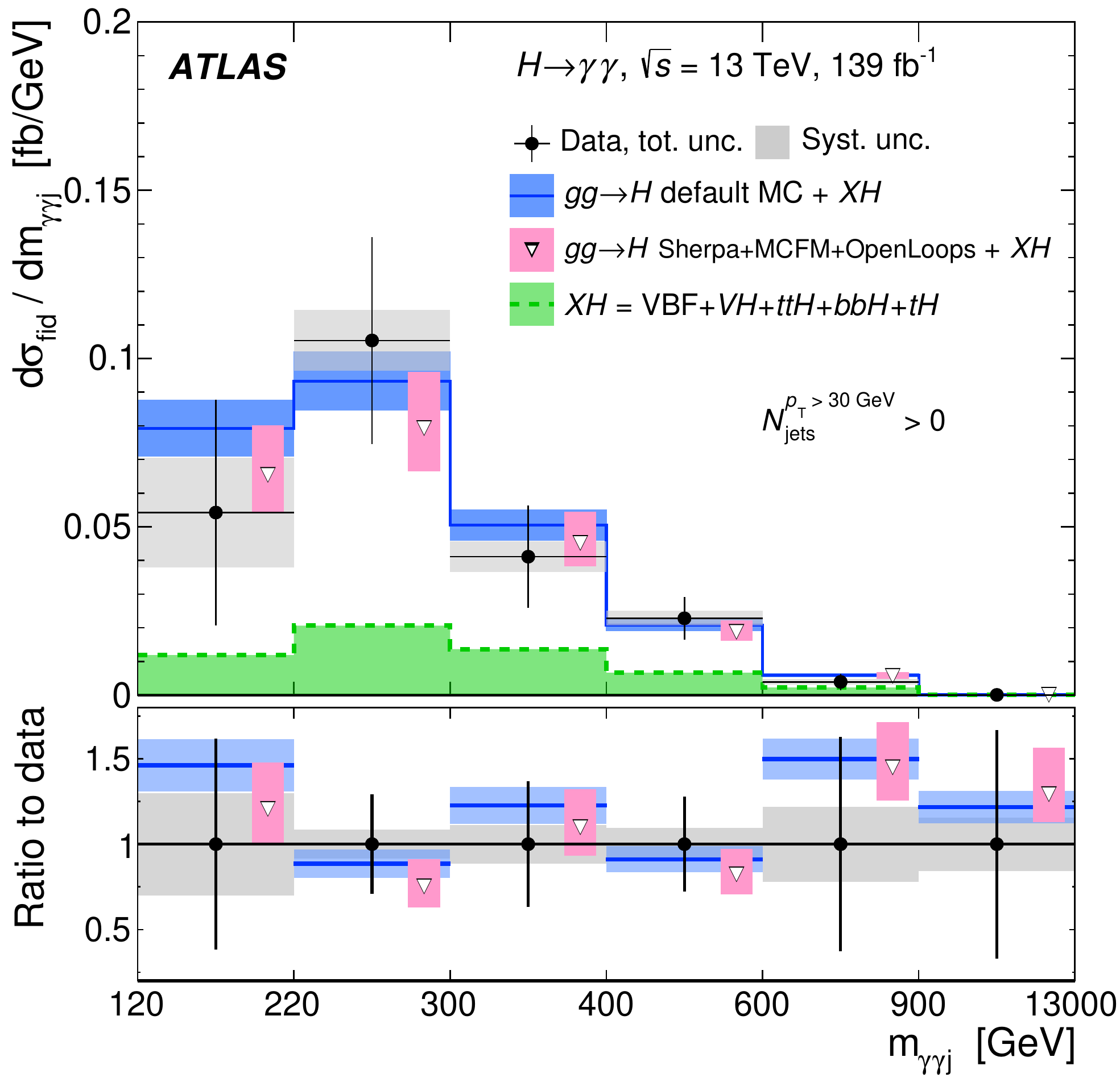}\label{appfig:data_unfolded_xsections_matrixinversion_1D_1jet_2_a}}
\subfloat[]{\includegraphics[width=0.5\textwidth]{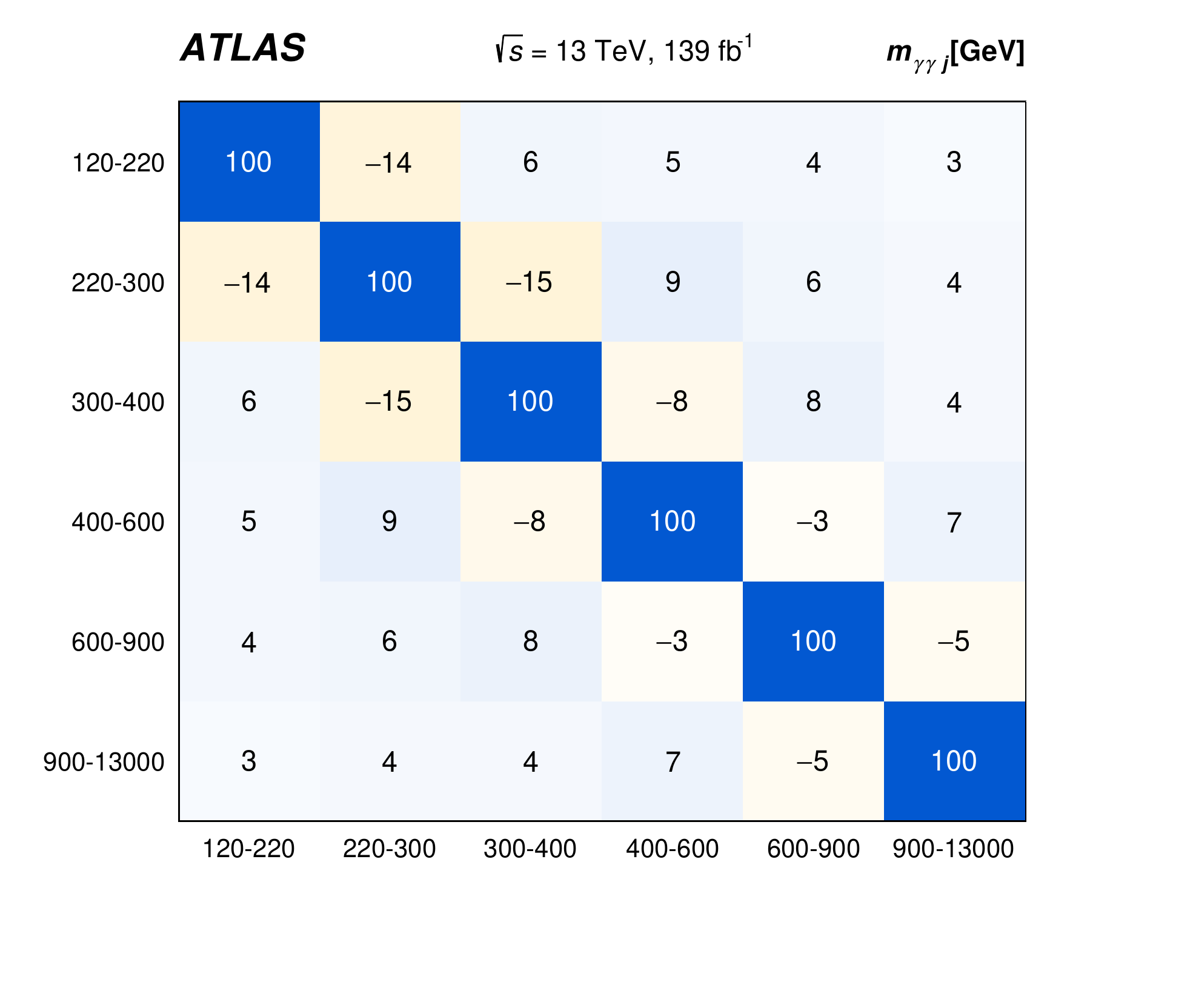}\label{appfig:data_unfolded_xsections_matrixinversion_1D_1jet_2_b}} \\
\subfloat[]{\includegraphics[width=0.5\textwidth]{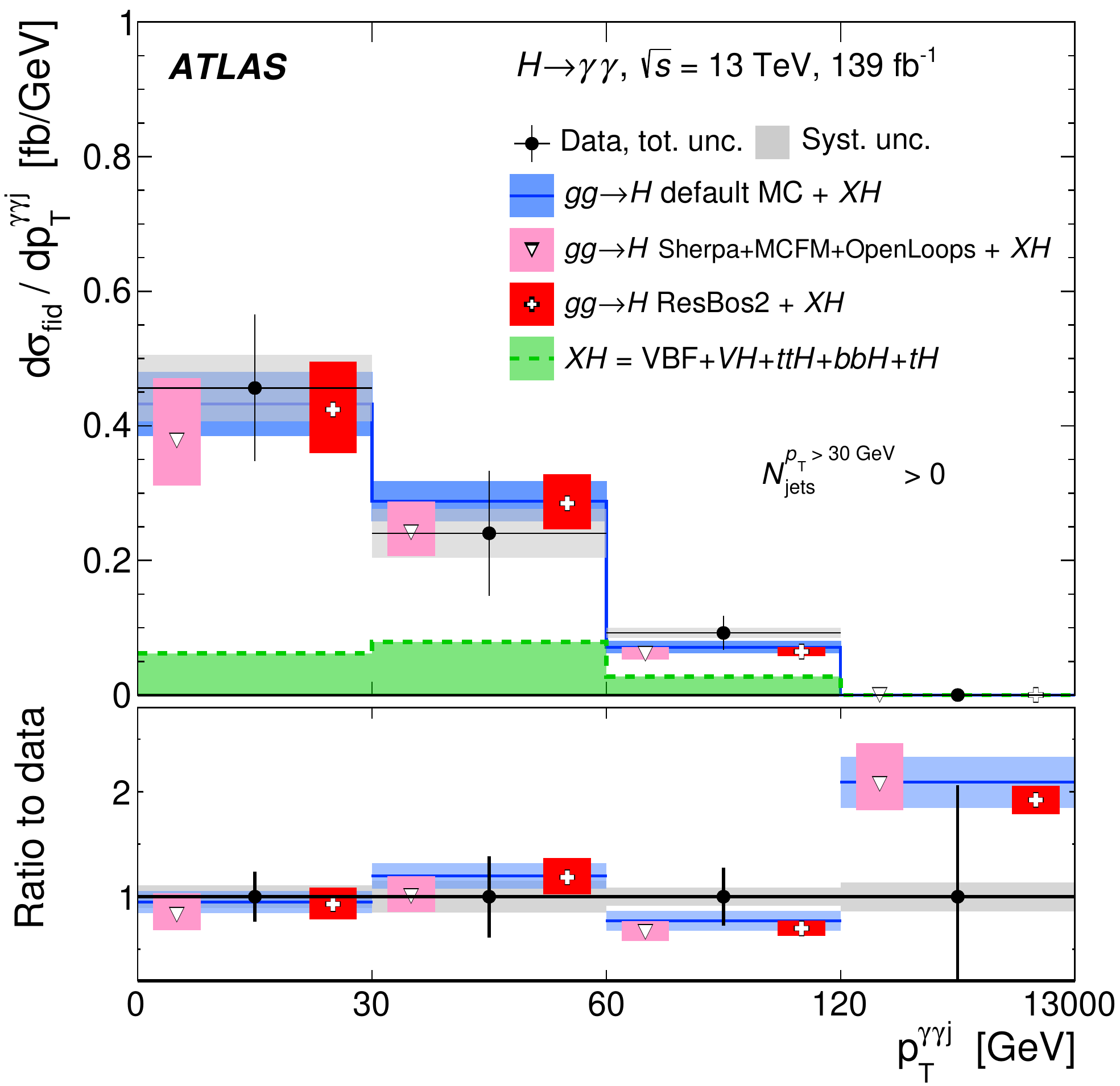}\label{appfig:data_unfolded_xsections_matrixinversion_1D_1jet_2_c}}
\subfloat[]{\includegraphics[width=0.5\textwidth]{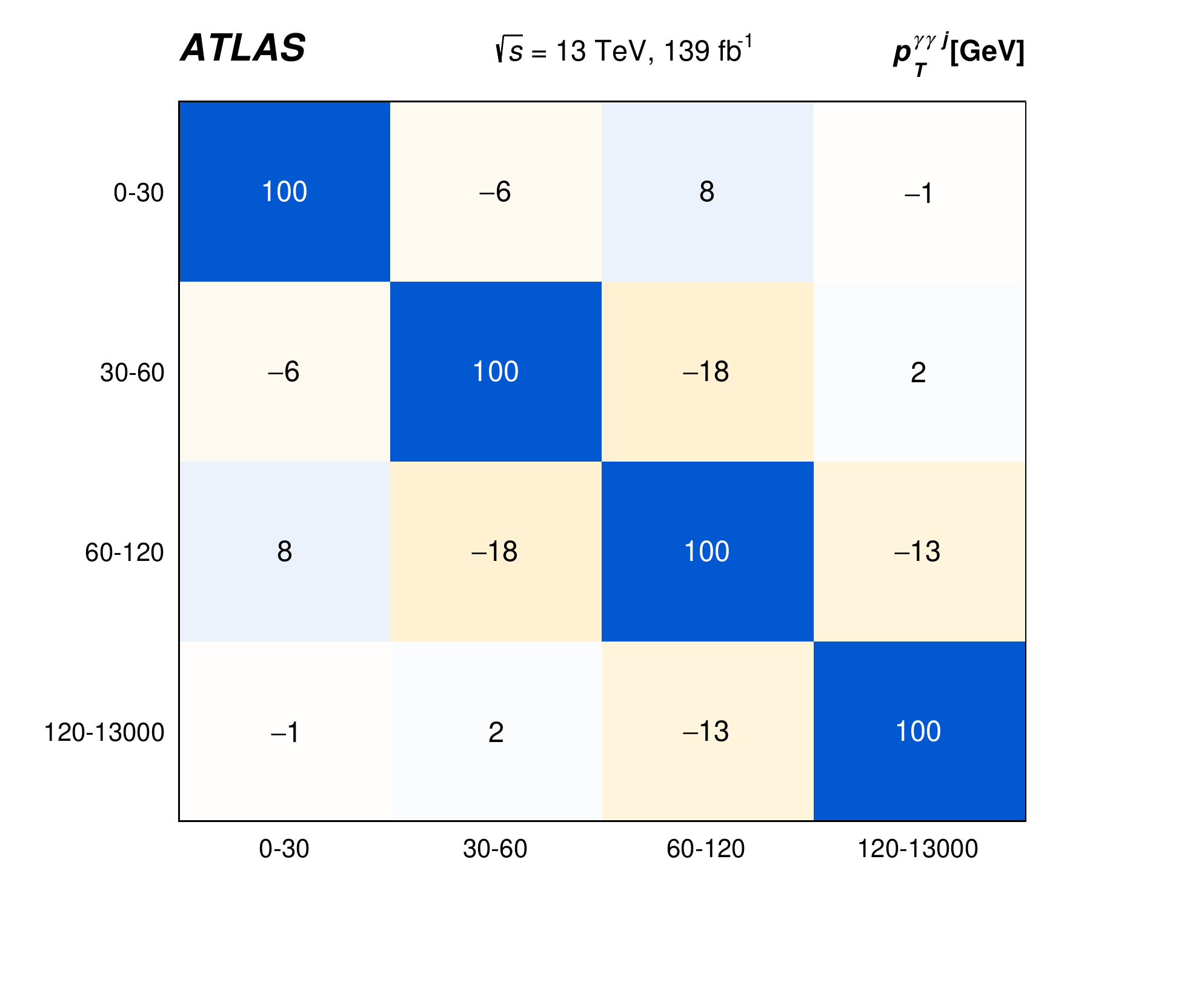}\label{appfig:data_unfolded_xsections_matrixinversion_1D_1jet_2_d}}
\caption{Particle-level fiducial differential cross-sections times branching ratio for the variables: \protect\subref{appfig:data_unfolded_xsections_matrixinversion_1D_1jet_2_a} \mggj\ and \protect\subref{appfig:data_unfolded_xsections_matrixinversion_1D_1jet_2_c} \ptggj\ together with the corresponding correlation matrices (\protect\subref{appfig:data_unfolded_xsections_matrixinversion_1D_1jet_2_b} and \protect\subref{appfig:data_unfolded_xsections_matrixinversion_1D_1jet_2_d}).}
\label{appfig:data_unfolded_xsections_matrixinversion_1D_1jet_aux_2}
\end{figure}

\begin{figure}[htb!]
\centering
\subfloat[]{\includegraphics[width=0.5\textwidth]{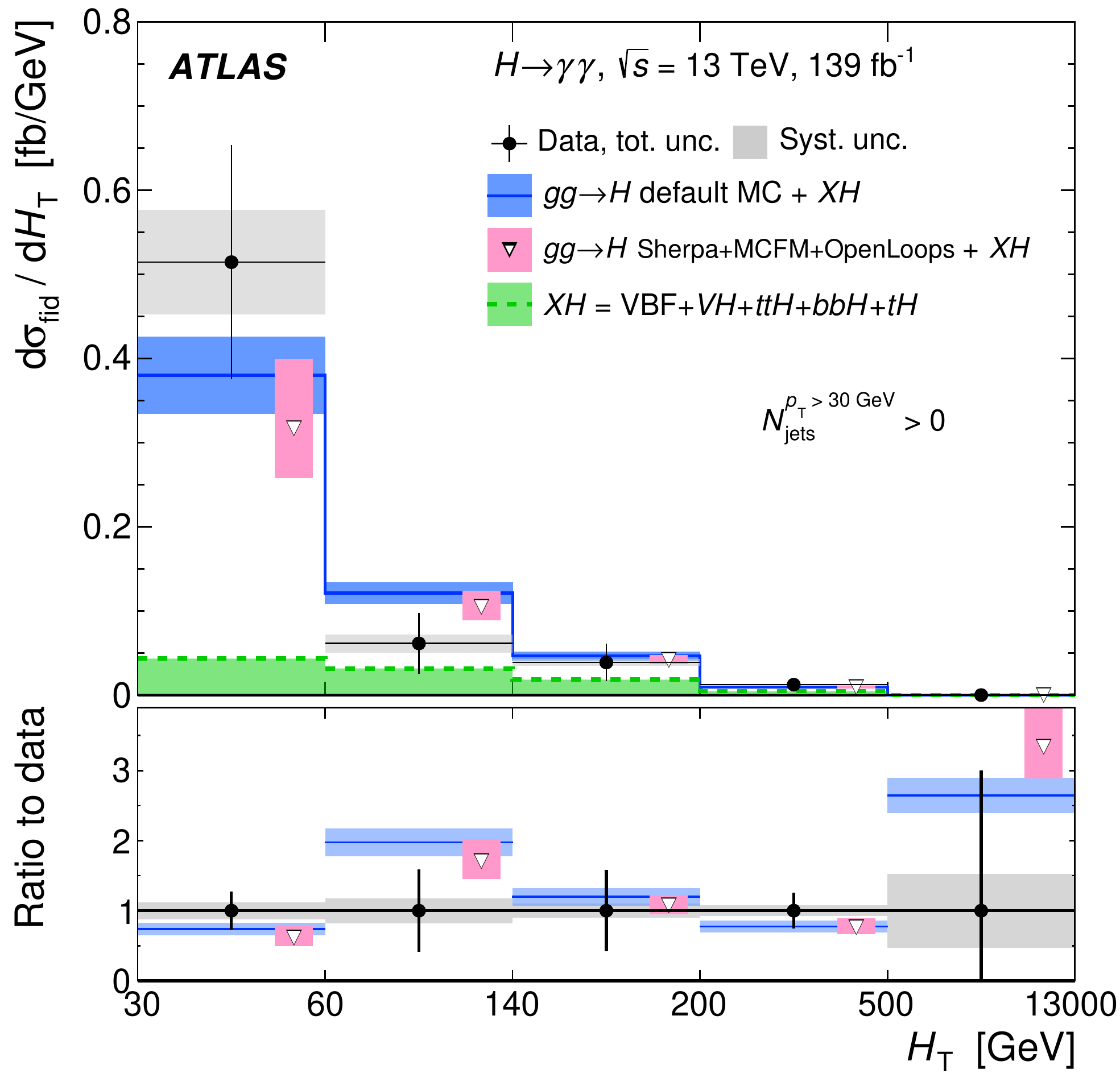}\label{appfig:data_unfolded_xsections_matrixinversion_1D_1jet_1_c}}
\subfloat[]{\includegraphics[width=0.5\textwidth]{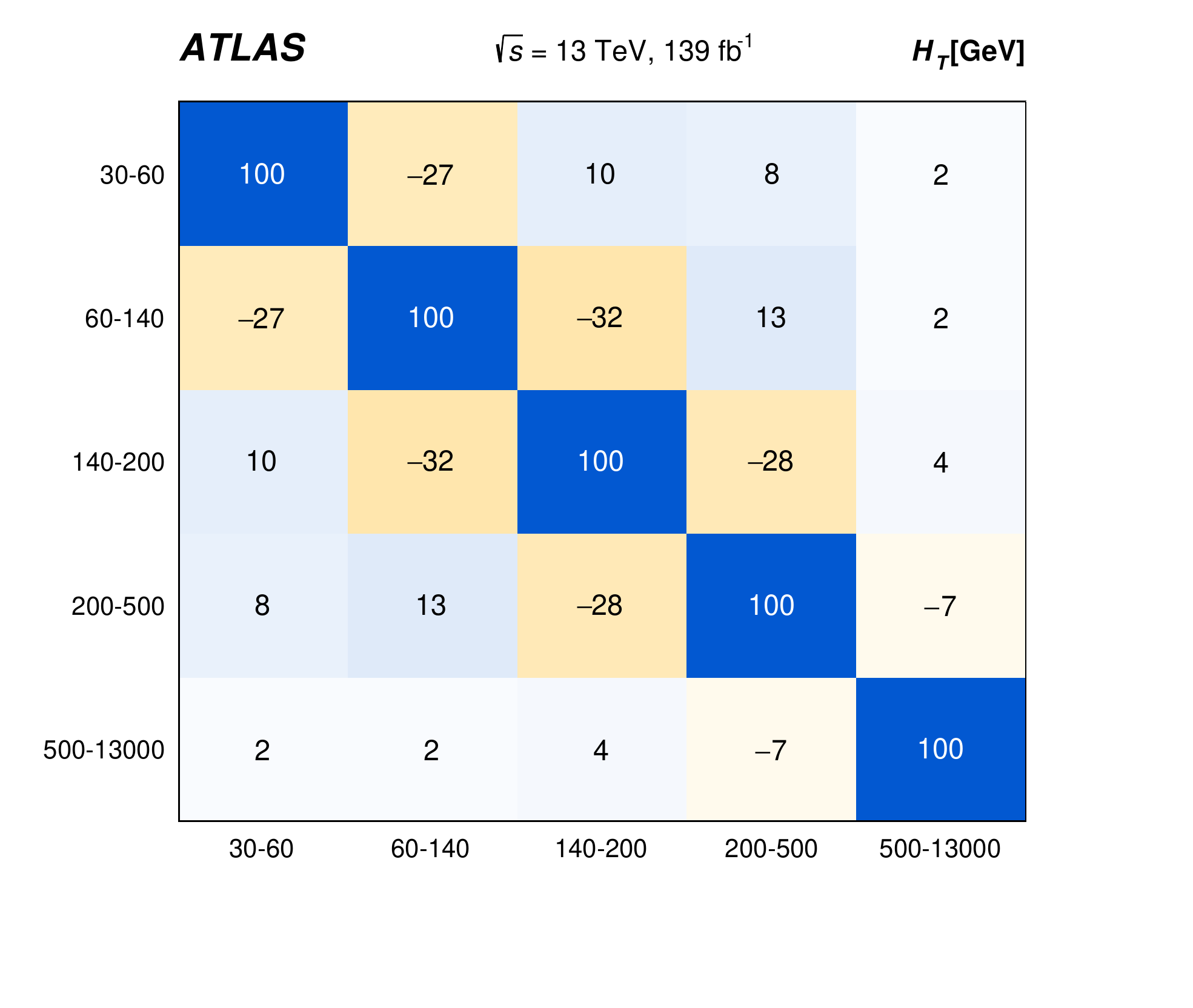}\label{appfig:data_unfolded_xsections_matrixinversion_1D_1jet_1_d}}
\caption{Particle-level fiducial differential cross-sections times branching ratio for \protect\subref{appfig:data_unfolded_xsections_matrixinversion_1D_1jet_1_c} \HT\ together with the corresponding correlation matrix  \protect\subref{appfig:data_unfolded_xsections_matrixinversion_1D_1jet_1_d}.}
\label{appfig:data_unfolded_xsections_matrixinversion_1D_1jet_aux_1}
\end{figure}

\begin{figure}[htbp]
\centering
\subfloat[]{\includegraphics[width=0.5\textwidth]{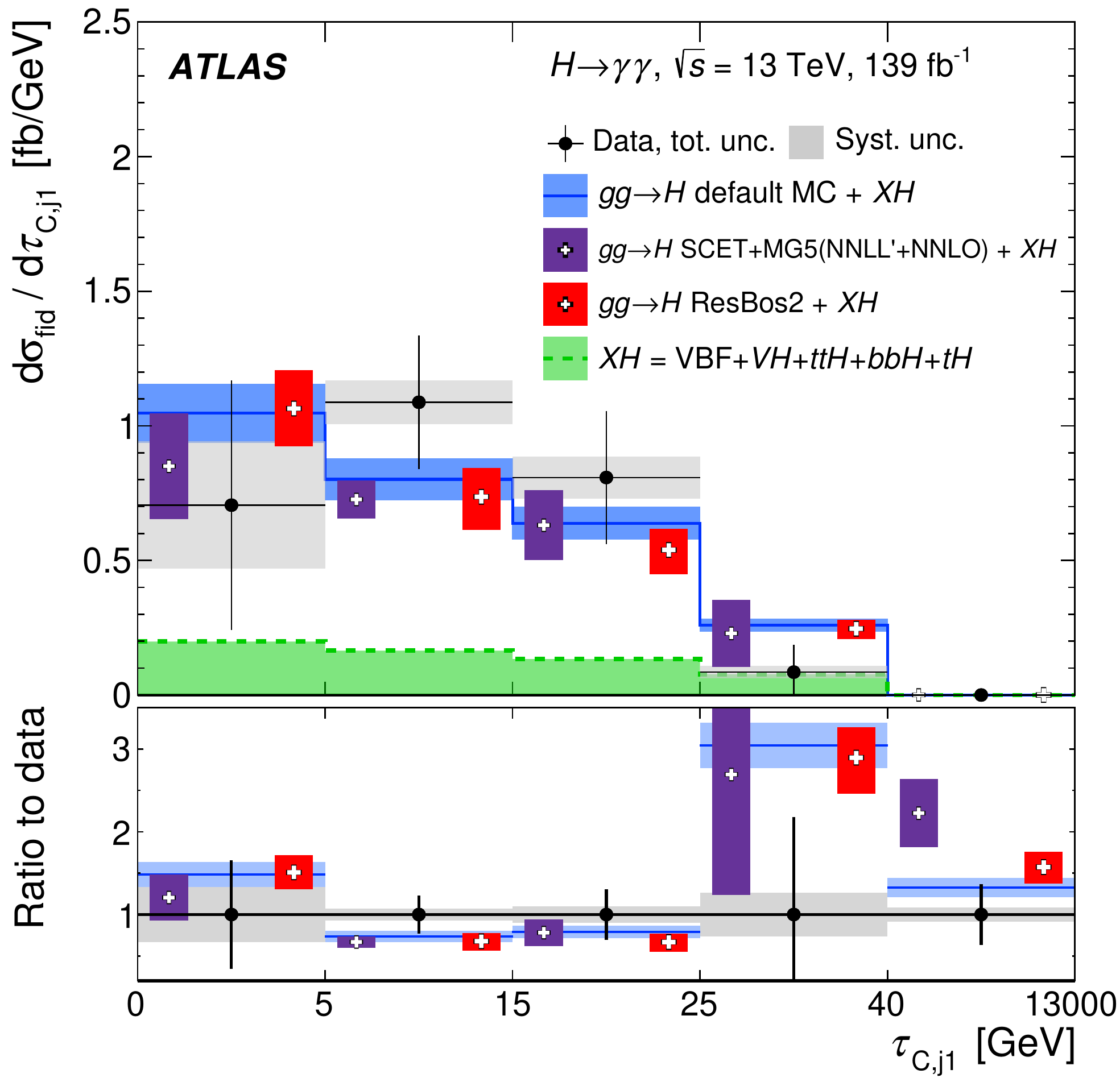}\label{appfig:data_unfolded_xsections_matrixinversion_1D_1jet_3_a}}
\subfloat[]{\includegraphics[width=0.5\textwidth]{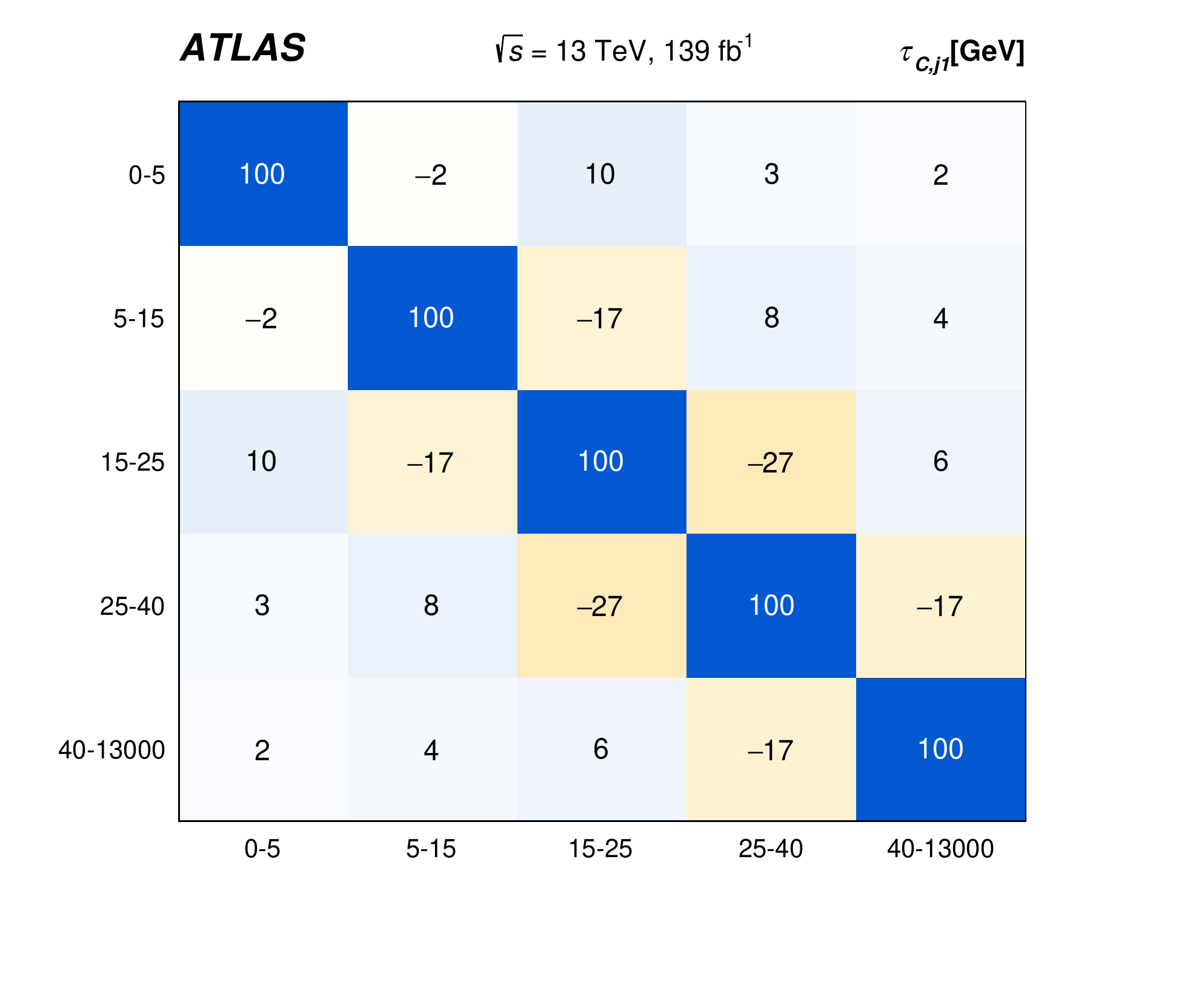}\label{appfig:data_unfolded_xsections_matrixinversion_1D_1jet_3_b}} \\
\subfloat[]{\includegraphics[width=0.5\textwidth]{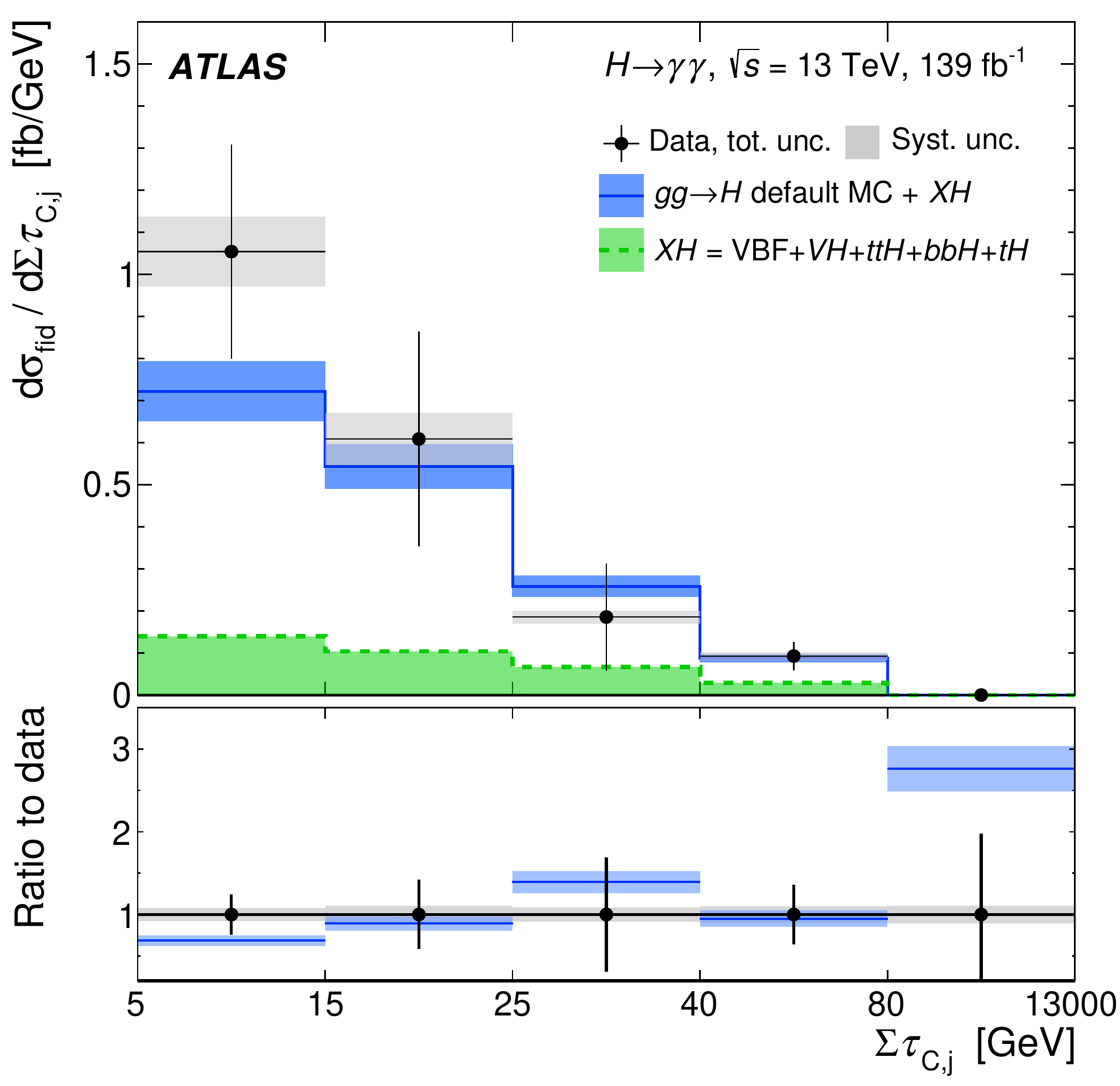}\label{appfig:data_unfolded_xsections_matrixinversion_1D_1jet_3_c}}
\subfloat[]{\includegraphics[width=0.5\textwidth]{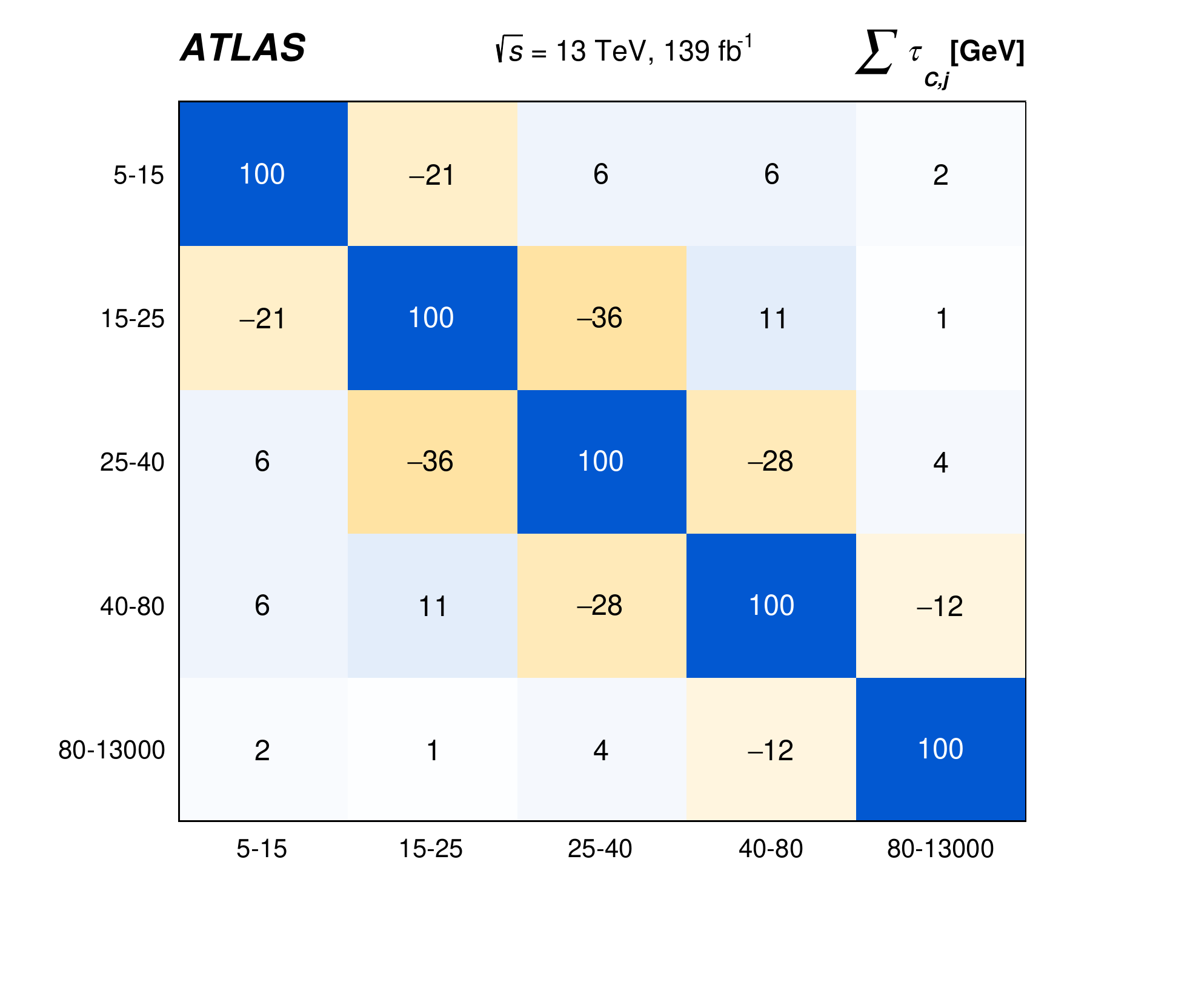}\label{appfig:data_unfolded_xsections_matrixinversion_1D_1jet_3_d}}\\
\caption{Particle-level fiducial differential cross-sections times branching ratio for the variables: \protect\subref{appfig:data_unfolded_xsections_matrixinversion_1D_1jet_3_a} \maxtau\ and \protect\subref{appfig:data_unfolded_xsections_matrixinversion_1D_1jet_3_c} \sumtau\ together with the corresponding correlation matrices (\protect\subref{appfig:data_unfolded_xsections_matrixinversion_1D_1jet_3_b} and \protect\subref{appfig:data_unfolded_xsections_matrixinversion_1D_1jet_3_d}).}
\label{appfig:data_unfolded_xsections_matrixinversion_1D_1jet_3}
\end{figure}

\begin{figure}[htbp]
\centering
\subfloat[]{\includegraphics[width=0.5\textwidth]{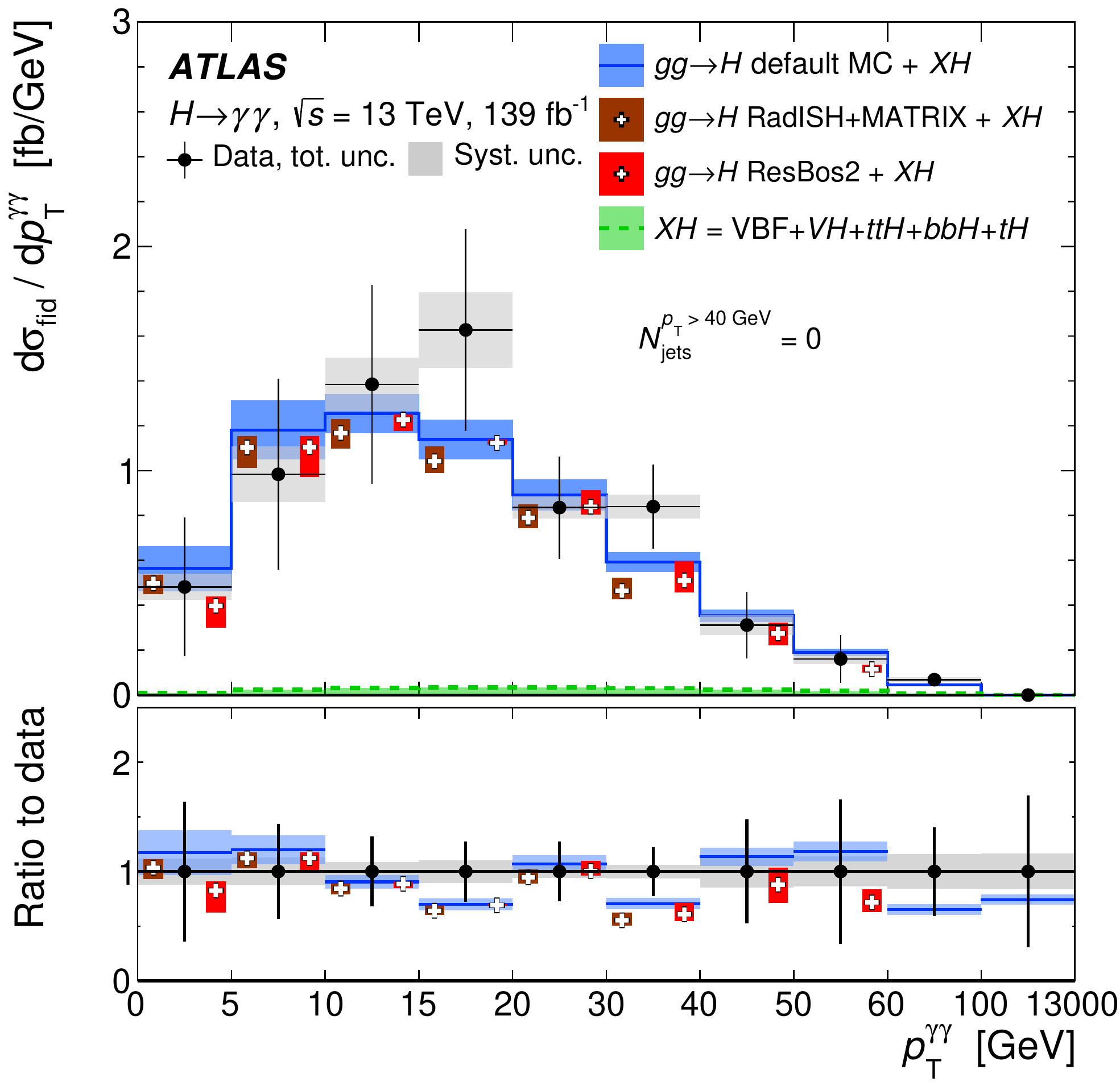}\label{appfig:data_unfolded_xsections_matrixinversion_1D_JV_1_a}}
\subfloat[]{\includegraphics[width=0.5\textwidth]{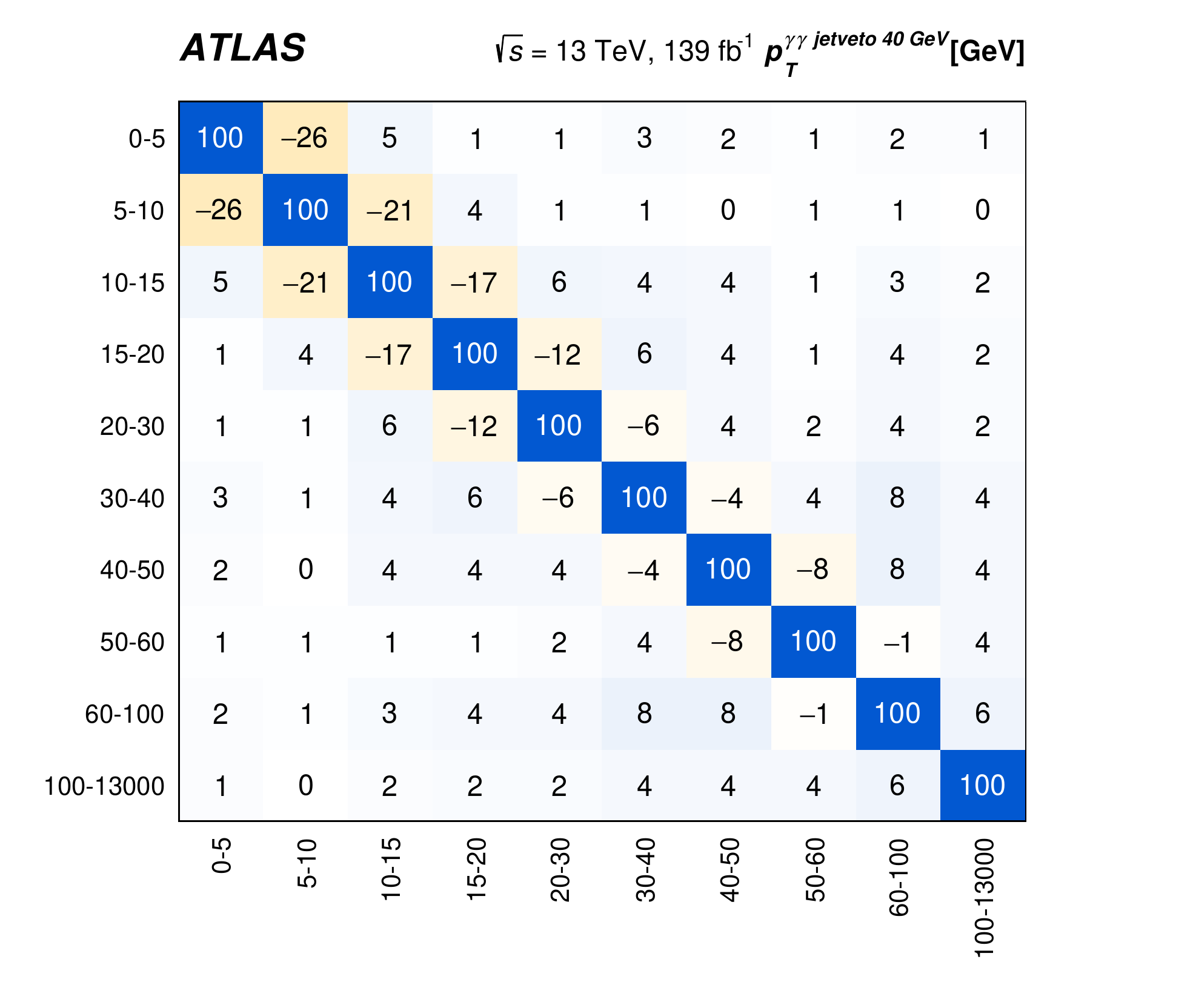}\label{appfig:data_unfolded_xsections_matrixinversion_1D_JV_1_b}} \\
\subfloat[]{\includegraphics[width=0.5\textwidth]{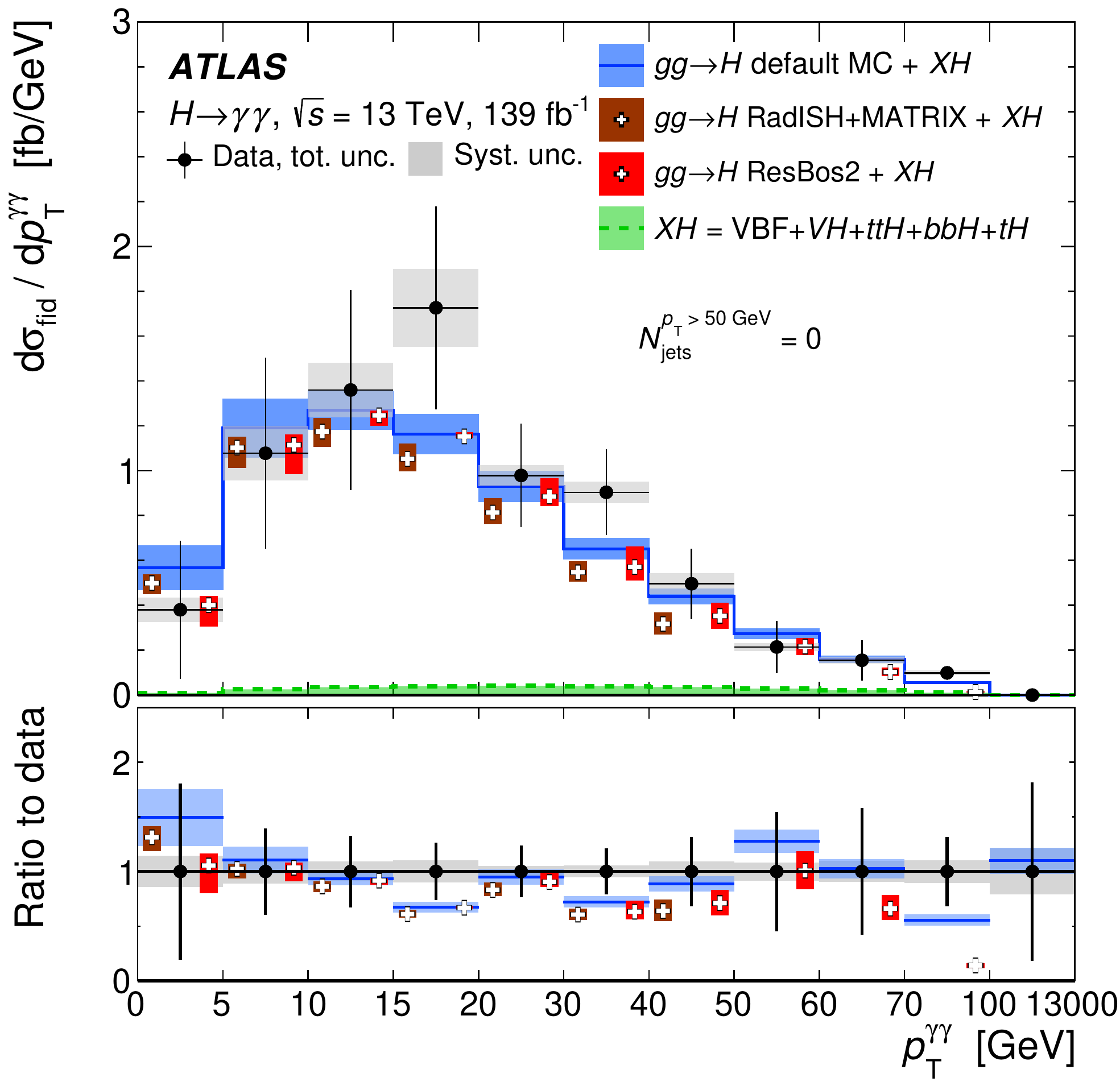}\label{appfig:data_unfolded_xsections_matrixinversion_1D_JV_1_c}}
\subfloat[]{\includegraphics[width=0.5\textwidth]{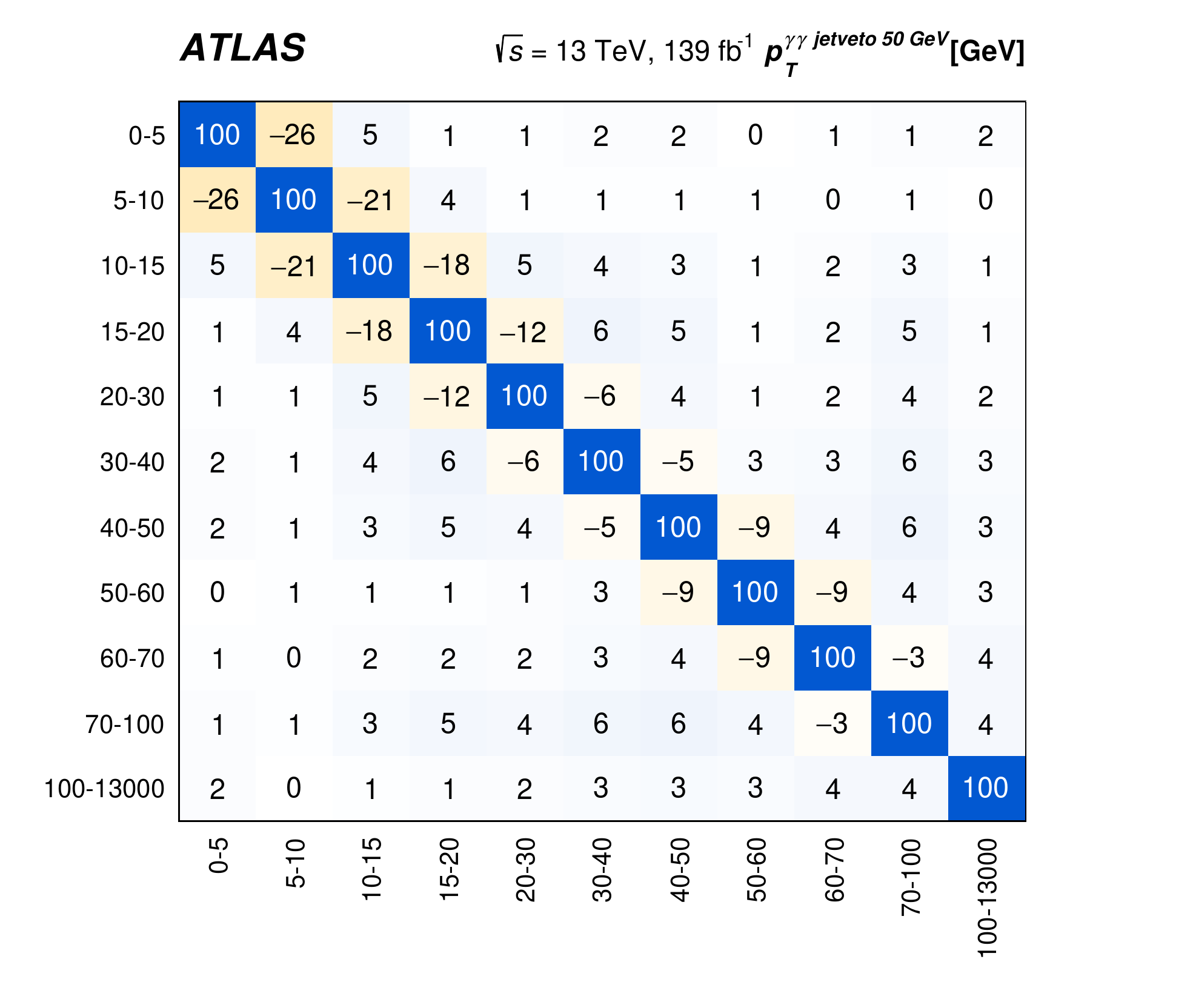}\label{appfig:data_unfolded_xsections_matrixinversion_1D_JV_1_d}}
\caption{Particle-level fiducial differential cross-sections times branching ratio for \ptgg\ with a jet veto at \protect\subref{appfig:data_unfolded_xsections_matrixinversion_1D_JV_1_a} \SI{40}{\GeV} and \protect\subref{appfig:data_unfolded_xsections_matrixinversion_1D_JV_1_c} \SI{50}{\GeV} together with the corresponding correlation matrices (\protect\subref{appfig:data_unfolded_xsections_matrixinversion_1D_JV_1_b} and \protect\subref{appfig:data_unfolded_xsections_matrixinversion_1D_JV_1_d}). The \textsc{ResBos2} predictions are available up to \SI{60}{\GeV} (\SI{100}{\GeV}) for \ptgg\ with a \SI{40}{\GeV} (\SI{50}{\GeV}) jet veto. The \textsc{RadISH+Matrix} predictions are available up to \SI{40}{\GeV} (\SI{50}{\GeV}) for \ptgg\ with a \SI{40}{\GeV} (\SI{50}{\GeV}) jet veto. }
\label{appfig:data_unfolded_xsections_matrixinversion_1D_JV_1}
\end{figure}

\begin{figure}[htb!]
\centering
\subfloat[]{\includegraphics[width=0.5\textwidth]{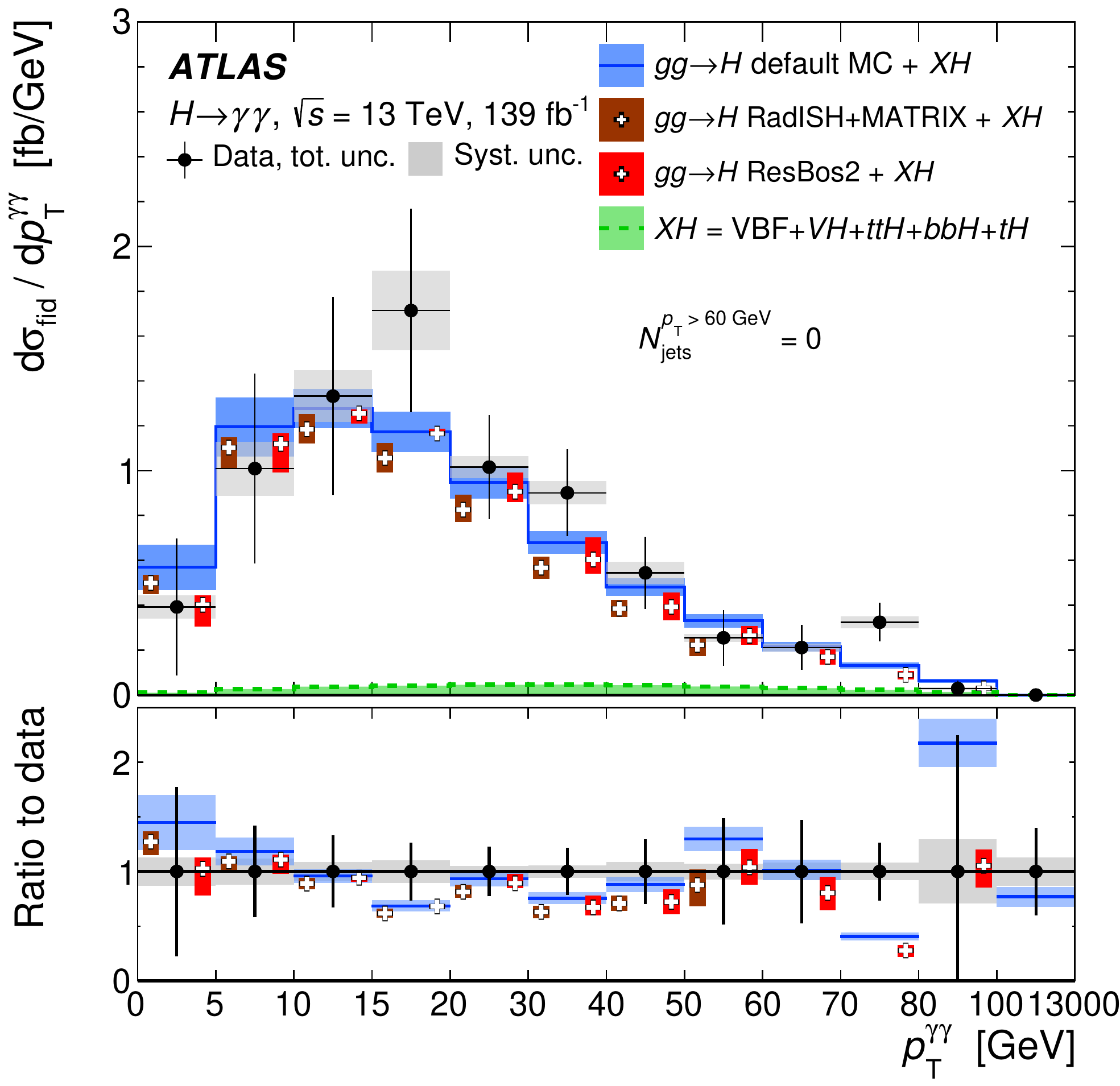}\label{appfig:data_unfolded_xsections_matrixinversion_1D_JV_2_c}}
\subfloat[]{\includegraphics[width=0.5\textwidth]{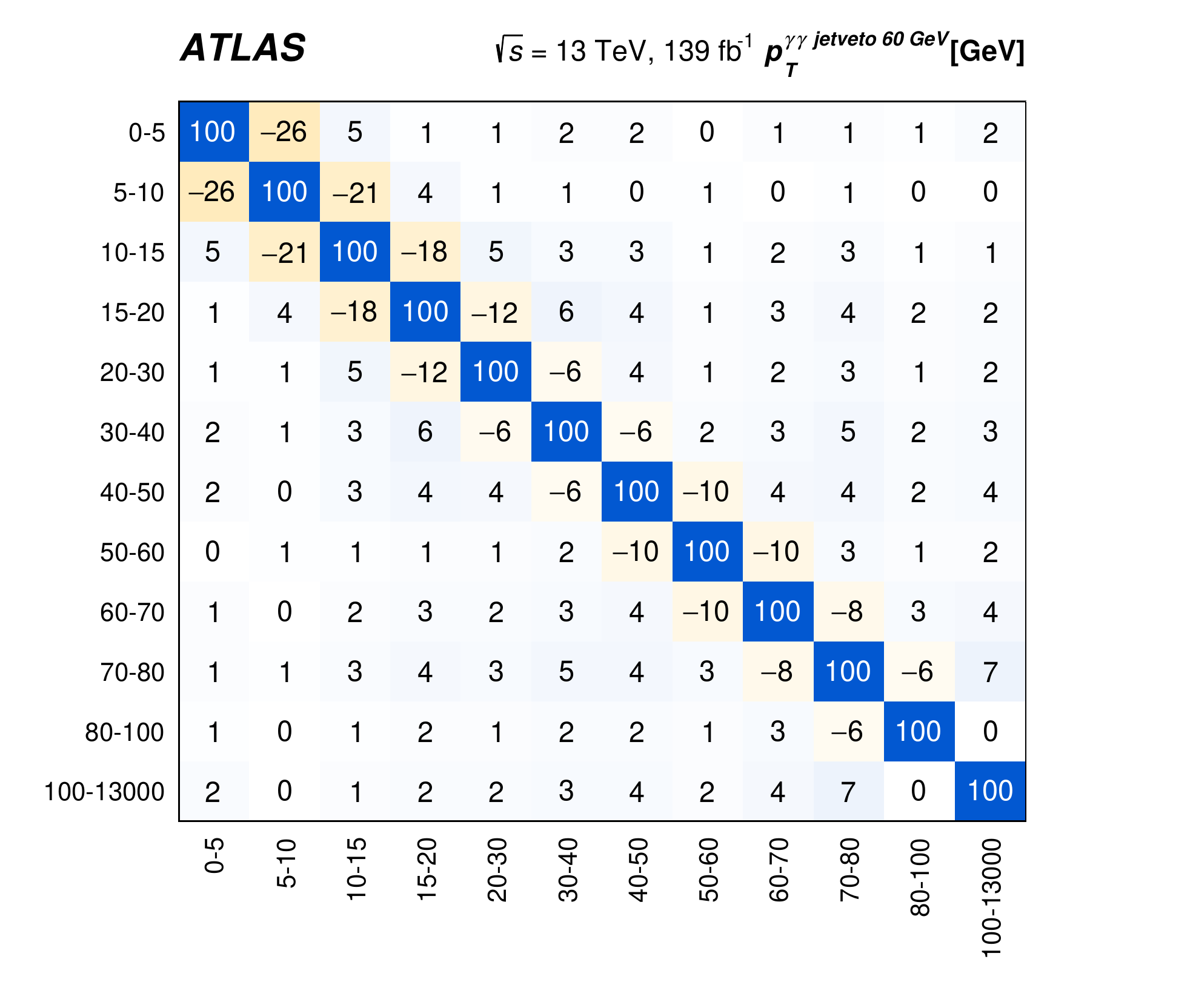}\label{appfig:data_unfolded_xsections_matrixinversion_1D_JV_2_d}}
\caption{Particle-level fiducial differential cross-sections times branching ratio for \ptgg\ with a \SI{60}{\GeV} jet veto \protect\subref{appfig:data_unfolded_xsections_matrixinversion_1D_JV_2_c} and the corresponding correlation matrix \protect\subref{appfig:data_unfolded_xsections_matrixinversion_1D_JV_2_d}. The \textsc{ResBos2} predictions are available up to \SI{100}{\GeV}. The \textsc{RadISH+Matrix} predictions are available up to \SI{60}{\GeV}.}
\label{appfig:data_unfolded_xsections_matrixinversion_1D_JV_2}
\end{figure}

Figures~\labelcref{appfig:data_unfolded_xsections_matrixinversion_1D_1jet_aux_2,appfig:data_unfolded_xsections_matrixinversion_1D_1jet_aux_1} show the measured differential cross-sections for the variables: \mggj, \ptggj and \HT. Figure~\ref{appfig:data_unfolded_xsections_matrixinversion_1D_1jet_3} shows the variables \maxtau\ and \sumtau.
Figures~\labelcref{appfig:data_unfolded_xsections_matrixinversion_1D_JV_1,appfig:data_unfolded_xsections_matrixinversion_1D_JV_2} show \ptgg\ with different jet vetoes. All predictions agree well with the data within uncertainties for the different predictions.
 
\paragraph{\(\geq\) 2-jet differential cross-sections}
 
\begin{figure}[htbp]
\centering
\subfloat[]{\includegraphics[width=0.5\textwidth]{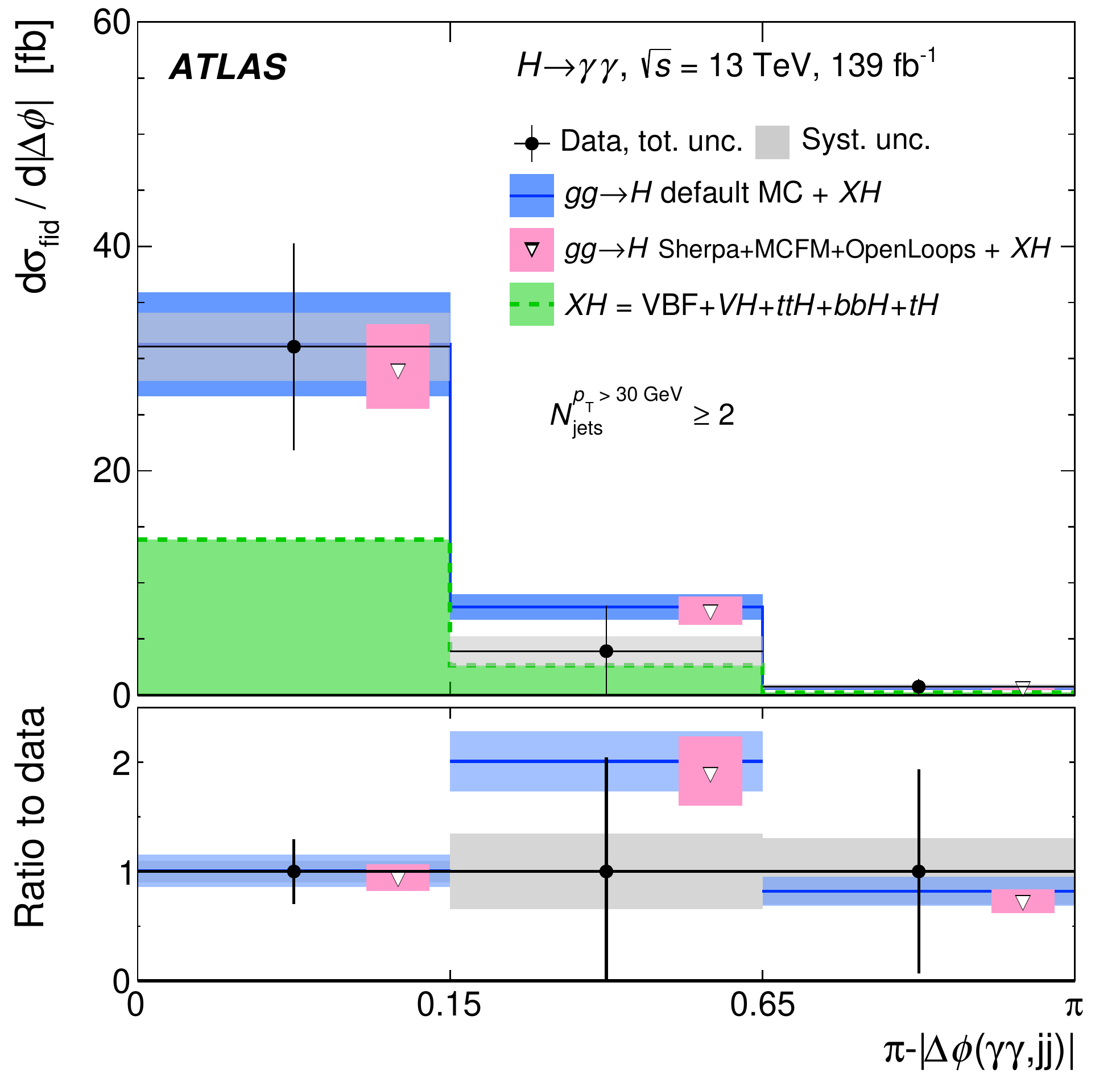}\label{appfig:data_unfolded_xsections_matrixinversion_1D_2jet_2_a}}
\subfloat[]{\includegraphics[width=0.5\textwidth]{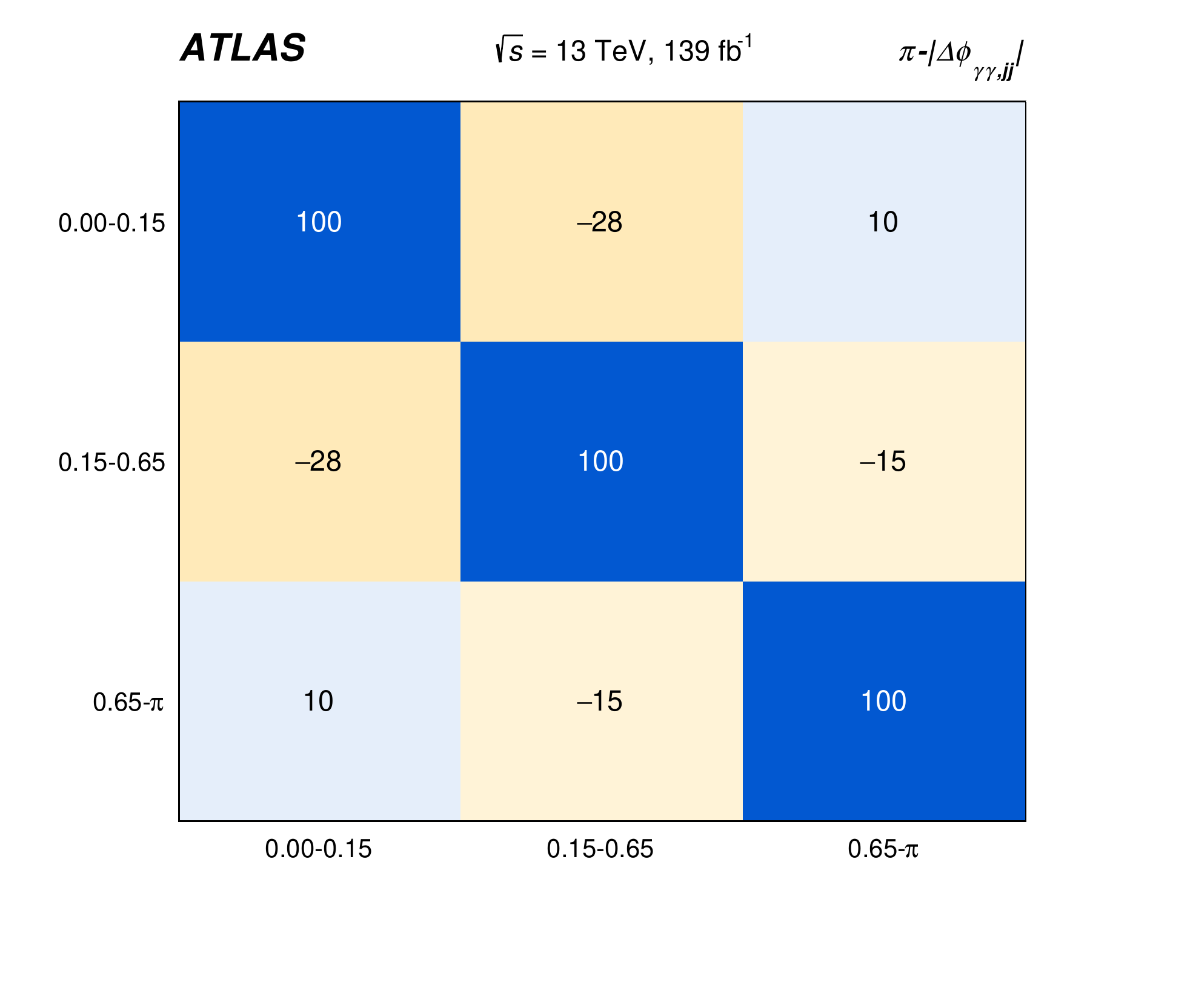}\label{appfig:data_unfolded_xsections_matrixinversion_1D_2jet_2_b}} \\
\subfloat[]{\includegraphics[width=0.5\textwidth]{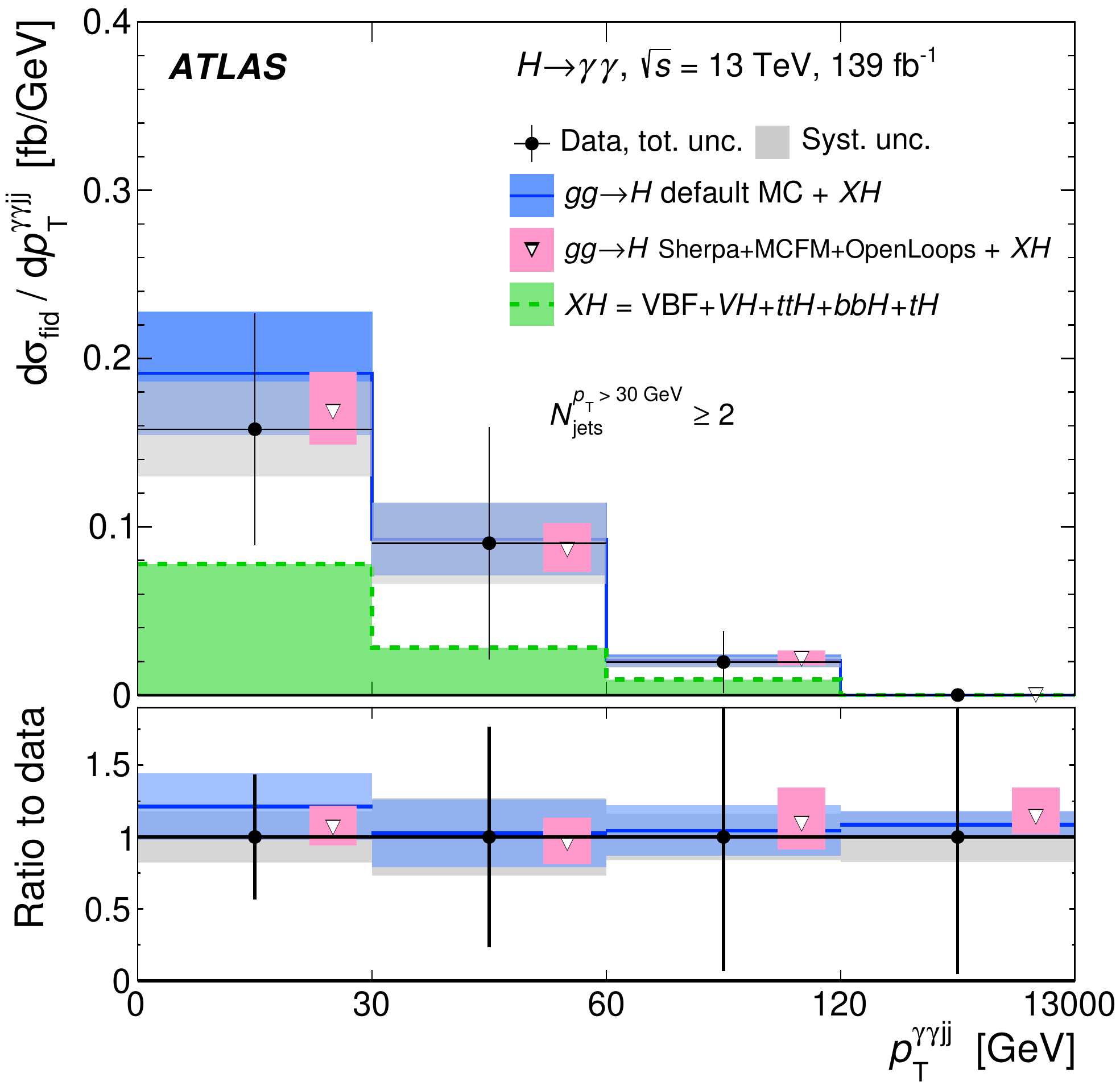}\label{appfig:data_unfolded_xsections_matrixinversion_1D_2jet_2_c}}
\subfloat[]{\includegraphics[width=0.5\textwidth]{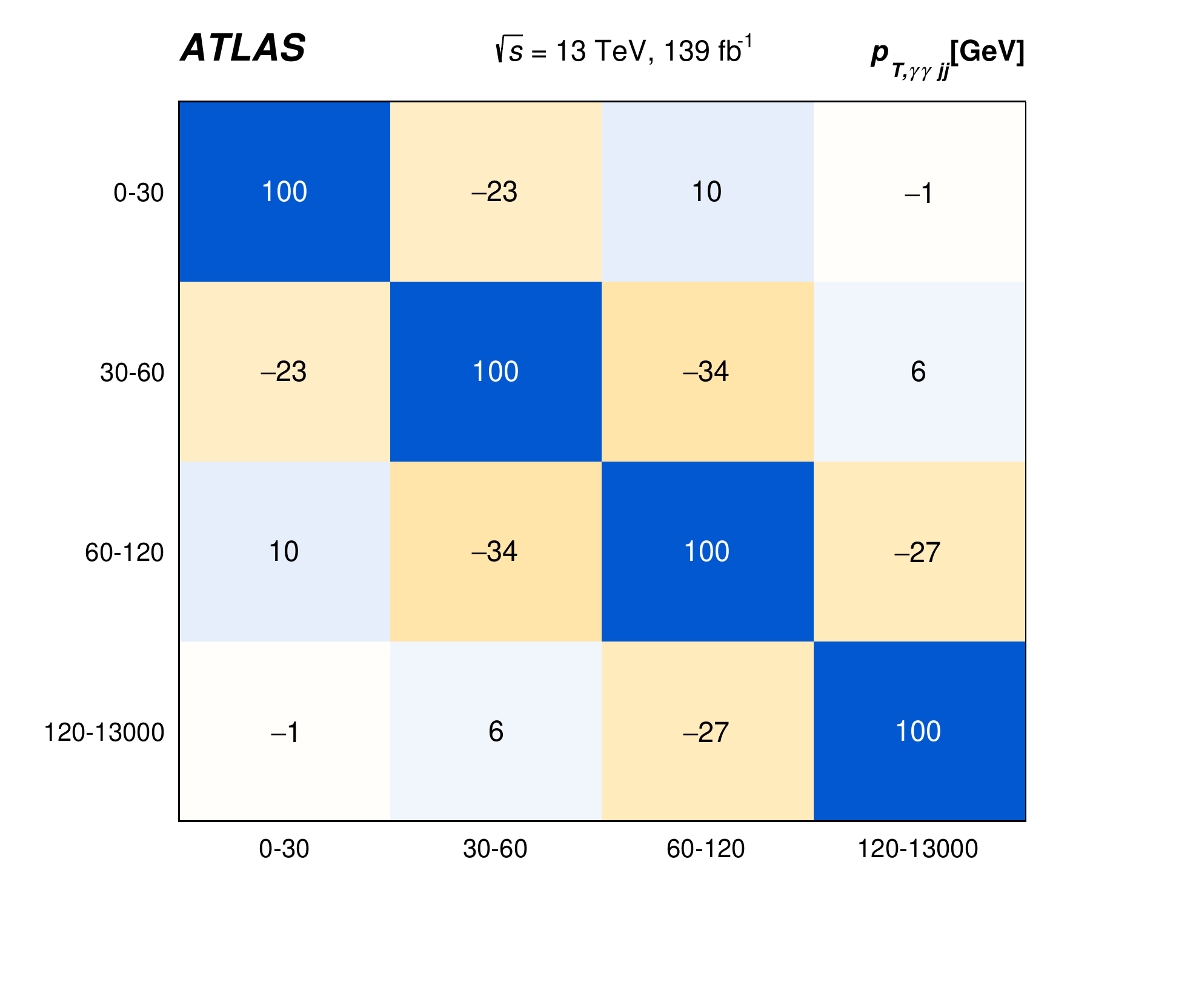}\label{appfig:data_unfolded_xsections_matrixinversion_1D_2jet_2_d}}\\
\caption{Particle-level fiducial differential cross-sections times branching ratio for the variables \protect\subref{appfig:data_unfolded_xsections_matrixinversion_1D_2jet_2_a} \(\pi-\absdphiggjj\) and \protect\subref{appfig:data_unfolded_xsections_matrixinversion_1D_2jet_2_c} \ptggjj together with the corresponding correlation matrices (\protect\subref{appfig:data_unfolded_xsections_matrixinversion_1D_2jet_2_b} and \protect\subref{appfig:data_unfolded_xsections_matrixinversion_1D_2jet_2_d}).}
\label{appfig:data_unfolded_xsections_matrixinversion_1D_2jet_2}
\end{figure}

Figure~\ref{appfig:data_unfolded_xsections_matrixinversion_1D_2jet_2} shows the differential cross-section for the variables \(\pi-\absdphiggjj\) and \ptggjj.
 
\begin{figure}[htbp]
\centering
\subfloat[]{\includegraphics[width=0.5\textwidth]{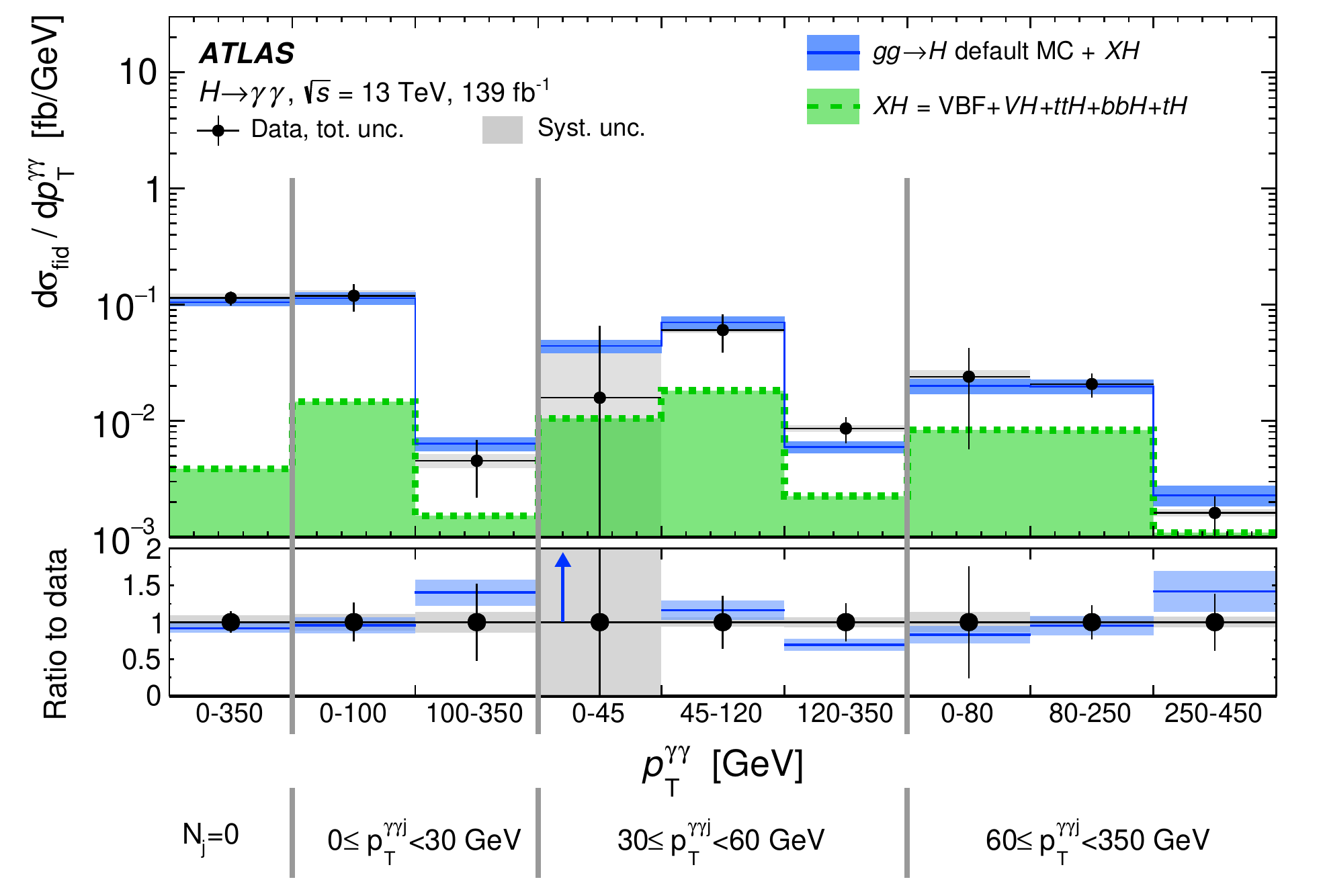}\label{appfig:data_unfolded_xsections_matrixinversion_2D_2_a}}
\subfloat[]{\includegraphics[width=0.5\textwidth]{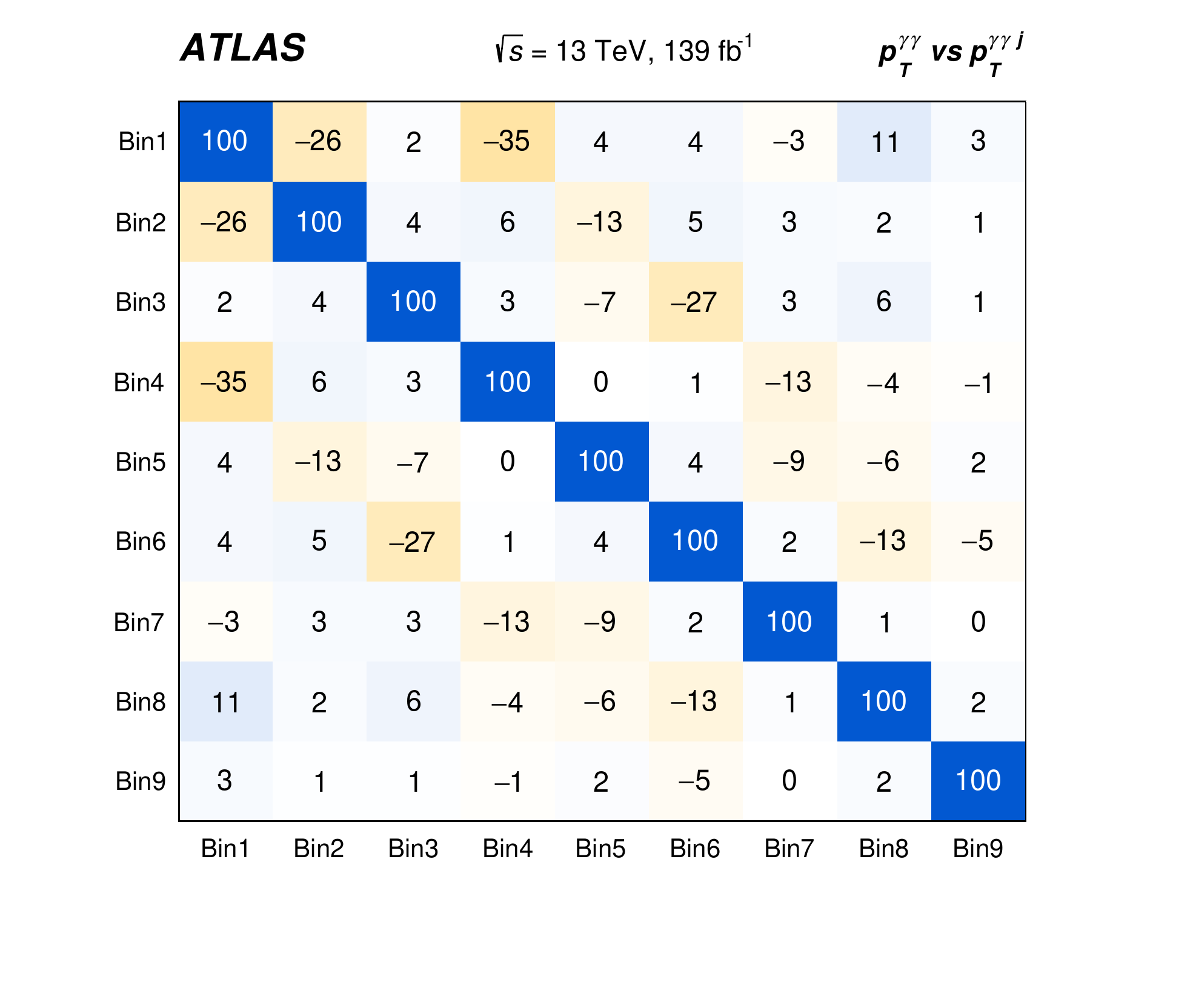}\label{appfig:data_unfolded_xsections_matrixinversion_2D_2_b}} \\
\subfloat[]{\includegraphics[width=0.5\textwidth]{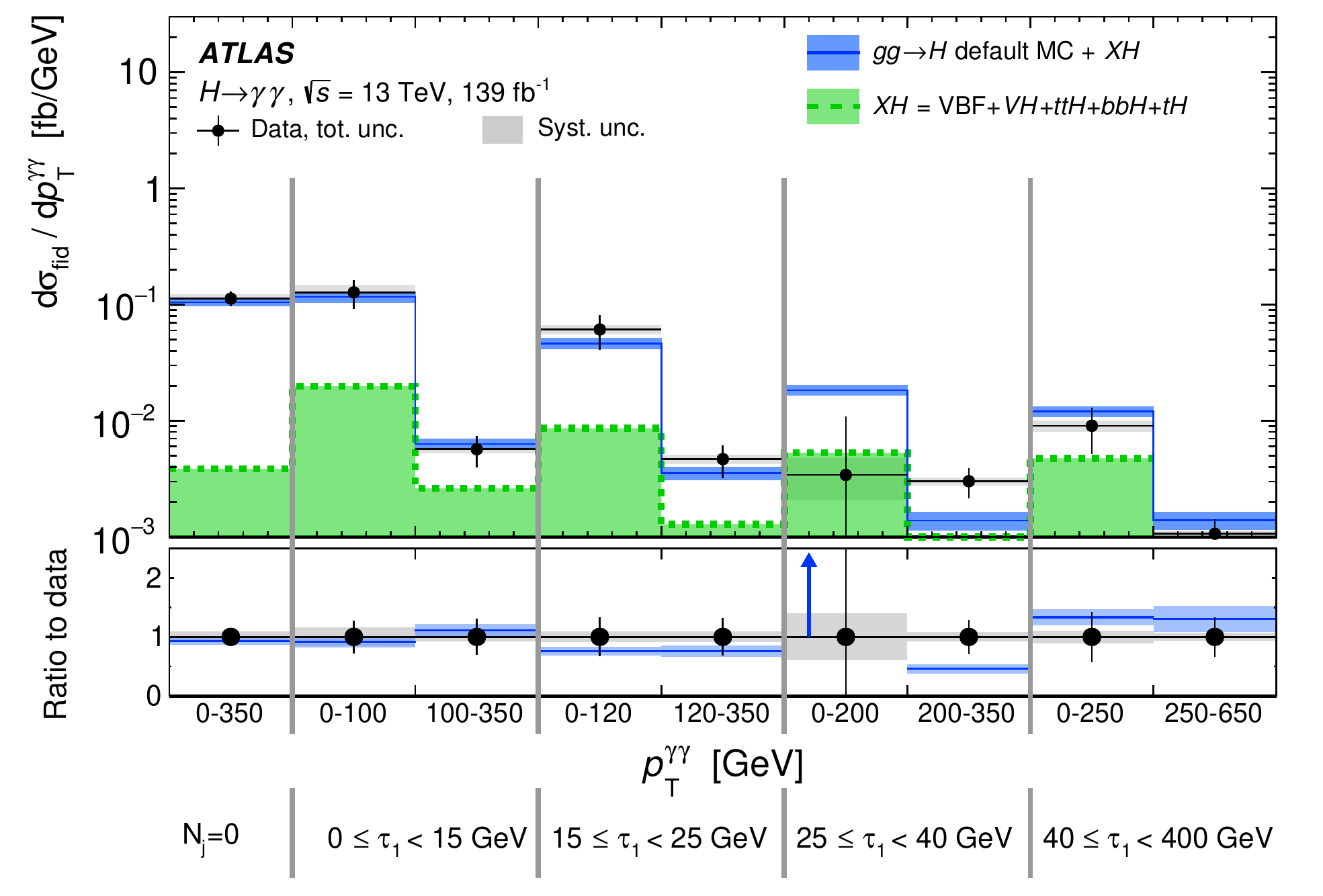}\label{appfig:data_unfolded_xsections_matrixinversion_2D_2_c}}
\subfloat[]{\includegraphics[width=0.5\textwidth]{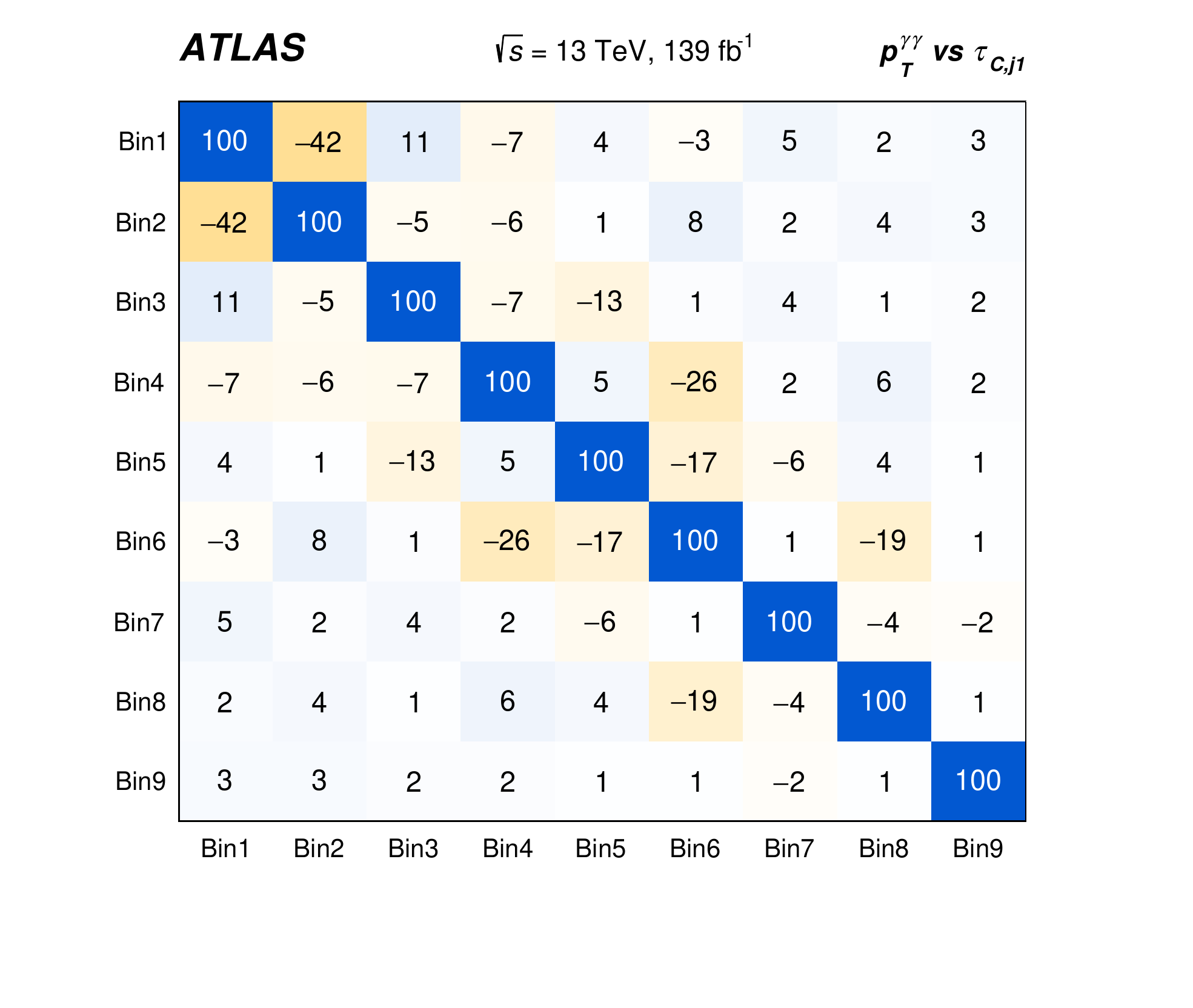}\label{appfig:data_unfolded_xsections_matrixinversion_2D_2_d}}\\
\caption{Double-differential particle-level fiducial cross-sections times branching ratio of \protect\subref{appfig:data_unfolded_xsections_matrixinversion_2D_2_a} \ptgg\ in bins of \ptggj\ and \protect\subref{appfig:data_unfolded_xsections_matrixinversion_2D_2_c} \ptgg\ in bins \maxtau\ together with the corresponding correlation matrices (\protect\subref{appfig:data_unfolded_xsections_matrixinversion_2D_2_b} and \protect\subref{appfig:data_unfolded_xsections_matrixinversion_2D_2_d}). The first bin corresponds to the case of \(\Njet=0\). The order of the bins in the correlation plots is the same as in the plots with the values of the cross-sections.}
\label{appfig:data_unfolded_xsections_matrixinversion_2D_2}
\end{figure}

\begin{figure}[htb!]
\centering
\subfloat[]{\includegraphics[width=0.5\textwidth]{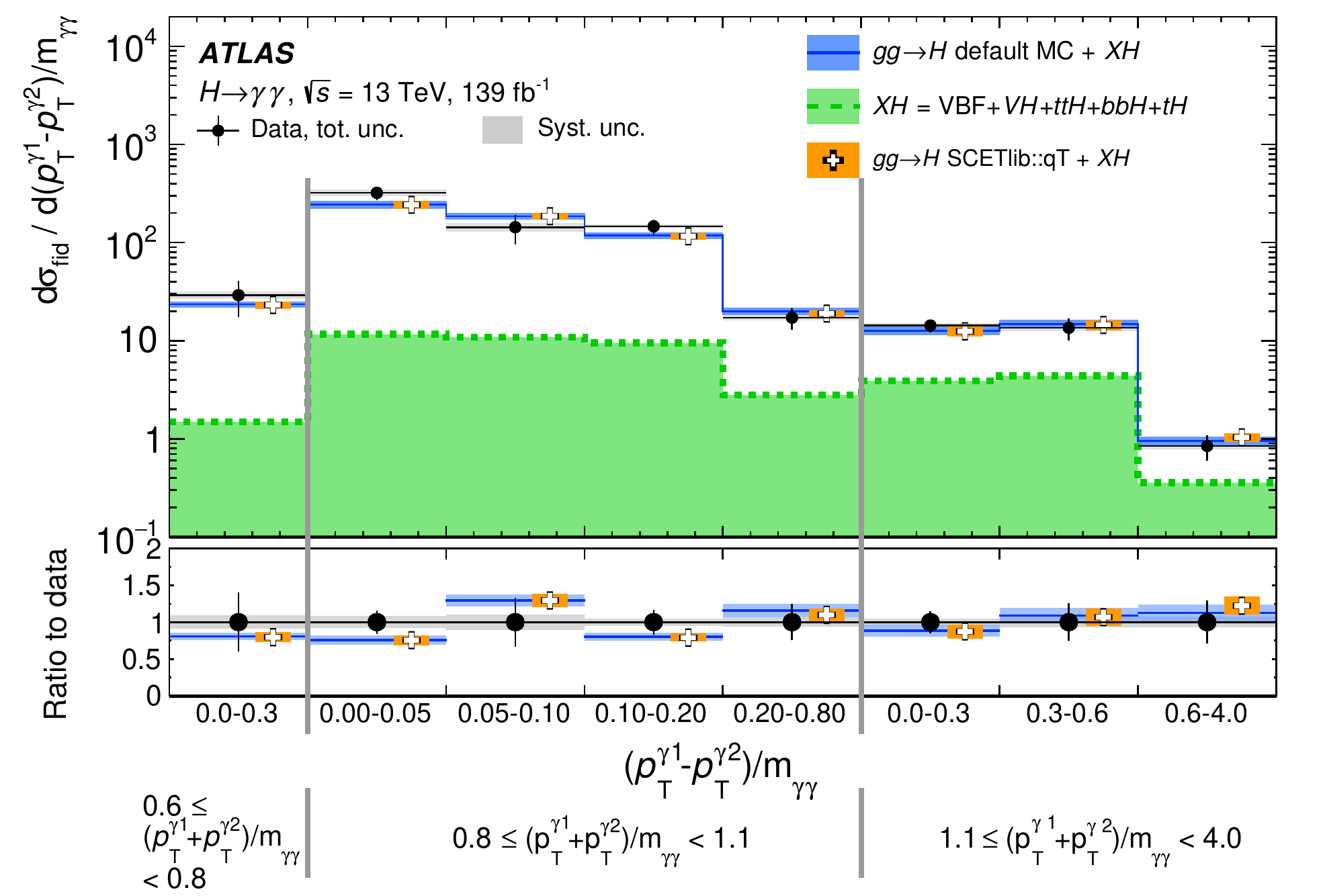}\label{appfig:data_unfolded_xsections_matrixinversion_2D_1_c}}
\subfloat[]{\includegraphics[width=0.5\textwidth]{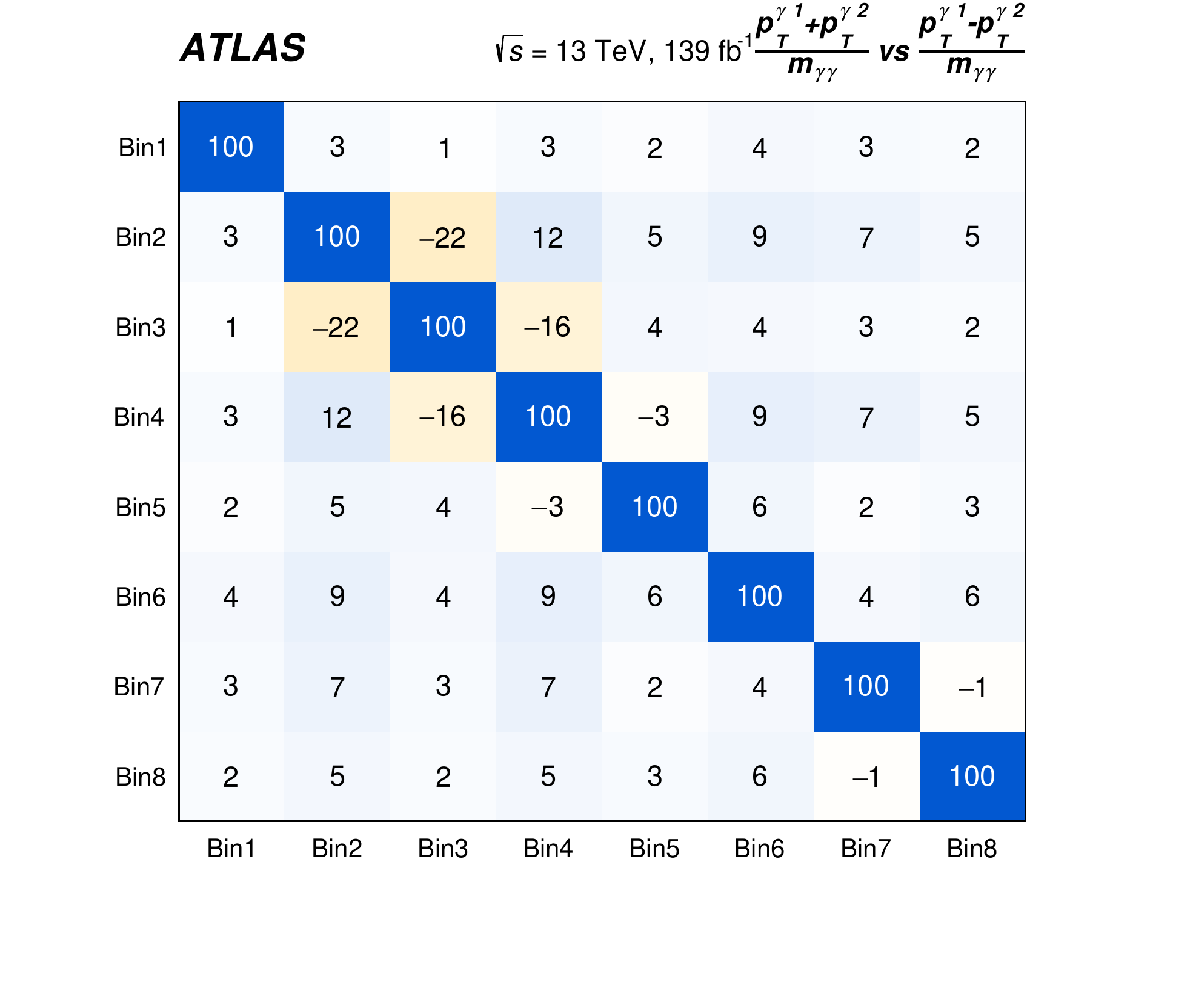}\label{appfig:data_unfolded_xsections_matrixinversion_2D_1_d}}
\caption{Double-differential particle-level fiducial cross-sections times branching ratio of \protect\subref{appfig:data_unfolded_xsections_matrixinversion_2D_1_c} \reldiffptgg\ in bins of \relsumptgg\ together with the corresponding correlation matrix \protect\subref{appfig:data_unfolded_xsections_matrixinversion_2D_1_d}. The order of the bins in the correlation plots is the same as in the plots with the values of the cross-sections.}
\label{appfig:data_unfolded_xsections_matrixinversion_2D_1}
\end{figure}

\paragraph{Double-differential cross-sections}
Figures~\labelcref{appfig:data_unfolded_xsections_matrixinversion_2D_2,appfig:data_unfolded_xsections_matrixinversion_2D_1} show the double-differential cross-sections for the variables: \ptgg\ vs \ptggj, \ptgg\ vs \maxtau\ and \relsumptgg\ vs \reldiffptgg. Overall, good agreement is observed for the double-differential cross-sections for the photon variables, with \scetlib providing a more accurate description. Slight shape differences between data and the default simulation are observed for some bins in the double-differential cross-sections with an additional jet, namely for \ptgg\ vs \maxtau.

\begin{figure}[htbp]
\centering
\subfloat[]{\includegraphics[width=0.5\textwidth]{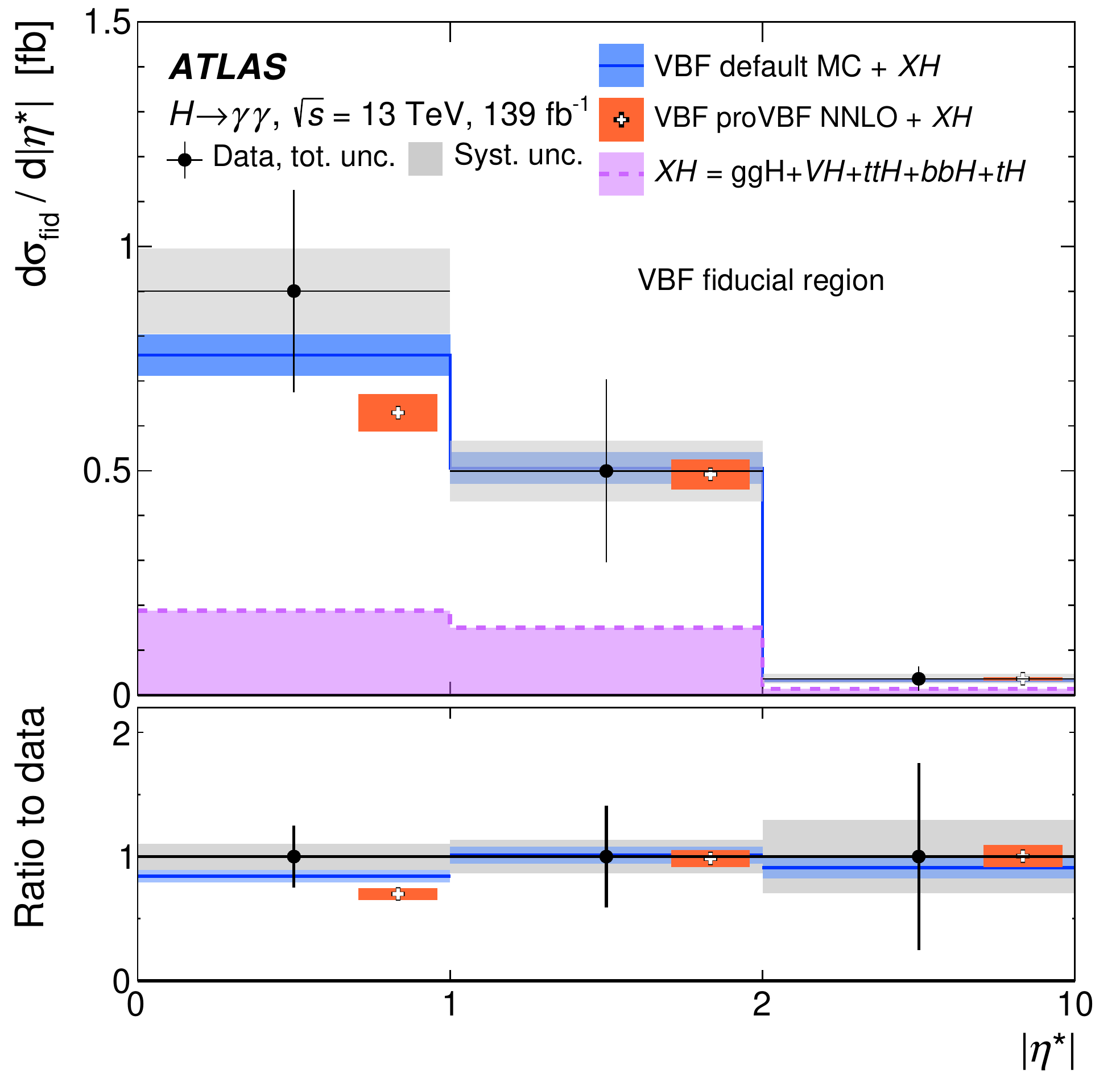}\label{appfig:data_unfolded_xsections_matrixinversion_VBF_2_a}}
\subfloat[]{\includegraphics[width=0.5\textwidth]{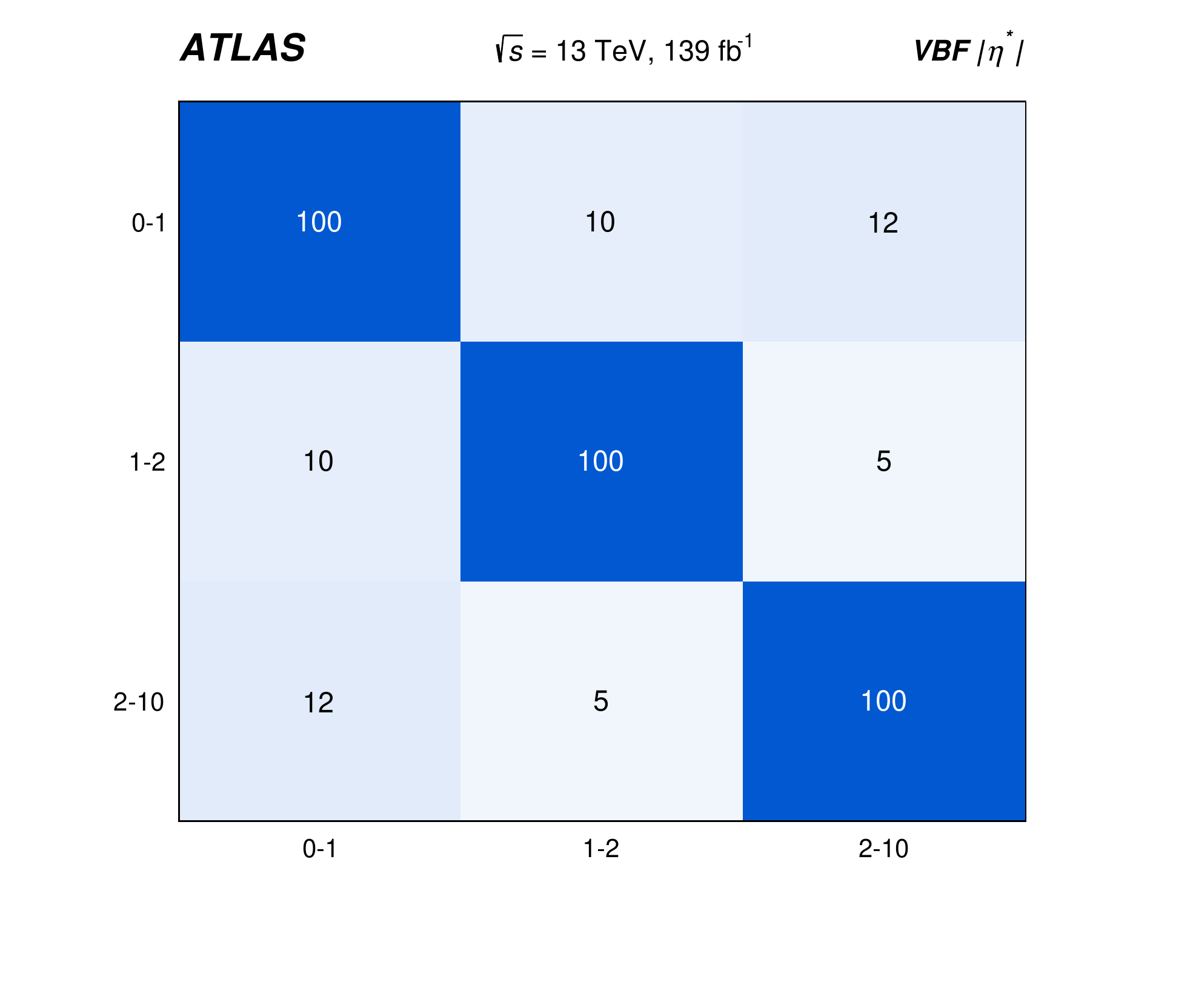}\label{appfig:data_unfolded_xsections_matrixinversion_VBF_2_b}} \\
\subfloat[]{\includegraphics[width=0.5\textwidth]{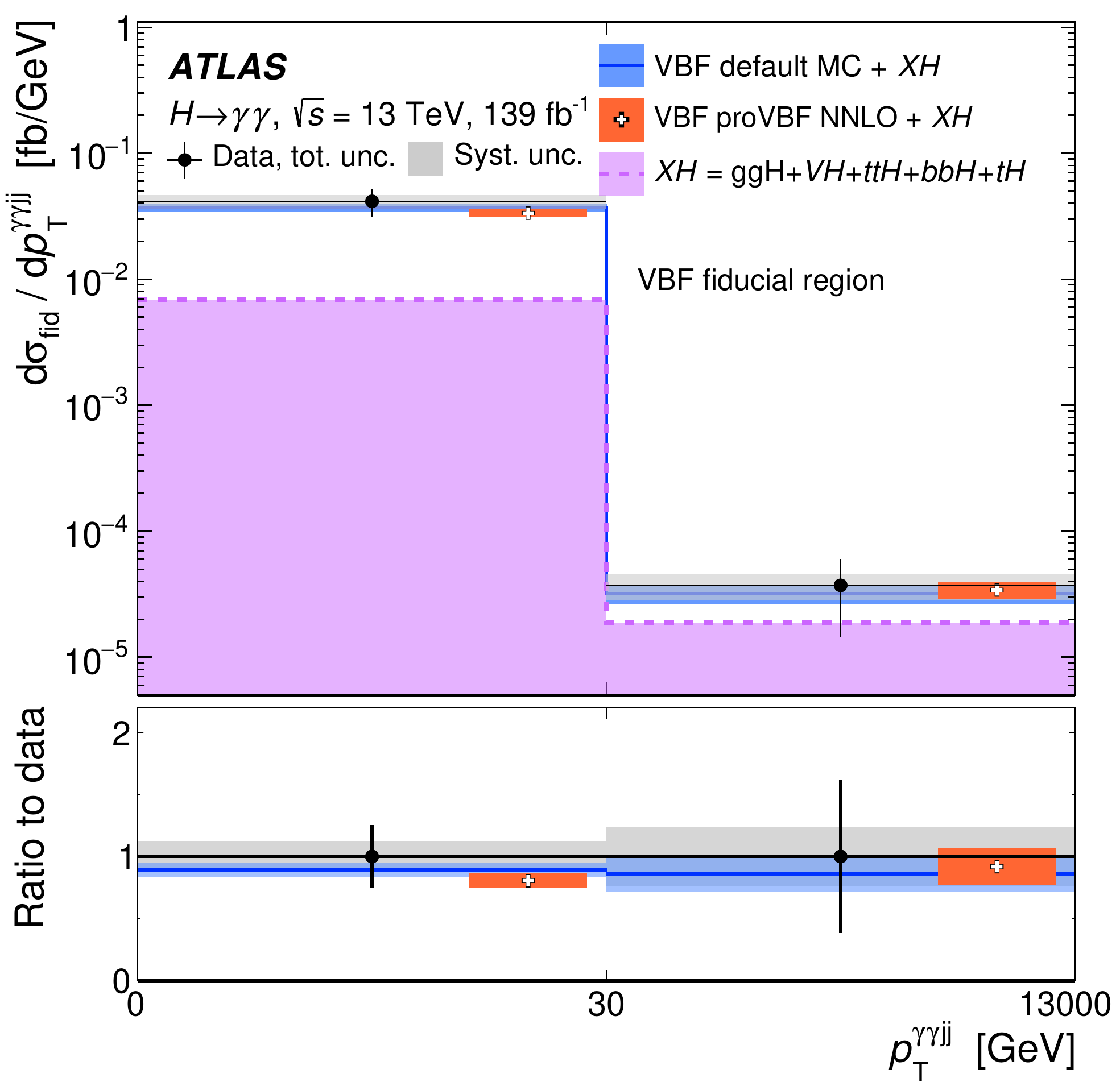}\label{appfig:data_unfolded_xsections_matrixinversion_VBF_2_c}}
\subfloat[]{\includegraphics[width=0.5\textwidth]{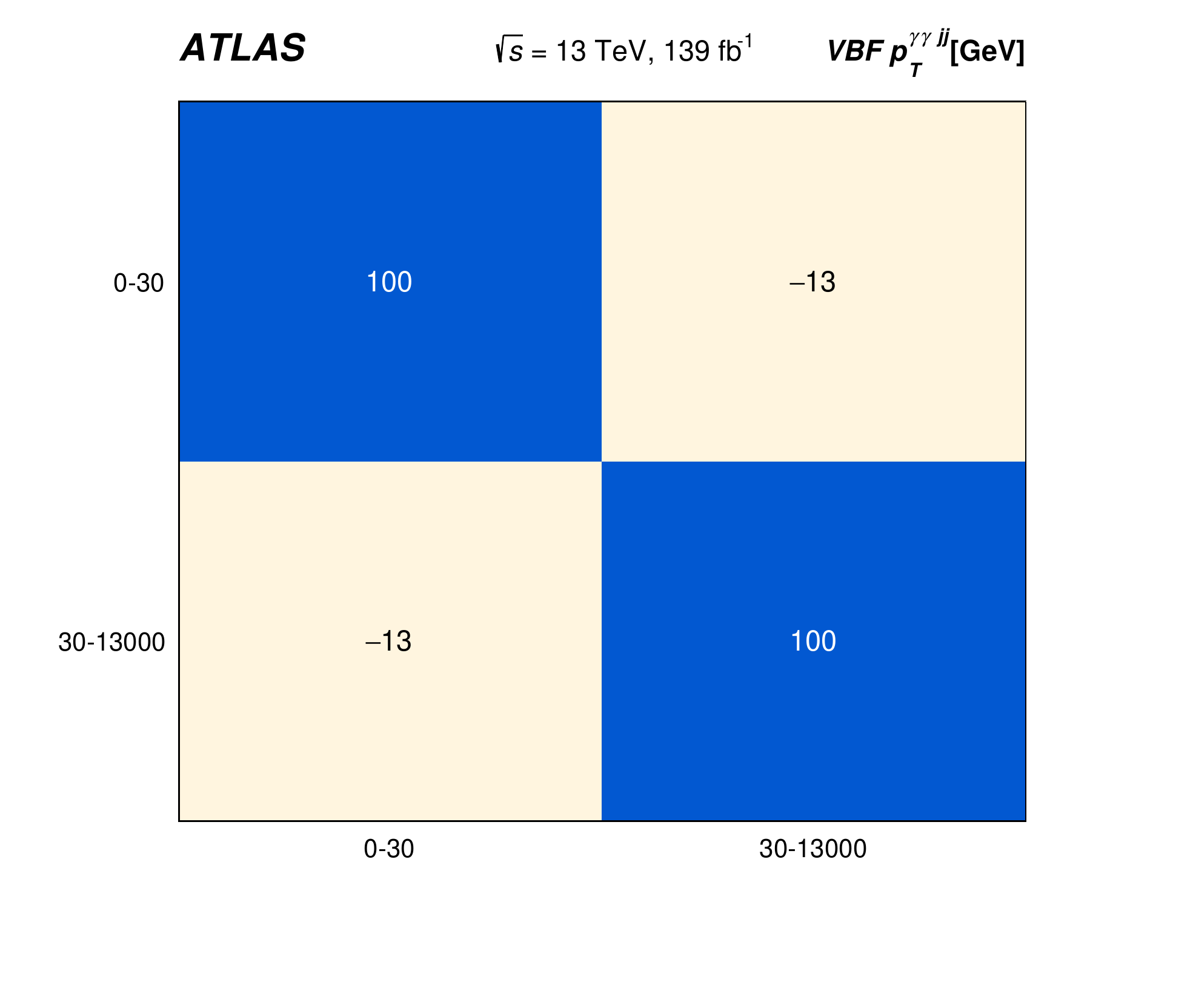}\label{appfig:data_unfolded_xsections_matrixinversion_VBF_2_d}}\\
\caption{Particle-level fiducial differential cross-sections times branching ratio for the variables \protect\subref{appfig:data_unfolded_xsections_matrixinversion_VBF_2_a} \(|\eta^*|\) and \protect\subref{appfig:data_unfolded_xsections_matrixinversion_VBF_2_c} \ptggjj\ together with the corresponding correlation matrices (\protect\subref{appfig:data_unfolded_xsections_matrixinversion_VBF_2_b} and \protect\subref{appfig:data_unfolded_xsections_matrixinversion_VBF_2_d}) in the VBF-enhanced fiducial region.}
\label{appfig:data_unfolded_xsections_matrixinversion_VBF_2}
\end{figure}

\begin{figure}[htb!]
\centering
\subfloat[]{\includegraphics[width=0.5\textwidth]{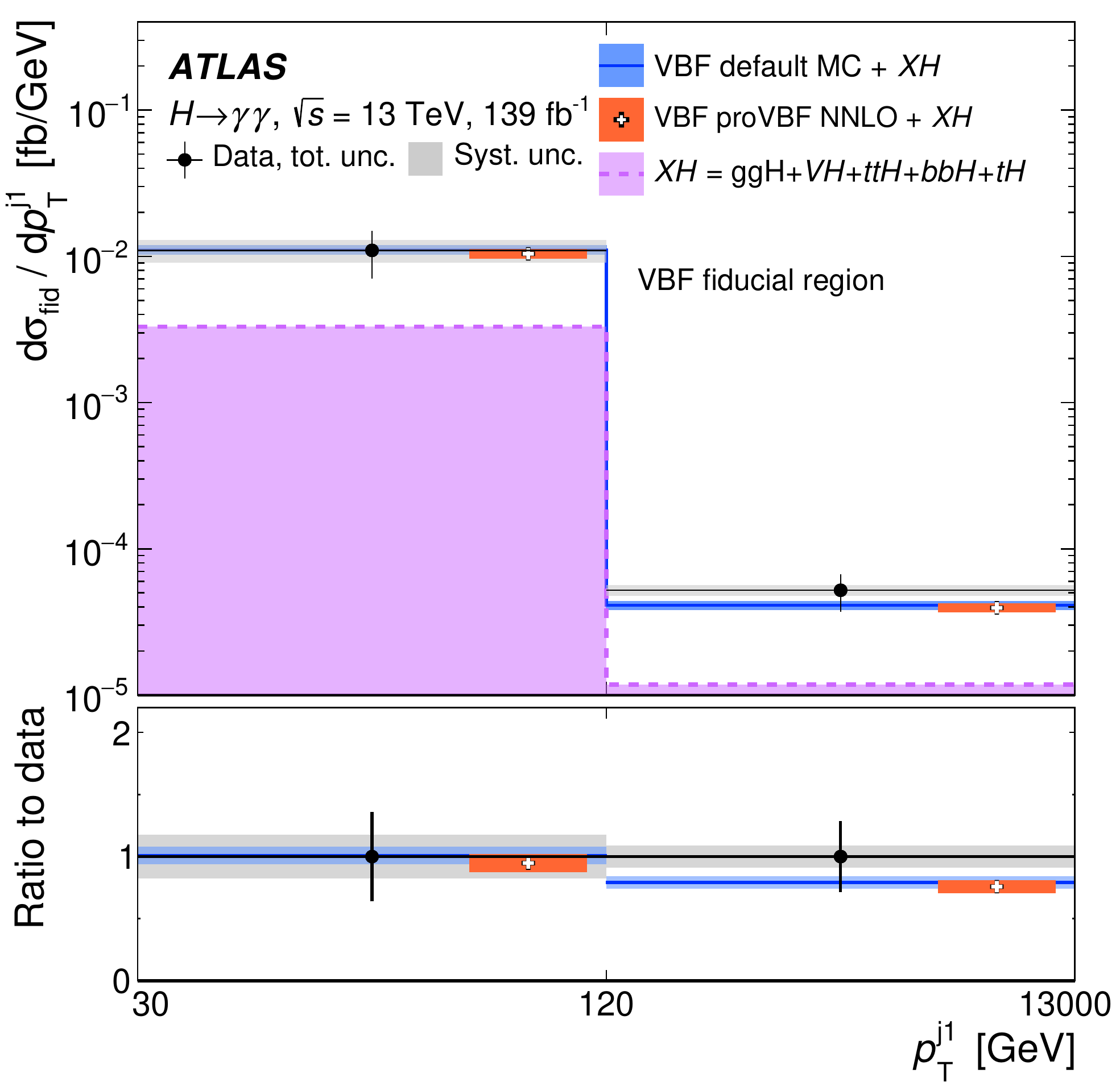}\label{appfig:data_unfolded_xsections_matrixinversion_VBF_1_a}}
\subfloat[]{\includegraphics[width=0.5\textwidth]{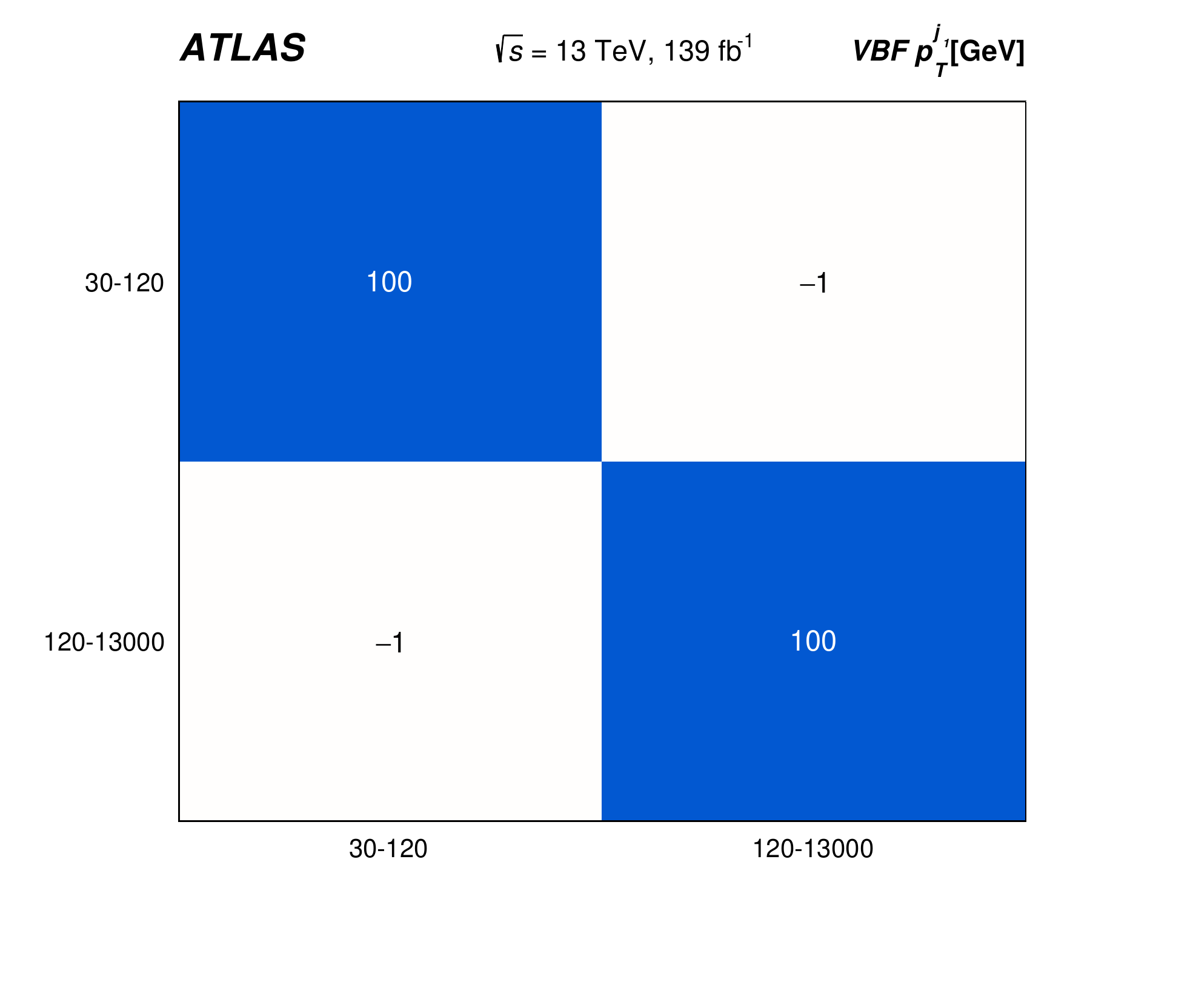}\label{appfig:data_unfolded_xsections_matrixinversion_VBF_1_b}} \\
\subfloat[]{\includegraphics[width=0.5\textwidth]{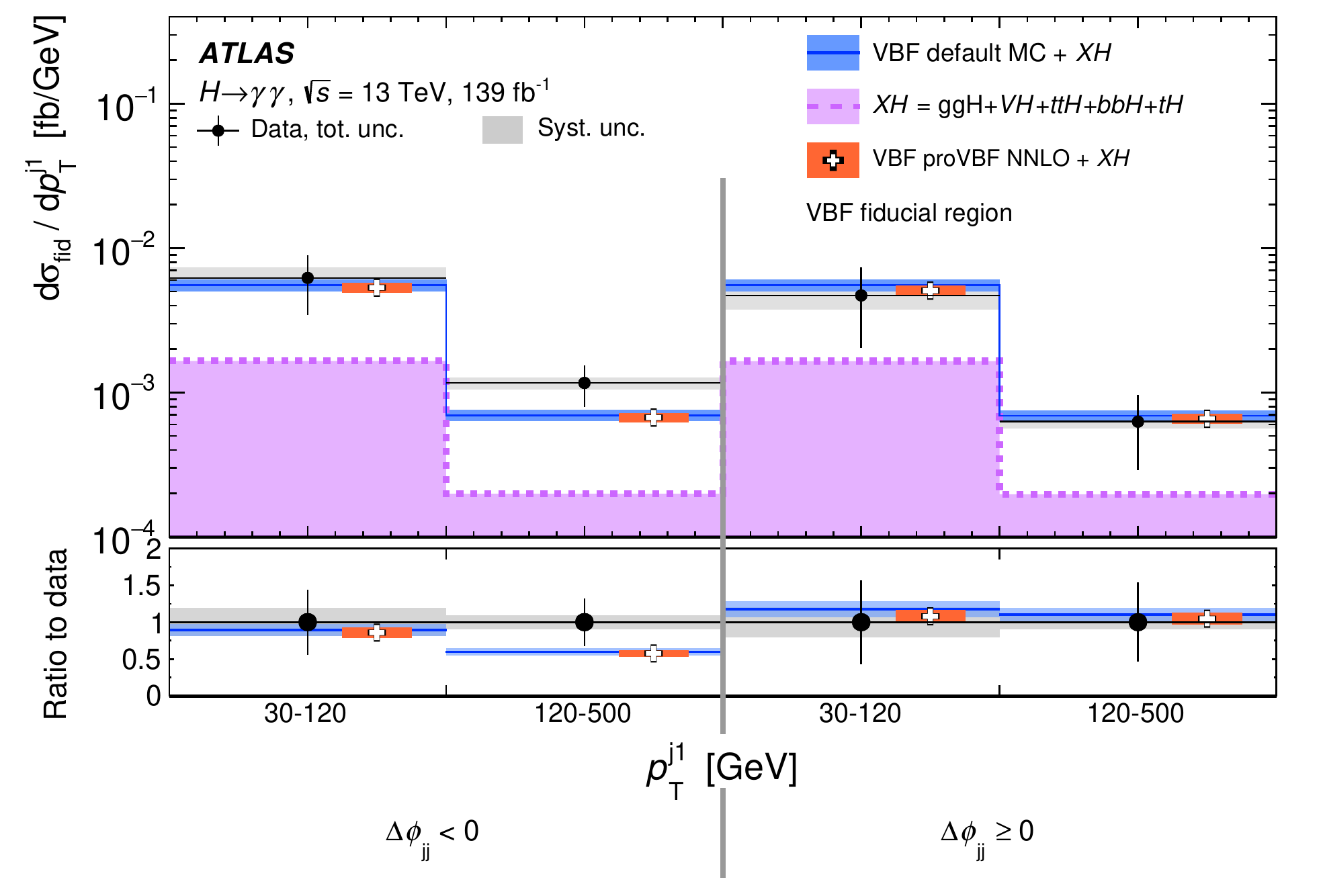}\label{appfig:data_unfolded_xsections_matrixinversion_VBF_3_a}}
\subfloat[]{\includegraphics[width=0.5\textwidth]{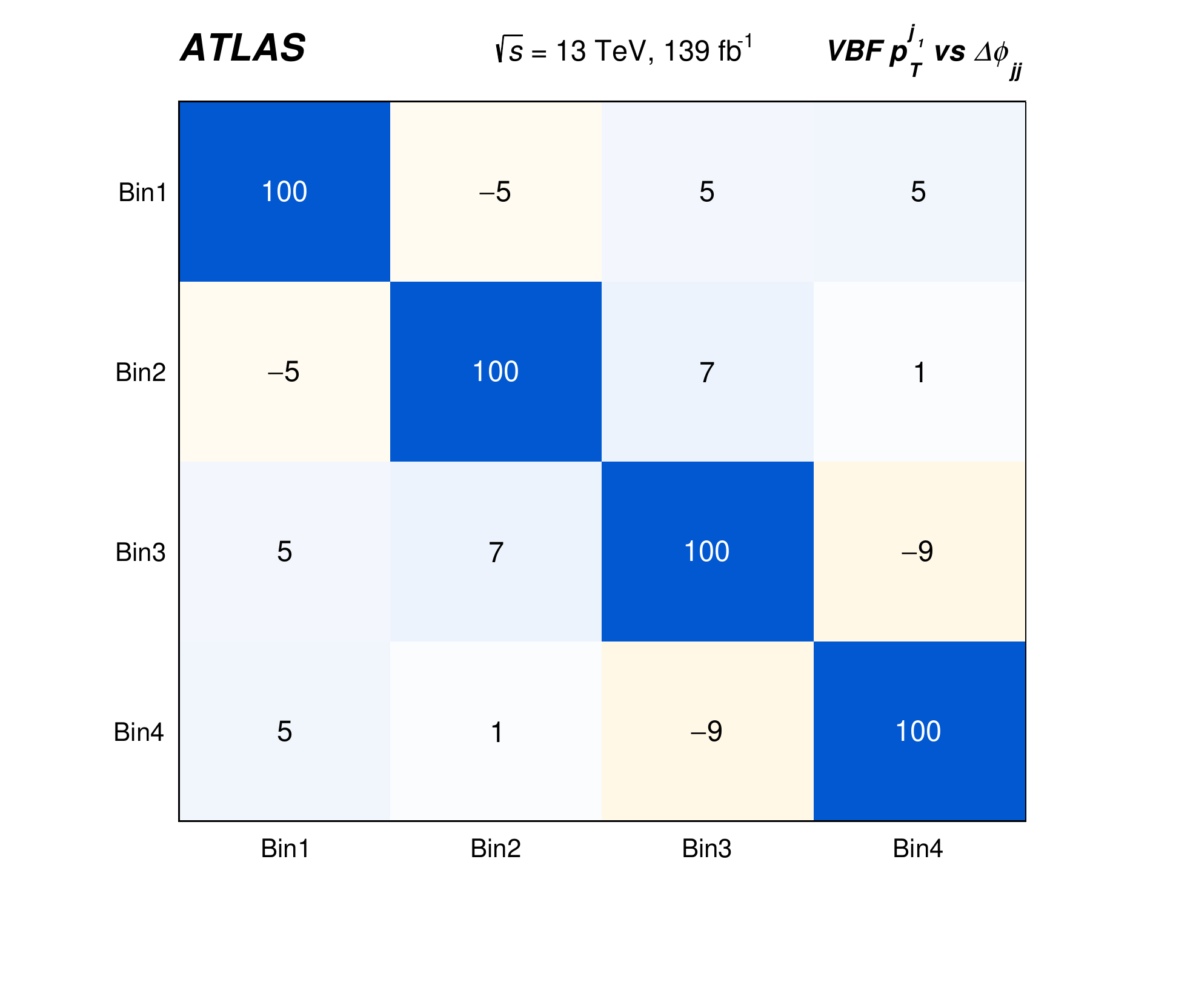}\label{appfig:data_unfolded_xsections_matrixinversion_VBF_3_b}}
 
\caption{Particle-level fiducial differential cross-section times branching ratio for the variable \protect\subref{appfig:data_unfolded_xsections_matrixinversion_VBF_1_a} \ptj[1] together with the corresponding correlation matrix in the VBF-enhanced fiducial region \protect\subref{appfig:data_unfolded_xsections_matrixinversion_VBF_1_b}. Double-differential particle-level fiducial cross-sections times branching ratio of \protect\subref{appfig:data_unfolded_xsections_matrixinversion_VBF_3_a} \ptj[1] in bins of \dphijj\ together with the \protect\subref{appfig:data_unfolded_xsections_matrixinversion_VBF_3_b} corresponding correlation matrices in the VBF-enhanced fiducial region. The order of the bins in the correlation plot is the same as in the plot with the values of the cross-sections. }
\label{appfig:data_unfolded_xsections_matrixinversion_VBF_1}
\end{figure}

\paragraph{Cross-sections in the VBF-enhanced phase space}
Figures~\labelcref{appfig:data_unfolded_xsections_matrixinversion_VBF_2,appfig:data_unfolded_xsections_matrixinversion_VBF_1} show the differential cross-sections in the VBF-enhanced phase space for the variables \(|\eta^*|\), \ptggjj, and \ptj[1], and the double-differential cross-section for \ptj[1] vs \dphijj\. The coarser binning for these measurements reflects the current statistical precision in the VBF-enhanced region.

\FloatBarrier
\section{Uncertainties in additional theory predictions}
\label{sec:aux theory preds unc}
In this section the uncertainties in the new additional theory predictions are summarised, along with the different scales used to produce them. For the inclusive predictions, the Higgs to diphoton branching ratio uncertainty from Ref.~\cite{deFlorian:2016spz} is included.
 
\paragraph{\textbf{ MATRIX+RadISH predictions}}
These predictions use the \NNPDF[3.1] PDF~\cite{Ball:2017nwa} set with \(\alphas(m_Z)=0.118\). The normalisation and factorisation scales were set to \mH\ and the resummation scale to \(\mH/2\). All scales were varied by a factor of 2 around their central values (but with the restriction \(1/2 \leq \muR/\muF\leq 2\)). This calculation was used to predict the fiducial differential cross-sections as a function of the diphoton transverse momentum in events passing a jet veto. The theoretical uncertainty is of the order of \(10\%\) for \(p_{\mathrm{T}}^{H}\leq p_{\mathrm{T}}^\mathrm{JV}\), where \(p_{\mathrm{T}}^\mathrm{JV}\) is the \pT\ used to define the jet veto.

\paragraph{\textbf{ RadISH+NNLOjet predictions}}
These predictions use the \PDFforLHC[15nnlo] PDF set and the normalisation, factorisation, and resummation scales were set to \(\mH/2\).
In addition, predictions for \ptj[1] are made at NNLL+NNLO QCD accuracy using \textsc{RadISH+NNLOjet} following Refs.~\cite{Banfi:2012jm,Banfi:2015pju}. In this case, the results are obtained for a stable Higgs boson and an acceptance correction is applied to account for the fiducial selection. These predictions use the NNPDF3.1 PDF set, the normalisation and factorisation scales were set to \mH\ and the resummation scale to \(\mH/2\).
In both cases, all scales were varied by a factor of 2 around the central values (but with the restriction \(1/2 \leq \muR/\muF\leq 2\)). In addition, finite top-quark mass corrections derived from the default simulation were applied to the \textsc{RadISH+NNLOjet} predictions.

\paragraph{\SHERPA[2.2.11]}
The \SHERPA\ predictions were produced using the \PDFforLHC[15nlo] PDF set~\cite{Butterworth:2015oua}. The uncertainties in the predictions are estimated from the six \(1/2\le  \muR, \muF \le 2\) scale variations (with the restriction \(1/2 \leq \muR/\muF\leq 2\)) and the 30 PDF eigen-variations.

\paragraph{ResBos2}
For the inclusive calculation, the renormalisation and factorisation scales were varied by a factor of 2 around the central value of \(\mH/2\) with the usual the restriction \(1/2 \leq \muR/\muF\leq 2\). In addition, the resummation scale is set to be the same as the renormalisation scale. Furthermore, finite top-quark mass corrections derived from the default simulation were applied to the inclusive Higgs production \textsc{ResBos2} predictions.
In the Higgs-plus-jet calculation, the resummation scale is fixed to be the jet transverse momentum as suggested in Ref.~\cite{Sun:2016kkh}, and the renormalisation and factorisation scales are varied by a factor of two around the central value of \(\mH/2\) (with the restriction \(1/2 \leq \muR/\muF\leq 2\)).
 
\paragraph{\scetlib}
The predictions are computed using the \PDFforLHC[15nnlo] PDF set.
All required contributions for matching to fixed-order NNLO calculations are included directly in
\scetlib. The matching to N\(^3\)LO calculations uses as inputs the known
N\(^3\)LO\(_0\) correction to the total inclusive cross-section~\cite{Mistlberger:2018etf}
and existing \textsc{NNLOjet} results for the NNLO\(_1\)
corrections to the \pT\ spectrum from Refs.~\cite{Chen:2018pzu, Bizon:2018foh}.
 
The top-quark Yukawa coupling \(y_t^2\) contributions at N\(^3\)LL\('+\)N\(^3\)LO accuracy are computed in the rEFT limit,
i.e.\ in the \(m_t\to\infty\) limit rescaled with the exact LO \(m_t\)-dependence.
This approximation is valid up to around \(\pt = \SI{200}{GeV}\).
In addition, \textsc{SCETlib} predictions were provided including the \(y_t^2\), \(y_t y_b\), \(y_t y_c\), \(y_b^2\), \(y_c^2\), \(y_b y_c\)
contributions for the \(gg\to H\) \pT\ spectrum to NNLL\(+\)NLO accuracy with
the exact dependencies on \(m_t\), \(m_b\), \(m_c\), and are used for the bottom- and charm-quark Yukawa coupling interpretations detailed in Section~\ref{subsec:YukawaInterpretation}.
 
Perturbative uncertainties from several sources were estimated
through appropriately chosen variations following Refs.~\cite{Stewart:2013faa, Lustermans:2019plv,Ebert:2020dfc}.
These include: (i) resummation uncertainties estimated as the maximum envelope of 36
combinations of upward/downward variations of the four involved resummation scales, resulting in a 15\% uncertainty for the lowest \ptgg\ bins, decreasing with increasing \ptgg; (ii) a matching uncertainty corresponding to the ambiguity in carrying out the
matching to fixed order; (iii) a fixed-order uncertainty which estimates the effect of missing higher-order corrections by varying the overall fixed-order scale \(\mu_\mathrm{FO}\) by a
factor of 2; this uncertainty dominates for \(\ptgg>\SI{45}{\GeV}\) and reaches 8\%; (iv) a non-perturbative uncertainty due to the sensitivity to non-perturbative effects below \(\pT \sim \SI{1}{GeV}\).
 
\paragraph{SCETlib predictions for \ptj}
 
The predictions are made using the \PDFforLHC[15nnlo] PDF set with \(\alphas(\mZ) = 0.118\).
All fiducial requirements are applied when performing the calculation, except for the photon isolation requirement, for which a dedicated correction computed from the full-simulation MC samples is applied.

\paragraph{proVBF predictions}
The renormalisation and factorisation scale were varied to estimate the residual theoretical uncertainties. A three-point scale variation $\muR=\muF$ with $\muR/\muF=\{\frac{1}{2}, 1, 2\}\mu_{0}$ was used to estimate the uncertainties, where $\mu_0$ is the $\pt$-dependent scale described in Ref.~\cite{proVBFNNLO}. The predictions were made using \NNPDF[3.0] PDF set. Corrections were applied to the predictions to account for the fiducial acceptance corrections obtained from the default simulation.
\FloatBarrier
\section{Two-dimensional limits on the Effective Field Theory couplings}
\label{sec:aux eft 2d}
\begin{figure}[!htbp]
\centering
\subfloat[]{\includegraphics[width=0.5\textwidth,page=1]{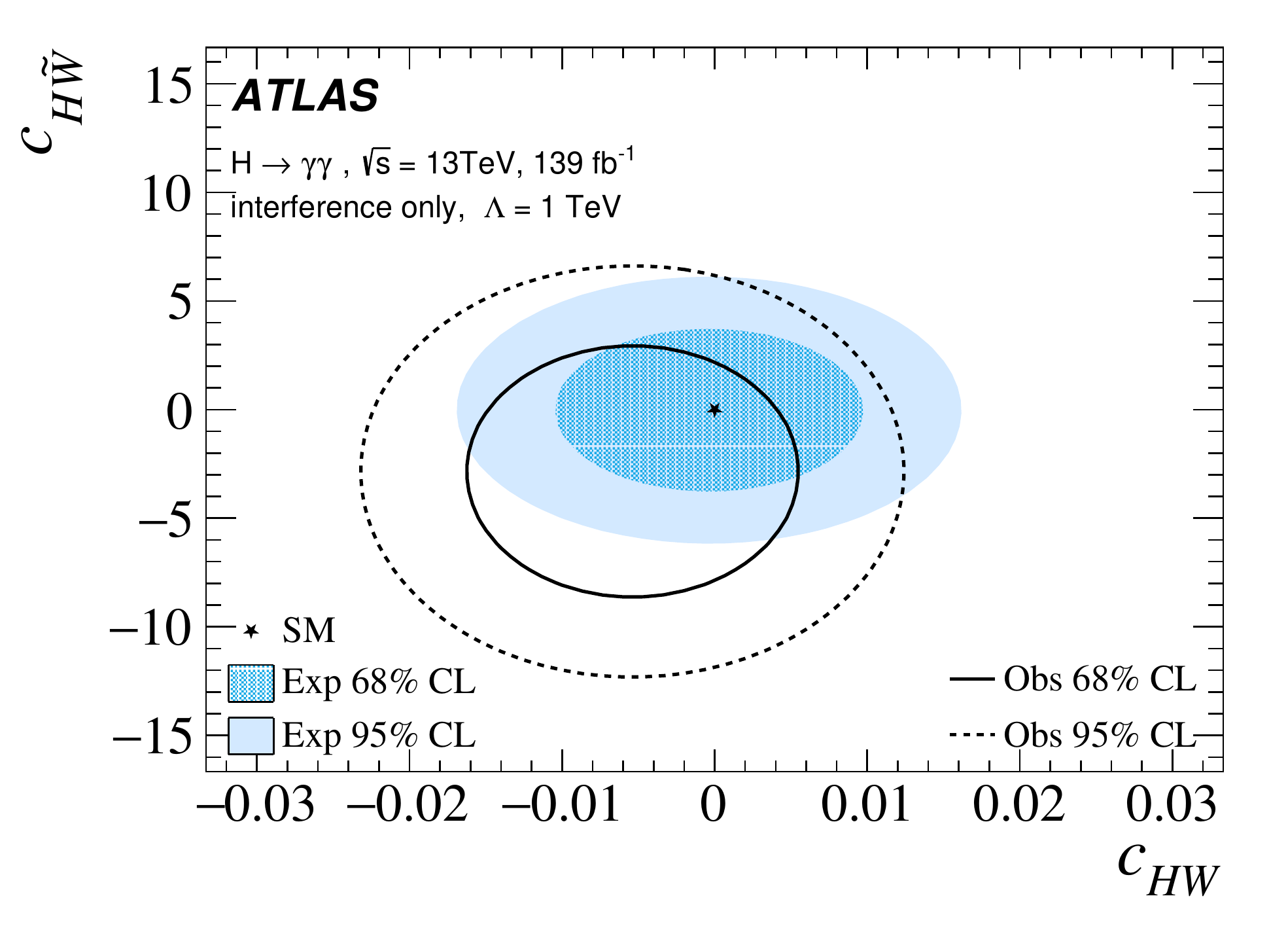}\label{fig:2dSMEFT_a}}
\subfloat[]{\includegraphics[width=0.5\textwidth,page=1]{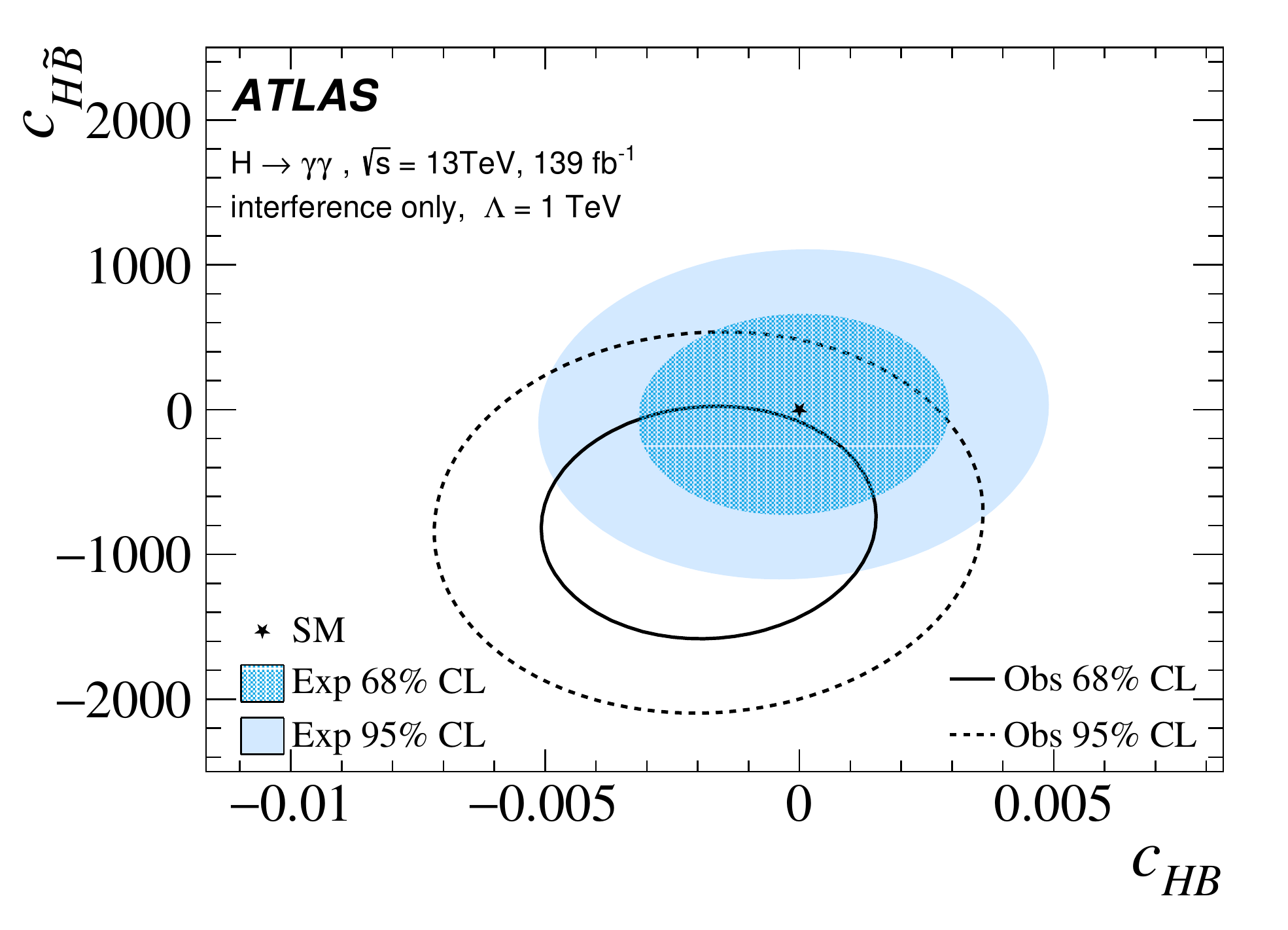}\label{fig:2dSMEFT_b}}\\
\subfloat[]{\includegraphics[width=0.5\textwidth,page=1]{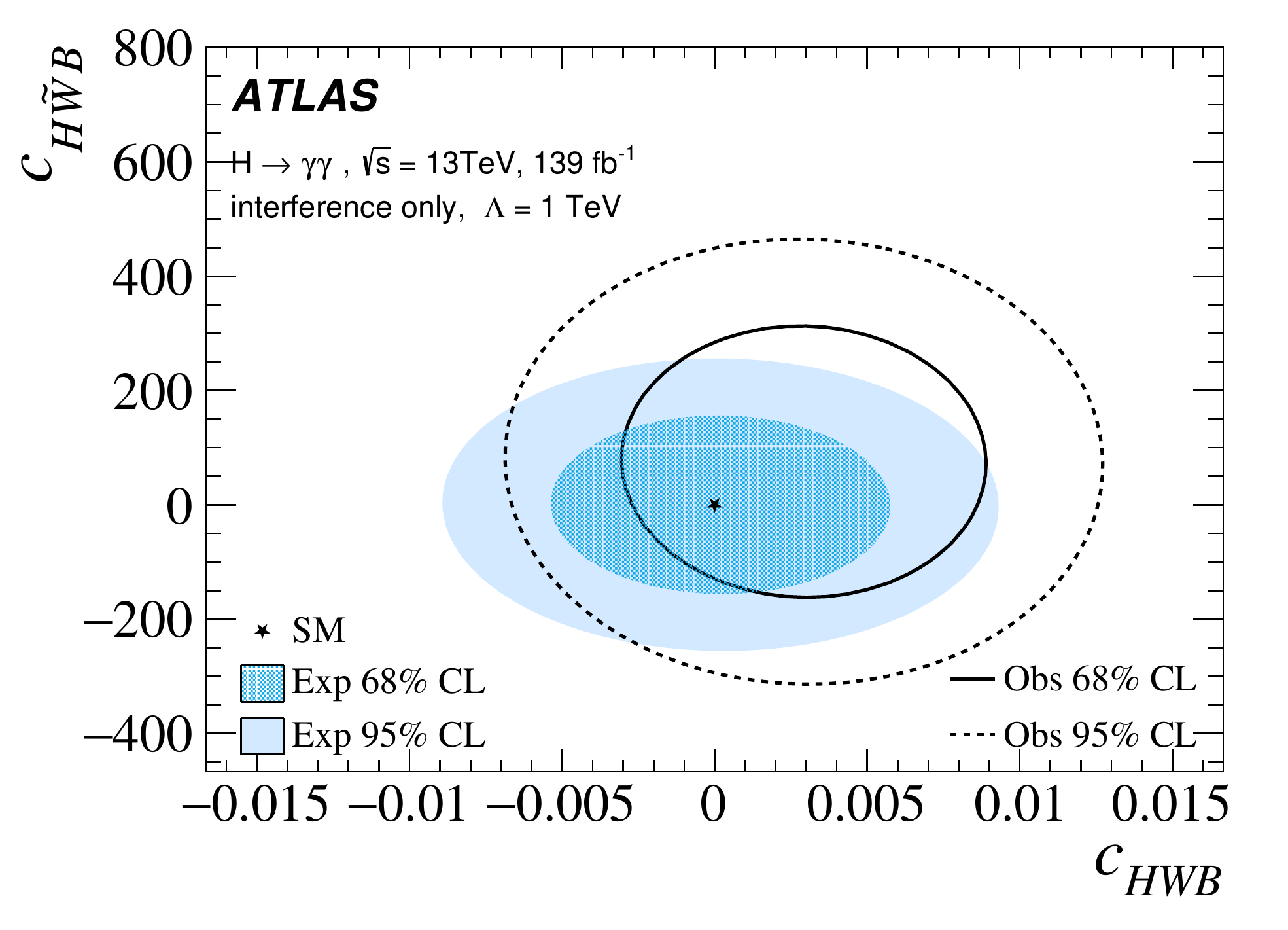}\label{fig:2dSMEFT_c}}
\subfloat[]{\includegraphics[width=0.5\textwidth,page=1]{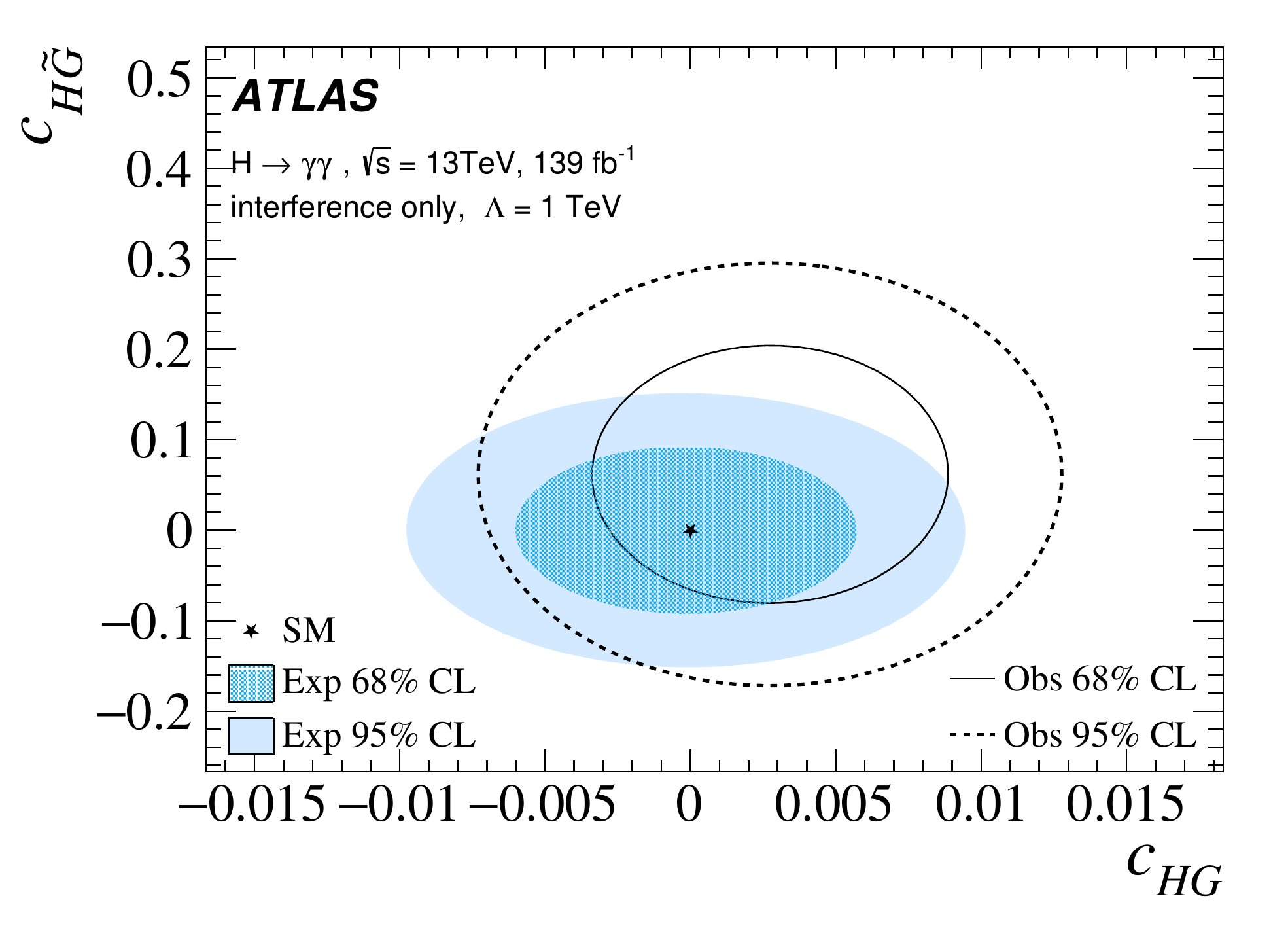}\label{fig:2dSMEFT_d}}\\
\caption{Plots showing the 2D 68\% and 95\% observed and expected limits obtained from various combinations of two Wilson coefficients using only the SM--dimension-6 interference in the SMEFT basis: \protect\subref{fig:2dSMEFT_a} \CHW\ vs \CHWt, \protect\subref{fig:2dSMEFT_b} \CHB\ vs \CHBt, \protect\subref{fig:2dSMEFT_c} \CHWB\ vs \CHWBt, \protect\subref{fig:2dSMEFT_d} \CHG\ vs \CHGt. The limits are computed at a new-physics scale \(\Lambda=\SI{1}{\TeV}\).}
\label{fig:2dSMEFT}
\end{figure}

\begin{figure}[!htbp]
\centering
\subfloat[]{\includegraphics[width=0.5\textwidth,page=1]{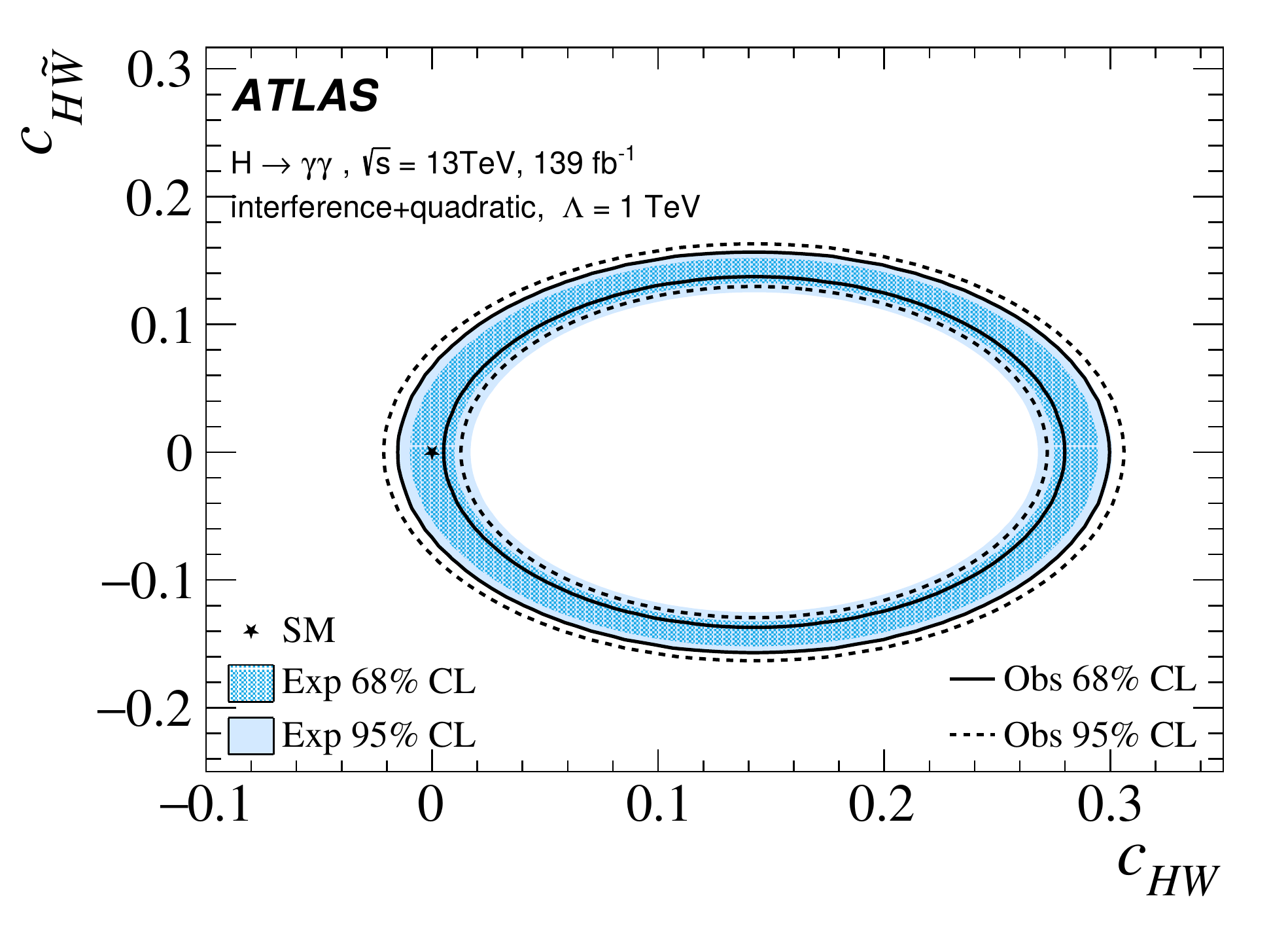}\label{fig:2dSMEFT_1_a}}
\subfloat[]{\includegraphics[width=0.5\textwidth,page=1]{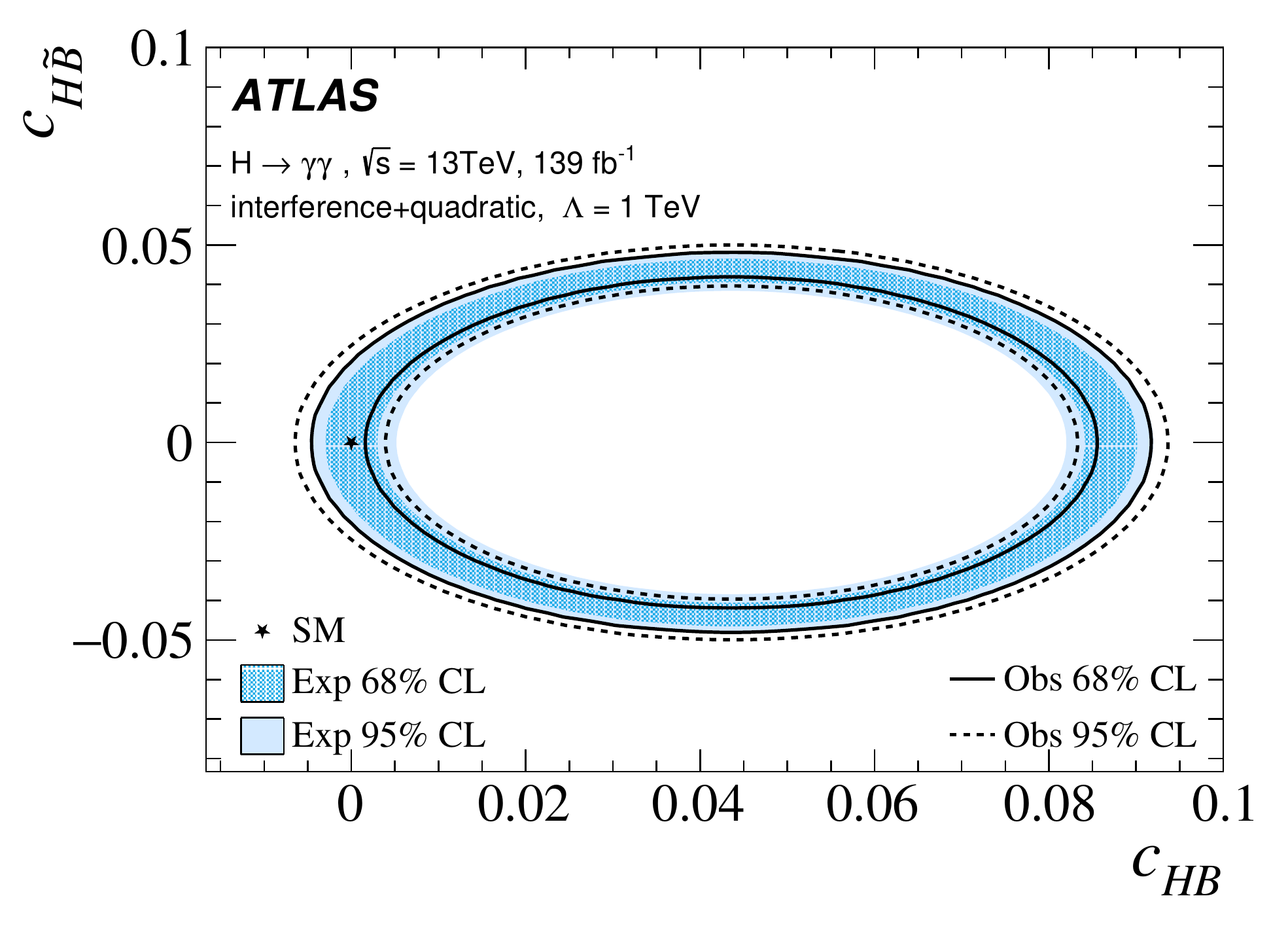}\label{fig:2dSMEFT_1_b}}\\
\subfloat[]{\includegraphics[width=0.5\textwidth,page=1]{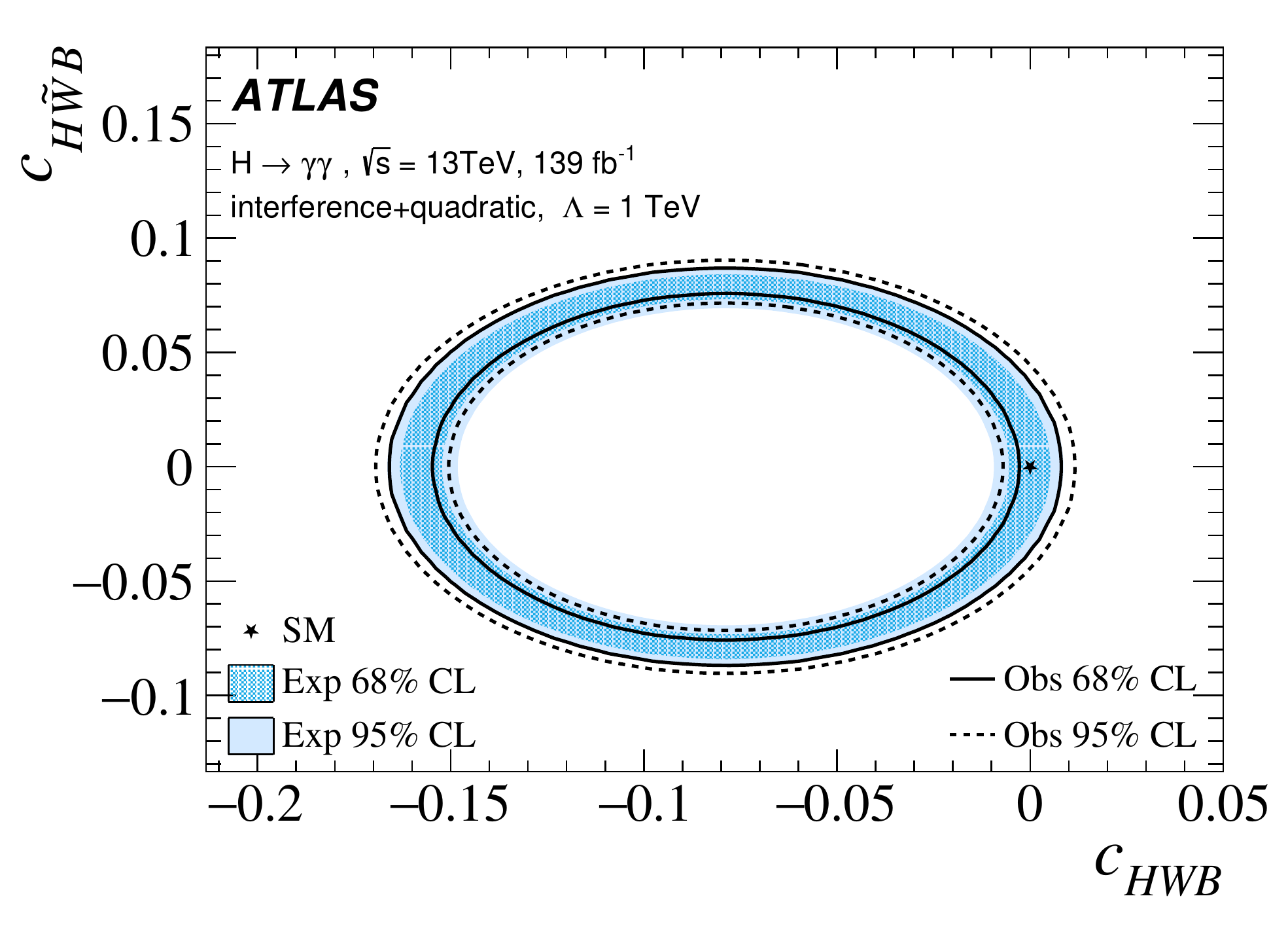}\label{fig:2dSMEFT_1_c}}
\subfloat[]{\includegraphics[width=0.5\textwidth,page=1]{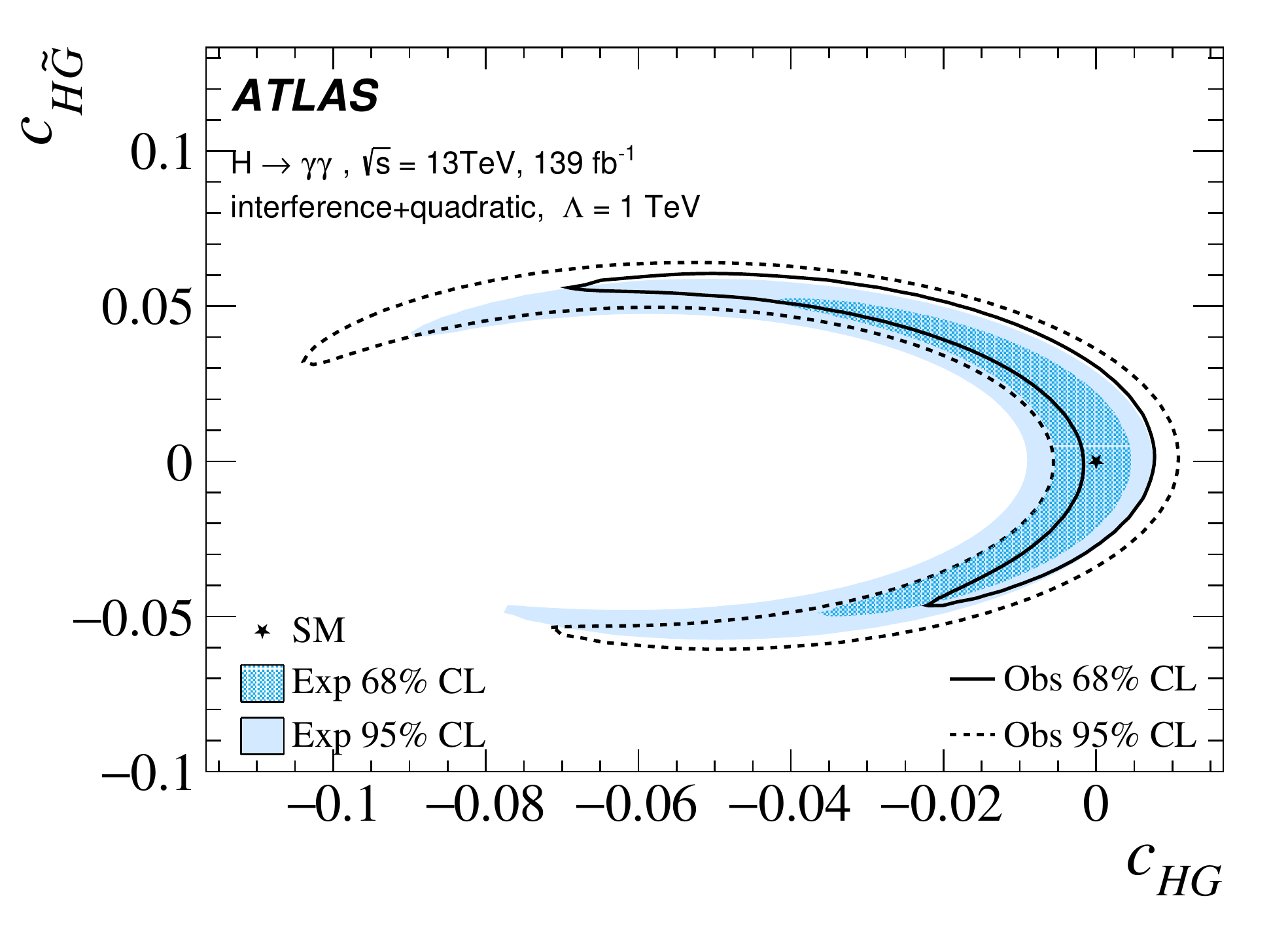}\label{fig:2dSMEFT_1_d}}\\
\caption{Plots showing the 2D 68\% and 95\% observed and expected limits obtained from various combinations of two Wilson coefficients including both the SM--dimension-6 interference and the quadratic dimension-6 terms in the SMEFT basis: \protect\subref{fig:2dSMEFT_1_a} \CHW\ vs \CHWt, \protect\subref{fig:2dSMEFT_1_b} \CHB\ vs \CHBt, \protect\subref{fig:2dSMEFT_1_c} \CHWB\ vs \CHWBt, \protect\subref{fig:2dSMEFT_1_d} \CHG\ vs \CHGt. The limits are computed at a new-physics scale \(\Lambda=\SI{1}{\TeV}\).}
\label{fig:2dSMEFT_1}
\end{figure}

In addition to the one-dimensional limits on EFT Wilson coefficients presented in Section~\ref{sec:eft}, two-dimensional limits are derived, allowing two Wilson coefficients (a CP-even coefficient and its CP-odd counterpart) to vary simultaneously using the interference-only cross-section, shown in Figure~\ref{fig:2dSMEFT}, and including the quadratic dimension-6 cross-section, shown in Figure~\ref{fig:2dSMEFT_1}.
The shape difference between the interference-only 2D limits and the interference-plus-quadratic limits is due to the fact that the interference-plus-quadratic cross-section affects the \Hgg\ branching-ratio for the both CP-even and CP-odd operators.
This is represented by the ring shape centred around zero for the CP-odd coefficient in the interference-plus-quadratic limits, since the interference-only cross-section vanishes for inclusive observables.
In contrast, the interference-only cross-section for CP-odd operators affects only the shape of the \dphijj\ distribution.
The 2D limits are compatible with the 1D limits due to the absence of significant correlation between the CP-even and CP-odd operators.
 
\FloatBarrier
 
\printbibliography
 
\clearpage
\input{atlas_authlist}

\end{document}

%% file: atlas_authlist.tex
 
\begin{flushleft}
\hypersetup{urlcolor=black}
{\Large The ATLAS Collaboration}

\bigskip

\AtlasOrcid[0000-0002-6665-4934]{G.~Aad}$^\textrm{\scriptsize 99}$,    
\AtlasOrcid[0000-0002-5888-2734]{B.~Abbott}$^\textrm{\scriptsize 125}$,    
\AtlasOrcid[0000-0002-7248-3203]{D.C.~Abbott}$^\textrm{\scriptsize 100}$,    
\AtlasOrcid[0000-0002-2788-3822]{A.~Abed~Abud}$^\textrm{\scriptsize 35}$,    
\AtlasOrcid[0000-0002-1002-1652]{K.~Abeling}$^\textrm{\scriptsize 52}$,    
\AtlasOrcid[0000-0002-2987-4006]{D.K.~Abhayasinghe}$^\textrm{\scriptsize 92}$,    
\AtlasOrcid[0000-0002-8496-9294]{S.H.~Abidi}$^\textrm{\scriptsize 28}$,    
\AtlasOrcid[0000-0002-9987-2292]{A.~Aboulhorma}$^\textrm{\scriptsize 34e}$,    
\AtlasOrcid[0000-0001-5329-6640]{H.~Abramowicz}$^\textrm{\scriptsize 158}$,    
\AtlasOrcid[0000-0002-1599-2896]{H.~Abreu}$^\textrm{\scriptsize 157}$
% ...
\bigskip
\\

% ...
$^{28}$Physics Department, Brookhaven National Laboratory, Upton NY; United States of America.\\
% ...
$^{34}$$^{(a)}$Facult\'e des Sciences Ain Chock, R\'eseau Universitaire de Physique des Hautes Energies - Universit\'e Hassan II, Casablanca;$^{(b)}$Facult\'{e} des Sciences, Universit\'{e} Ibn-Tofail, K\'{e}nitra;$^{(c)}$Facult\'e des Sciences Semlalia, Universit\'e Cadi Ayyad, LPHEA-Marrakech;$^{(d)}$LPMR, Facult\'e des Sciences, Universit\'e Mohamed Premier, Oujda;$^{(e)}$Facult\'e des sciences, Universit\'e Mohammed V, Rabat;$^{(f)}$Mohammed VI Polytechnic University, Ben Guerir; Morocco.\\
$^{35}$CERN, Geneva; Switzerland.\\
% ...
$^{52}$II. Physikalisches Institut, Georg-August-Universit\"{a}t G\"ottingen, G\"ottingen; Germany.\\
% ...
$^{92}$Department of Physics, Royal Holloway University of London, Egham; United Kingdom.\\
% ...
$^{99}$CPPM, Aix-Marseille Universit\'e, CNRS/IN2P3, Marseille; France.\\
$^{100}$Department of Physics, University of Massachusetts, Amherst MA; United States of America.\\
% ...
$^{125}$Homer L. Dodge Department of Physics and Astronomy, University of Oklahoma, Norman OK; United States of America.\\
% ...
$^{157}$Department of Physics, Technion, Israel Institute of Technology, Haifa; Israel.\\
$^{158}$Raymond and Beverly Sackler School of Physics and Astronomy, Tel Aviv University, Tel Aviv; Israel.\\
% ...
$^{a}$ Also at Borough of Manhattan Community College, City University of New York, New York NY; United States of America.\\
$^{b}$ Also at Bruno Kessler Foundation, Trento; Italy.\\
$^{c}$ Also at Center for High Energy Physics, Peking University; China.\\
$^{d}$ Also at Centro Studi e Ricerche Enrico Fermi; Italy.\\
$^{e}$ Also at CERN, Geneva; Switzerland.\\
$^{f}$ Also at D\'epartement de Physique Nucl\'eaire et Corpusculaire, Universit\'e de Gen\`eve, Gen\`eve; Switzerland.\\
% ...

\end{flushleft}
